\documentclass[review]{elsarticle}
\usepackage{lineno,hyperref,amsmath,amssymb}

\usepackage{graphicx, subcaption, gensymb}
\usepackage{esvect}
\usepackage{enumitem,textcomp}

\biboptions{numbers,sort&compress}

\begin{document}

\begin{frontmatter}

\title{The physics and applications of strongly coupled plasmas levitated in electrodynamic traps}

\author{Bogdan M. Mihalcea}
\ead{bogdan.mihalcea@inflpr.ro}
\address{National Institute for Laser, Plasma and Radiation Physics (INFLPR), \\ Atomi\c stilor Str. 407, 077125 M\u agurele, Ilfov County, Romania}
\author{Vladimir S. Filinov}
\ead{vladimir\_filinov@mail.ru}
\author{Roman A. Syrovatka}
\author{Leonid M. Vasilyak}
\address{Joint Institute for High Temperatures (JIHT), \\ Russian Academy of Sciences, Izhorskaya 13, Bld. 2, 125412 Moscow, Russia}




\begin{abstract}
 
Charged (nano)particles confined in electrodynamic traps can evolve into strongly correlated Coulomb systems which are the subject of current investigation. Exciting physical phenomena associated to Coulomb systems are reported such as autowave generation, phase transitions, defect formation, system self-locking at the edges of a linear Paul trap, self-organization in layers, or pattern formation and scaling. We investigate the dynamics of ordered structures consisting of highly nonideal similarly charged nanoparticles with coupling parameter of the order $\Gamma = 10^8$. This approach enables us to study the interaction of nanoparticle structures in presence and in absence of the neutralizing plasma background, as well as to investigate various types of phenomena and physical forces experienced by these structures. We review applications of electrodynamic levitation for mass spectrometry including containment and study of single aerosols and nanoparticles, with an emphasis on state of the art experiments and techniques, while also focusing on future trends and directions of investigation. Late experimental data suggest that inelastic scattering can be successfully applied to the detection of biological particles such as pollen, bacteria, aerosols, traces of explosives or synthetic polymers. 

Brownian dynamics is used to characterize charged particle evolution in time and thus identify regions of stable trapping. An analytical model is used to explain the experimental results. Numerical simulations take into account the stochastic forces of random collisions with neutral particles, the viscosity of the gas medium, regular forces produced by the a.c. trapping voltage and the gravitational force. We show that microparticle dynamics is characterized by a stochastic Langevin differential equation. Laser plasma acceleration of charged particles is also discussed, with an emphasize on dielectric capillaries and Paul traps employed for target micropositioning. 

\end{abstract}


\end{frontmatter}

\tableofcontents


\section{Introduction}\label{Intro}

Until the 1950s experimental investigations performed in atomic and molecular physics were limited in sensitivity, as the spectral resolution achievable was severely limited by the thermal motion of atoms that is responsible for the Doppler broadening of the spectral emission lines. The occurrence of Doppler-free spectroscopy techniques enables scientists to mitigate Doppler non-relativistic effects. Nevertheless, time-of-flight (ToF) broadening of spectral lines caused by the limited period of time that atoms spend in the interaction region with a laser beam also imposes restrictive boundaries to the spectral resolution of resonance microwave and optical lines \cite{Dem15, Dem18, Van15, Tome18}. In order to enhance the precision of such spectroscopic measurements ingenious methods are required to maintain atoms within the interaction region as long as possible, under minimal perturbations produced by collisions with other atoms or molecules or due to the interaction with the containing vessel walls.        

In a perpetual quest aimed at disclosing nature's inner secrets, physicists and chemists always aspired towards the ideal of isolating and investigating single atoms in a perturbation free, pristine environment. Such an objective seemed to be quite far-fetched as E. Schr{\"o}dinger stated in 1952: “We never experiment with just one electron or atom or (small) molecule. In thought experiments we sometimes assume that we do, this invariably entails ridiculous consequences.   ...  We are scrutinising records of events long after they have happened” \cite{Schro52}. A few years later Schr{\"o}dinger's statement was invalidated by W. Paul who invented the radiofrequency (RF) trap (also called Paul trap) \cite{Paul90, Ghosh95}, that became an invaluable instrument for modern spectroscopy, opening new pathways towards investigating atomic properties and testing quantum physics laws with unbeatable accuracy \cite{Major05, Kno14}.

Another ion trap pioneer is H. Dehmelt whose contributions are mainly linked to the development and use of Penning traps \cite{Bla08}. He invented ingenious methods of cooling, perturbing, trapping (one single electron was trapped for more than 10 months \cite{vanDyck87}) and probing levitated particles, thus forcing them to reveal their intrinsic properties \cite{Dehm90}. In the combined electric and magnetic fields in a Penning trap charged particles describe a complex motion \cite{Major05, Vogel18}. Penning traps exhibit unique features that make them suited to perform tests of fundamental physics. For example, ultra-fine Penning trap measurements of masses and magnetic moments for elementary particles (electrons, positrons, protons and antiprotons) (a) validate charge, parity, and time reversal (CPT) conservation; (b) enable accurate determination of the fine-structure constant $\alpha$ and of other fundamental constants \cite{Bla06, Bla10, Quint14}. 

Laser cooling of ions was firstly proposed by H. Dehmelt and D. J. Wineland in 1975 \cite{Wine87} and then achieved in 1978 \cite{Esch03, Haro10, Wine11}. Both W. Paul and H. Dehmelt were awarded the Nobel Prize in 1989 "for the development of the ion trap technique". Progress in the development of extremely stable laser sources allows scientists to manipulate and control the inner as well as the outer degrees of freedom of atoms and even molecules with ever increasing precision, right down to the quantum limit. Trapping and cooling techniques for electrically charged particles have expanded the frontiers of physics, leading to the emergence of new areas such as quantum engineering or quantum metrology \cite{Gobel15}. The Nobel Prize in Physics in 1997 was jointly awarded to S. Chu, C. Cohen-Tannoudji and W. D. Phillips, “for the development of methods to cool and trap atoms with laser light” \cite{Haro10}. J. L. Hall and T. W. H{\"a}nsch were awarded the Nobel Prize in Physics in 2005 “for their contributions to the development of laser-based precision spectroscopy, including the optical frequency comb technique” \cite{Poli13}. Finally, the Nobel Prize in Physics in 2012 was jointly awarded to S. Haroche and D. J. Wineland “for ground-breaking experimental methods that enable measuring and manipulation of individual quantum systems” \cite{Wine13}. Thus, impressive progress was achieved in quantum metrology by enabling measurements of unprecedented accuracy on space and time \cite{Haro10, Peik06, Bush13}. Both the plasma state of matter as well as systems of charged dust particles in plasma exhibit a great interest for modern physics \cite{Ichi82, Fort05, Fort10a, Bouf11, Ivlev12, Vasi15, Usach16, Ramaz16, Myas17, Ebel17}. When the potential energy associated to the Coulomb interaction between charged dust particles significantly exceeds their kinetic energy, they can evolve into ordered structures called Coulomb or plasma crystals that represent the subject of recent investigations performed onboard the International Space Station (ISS) \cite{Khra16, Pust16, Lipa20}. 

The paper is a review on strongly coupled Coulomb systems (SCCS) of finite dimensions such as electrically charged particles confined in quadrupole or multipole electrodynamic traps, either under Standard Ambient Temperature and Pressure (SATP) conditions or in vacuum \cite{Boll84}. An extended review of the underlying physical mechanisms is also performed. Microparticle electrodynamic ion traps (MEITs) are versatile instruments than can be used to investigate properties of individual charged particles (with dimensions ranging from 10 nm to 100 $\mu$m) such as aerosols, liquid droplets, solid particles, nanoparticles, DNA sequences, and even microorganisms. When the background gas is weakly ionized, the associated dynamics exhibits strong coupling regimes characterized by collective motion as damping is low for such systems. Late experimental evidence suggests that multipole trap geometries present certain advantages, among which better stability due to the existence of multiple regions of stable trapping and less sensibility to environment fluctuations. The mathematical models used and the numerical simulations performed simply create a clear and thorough picture on the phenomena that are investigated. We study particle oscillations around equilibrium positions where gravity is balanced by the trapping potential. Close to the trap centre the particles are almost frozen, which means they can be considered motionless due to their very low amplitude of oscillation.

Particular examples of SCCS would be electrons and excitons in quantum dots \cite{Boni14} or laser cooled ions levitated in Paul or Penning type traps \cite{Major05, Boll03, Thomp16}. We can also mention ultracold fermionic or bosonic atoms confined in atom traps \cite{Baly00} or in the periodic potential of an optical lattice \cite{Boni10b}. Experimental investigations of charged particles in external potentials have largely profited from the invention of ion traps, as they have greatly influenced the future of modern physics and state of the art technology \cite{Quint14, Werth09, Kno16}.

\section{Particle levitation techniques}\label{Sec2}

\subsection{The electrodynamic (Paul) trap. The underlying physics. Electrodynamic levitation}\label{EDB}

\begin{figure}[!ht] 
	\begin{center}
		\includegraphics[scale=1]{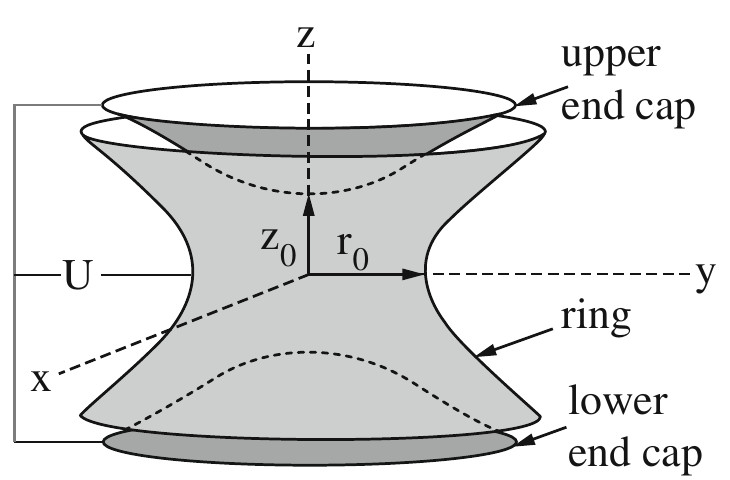}
		\caption{Basic design characteristic of three dimensional (3D) Paul and Penning traps. The inner electrode surfaces are hyperboloids. Dynamic stabilization in the Paul trap is achieved by supplying a RF voltage $V_0 \cos \left(\Omega t\right)$. Static stabilization in the Penning trap is ensured by means of a d.c. voltage $U = U_0$ and an axial magnetic field liable for radial confinement. Picture reproduced from \cite{Werth09} by courtesy of Prof. G. Werth.}
		\label{PaulWerth}
	\end{center}
\end{figure}

Electrically charged particles such as ions or microparticles can be levitated by using electromagnetic fields. Trapping of particles in a harmonic potential requires generating an electric restoring potential ${\vec F} = -q {\vec E}$ which increases linearly with the distance from the trap centre, such as $\vec F \propto -{\vec r}$. The corresponding trapping forces are characterized by a quadrupole potential $\Phi = \Phi_0 \left( \alpha x^2 + \beta y^2 + \gamma z^2 \right)/r_0^2$, where $\Phi_0$ denotes the electric potential applied to a quadrupole electrode configuration, $r_0$ is the trap radius, while $\alpha, \beta, \gamma$ represent weighing factors that determine the shape of the electric potential given by the solution of the Laplace equation $\Delta \Phi = 0$. For a three dimensional (3D) electric field  we infer $\alpha = \beta = -2\gamma$. The potential is attractive in the $x$ and $y$ directions and repulsive along the $z$ direction. According to the Earnshaw theorem a static electric field cannot achieve 3D binding \cite{Ghosh95, Major05}. In order to overcome this problem two main types of traps have been developed: (1) the Penning trap which employs a static electric field to achieve axial confinement and a superimposed static magnetic field that provides radial confinement, and (2) the Paul or RF trap which relies on an oscillating, inhomogeneous electric field \cite{Bla08, Kal09}, that creates a dynamic pseudopotential \cite{Berd19a, Berd19b}. A sketch of a quadrupole ion trap (QIT) is shown in Fig.~\ref{PaulWerth}. This review mainly considers levitation of particles in RF traps as the subject is extremely vast.    

We return to the trapping potential to emphasize that if an oscillating electric field is applied to the trap electrodes (see Fig.~\ref{PaulWerth}), then a saddle shaped potential is created \cite{Kiri16} that harmonically confines ions in the region where the field exhibits a minimum, under conditions of dynamical stability. The resulting potential is attractive in the $x$ and $y$ (radial) directions for the first half-cycle of the field and attractive in the $z$ (axial) direction during the second half-cycle. An adequately chosen amplitude and frequency $\Omega$ of the oscillating RF field ensures trapping of charged particles of mass $M$ and charge $Q$ in all three dimensions by means of a ponderomotive force oriented towards the trap centre. The 3D Paul trap provides a confining force with respect to a single point in space called the node of the RF field, which recommends its use for single ion experiments or to achieve levitation of 3D crystalline ion structures \cite{Werth08}.

Considering the electric field in a two dimensional geometry along the $x$ and $y$ axis only (linear Paul trap), we find $\alpha = -\beta, \gamma = 0$. In such case confinement of a charged particle is accomplished only in the (radial) $x$ and $y$ direction. An additional static dc-potential applied along the $z$ direction is required to confine the particle both radially and axially \cite{Kal09}. 

The electrodynamic (Paul) trap was invented in the 1950s by Prof. Wolfgang Paul and his co-workers at the University of Bonn \cite{Paul58}. The device was the subject of a patent awarded in 1954, for a 3D quadrupole ion trap (QIT) which consists of a ring electrode and two endcap electrodes \cite{Werth08, March09}. By applying a RF potential to the ring electrode, a quadrupole trapping field is produced in all three dimensions. A sketch of a QIT is shown in Fig.~\ref{PaulWerth}. Many other kinds of QITs exist, most notably the 2D linear ion trap with a quadrupole potential in the radial plane and a d.c. field for axial trapping. The quadrupole ion trap is a remarkably versatile mass spectrometer that is capable of multiple stages of mass selection (tandem mass spectrometry), high sensitivity, and moderate mass resolution and mass range. In combination with electrospray ionization (ESI), the QIT is widely used to investigate polar molecules such as peptides and proteins \cite{Sny19}. 

Electrically charged particles and ions levitated in oscillating RF fields represent a special class of physical systems, where classical and quantum effects \cite{Haro10, Werth09} compel to establish a dynamical equilibrium as they weight each other. Classical particle dynamics in a Paul trap is characterized by Mathieu equations \cite{Baril74, March05}. The stable solutions of the Mathieu equation are found by employing the Floquet theory \cite{Major05}. Rapid trapping field oscillations generate an effective Kapitza-Dirac potential that restrains the motion of electrically charged ions to a well defined region of space, which leads to a quasi-equilibrium distribution \cite{Aku14}. A picture of a 3D Paul or Penning trap is given in Fig.~\ref{PaulWerth} \cite{Bla08, Werth09}. We do not discuss stability and solutions of the Mathieu equation because the subject is widely covered in literature \cite{Major05, Werth08, March05}.  

The method of trapping charged particles using time-varying electric fields is not restricted to atomic or molecular ions, it equally applies to charged microparticles. The first demonstration of microparticle trapping was performed by Straubel \cite{Stra55} who confined charged oil drops in a simple quadrupole trap operated in air. Wuerker {\em et al} have obtained stable arrays of positively charged aluminium particles using a 3D Paul trap operated in vacuum \cite{Wuerk59}. The particles repeatedly crystallize into a regular array and then melt, owing to the dynamical equilibrium between the trapping potential and the inter-particle Coulomb repulsion. A similar experiment was performed by Winter, thus demonstrating storage of macroscopic dust particles (anthracene) in a 3D Paul trap operating under SATP conditions \cite{Wint91}, as friction in air is proven to be an efficient mechanism to {\em cool} the microparticles. The mechanism is similar with cooling of ions in ultrahigh vacuum conditions due to collisions with buffer gas molecules \cite{Major05}. The dynamics of a charged microparticle in a Paul trap (near SATP conditions) was studied by Izmailov \cite{Izma95} using a Mathieu differential equation with damping term and stochastic source, under conditions of combined periodic parametric and random external excitation. Further research on electrodynamic traps is presented in \cite{Tak04, Kja05}. 

The combination between a Paul and Penning trap, which uses both a RF voltage and an axial magnetic field, is called a combined trap \cite{Vogel18, Naka01}. Particle dynamics in nonlinear traps has also been intensively explored. It was found that ion motion is well described by the Duffing oscillator model \cite{Nay04, Kova11} with an additional nonlinear damping term \cite{Mih10b, Ake10}. Regions of stable (chaos) and unstable dynamics were illustrated, along with the occurrence of strange attractors and fractal properties for the associated dynamics. The dynamics of a parametrically driven levitated nanoparticle is investigated in \cite{Gies15}.

Ion traps have also opened new horizons towards performing investigations on the physics of few-body phase transitions \cite{Fort10a, Shuk02a, Tsyto08, Fort06a, Died87, Blu90, Schli96} or the study of nonlinear dynamics and (quantum) chaos. The issue of accurate control of the internal and motional quantum degrees of freedom of laser-cooled trapped ions is the subject of intense theoretical and experimental investigations \cite{Leibf03, Johan09}. Quantum engineering has opened new horizons in quantum optics and quantum metrology \cite{Haro10, Zago11, Jord20}. Applications of ion traps span mass spectrometry, very high precision spectroscopy, quantum physics tests, study of non-neutral plasmas \cite{David01, Dub99}, quantum information processing (QIP) \cite{Haff08, Brown16, Baut19, Bruz19} and quantum metrology \cite{Wine11, Sinc11, McCor19}, use of optical transitions in highly charged ions for detection of variations in the fine structure constant \cite{Koz18} or very accurate optical atomic clocks \cite{Van15, Poli13, Ved09, Lud15, Kel15}. A phase-insensitive Fock-state-based protocol implemented using trapped ions that helps in overcoming the quantum noise limit is presented in \cite{Wolf19}. 

The ability to confine single particles or nanoparticles (NPs) under conditions of minimal perturbations and then expose them to laser beams, makes it possible to perform light scattering measurements without the presence of parasitic effects that might result as an outcome of the interaction with other particles that may be of different sizes and morphology, and different chemical and/or optical properties. This can be achieved by trapping a charged microparticle using an electrodynamic balance (EDB) or ion trap, by means of optical levitation, or using acoustic levitation. The technique called electrodynamic levitation is used to isolate and analyze aerosols and microparticles, with an aim to study the mechanisms from the Earth atmosphere and supply information about the optical properties of the species of interest \cite{Davis10}. The EDB is an outcome of the quadrupole electric mass filter of Paul \cite{Paul90, Major05} and of the bihyperboloidal electrode configuration introduced by Wuerker {\it et al} \cite{Wuerk59}. 

\begin{figure}[!ht] 
	\begin{center}
	\includegraphics[scale=1.6]{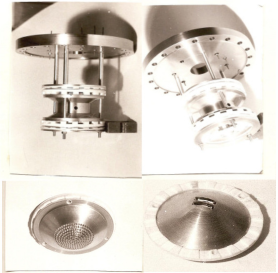}
	\caption{Classical Paul 3D trap with hyperbolic geometry used at INFLPR to trap $^{137}$Ba$^+$ ions in vacuum. The upper images show the 3D hyperbolic trap attached to the vacuum flange, while the lower images show the upper and lower endcap (slit) electrode, respectively. The lower endcap electrode exhibits 2 slits with dimensions $16 \times 3$ mm, corresponding to the location of two platinum belts covered with BaCl$_2$ solution \cite{Giu94}.}
	\label{PaulTrapINFLPR}
	\end{center}
\end{figure}

\begin{figure}[!ht] 
	\begin{center}
		\includegraphics[scale=0.28]{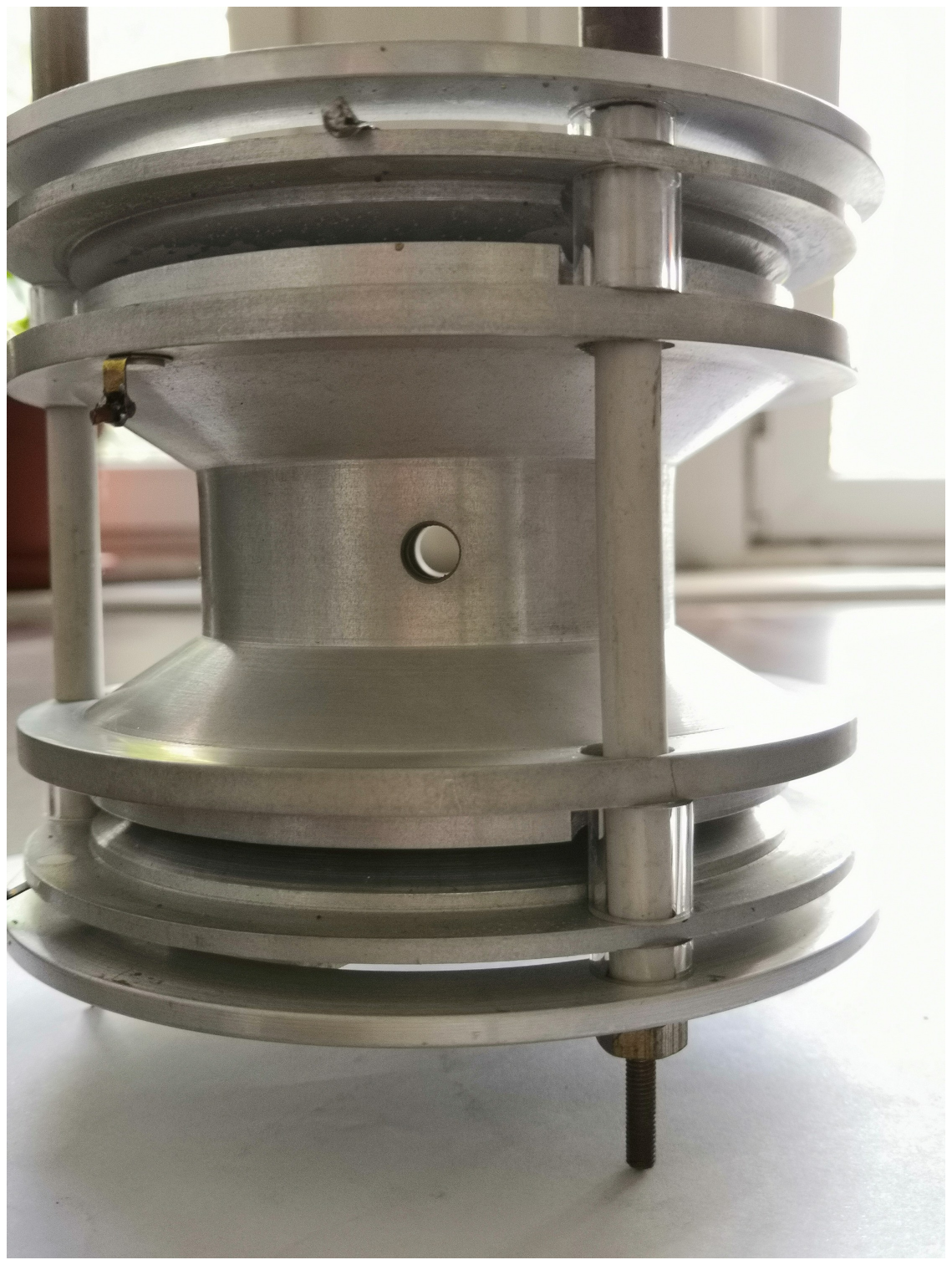}
		\includegraphics[scale=0.28]{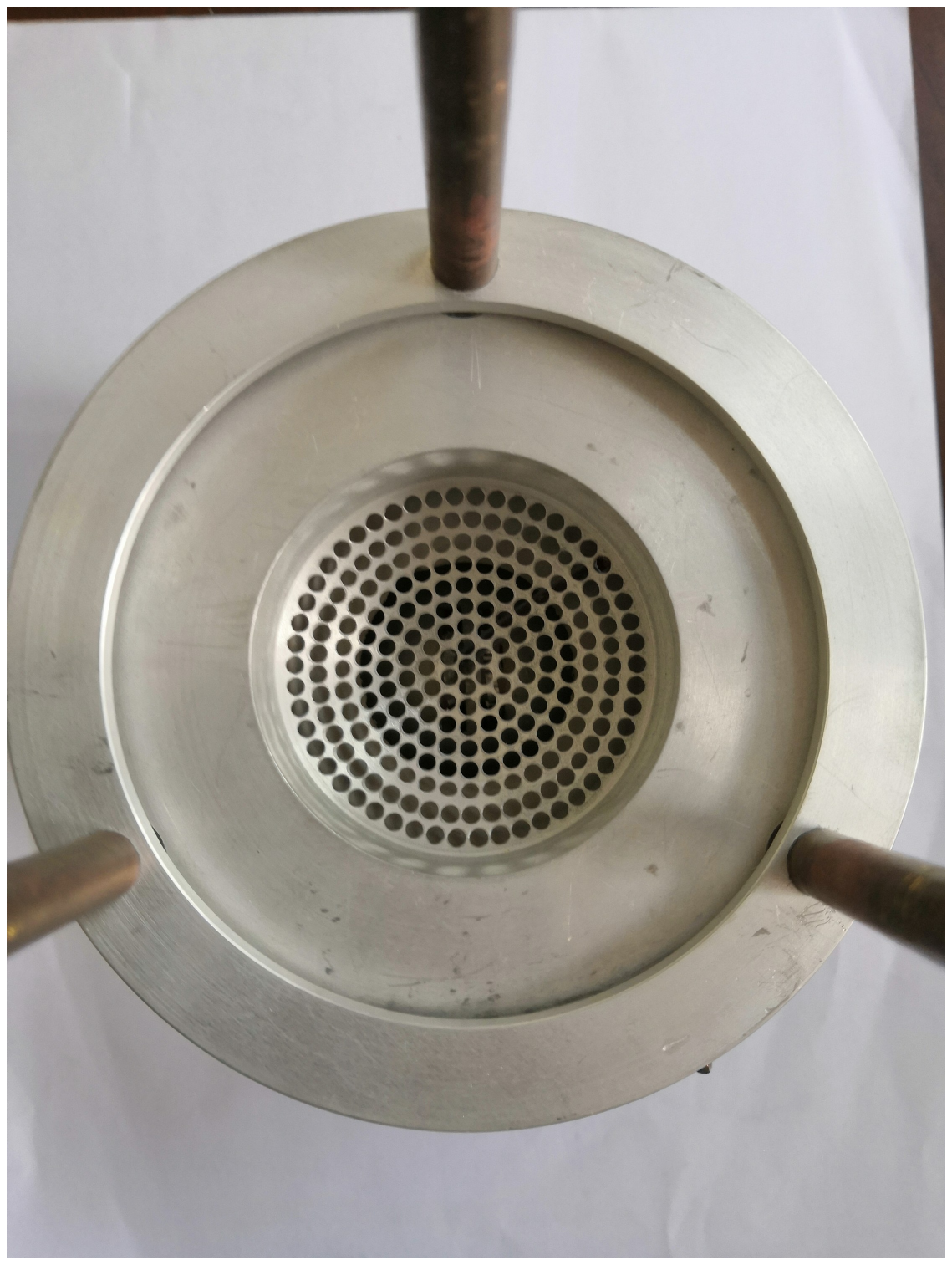}
		\caption{Classical Paul 3D trap with hyperbolic geometry used at INFLPR to trap $^{137}$Ba$^+$ ions in vacuum. The left image shows the 3D hyperbolic trap with the ring electrode pierced to allow radiation access. The right image shows the upper endcap (sieve) electrode with 256 holes of 2.4 mm diameter, equally spaced at 0.6 mm, that achieves a transparency of 56 \% in the $2.5 \div 25 \ \mu$m range. This particular design of the sieve electrode facilitates collection of fluorescence radiation. The trap dimensions are $r_0 = 1.75$ cm and $z_0 = 1.24$ cm \cite{Ghe76, Ghe97}.}
		\label{PaulTrapINFLPR2}
	\end{center}
\end{figure}

A hyperbolic geometry classical 3D Paul trap used at INFLPR in the experiments performed by Prof. V. Gheorghe with an aim to levitate $^{137}$Ba$^+$ ions in vacuum $\left( 10^{-4} \div 10^{-5} \ \text{Torr} \right)$, is shown in Fig.~\ref{PaulTrapINFLPR} and Fig.~\ref{PaulTrapINFLPR2} \cite{Ghe76, Giu94, Ghe97}.

\begin{figure}[bth]
	\centering	
	\includegraphics[scale=1.3]{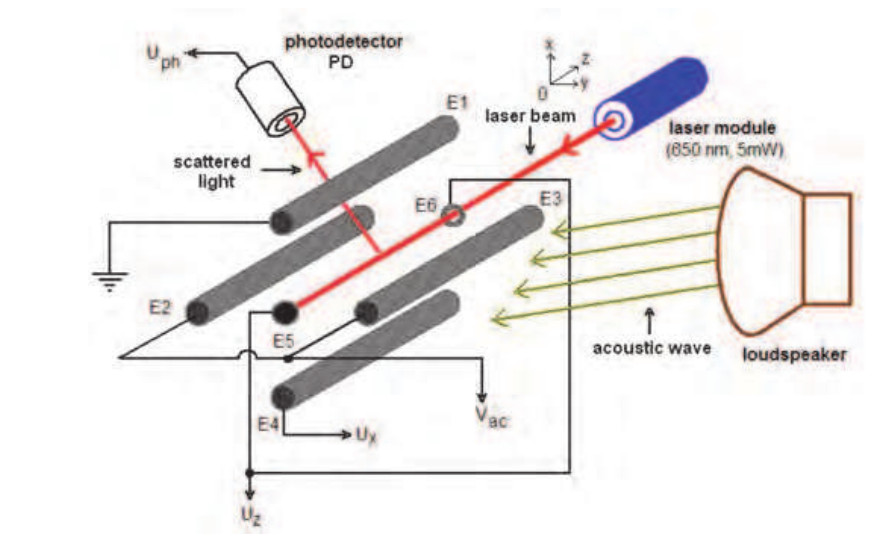}
	\caption{Experimental setup for acoustic wave excitation. Picture reproduced from \cite{Sto11} with kind permission from O. Stoican}
	\label{ac1}
\end{figure}

In the conventional operation mode of an EDB an alternating current (a.c.) potential is applied to the ring electrode, and a direct current (d.c.) field is generated by applying equal but opposite polarity d.c. potentials to the endcap electrodes. The a.c. field is used to trap the particle in an oscillatory mode, while the d.c. field serves the purpose of the Millikan condenser to balance the gravitational force and any other vertical forces that act upon the particle. A photodiode array can be mounted on the ring electrode as shown in Fig. \ref{ac1}, with an aim to measure the irradiance of the scattered light as a function of angle. In addition, a photomultiplier tube (PMT) is located at right angles to the incident laser beam to record the irradiance at a single angle as a function of time. The appropriate a.c. trapping frequency depends on the particle mass and drag force on the particle. When the vertical forces ({\em e.g.}, gravity) are balanced by the additional d.c. electric field applied between the upper and lower trap electrodes, the particle can be maintained at the midpoint of the EDB provided that the a.c. potential is not too large \cite{Sto11}. If the a.c. field is too large, the particle ends up in being expelled from the balance chamber. This effect is used to isolate single particles by varying the a.c. potential and frequency, which eliminates undesired particles. The principles and stability characteristics of the EDB were firstly analyzed by Frickel {\em et al.} \cite{Kul11}, while other aspects of the EDB are discussed by Davis and Schweiger \cite{Davis02, Hart92}.

\subsection{Anharmonic contributions in RF linear quadrupole traps}

As the RF linear Paul trap is widely used in physics and chemistry, an important feature is the trap capacity as there are situations that impose levitation of large ion clouds. Due to the presence of anharmonic terms in the series expansion of the trap potential \cite{Mih18}, there is a limit on the trap capacity (the number of ions that may be confined). Ref. \cite{Home11} investigates the effects of anharmonic terms in the trapping potential for linear chains of levitated ions. Two paramount effects can be distinguished. The first effect consists in a modification of the oscillation frequencies and amplitudes of the ions’ normal modes of vibration for multi-ion crystals, which is due to the fact that each ion experiences a different potential curvature. The second effect reported are amplitude-dependent shifts of normal-mode frequencies as an outcome of elevated anharmonicity or higher excitation amplitude \cite{Home11}. The issue of anharmonic contributions to the electric potential generated by a linear RF trap is investigated in Refs. \cite{Pedre10, Pedre18a, Pedre18b}. Although the anharmonic part is quite comparable in the trap centre region for all configurations investigated, the behaviour is quite different as one moves further away from the trap centre.

\subsection{Particle dynamics in the trap. Equations of motion. Stability analysis}

Particle dynamics in an electrodynamic trap is described by a Mathieu equation which represents a particular case of the Hill equation \cite{McLac64, Mag66}. Paul traps are able to levitate charged microparticles over a wide range of charge-to-mass ratios, by modifying the a.c. voltage amplitude and frequency \cite{Major05, Davis97}. However, the spatial charge potential caused by the presence of other trapped particles changes the limits of the stability diagram, depending on the type of particle ordering, particle size, and particle species used. The RF and d.c. endcap voltages also influence the shape of trapped particles. By increasing the RF voltage and the specific charge-to-mass ratio $Q/M$, the operating point of the trap can shift towards the border of the Mathieu stability diagram, case when the particle species can end up by being expelled out of the trap. 

A harmonically trapped and laser-cooled ion represents a key system for quantum optics \cite{Orsz16} used in order to investigate pure quantum phenomena \cite{Wine13}. Ion traps enable scientists to perform tests of fundamental physics such as investigations on parity violation, tests of the CPT theorem or Lorenz symmetry, searches for spatio-temporal variations of the fundamental constants in physics at the cosmological scale, tests of the General Relativity (GR) and of the equivalence principle, or searches for dark matter, dark energy and extra forces   \cite{Safro18, Koz18}. To these we also add applications such as high-resolution spectroscopy, quantum computation \cite{Haff08, Bruz19, Har14, Humb19, Kau20}, studies of quantum chaos and integrability \cite{Ber00, Gra13}, nonequilibrium quantum dynamics of strongly coupled many-body systems \cite{San18} and quantum sensing \cite{Rei17}. One of the most promising systems that reduces decoherence effects \cite{Haro10} down to a level where quantum state engineering (state preparation) can be achieved, consists of one or more trapped ions \cite{Wine13, Cler16}. In a good approximation the centre of mass (CM) of a trapped ion experiences a harmonic external trapping potential. Hence, the ion trap can be assimilated with a quantum harmonic oscillator \cite{Pli08a, Mih09, Mih10a, Kie15, Tiba20a, Witt20, Kel21, Chak21, Sai21}. 

We shortly present a classical model proposed to characterize regular and chaotic orbits for a system of two ions (charged particles) in a 3D Paul trap, depending on the chosen control parameters. We consider two ions of mass $M_1$ and $M_2$ levitated in a quadrupole (3D) Paul trap, characterized by the force constants $k_1$ and $k_2$, respectively. The reduced equations of motion restricted to the $xy$ - plane can be cast into \cite{Mih21}:

\begin{equation}
\begin{cases}\label{cla1}
M_1 \ddot x = -k_1 x + a \left( x - y \right) \\ 
M_2 \ddot y = -k_2 y - a \left( x - y \right) \ , \\
\end{cases}
\end{equation}
where $a$ represents the constant of force that characterizes the Coulomb repelling between the two trapped ions. The trap control parameters are: 
\begin{equation}\label{}
\label{cla2}k_i = \frac{M_i q_i^2 \Omega}8 , \ q_i = 4 \frac {Q_i}{M_i} \frac{V_0}{\left( z_0 \Omega \right)^2} \ , \ i = 1, 2 \ ,
\end{equation}
which implies a time average on the micromotion at frequency $\Omega$, where the higher order terms in the Mathieu equation that describes ion dynamics are discarded. $V_0$ stands for the RF voltage supplied to the 3D Paul trap electrodes and $Q_i$ represents the ion electric charge. The Coulomb force constant is $ a \equiv{2Q^2}/{r^3} < 0 $, as it results from the series expansion of ${Q^2}/{r^2}$ around an average deviation of the trapped particle with respect to the trap centre $r_0 \equiv \left( x_0 - y_0 \right) < 0 $, determined by the initial conditions. The kinetic and potential energies for the system of ions are:

\begin{equation}
\label{cla4}T = \frac{M_1 \dot x^2}2 + \frac{M_2 \dot y^2}2 ,\  U = \frac{k_1 x^2}2 + \frac{k_2 y^2}2 + \frac 12 a \left( x - y\right)^2 \  .
\end{equation}
We choose
\begin{equation}
\label{cla6}\frac{k_1}2 = Q_1 \beta_1 , \; x = z_1, y = z_2 \ ,
\end{equation}
We assume the two ions possess equal electric charges, $Q_1 = Q_2$. The Lagrange function is written in the standard form $ L\left( \zeta_i,\dot \zeta_i\right) = T - U $ and the Lagrange equations are 
\begin{equation}
\label{cla7}\frac d{dt} \left( \frac{\partial L}{\partial \dot \zeta_i} \right)  - \frac{\partial L}{\partial \zeta_i} = 0 , \; i = 1,2 \  , 
\end{equation}
where $\zeta_i$ are the generalized coordinates and ${\dot \zeta}_i$ represent the generalized velocities. In case of a one-dimensional system of $s$ particles (ions) or for a system with $s$ degrees of freedom, the potential energy can be expressed as:
\begin{equation}
\label{cla11}U = \sum_{i = 1}^s \frac{k_i \zeta_i^2}2 + \frac 12 \sum_{1 \leq i < j \leq s} a_{ij} \left( \zeta_i - \zeta_j \right)^2  \ .
\end{equation}

By performing a series expansion of the Coulomb potential in spherical coordinates and assuming a diluted medium, the interaction potential is
\begin{equation}
\label{cla13}V_{int}=\frac 1{4\pi \varepsilon_0} \sum_{1 \leq i < j \leq s} \frac{Q_iQ_j}{\left| {\vec r}_i - {\vec r}_j \right| }   \ . 
\end{equation}
We denote \cite{Wine88}
\begin{equation}
\label{cla14}k_1 = 2 Q_1 \beta_1 \ , \beta_1 = \frac{4Q_1 V_0^2}{ M_1 \Omega ^2 \xi^4} - \frac{2U_0}{\xi^2} \ ,
\end{equation}
where $\xi^2 = r_0^2 + 2 z_0^2 $. We consider $ U_0 = 0 $ (the d.c. trapping voltage) and take $r_0$ as negligible, with $r_0$ and $z_0$ the trap semiaxes. The trap control parameters are  $ U_0, V_0, \xi_0 $ and $ k_i $ \cite{Hoff95}. We choose an electric potential $V = \frac 1{|z|} $ which we expand in series around $z_0 > 0$
\begin{multline}\label{cla18}
V\left( z \right)  = \frac 1{z_0} - \frac 1{z_0^2} \left( z - z_0 \right) + \frac 1{z_0^3} \left( z - z_0 \right)^2 - \ldots \ ,
\end{multline} 
with $z - z_0 = x - y$. Then, the expression of the potential energy in eq. \ref{cla4} for the system of ions can be cast into: 
\begin{equation}
\label{cla20}U = \frac{k_1 x_1^2}2 + \frac{k_2 x_2^2}2 + \frac 1{4\pi \varepsilon_0} \frac{Q_1 Q_2}{\left| x_1 - x_2\right| } \ ,
\end{equation}
and it represents the sum of the potential energies of the ions assimilated with two harmonic oscillators, while the third term in the relationship \ref{cla20} describes the Coulomb interaction between the ions. As Hamilton's principle requires the potential energy to be minimum in order for the system to be stable, we search for a minimum of $U$. A system of equations results, then after calculus we obtain 
\begin{equation}\label{cla25}
\lambda = \sqrt[3]{\frac 1{4 \pi \varepsilon_0} \frac{k_1^2 k_2^2}{\left( k_1 + k_2 \right)^2}\, Q_1 Q_2}   \  ,
\end{equation}
which enables us to determine the points of minimum $x_1$ and $x_2$ of the system in an equilibrium state. We choose
\begin{equation}
\label{cla26}z_0 = x_{1\min} - x_{2\min} = \lambda \frac{k_1 k_2}{k_1 + k_2} \ ,
\end{equation}
and denote
\begin{equation} 
\label{cla27}z = x_1 - x_2 \ ,\; x_1 = x_{1\min} + x  \ ,\; x_2 = x_{2\min} + y \  \Rightarrow z = z_0 + x - y  \ .
\end{equation}
We revert to the expansion of the trap electric potential in eq. (\ref{cla18}) then express the potential energy as  
\begin{equation}
\label{cla31} U = \frac{k_1}2x_{1\min }^2 + \frac{k_2}2x_{2\min }^2 + \frac{k_1}2x^2 + \frac{k_2}2y^2 + \lambda{z_0} + \frac \lambda{z_0} \left( z - z_0 \right)^2 - \ldots \ ,
\end{equation}
where $x_{1\min} = \lambda / k_1$, $x_{2\min} = - \lambda / k_2$. From eqs. (\ref{cla4}) we obtain $a/2 = \lambda / z_0$. According to the principle of least action, the system is stable when the potential energy reaches a minimum. 

\subsubsection{Solutions of the equations of motion}
We search solutions such as $x = A \sin \omega t$ and $y = B \sin \omega t$, which we introduce in the equations of motion \ref{cla1}. The resulting determinant of the system of equations is:
\begin{equation} 
\label{cla35}\left| 
\begin{array}{cc}
a - k_1 + M_1 \omega^2 & -a \\ 
-a & a - k_2 + M_2 \omega^2 
\end{array}
\right| =0  \ .
\end{equation}
The stability requirement for the system demands for the determinant of this equation to be zero
\begin{equation}
\label{cla36}\left( a - k_1 + M_1 \omega^2 \right) \left( a - k_2 + M_2 \omega^2 \right) - a^2 = 0 \ .
\end{equation}
The discriminant of eq. (\ref{cla36}) can be cast as 
\begin{equation}
\label{cla40} \Delta = \left[ M_1 \left( a - k_2 \right) - M_2 \left( a - k_1 \right) \right]^2 + 4 M_1 M_2 a^2 \ .
\end{equation}
As it is well known a linear homogeneous system of equations admits nonzero solutions only if the determinant associated to the system is zero. Hence, the solution of eq. (\ref{cla36}) can be expressed as:
\begin{equation}
\label{cla41}\omega_{1,2} = \frac{M_1 \left( k_2 - a \right) + M_2 \left( k_1 - a \right) \pm \sqrt{\Delta}}{2 M_1 M_2} \ .
\end{equation}
A solution for this system would be
\begin{subequations}\label{subeqcla42}
	\begin{eqnarray}
	x = C_1 \sin \left( \omega_1 t + \varphi_1 \right) + C_2 \sin \left( \omega_2 t + \varphi_2 \right) \label{subeqcla42a} \ ,\\
	y = C_3 \sin \left( \omega_1 t + \varphi_3 \right) + C_4 \sin \left( \omega_2t + \varphi_4 \right) \label{subeqcla42b} \ .
	\end{eqnarray}
\end{subequations}
In fact, such a solution represents a superposition of two oscillations with secular frequencies $\omega_1$ and $\omega_2$, that denote the eigenfrequencies of the system investigated. $\varphi_1 \div \varphi_4$ stand for the initial phases while $C_1 \div C_4$ are a set of constants $\left( 0 < C_i \leq 1\right)$.  Assuming that $a\ll k_{1,2}$ (a requirement that is frequently met in practice) in eq. (\ref{cla36}), we can define the condition of strong coupling as
\begin{equation}
\label{coup1}\left| {\dfrac{a}{k_i}} \right| \gg \left| \dfrac{M_1 - M_2}{M_2} \right| \ ,
\end{equation}
where the modes of oscillation are 
\begin{equation}
\label{coup2} \omega_1^2 = \dfrac{1}{2} \left( \dfrac{k_1}{M_1} + \dfrac{k_2}{M_2} \right) , \;\ \omega_2^2 = \dfrac{1}{2} \left( \dfrac{k_1 - 2a}{M_1} + \dfrac{k_2 - 2a}{M_2} \right) \ .
\end{equation}

By studying the phase relations between the solutions of eq. (\ref{cla1}) we notice that the $\omega_1$ mode describes a translation of the two ions, where the distance $r_0$ between them is constant. The Coulomb repulsion force remains unchanged, a fact that is indicated by the absence of the term $a$ in the first eq. (\ref{coup2}). This mode of translation generates an axial current which can be detected. In case of the $\omega_2$ mode the distance between the ions increases and then decreases around the location of a fixed centre of mass (CM), situation when the electric current and implicitly the signal are zero. Although this mode can not be electronically detected, optical detection is possible. Therefore, for collective modes of motion the theory predicts that only a peak of the mass will be detected,  corresponding to the average mass of the two ions. In case of weak coupling the inequality in eq. (\ref{coup1}) reverses, while from eq. (\ref{cla41}) we infer
\begin{equation}
\label{coup4} \omega_{1, 2}^2 = \left( k_{1, 2} - a \right) /M_{1, 2} \ ,
\end{equation}
that is every mode of the motion corresponds to a single mass and the resonance is shifted by the term $a$. Moreover, within the limit $m_1 = m_2$ the strong coupling requirement in eq. (\ref{coup1}) is always satisfied no matter how weak the Coulomb coupling, which renders the weak coupling condition inapplicable in practice.  

\begin{figure}[!ht]
	\begin{minipage}[h]{0.45\linewidth}
		\center{\includegraphics[width=\linewidth]{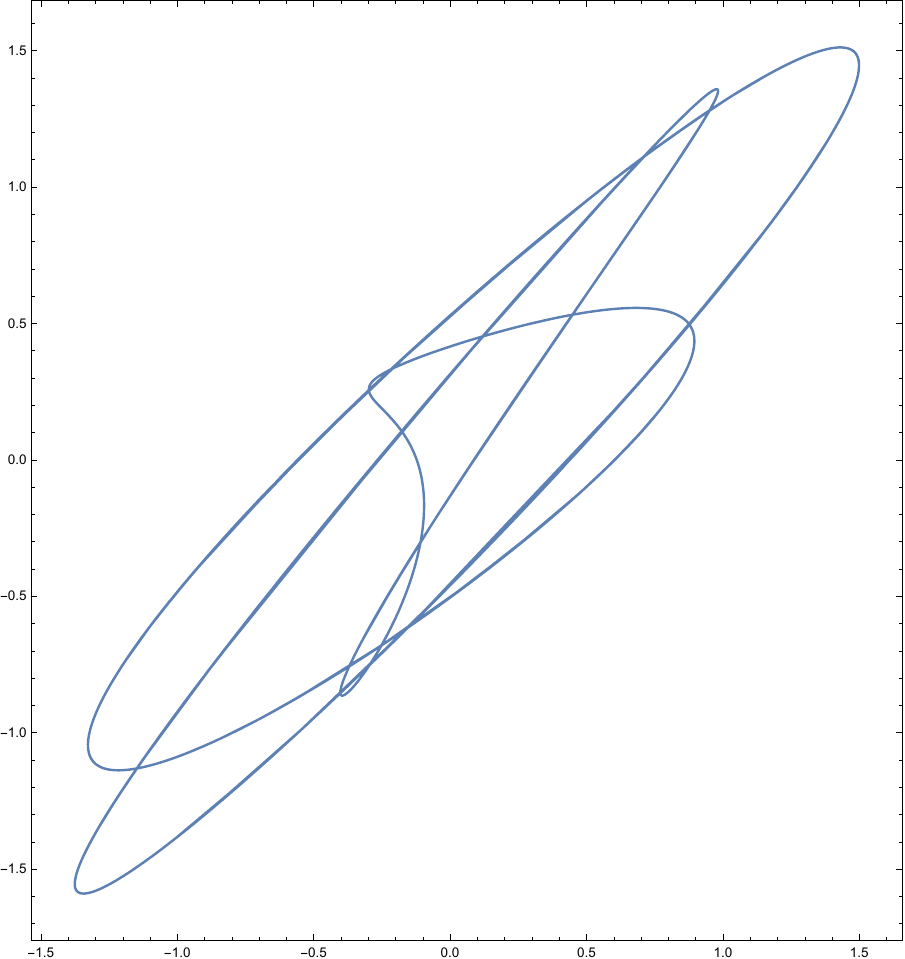}\\a)}
	\end{minipage}
	\begin{minipage}[h]{0.45\linewidth}
		\center{\includegraphics[width=\linewidth]{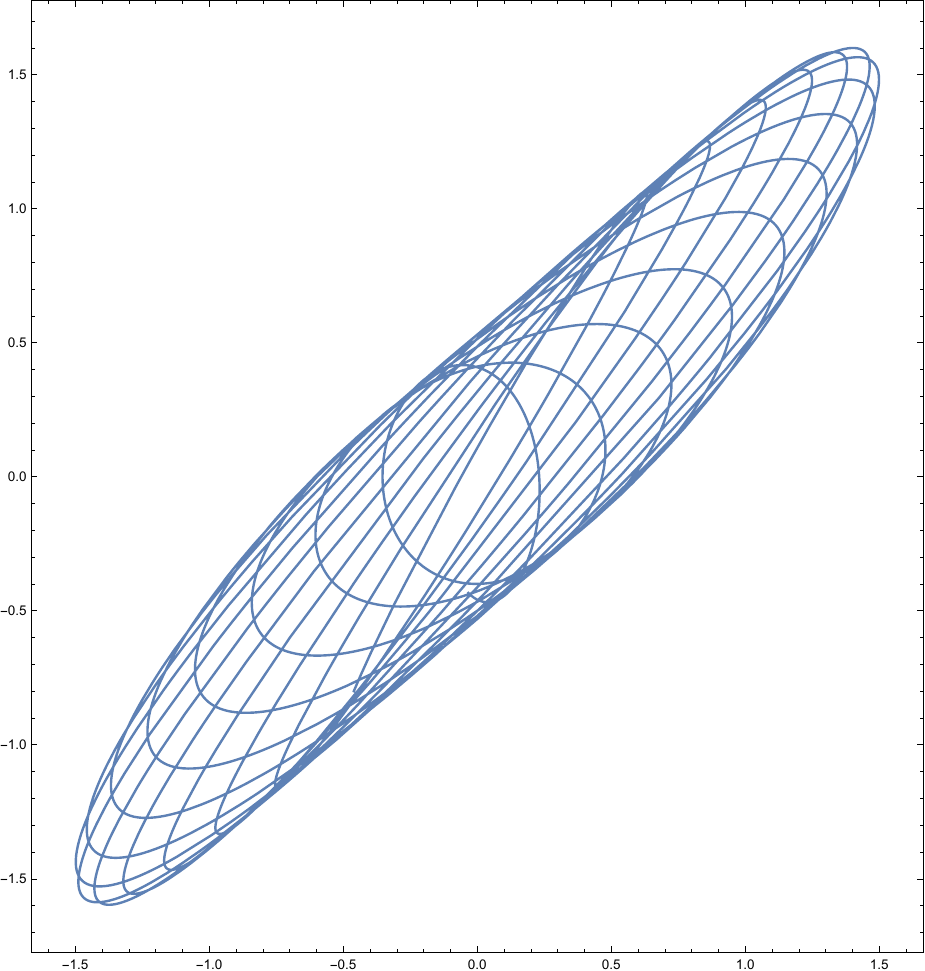}\\b)}
	\end{minipage}
	\begin{minipage}[h]{0.5\linewidth}
		\center{\includegraphics[width=\linewidth]{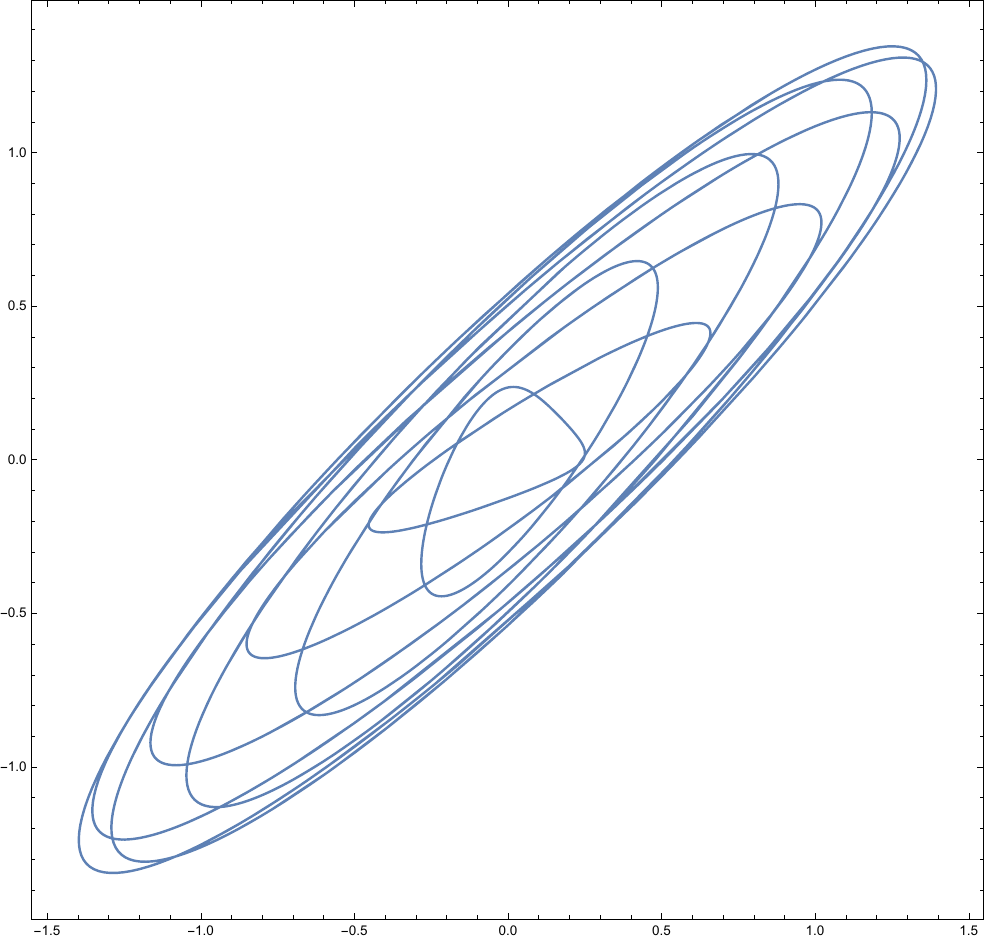}\\c)}
	\end{minipage}
	\begin{minipage}[h]{0.5\linewidth}
	\center{\includegraphics[width=\linewidth]{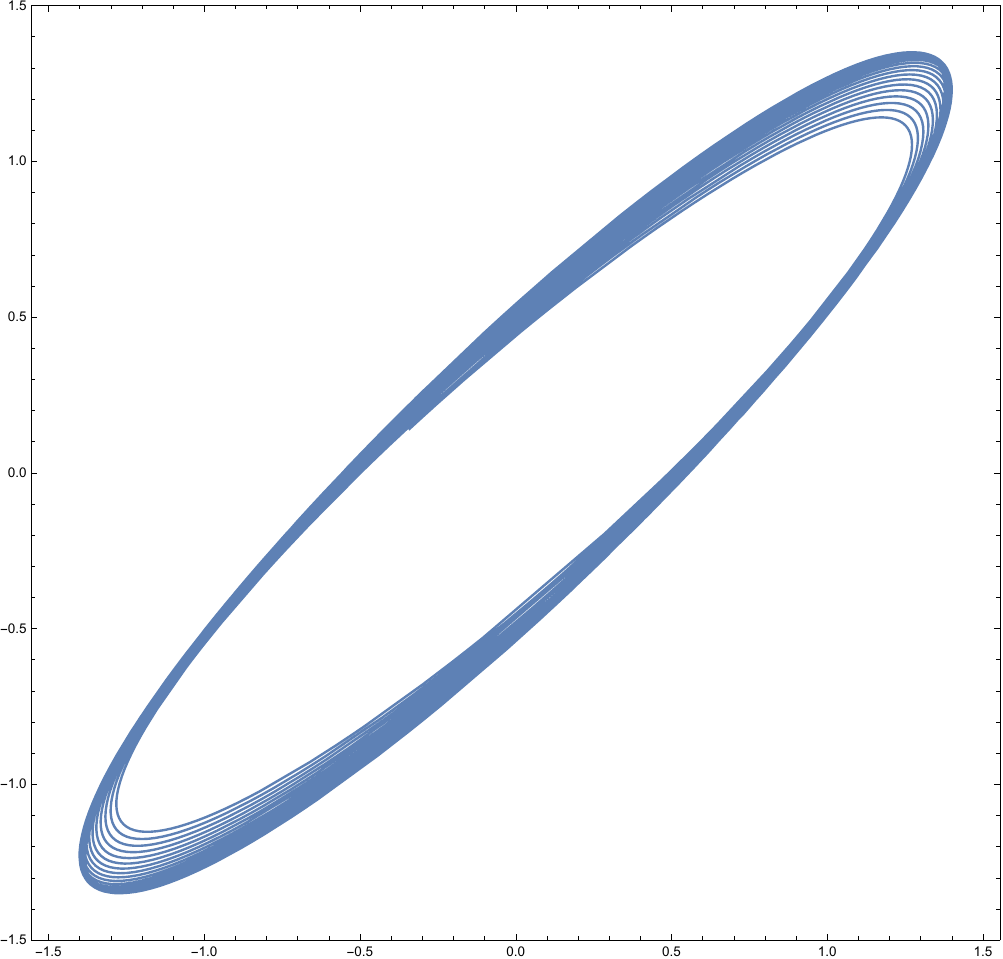}\\d)}
\end{minipage}
	\caption{Phase portrait for the two ion system with initial conditions: (a) $C_1 = 0.8, C_2 = 0.6, C_3 = 0.75, C_4 = 0.6, \varphi_1 = \pi/3, \varphi_2 = \pi/4, \varphi_3 = \pi/2, \varphi_4 = \pi/4,  \omega_1/\omega_2 = 1.5/2$. The dynamics is periodic; (b) $\omega_1/\omega_2 = 1.5/\sqrt{2}$. Ion dynamics is quasiperiodic; (c) $C_1 = 0.8, C_2 = 0.6, C_3 = 0.76, C_4 = 0.6, \varphi_1 = \pi/3, \varphi_2 = \pi/4, \varphi_3 = \pi/2, \varphi_4 = \pi/3,  \omega_1/\omega_2 = 1.8/2$. Ion dynamics is ergodic, the motion is periodic; (d) case $\omega_1/\omega_2 = 1.83/1.85$. Regular motion (quasiperiodic). Pictures reproduced from \cite{Mih21}.} 
	\label{pha1}
\end{figure}

\begin{figure}[!ht]
	\vspace{-2.5cm}
	\begin{minipage}[h]{0.45\linewidth}
		\center{\includegraphics[width=\linewidth]{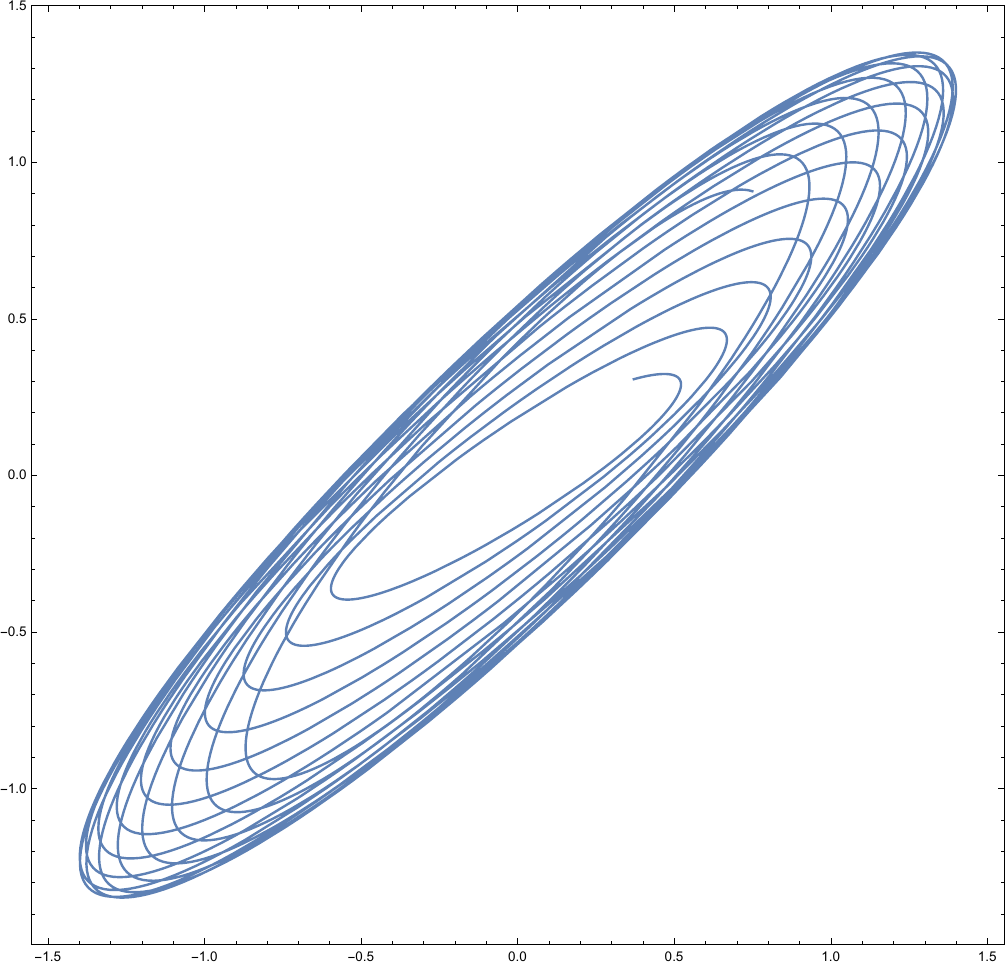}\\a)}
	\end{minipage}
	\begin{minipage}[h]{0.45\linewidth}
		\center{\includegraphics[width=\linewidth]{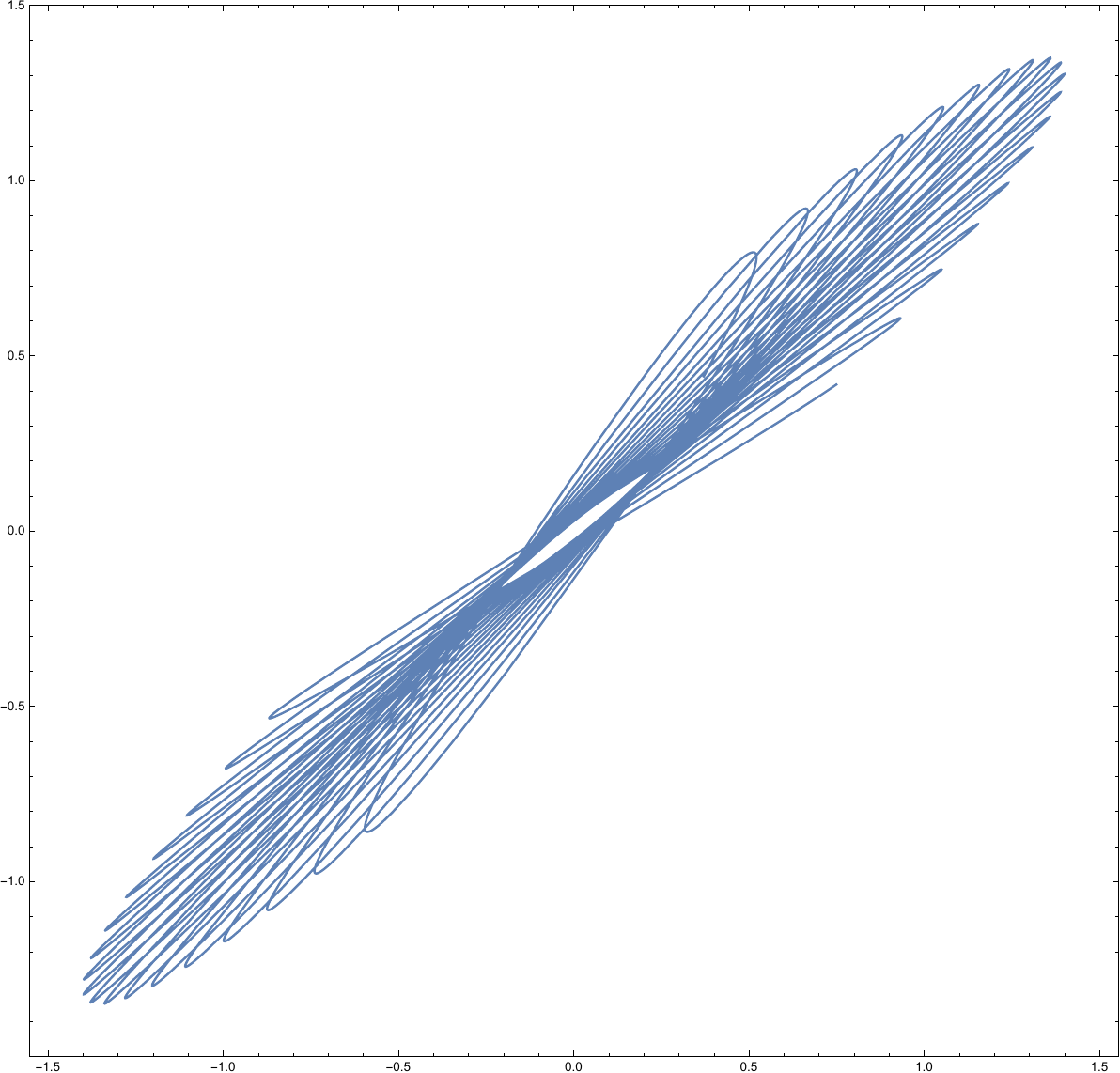}\\b)}
	\end{minipage}
	\caption{Phase portrait for the two ion system with initial conditions: (a) quasiperiodic motion $\omega_1/\omega_2 = 1.78/1.85$; (b) quasiperiodic motion but without chaos present  $\varphi_1 = \pi/3, \varphi_2 = \pi/4, \varphi_3 = \pi/4, \varphi_4 = \pi/3,  \omega_1/\omega_2 = 1.83/1.85$. Pictures reproduced from \cite{Mih21}.}
	\label{pha2}
\end{figure}

Fig.~\ref{pha1} (a) illustrates the phase portrait for a system of two ions (particles) when the ratio between the frequencies is $\omega_1 / \omega_2 = 1.5 / 2$. The theory of differential equations states that when the ratio of the eigenfrequencies is a rational number $\omega_1 / \omega_2 \in {\mathbb Q}$, the solutions of the equations of motion (\ref{subeqcla42}) are periodic trajectories. Ion dynamics is then characterized as stable. If the ratio between the eigenfrequencies is an irrational number $\omega_1 / \omega_2 \notin {\mathbb Q}$, iterative rotations at irrational angles occur around a  point \cite{Gutz90}. Such iterative irrational rotations are called {\em ergodic}, which is stated by the theorem of Weyl \cite{Buni00, Berg03}. The angular frequencies between points located on the circles take discrete values. Fig.~\ref{pha1} (b) and (c) illustrate ergodic dynamics \cite{Kras16} of the system under investigation. In this case ion dynamics is unstable while also showing traces of chaos. Fig.~\ref{pha2} (a) - (b) illustrate the onset or presence of chaos. According to Gutzwiller: "Ergodicity implies that the phase space $M$ cannot be decomposed into subsets with non-vanishing measure each of which remains invariant" \cite{Gutz90}.

\subsection{Dynamics of a system of two ions in a Paul trap}\label{ham}

\subsubsection{Hamilton function associated to the system. Hessian matrix}
The relative motion of two ions of equal electric charges can be expressed as \cite{Moore93, Far94a, Blu95}

\begin{equation}\label{ham1}
\frac{d^2}{dt^2} \begin{bmatrix}
x \\
y \\
z \\
\end{bmatrix} 
+ \left[ a + 2q \cos\left( 2 t \right) \right] \begin{bmatrix}
x \\
y \\
-2 z \\
\end{bmatrix} = \frac{\mu_x^2}{|{\bf r}|^3} \begin{bmatrix}
x \\
y \\
z \\
\end{bmatrix} \ ,
\end{equation}
where $\mu_x = \sqrt{ a + \frac 12 q^2 }$ represents the radial secular characteristic frequency while $a$ and $q$ stand for the adimensional trap parameters in the Mathieu equation, namely 

$$
a = \frac{8 Q U_0}{M \Omega^2 \left( r_0^2 + 2z_0^2\right)}  \ , \;\; q = \frac{4 Q V_0}{M \Omega^2 \left( r_0^2 + 2 z_0^2\right)} \ .
$$

$U_0$ and $V_0$ stand for the d.c and RF trap voltage respectively, $Q$ is the particle electric charge and $\Omega$ represents the RF voltage frequency. For $ a, q \ll 1$,  eq. (\ref{ham1}) can be investigated in the pseudopotential approximation in which the motion is averaged over the driving terms that induce a high frequency oscillating motion (micromotion) \cite{Far94a, Blu98, Blu21a}. Then, an autonomous Hamilton function results which can be expressed in scaled cylindrical coordinates $\left( \rho, \phi, z\right)$ as

\begin{equation}
\label{ham2} H = \frac 12 \left( {p_\rho^2} + {p_z^2} \right) + U\left( \rho, z \right) \ ,
\end{equation}
where 
\begin{equation}
\label{ham3} U\left( \rho, z\right) = \frac 12 \left( \rho^2 + \lambda^2 z^2 \right) + \frac{\nu^2}{2 \rho^2} + \frac 1r \ ,
\end{equation}
and $r = \sqrt{\rho^2 + z^2} \ , \; \lambda = \mu_z/\mu_x \ , \; \mu_z = \sqrt{2\left( q^2 - a\right) }$. $\nu $ is the scaled axial ($z$) component of the angular momentum $L_z$ and it represents a constant of motion, while $\mu_z$ represents the second secular frequency \cite{Far94b}. Both $\lambda$ and $\nu$ are positive control parameters. For arbitrary $\nu $ and $\lambda = 1/2, 1, 2$, eq. (\ref{ham3}) is integrable and even separable, except the case when $\lambda = 1/2$ and $ \left| \nu \right| > 0$ $\left( \nu \neq 0 \right) $. After calculus we obtain

\begin{equation}
\label{ham4}\lambda^2 = 4 \frac{q^2 - a}{q^2 + 2 a}, \; \; \nu^2 = \frac{2L_z^2}{q^2 + 2a} \  ,  
\end{equation}
and we distinguish between three cases: (a) $\lambda = 1/2$, (b) $\lambda = 1$, and (c) $\lambda = 2$ \cite{Moore93, Blu95}. By using the Morse theory \cite{Mil63, Chang93}, the critical points of the $U$ potential can be determined as solutions of the system of equations: 
\begin{subequations}\label{ham5}
	\begin{eqnarray}
	\frac{\partial U}{\partial \rho} = \rho - \frac{\nu^2}{\rho^3} - \frac 1{r^2}\frac{\rho}r = 0 \ , \\
	 \frac{\partial U}{\partial z} = \lambda^2 z - \frac 1{r^2} \frac zr = 0  \ ,
	\end{eqnarray} 
\end{subequations}
with $\partial r/ \partial \rho = \rho /r $ and $\partial r/ \partial z = z / r $. Furthermore, we find that  
\begin{equation}
\label{ham6} z \left( \lambda^2 - \frac 1{r^3} \right) = 0  \ , 
\end{equation}
which obviously leads to two possible cases: (i) $z = 0$, and (ii) $ r^3 = 1/ \lambda^2$. An extended treatment of the problem is found in \cite{Mih21}. By studying the sign of the Hessian matrix eigenvalues one can analyze and classify the critical points, while also characterizing system stability. The degenerate critical points (characterized by $\det {\mathsf H} = 0 $) compose the bifurcation set, whose image in the control parameter space (namely the $\nu - \lambda $ plan) establishes the catastrophe set of equations which defines the separatrix:

\begin{equation}
\label{ham21}\nu = \sqrt{\lambda^{-8/3} - \lambda^{-2/3}}\ \mbox{ or } \   \lambda = 0
\end{equation}

\begin{figure}[bth]
	\begin{center}
		\includegraphics[scale=0.35]{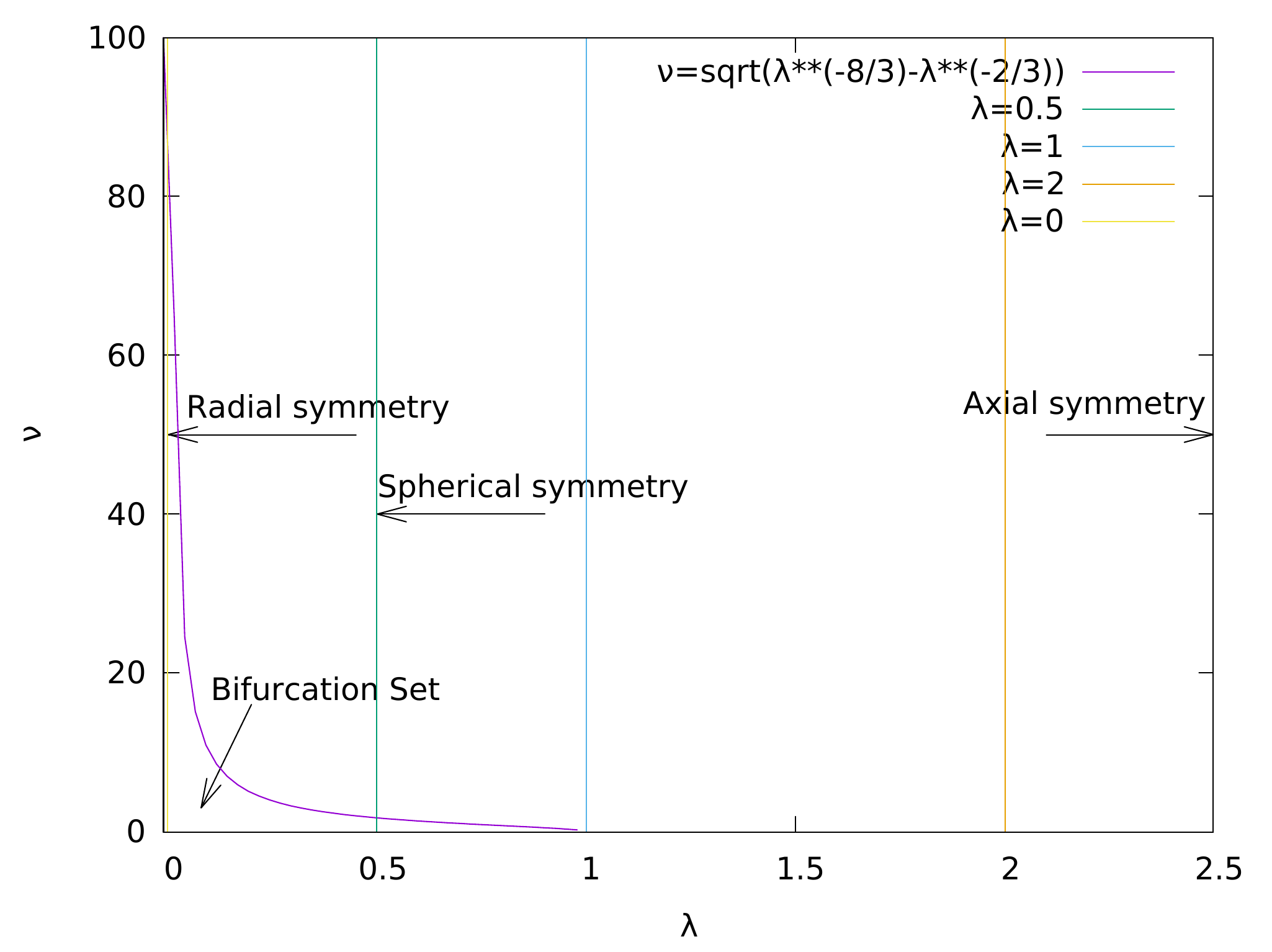} 
	\end{center}
	\vspace{.15cm}
	\caption[]{The bifurcation set for a system of two ions confined in a Paul trap. Picture reproduced from \cite{Mih21}.}
	\label{bifurc}
\end{figure}

Fig. \ref{bifurc} shows the bifurcation diagram for a system of two ions levitated in a Paul trap. The relative motion of the ions is governed by the Hamiltonian function described by eqs. (\ref{ham2}) and (\ref{ham3}). The diagram illustrates the stability and instability regions, where ion dynamics is integrable and non-integrable, respectively. It can be noticed that chaos prevails. Ion dynamics is integrable when $\lambda = 0.5 , \  \lambda = 1 $, and $\lambda = 2$.

Thus, when the electric potential is time independent or in the case of the pseudopotential approximation for RF traps, an autonomous Hamilton function can be associated to the ion system. In this case the family of electric potentials can be classified according to the Morse theory \cite{Nico11}, the bifurcation theory \cite{Dessup16, Alb18}, the ergodic theory, and the catastrophe theory \cite{Gutz90, Stee14}. This classification qualitatively determines the structure of the phase space, the minimum, maximum, as well as the critical points, the periodic orbits and the separatrices. Besides the case of axial motion ion (particle) dynamics in a nonlinear trap is generally non-integrable \cite{Mih02} and chaotic orbits prevail in the phase portrait, except specific neighbourhoods around the points of minimum which consist of regular orbits, according to the Kolmogorov-Arnold-Moser (KAM) theory \cite{Broer10, Dum14}. It is exactly these points of minimum that exhibit a particular interest with an aim to implement quantum logic operations for systems of trapped ions \cite{War20}. Dynamical systems of trapped ions (particles) can be investigated by using the dynamical group theory to describe regular and chaotic dynamics \cite{Ghe92, Mih11, Mih18, Mih22a}, depending on the control parameters determined by the RF trapping field and the trap geometry \cite{Mih21}. 

\subsection{Trap geometries. Nonlinear traps}

One of the major concerns for scientists lies in the design of ion traps with considerably reduced dimensions, required by applications such as optical clocks \cite{Koz18, Kel19, Pyka14, Dele18}. Downsizing the trap dimensions comes at the expense of some unwanted drawbacks, such as increased sensitivity to trap design imperfections or the occurrence of stray potentials \cite{Champ01}. An experiment focused on levitating single ions in a novel RF QIT with spherical shape is described in \cite{Nosh14}. By optimizing the spherical ion trap (SIT) geometry, nonlinearity is sensibly reduced by eliminating the electric octopole moment. 

Ref. \cite{Kassa17} reports on a novel miniature Paul ion trap design that exhibits an integrated optical fibre cavity. Optimal coupling of the ions to the cavity mode is achieved by moving the ion with respect to the cavity mode. The trap features a novel design that includes extra electrodes to which additional RF voltages are applied, in order to fully control the pseudopotential minimum in three dimensions. This method lifts the need to use 3D translation stages for moving the fibre cavity with respect to the ion and achieves high integrability, mechanical rigidity, and scalability. As it does not rely on modifying the capacitive load of the trap, the method leads to precise control of the pseudopotential minimum \cite{Kassa17}. 

Cavity quantum electrodynamics relies on achieving a firm coupling between an atom and an electromagnetic resonator, a requirement that is very difficult to fulfill in experiments. Ref. \cite{Taka20} reports on strong coupling between a single $^{40}$Ca$^+$ ion and an optical cavity, that exceeds both atomic and cavity decay rates. The ion-cavity coupling strength is precisely characterized by means of cavity assisted Raman spectroscopy.

\subsection{Optical levitation and optical tweezing}\label{OpticLev}

\begin{figure}[bth]
	\begin{center}
		\includegraphics[scale=1.5]{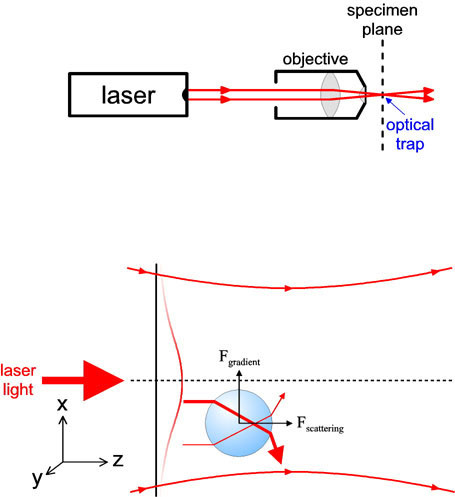} 
	\end{center}
	\caption[]{Simplified view of an optical trap. Photo reproduced from Block lab at Stanford University  https://blocklab.stanford.edu/optical\_tweezers.html}
	\label{OptTrap}
\end{figure}

The radiation pressure concept was probably first suggested by Kepler in 1619, when he advanced this hypothesis to elucidate why comet tails always point away from the sun \cite{Bra89}. A fundamental concept that emerged from the work of Maxwell was the idea to use electromagnetic radiation to manipulate matter using light pressure, or what is commonly known as radiation pressure. The first step in this direction is the pioneering work of Lebedev at the beginning of the 1900s, where a proof of concept is made with respect to the pressure force exerted by the electromagnetic radiation \cite{Lebe1901, Bere16}. Optical trapping was first reported in 1970 when Arthur Ashkin demonstrated that micron-sized dielectric particles could be accelerated and manipulated (trapped) in stable optical potentials, produced by counter-propagating laser beams \cite{Ash70}. Over the next 15 years Ashkin and others experimented with various optical systems that could trap microscale objects, as well as atoms and molecules. In 1986 Ashkin and his colleagues developed a single-beam optical trap \cite{Ash86} that was later given the name of {\em optical tweezers} \cite{Gies15}. This beam is so sharply focused that it generates an optical gradient force that balances the radiation pressure and establishes a stable optical trap in three dimensions (3D) \cite{Leh15}. A simplified sketch of an optical trap is given in Fig.~\ref{OptTrap}.

Noncontact and noninvasive optical trapping was mainly applied to manipulate transparent materials such as biological cells, microorganisms or polystyrene latex spheres \cite{Sato96}. Many experiments have been performed in an attempt to reveal the features of optical trapping and its applicability to less transparent materials. For instance, optical trapping of micron-sized \cite{Sato94} and nanometre-sized \cite{Svobo94} metal particles has been demonstrated in the 1990s. The optical forces generated are in the femto (fN) to tenths of picoNewton (pN) range, which renders them extremely suited to investigate physics at the mesoscopic scale \cite{Dhola08}. Optical tweezers represent an extremely sensible nano-manipulation technique.  

Optical levitation is successfully implemented in two size ranges: (a) in the subnanometre scale where light–matter mechanical coupling enables cooling of atoms, ions and molecules down to the quantum limit, and (b) in the micron scale where the momentum transfer resulting from light scattering \cite{Berg17} allows manipulation of microscopic objects such as cells. Nevertheless, it becomes problematic when one tries to employ these techniques to the intermediate — nanoscale range that encompasses structures such as quantum dots, nanowires, nanotubes, graphene, and 2D crystals. These structures are of critical importance for nanomaterials-based applications. Recently, new approaches have been developed and tested with respect to trapping of plasmonic nanoparticles, semiconductor nanowires and carbon nanostructures. A detailed review on recent advances in state-of-the-art optical trapping at the nanoscale is performed in Ref. \cite{Mara13}. The review focuses on some of the most remarkable advances, among which we enumerate controlled manipulation and aggregation of single and multiple nanostructures, force measurement under conditions of femtonewton resolution and biosensors. Thus, the ability to manipulate nanoparticles is vital in nanoscale science and technology. As object dimensions scale down to the sub-10 nm regime, such a requirement imposes a great challenge for conventional optical tweezers. This explains why a lot of effort has been invested to explore alternative manipulation methods, including using nanostructures, electron beams, scanning probes, etc. \cite{Zheng13}. 

Computational modelling can be applied, for example, to simulate particle dynamics in optical tweezers. This approach is effective when dealing with systems that exhibit many degrees of freedom or to perform experimental simulations. Ion dynamics simulation in optical traps is centred on modelling the optical force, but to enhance precision it should also consider non-optical forces among which the most prominent are Brownian motion \cite{Chav11} and viscous drag. We mention Ref. \cite{Bui17} as a review that focuses on the theory and practical principles of simulation of optical tweezers, while also discussing the challenges and mitigation of known issues.

A comprehensive review of optical trapping and manipulation techniques for single particles in air, including different experimental setups and applications, can be found in Ref. \cite{Gong18}. An experimental method that relies on employing dynamic split-lens configurations with an aim to achieve trapping and spatial control of microparticles by means of the photophoretic force, is reported in \cite{Liz18}. 

It is known that nonspherical particles can severely influence processes that occur in the atmosphere such as radiative forcing, photochemistry, new particle formation and phase transitions \cite{Rama18}. Hence, in order to further investigate and characterize the properties associated with these particles and thus quantify their effect on global climate, it is of utmost importance to perform measurements on single particles. Such step requires continued refinement of the experimental setup and a more detailed understanding of the physical processes associated with optical levitation \cite{Arts19}. An investigation of optical trapping of nonspherical particles in air is performed in Ref. \cite{David18}, where an analytical model aimed at better explaining the physical mechanisms is introduced. The 6D motion of a trapped peanut-shaped particle (3D for translation and 3D for rotation) is analyzed by means of a holographic microscope. In addition, optical forces and torques exerted by the optical trap on the peanut-shaped particle are calculated using Finite Difference Time Domain (FDTD) simulations. A good agreement with experimental results is obtained for the particle motion, while there are still some specific aspects of particle dynamics that cannot be explained \cite{David18}. 

Levitation of nanoparticles in optical or quadrupole traps is expected to result into new advances towards better characterizing quantum decoherence at the mesoscopic scale or into novel ultra-sensitive sensing schemes \cite{Libb04}, as shown in \cite{Peder20, Schel19}. Recent experimental advances indicate that an optically levitated silica microsphere can serve as a spinning-rotor vacuum gauge \cite{Blake20}. Using state-of-the art techniques to manipulate the silica microparticle, absolute vacuum measurements can be performed in space together with gas species identification by measuring the effective molecular mass of a gas mixture at very low pressures (up to $\sim 1$ mbar). Ref. \cite{Deli20} reports on quantum ground state cooling of a levitated silica nanoparticle (NP) in a room temperature environment. As an outcome of the interaction between acoustic phonons and the rotation of a NP around its own axis, the acoustic and optical properties of the NP change when it rotates at a fixed frequency \cite{Humm20}. In addition, optically levitated rotors are excellent candidates for high-precision torque sensors \cite{Peder20, Laan20}.

Ref. \cite{Seb20a} discusses a first experimental demonstration of an optically levitated Yttrium Iron Garnet (YIG) NP both in air and in vacuum, together with a scheme to achieve ground state cooling of the translational motion. The theoretical cooling scheme suggested is based on performing sympathetic cooling of a ferromagnetic YIG nanosphere, using a spin-polarized atomic gas. Finally, state of the art research methods and results in the field of optical trapping of ions are provided in Ref. \cite{Karp19}, while optical pulling forces and their area of applications are presented in Ref. \cite{Li20}. An interesting approach focused on optical manipulation of micro/nano-sized objects is presented in \cite{Sab20}. 

Section \ref{RamanFl} performs a more detailed review of Raman spectroscopy and optical trapping techniques with respect to state-of-the-art experimental results and methods. 
 
\subsection{Acoustic levitation}\label{AcoustLev}

Both electromagnetic and acoustic waves exert radiation forces upon an obstacle placed in the path of the wave \cite{Bra89}. These forces are in a direct relationship with the mean energy density of the wave motion \cite{Borg53, Live81}. Acoustic levitation uses acoustic radiation pressure associated with plane acoustic (sound) waves to confine solids, fluids and heavy gases \cite{Bra01}. Acoustic standing waves are widely used to trap, pattern and manipulate particles. An overview of dust acoustic waves can be found in Ref. \cite{Chop14}. Nevertheless, there is still one major obstacle to cross: there is little knowledge about force conditions on particles which mainly include acoustic radiation force (ARF) and acoustic streaming (AS) \cite{Sepe15, Liu17}. Recently, a numerical simulation model was suggested with an aim to evaluate the acoustic radiation force and streaming for spherical objects located in the path of a standing wave. The model was verified by comparing the results with those obtained from an analytical solution, proposed by Doinikov \cite{Doi94}. Unlike analytical solutions, the proposed numerical scheme is applicable to the case of multiple spheres in a viscous fluid \cite{Sepe15}.

Acoustic levitation uses acoustic radiation pressure forces to compensate the gravity force and levitate objects in air. Although acoustic levitation was firstly investigated almost a century ago, levitation techniques have been limited until recently to particles that are much smaller than the acoustic wavelength. Late experiments show that acoustic standing waves can be employed to steady levitate an object whose dimensions are larger with respect to the acoustic wavelength, under SATP conditions. Levitation of a heavy object (that weighs 2.3 g) is demonstrated by using a setup consisting of two 25 kHz ultrasonic Langevin transducers, connected to an aluminium plate \cite{Andra17}. The sound wave emitted by the experimental setup generates both a vertical acoustic radiation force that compensates gravity and a sideways restoring force that provides horizontal stability to the levitated object. An analysis of levitation stability is achieved by means of using a Finite Element Method (FEM) to determine the acoustic radiation force exerted upon the object. Consequently, recent advances in acoustic levitation allow not only suspending, but also rotating and translating objects in space (3D) \cite{Andra17, Andra18}. In addition, ultrasonic Langevin transducers represent extremely sensible contactless pick-and-place devices that enable one to manipulate delicate objects such as microelectromechanical (MEMS) devices or micro-optical devices \cite{Andra20a}. 

In-depth informations about the acoustic radiation pressure and its applications can be found in Refs. \cite{Borg53, Live81, Andra18}. Investigation of force conditions on micrometer size polystyrene microspheres in acoustic standing wave fields is performed in Ref. \cite{Liu17}, using the COMSOL Multiphysics particle tracing module. The velocity of the particle movement is experimentally measured by employing particle imaging velocimetry (PIV). Use of acoustic levitation to confine microparticles in an electrodynamic trap is presented in Section \ref{AcoustWave}.

\subsection{Instability in electrodynamic traps. Solutions}

For high values of the trapping voltage frequency, microparticles are confined in well defined regions in space and perform oscillations around points of dynamic equilibrium \cite{Lan13, Mih16a}. Particles located close to the trap centre can be practically considered as motionless. The amplitude of oscillation for trapped particles increases as they are located farther with respect to the trap centre. In addition, particle trajectories can be used to map the electric field within the trap. When the a.c. voltage frequency is reduced, particles that remain confined start to move chaotically. A structure consisting of charged particles that strongly interact via the Coulomb force is called {\em quasicrystal} \cite{Boll84}. Transition of the particles from regular motion towards a chaotic regime in case of a low frequency of the a.c. trapping voltage, is called {\em melting} \cite{Wuerk59}. A Coulomb cloud of atomic ions is then obtained, as described in Refs. \cite{Kja05, Blu89}.

Detailed measurements of instabilities of ion trajectories in a Paul trap, caused by distortions of the trap potential with respect to an ideal trap, are presented in \cite{Alhe96a}. Storage instabilities of electrons levitated in a Penning trap at low magnetic fields are described in \cite{Paas03}. Occurrence of these instabilities is attributed to the presence of higher order static perturbations in the trapping potential. The stability properties of a linear RF ion trap with cylindrical electrodes are explored in \cite{Drako06}. Instabilities analogous to those experimentally reported in case of 3D ion traps are also identified for an ideal linear trap. These instabilities are caused by higher order contributions in the series expansion of the trapping electric potential, which deviates from an ideal one.    

Experimental investigations of the stability region of a Paul trap characterized by very strong damping, demonstrate that the adimensional trapping parameter $q$ can reach values up to 25 times greater than the maximum value in absence of damping \cite{Nas01}. The considerable increase in size of the stability region that is observed at high pressures, can be explained by the numerical solution of the Mathieu differential equation with damping term. Nonlinear dynamics in Penning traps with hexapole and octopole terms of the electric potential is explored in \cite{Salas02, Lara04}. 

To confine charged dust particles in a dynamical trap, under SATP conditions, friction between charged particles and air must be considered \cite{Wint91}. Moreover, the trap parameters must be chosen carefully as the stability region is shifted \cite{Major05}. One of the most difficult tasks to perform in experiments consists in charging the aerosol particles with electric charges of identical sign \cite{Stern01}. In order to achieve similar charge-to-mass ratios,  electrically charged oil drops are used \cite{Rob99}. The drops assemble into a Coulomb structure, under SATP conditions. To confine charged particles of silicon carbide (SiC) and alumina (Al$_2$O$_3$), classical quadrupole traps \cite{Ghe95a, Ghe98, Sto01} and multi-electrode traps consisting of 8, 12 or 16 cylindrical brass electrodes (with 3 or 4 mm diameter) can be used \cite{Mih16a, Mih16b}. Experiments confirm that a 12 electrode trap geometry yields to higher stability and smaller amplitude of oscillation of levitated microparticles. Confinement conditions prove to be very sensitive with respect to the trap geometry. Quadrupole traps are used more frequently as they provide better control of the trapping parameters. The radius of the multipole traps designed and tested in INFLPR ranges between $1 \div 1.5$ cm \cite {Mih16a, Mih16b, Mih08}. Negatively charged particles of Al$_2$O$_3$ with diameters ranging between $60 \div 200$ microns (with mass of $ 4.18 \times 10^{-10}$ kg and $1.55 \times 10^{-8}$ kg, respectively), are confined close to the trap axis line if their number is small. If the particle number is larger, they start to move chaotically yielding a dusty cloud. To estimate the electric charge, the levitated cloud is subjected to the action of acoustic waves that induce parametric oscillations of the particles \cite{Sto11, Ghe96b, Sto08}, as shown in Fig.~\ref{ac1}. The value of the experimentally determined charge-to-mass ratio is around $5.4 \times 10^{-4}$ C/kg. 

Ion and particle traps represent versatile instruments to explore nonequilibrium statistical physics, where dissipation and nonlinearity can be very well controlled. In case of ion chains dissipation is achieved by means of laser heating and cooling, while nonlinearity is the outcome of trap anharmonicity and beam shaping. Ion dynamics is governed by an equation similar to the complex Ginzburg-Landau equation. As an exotic feature, the system can be described as both oscillatory and excitable at the same time. The experiment is presented in Ref. \cite{Lee11}, while the approach also allows controllable experiments with noise and quenched disorder.

\begin{figure*}[!ht]
	\centering
	\includegraphics[scale=0.95]{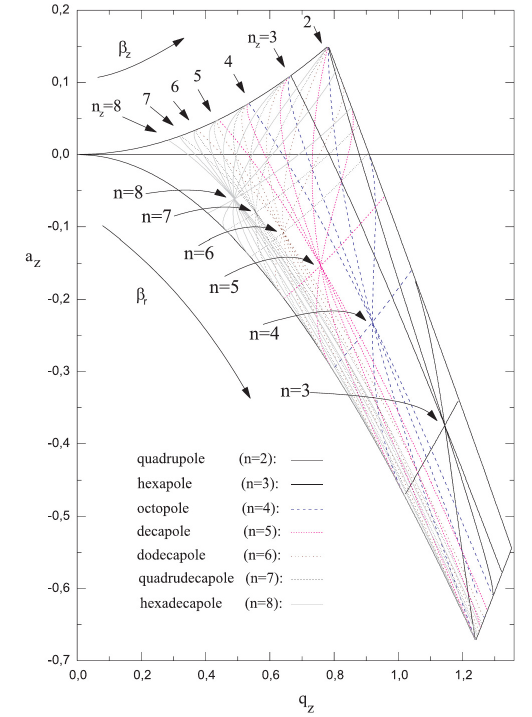}	
	\caption{Theoretical lines of instabilities in the first region of stability of the Paul trap, for perturbations of order $n = 3$ to $n = 8$. Picture reproduced from \cite{Major05} by courtesy of Prof. G. Werth.}
	\label{InstabPaul} 
\end{figure*}

A theoretical model used to describe ion transient response to a dipolar a.c. excitation in a QIT is presented in \cite{Xu11}, where high frequency ion motion components are also considered. The electric potential is a pure quadrupole one in case of an ideal ion trap and ion dynamics is described by a Hill (Mathieu) equation. Real traps used in experiments do not generate a quadrupole field due to geometric imperfections (electrode truncation) or misalignment of electrodes. The nonlinearity that real traps exhibit is the outcome of weak multipole fields (e.g., hexapole, octopole, decapole, and higher order fields \cite{Micha05, Zhao08, Pedre10, Marche21}). Ion dynamics in such a trap is governed by a nonlinear Mathieu equation that can not be solved analytically. The theoretical lines of instabilities in the first region of stability of the Paul trap, for perturbations of order $n = 3$ to $n = 8$, are illustrated in Fig~.\ref{InstabPaul}.

In Ref. \cite{Doro12} a technique is used to calculate axial secular frequencies of a nonlinear ion trap with hexapole, octopole, and decapole terms of the electric potential, based on the modified Lindstedt–Poincar{\'e} method. The equations of ion motion in the resulting effective potential are shown to be similar to a Duffing equation \cite{Kova11}. 

Ref. \cite{Wang13} introduces the harmonic balance method, used to examine the coupling effects induced by hexapole and octopole fields on ion motion in a QIT. Ion motion characteristics and buffer gas damping effects are investigated in presence of both hexapole and octopole fields. It is suggested that hexapole fields yield to higher impact on ion motion centre displacement, while octopole fields prevail on ion secular frequency shift. In addition, nonlinear features caused by hexapole and octopole fields could add or cancel each other. The method is further used to devise an ion trap with improved performance, based on a particular combination of hexapole and octopole fields \cite{Wang13}.

A detailed analysis based on numerical simulations of ion dynamics in case of a nonlinear RF (Paul) trap with terms of the electric potential up to the order 10, is performed in Ref. \cite{Herba14}. The Hill-Mathieu equations of motion that characterize the associated dynamics are numerically solved by employing the power series method. The stability diagram is qualitatively discussed and buffer gas cooling is implemented using a Monte Carlo method that takes dipole excitation in consideration. The method is then demonstrated in case of a real trap, where a good agreement with experiments and numerical simulations is obtained. RF ion traps encompass complex electric field shape which yields to complex ion motion. The paper of Li \cite{Li17} reports on SIMION simulations that show classical chaotic behaviour of ions levitated in a toroidal ion trap. Fractal-like patterns are illustrated in a series of zoomed-in regions of the stability diagram. 

Refs. \cite{Saxe18, Foot18} investigate the dynamics of a collection of charged particles (ions) in a dual-frequency Paul trap. In Ref. \cite{Saxe18} the authors infer the analytical expressions for the single particle trajectory and the plasma distribution function assuming a Tsallis statistics, by emphasizing on the difference between the trap configuration used and a conventional Paul trap. Use of a secondary RF frequency in a dual-frequency Paul trap allows one to modify the spatial extent of charged particles confinement. The plasma distribution function and temperature are periodic provided that the ratio of the eigenfrequencies is a rational number $\omega_1 / \omega_2 \in {\mathbb Q}$, case when the solutions of the equations of motion are periodic trajectories \cite{Mih21}. Then, the resulting period of plasma oscillation is given by the Least Common Multiple (LCM) of the time periods corresponding to the two driving frequencies and linear combinations of them. The plasma temperature varies spatially and the time-averaged plasma distribution is found to be double humped beyond a certain spatial threshold, indicating the presence of certain instabilities \cite{Syr19b, Syr19c}. The dual frequency effect on the energy level shifts of atomic orbitals is investigated, whilst uncertainties in the second order Doppler and Stark shifts are found to be of the same order as those of a single-frequency Paul trap \cite{Saxe18}.

In addition, the effective rotational potential can be used to describe dynamics of various diatomic particles with different centres of mass (CM) and charges, levitated in a plane quadrupole RF trap. By comparison between "the model of pseudopotential for localization of a single ion and the proposed model of the effective rotational potential for diatomic structure", one can determine "additional positions of quasi-equilibrium for the CM of the diatomic particle and orientation angle of the molecule" \cite{Vasi19}. 

Electrodynamic (ED) levitation of one or a few charged droplets using Paul traps opens new pathways towards explaining the Rayleigh instability \cite{Singh19, Singh20} and the interplay between various inter-particle forces that yield to pattern formation \cite{Blu89, Lee11}. Very recent papers \cite{Singh17, Singh18a} focus on developing a theory for two-dimensional patterns formed during levitation of two to a few droplets in an EDB. The theory is based on an extension of the classical Dehmelt approximation to interacting particle systems. Among other original contributions the theory demonstrates solutions to the inter-particle separations, secular frequencies, stabilities of two-drop systems, and collapse of droplets onto the $X-Y$ plane. The theory is also applied to predict the size and shape of closed structures that occur for few-drop systems. 

Spontaneous wave-function collapse models have been proposed in an attempt to reconcile deterministic quantum mechanics with the nonlinearity and stochasticity of the measurement operation. According to these models random collapses in space of the wave function of any system occur spontaneously, leading to a progressive spatial localization free of the measurement process. Noninterferometric tests of spontaneous collapse models for a nanoparticle levitated in a Paul trap, in ultrahigh cryogenic vacuum, are reported in Ref. \cite{Vin19}.

\subsection{Chaos}\label{chaos}

Quadrupole ion traps (QITs) \cite{Ghosh95, Major05, March09} are extremely versatile tools to perform studies of chaos and integrability for dynamical systems \cite{Aku14, Ber00, Gra13, Blu95, Blu98, Rozh17}, fundamental tests on quantum mechanics concepts \cite{Leibf03, Orsz16}, and investigations of non-neutral plasmas \cite{Haro10, Werth09, Dub99, Wine88, Meni07}. Deterministic chaos deals with long-time evolution of a system in time. A dynamical system represents a system that evolves in time. Chaos is related to the study of the dynamical systems theory \cite{Gutz90, Meiss17} or of nonlinear dynamics \cite{Mih10b, Blu95, Stro15, Schne17}. Dynamical systems can be either conservative when friction is absent and the system does not give up energy in time, or dissipative when we frequently encounter a behaviour called {\em limit cycle} in which the system approaches some asymptotic or limiting condition in time \cite{Nay04, Gutz90, Hilb10}. Under certain conditions, the asymptotic or limiting state is where chaos occurs. Hence, chaos develops in deterministic, nonlinear, dynamical systems. Other chaos-related geometric objects such as the boundary between periodic and chaotic motions in phase space may also exhibit fractal properties. 

The dynamics associated to a single charged particle in a Paul trap, in presence of a damping force, is theoretically and experimentally investigated in \cite{Izma95, Hase95}. Izmailov {\em et al} explore the damped microparticle motion in presence of combined periodic parametric and random external excitations, using the singular perturbation theory \cite{Izma95}. The modified stability diagrams in the parameter space are calculated in \cite{Hase95}, illustrating that stable regions in the $a - q$ parameter plane are not only enlarged but also shifted. It is also emphasized that in certain cases, the damping force may induce instability in particle dynamics. Later, Sevugarajan has developed a model relating perturbation in the ion axial secular frequency to geometric aberration, space charge, dipolar excitation, and collisional damping in nonlinear Paul trap mass spectrometers \cite{Sevu00}. A multipole superposition model that considers both hexapole and octopole superposition is introduced to account for field inhomogeneities. Dipolar excitation and damping are also considered in the equation of ion motion. The perturbed secular frequency of the ion is then inferred by use of a modified Lindstedt–Poincar{\'e} perturbation technique. Moreover, the shift in the ion secular frequency with the axial distance from the centre of the trap exhibits quadratic variation. Ref. \cite{Sevu02} investigates the role of field inhomogeneities in altering the stability boundaries in nonlinear Paul traps mass spectrometers, taking into account higher order terms in the equation of motion. The contribution of hexapole and octopole superposition in shifting the stable trapping region along with ion secular frequencies in nonlinear Paul traps are also explored. Ref. \cite{Zhou10} uses the Poincar{\'e}-Lighthill-Kuo (PLK) method to infer an analytical expression on the stability boundary and ion trajectory, while it also suggests a multipole superposition model with an aim to model electric field inhomogeneities. Using the method described both the dynamic shift and the ion secular frequency are expanded to asymptotic series. 

The presence of space charge and its outcome on ion dynamics in case of a LIT is nicely investigated in Ref. \cite{Mand14} where the authors emphasize on two distinguishable effects: (i) modification of the individual ion oscillation frequency due to the modification of the trap potential (similar to the case of a 3D QIT described in \cite{Alhe96a, Alhe96b}), and (ii) for particular experimental conditions of high charge density, ions might perform as a single collective body and exhibit motional frequency independent of the ion number. Collective oscillations are reported for both 3D Paul \cite{Alhe96b} and Penning \cite{Wine75} QIT, but Mandal {\em et al} report for the first time on criticality of the collective oscillation on ion number and collective oscillation of the radial motion \cite{Mand14}. The model introduced by the authors accounts for experimental observations. 

A nonlinear chaotic system, namely the parametrically kicked nonlinear oscillator, can be achieved in the dynamics of a trapped, laser-cooled ion, interacting with a sequence of standing wave pulses \cite{Mih10b, Ake10, Bres97, Adam01, Qing17}. RF ion traps are usually associated with complex electric field shape and sensibly intricate ion motion. Classical chaotic behaviour of ions in a toroidal ion trap is explored in \cite{Li17} by employing SIMION simulations. It is demonstrated that chaotic motion occurs due to the presence of nonlinear terms of the electric fields generated by the trap electrodes.

The quasistationary distribution of Floquet-state occupation probabilities for a parametrically driven harmonic oscillator coupled to a thermal bath is investigated in \cite{Dier19}. Recent investigations on the stochastic dynamics of a particle levitated in a periodically driven potential are presented in \cite{Mai19a}. Results show that in case of atomic ions confined in RF Paul traps, noise heating and laser cooling typically respond slowly with respect to the unperturbed motion. These stochastic processes can be characterized in terms of a probability distribution defined over the action variables. In addition, Ref. \cite{Mai19a} presents a semiclassical theory of low-saturation laser cooling applicable from the limit of low-amplitude motion to the extent of large-amplitude motion, that fully describes the time-dependent and anharmonic trap. Thus, a detailed study of the stochastic dynamics of a single ion is achieved. A single atomic ion confined in a time-dependent periodic anharmonic potential, in presence of stochastic laser cooling, is explored in Ref. \cite{Mai19b}. The paper shows how the competition between the energy gain from the time-dependent drive and damping leads to the stabilization of stochastic limit cycles. Such distinct nonequilibrium behaviour can be observed in experimental RF traps loaded with laser-cooled ions. 

Further on we present ion (particle) dynamics in a nonlinear quadrupole Paul trap with octopole anharmonicity. The system can be characterized as dissipative. Particle dynamics is described by a nonlinear Mathieu equation \cite{Brou11, Rand16}. All perturbing contributions such as: (i) damping, (ii) multipole terms of the potential, and (iii) harmonic excitation force, are considered in this approach. In order to complicate the picture and approximate real conditions it is also considered that the particle (ion) undergoes interaction with a laser field \cite{Mih10b}. The resultant equation of motion is shown to be similar to a perturbed Duffing-type equation, which is a generalization of the linear differential equation that describes damped and forced harmonic motion \cite{Kova11}. 

\subsubsection{Dynamics of a particle confined in a nonlinear trap}
We investigate the dynamics of a charged particle (ion) confined in a quadrupole nonlinear Paul trap \cite{Lan13}, which we treat as a time-periodic differential dynamical system. Dissipation in such system is very low which leads to a number of interesting phenomena. The equation of motion along the $x$-direction for a particle of electrical charge $Q$ and mass $M$ which undergoes interaction with a laser field in a quartic potential $V\left( u \right) = \mu u^4$, $\mu > 0$, in presence of damping \cite{Mih10b}, can be expressed as

\begin{equation}\label{cha1}
\frac{d^2u}{d\tau^2} + \gamma \frac{du}{d\tau} + \left[ a - 2q\cos\left(2 \tau \right)\right] u + \mu u^3 + \alpha \sin u = F \cos \left(\omega_0 t\right) \, 
\end{equation}
where $u = kx$, with $k$ a constant, $\tau = \Omega t /2$, $\alpha = 2k^2\Omega_0 \cos \theta/M\Omega^2$, and $\gamma$, $\mu$ stand for the damping and anharmonicity coefficient, respectively. The adimensional parameters are 

\begin{equation}\label{cha2}
a = \frac{-8QU_0}{M\Omega^2 d} , \;\; q = \frac{4QV_0}{M\Omega^2 d} , \;\; d = r_0^2 + 2 z_0^2 \ ,
\end{equation}
where the geometrical parameters $r_0$ and $z_0$ stand for the trap semiaxes. Generally, for a typical 3D Paul trap $a = 0.1$ and $q = 0.7$. The frequency of the applied a.c. voltage is denoted by $\Omega$, $U_0$ and $V_0$ are the static and time-varying trapping voltages, whereas $\Omega_0$ is the Rabi frequency for the ion–laser interaction, and $\cos \theta$ represents the expectation value of the $x$ projection spin operator for a two-level system with respect to a Bloch coherent state \cite{Dodon03, Gryn10, Ortiz19, Zela20a, Robe21}. The expression $F \cos \left( \omega_0 t\right)$ stands for the driving force, an external excitation at frequency $\omega_0$.

Equation (\ref{cha1}) can also be regarded in a good approximation as Newton's law for a particle located in a double-well potential. The force $F \cos\left( \omega t\right)$ represents an inertial force that arises from the oscillation of the coordinate system. The mathematical analysis of equation (\ref{cha1}) (which is dimensionless) requires using techniques from the global bifurcation theory \cite{Nay04, Gutz90, Alb18, Stro15}. We investigate equation (\ref{cha1}) by means of numerical simulations, in an attempt to gain more insight on the behaviour of such strongly nonlinear systems. The dynamical behaviour of the equation of motion is studied numerically by varying the damping and driving frequency parameters, as well as the amplitude parameter. We finally discuss the possibility of observing chaos in such a nonlinear system \cite{Bres97, Qing17, Gard97}. Chaotic regions in the parameter space can be identified by means of Poincar{\'e} sections \cite{Gutz90, Buni00, Stro15, Hilb10}, as illustrated in \cite{Robe18} where integrable and chaotic motion of a single ion is investigated for five-wire surface-electrode Paul traps. A late experiment reports on a monopole ion trap used to investigate nonlinear dynamics of a single charged polyethylene microsphere \cite{Per21}. Phase space portraits and power spectra show a period-doubling route to chaos as well as a period-3 window within the chaotic regime.

\subsubsection{Nonlinear parametric oscillator in a Paul trap. Phase-space orbits, Poincar{\'e} sections and bifurcation diagrams. The onset of chaos. Attractors} 
	
We consider the trapped particle as a forced harmonic oscillator described by a non-autonomous or time-dependent equation of motion \cite{Cruz20, Zela20b}. Forced oscillators exhibit many of the properties associated with nonlinear systems. Most nonlinear systems are impossible to solve analytically, a fact which makes numerical modelling a powerful tool to investigate the associated dynamics. The trajectory represents the solution of the differential equation starting from a set of initial conditions. A picture that illustrates all the qualitatively different trajectories of the system is called a phase portrait. The appearance of the phase portrait is controlled by fixed points. In terms of the original differential equation, fixed points represent equilibrium solutions. An equilibrium is considered as stable if all sufficiently small disturbances away from it damp out in time \cite{Nay04, Hilb10, Cvita22}. 

\begin{figure}[!ht]
	\centering
	\subcaptionbox{\label{sfig:a}}{\includegraphics[width=.3\textwidth]{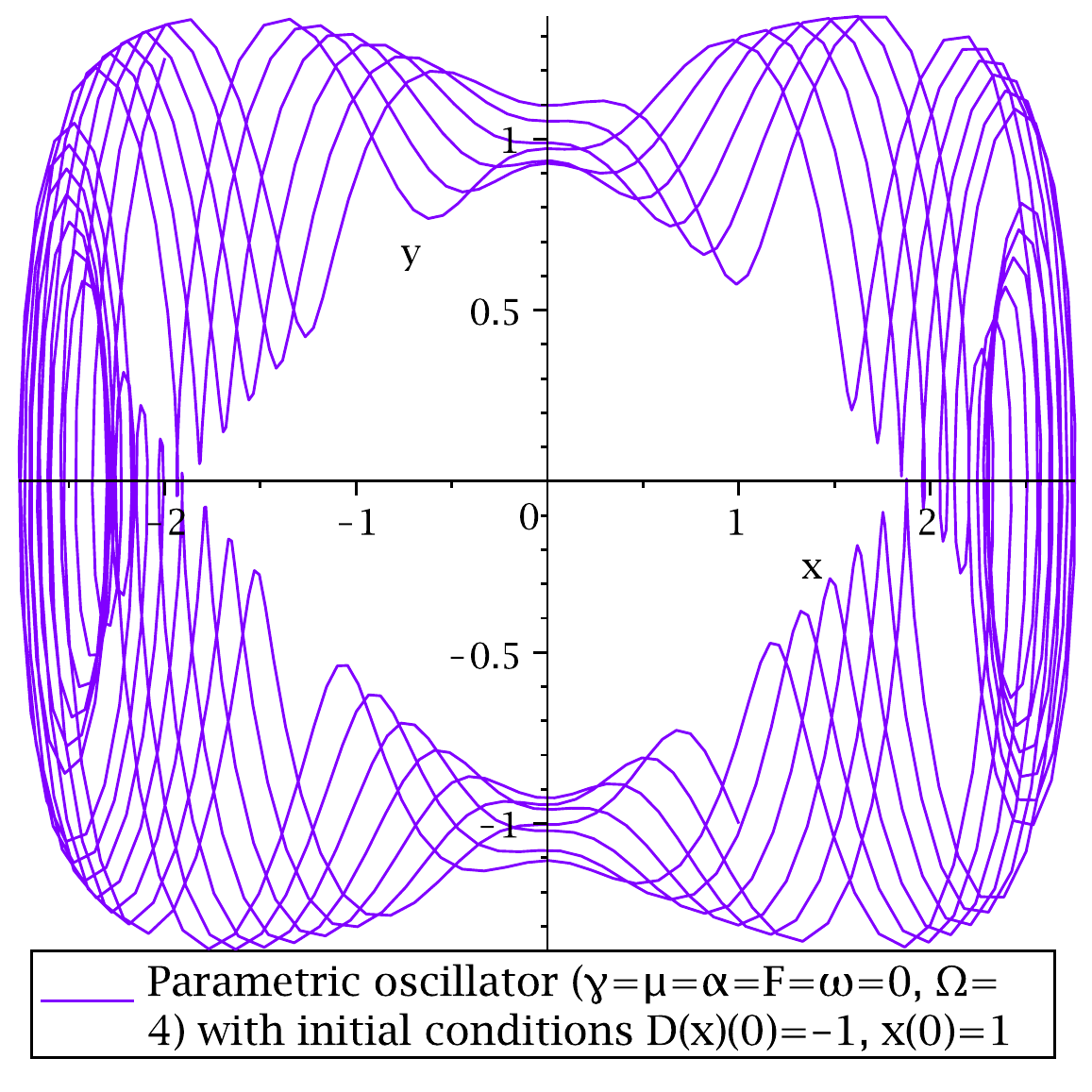}}
	\subcaptionbox{\label{sfig:b}}{\includegraphics[width=.3\textwidth]{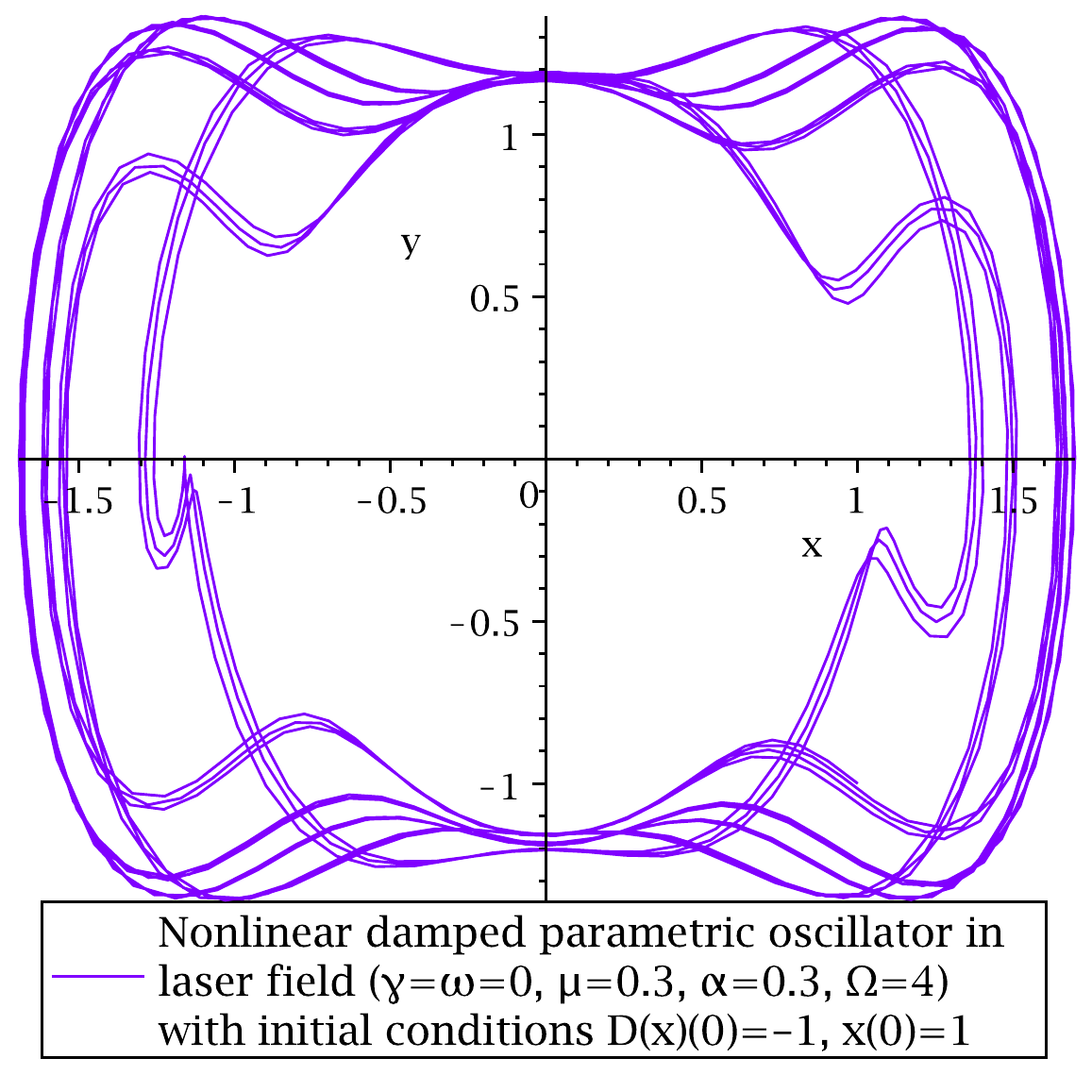}}
	\subcaptionbox{\label{sfig:c}}{\includegraphics[width=.3\textwidth]{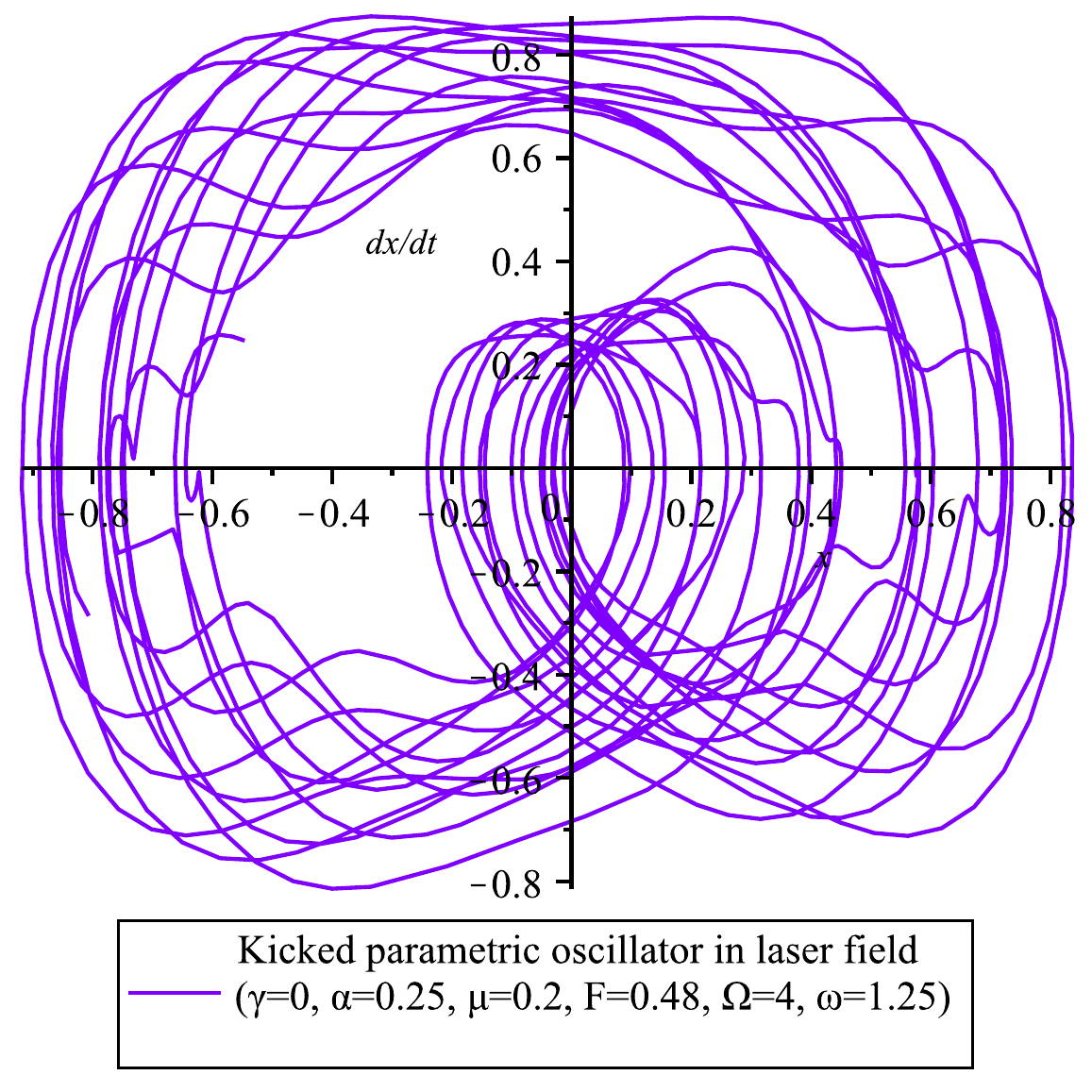}}
	\\
    \subcaptionbox{\label{sfig:d}}{\includegraphics[width=.3\textwidth]{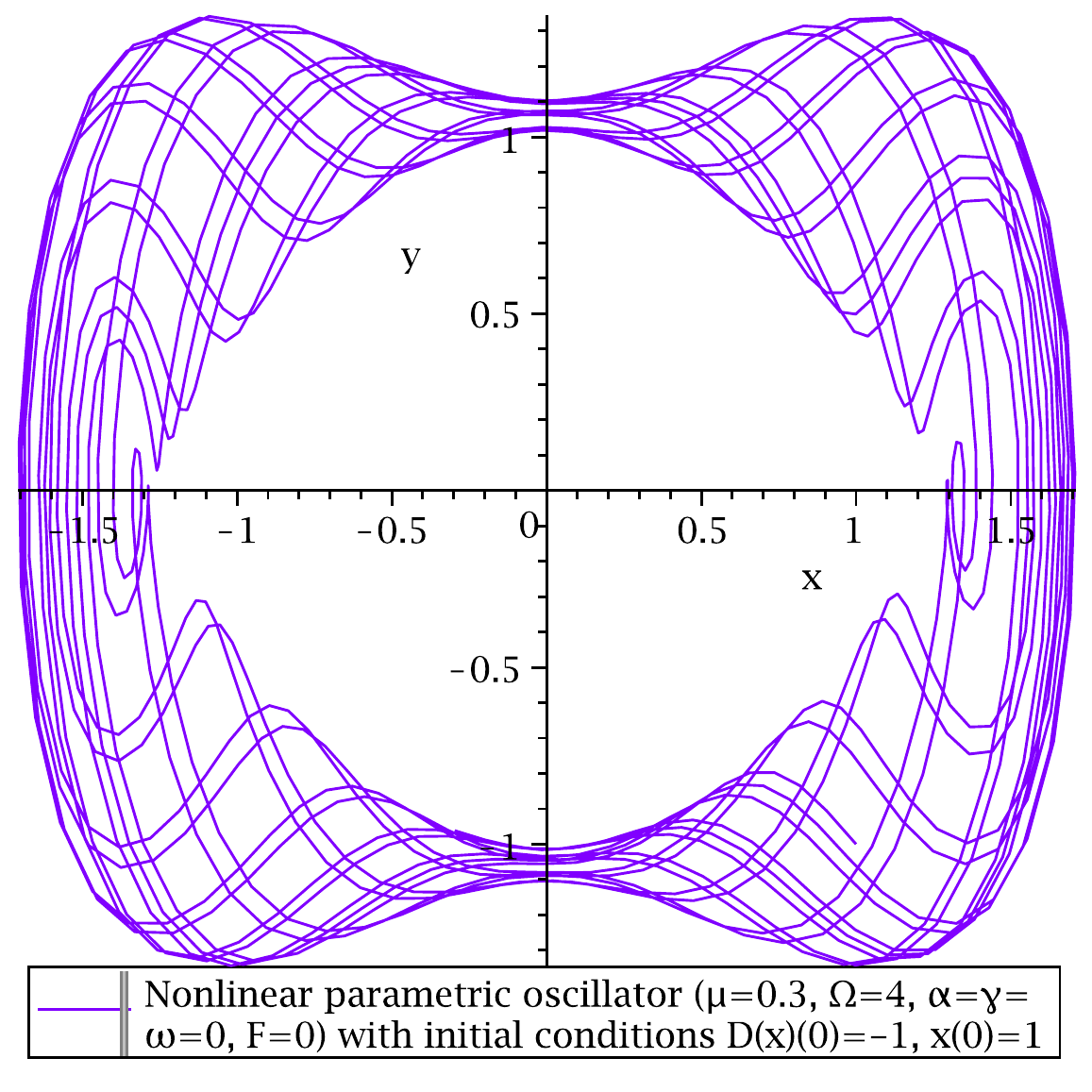}}
    \subcaptionbox{\label{sfig:e}}{\includegraphics[width=.3\textwidth]{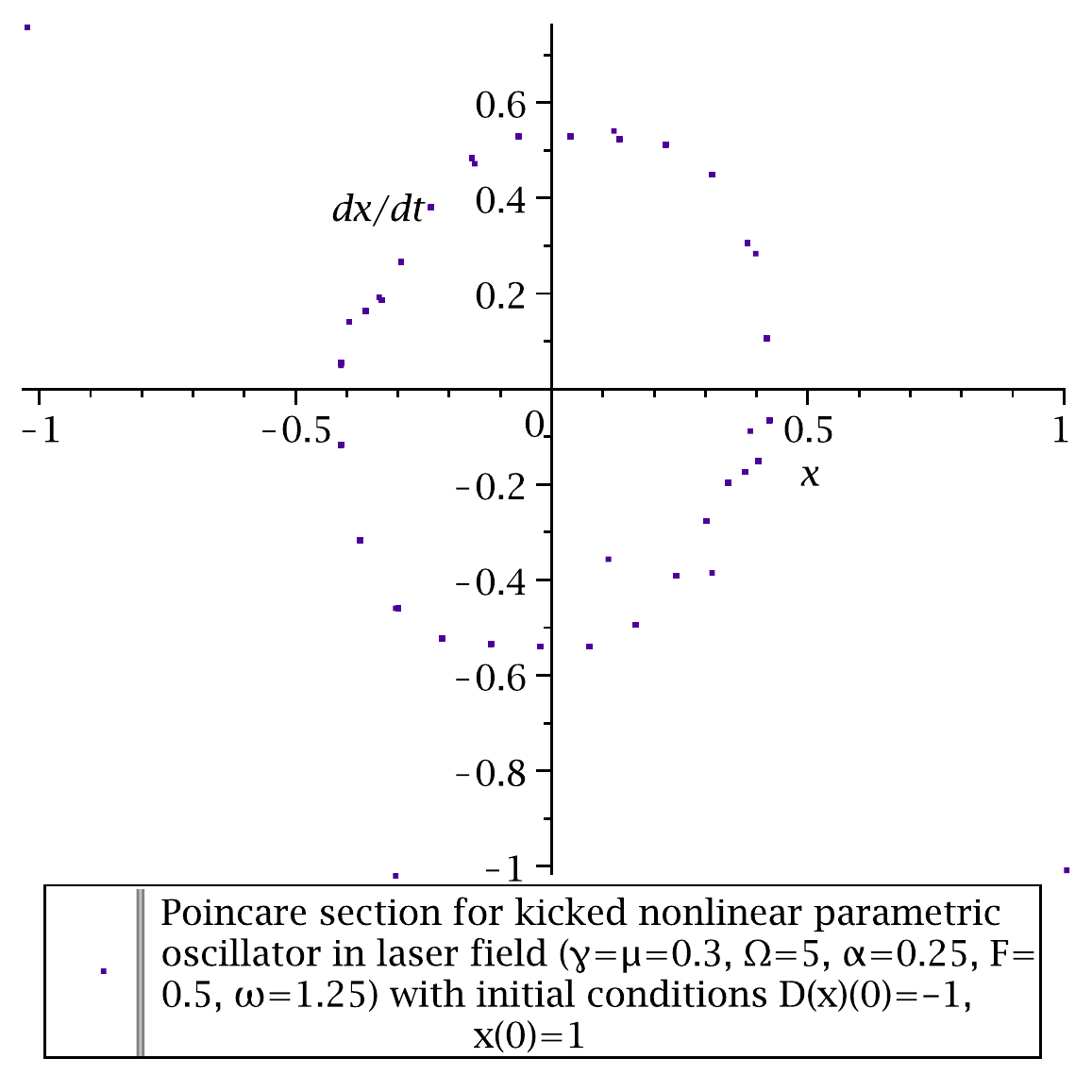}}
    \subcaptionbox{\label{sfig:f}}{\includegraphics[width=.3\textwidth]{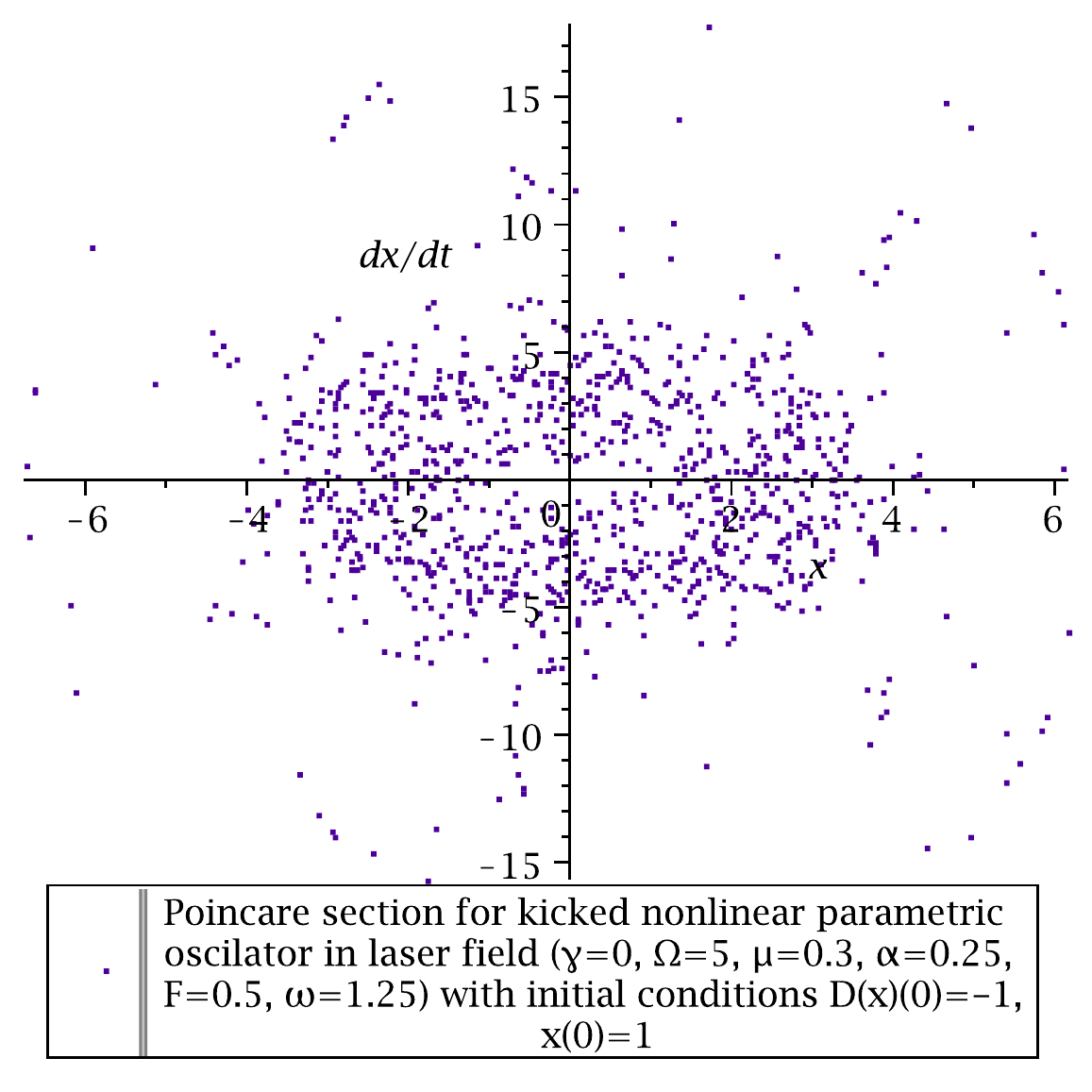}}
   	\\
	\subcaptionbox{\label{sfig:g}}{\includegraphics[width=.3\textwidth]{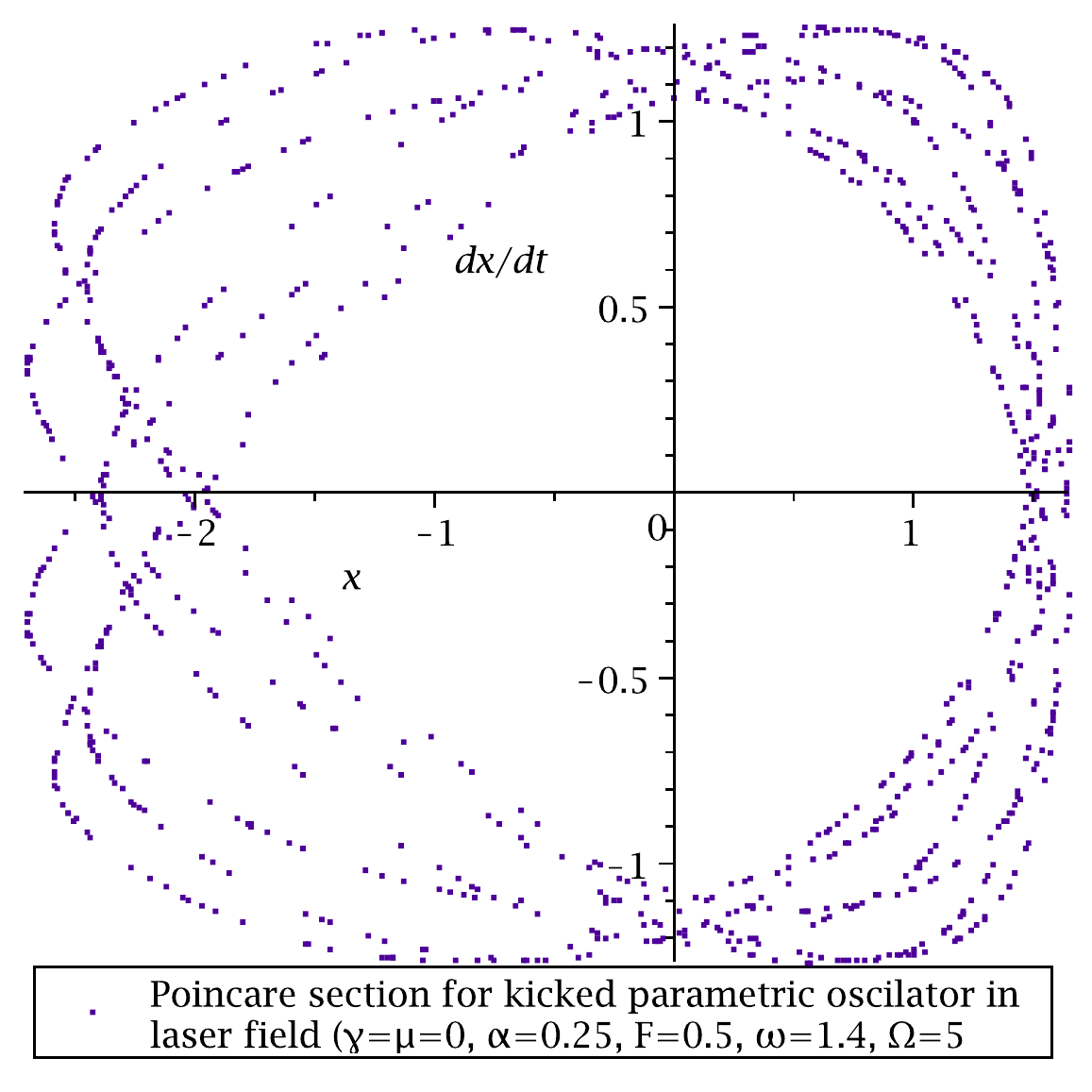}}
	\subcaptionbox{\label{sfig:h}}{\includegraphics[width=.3\textwidth]{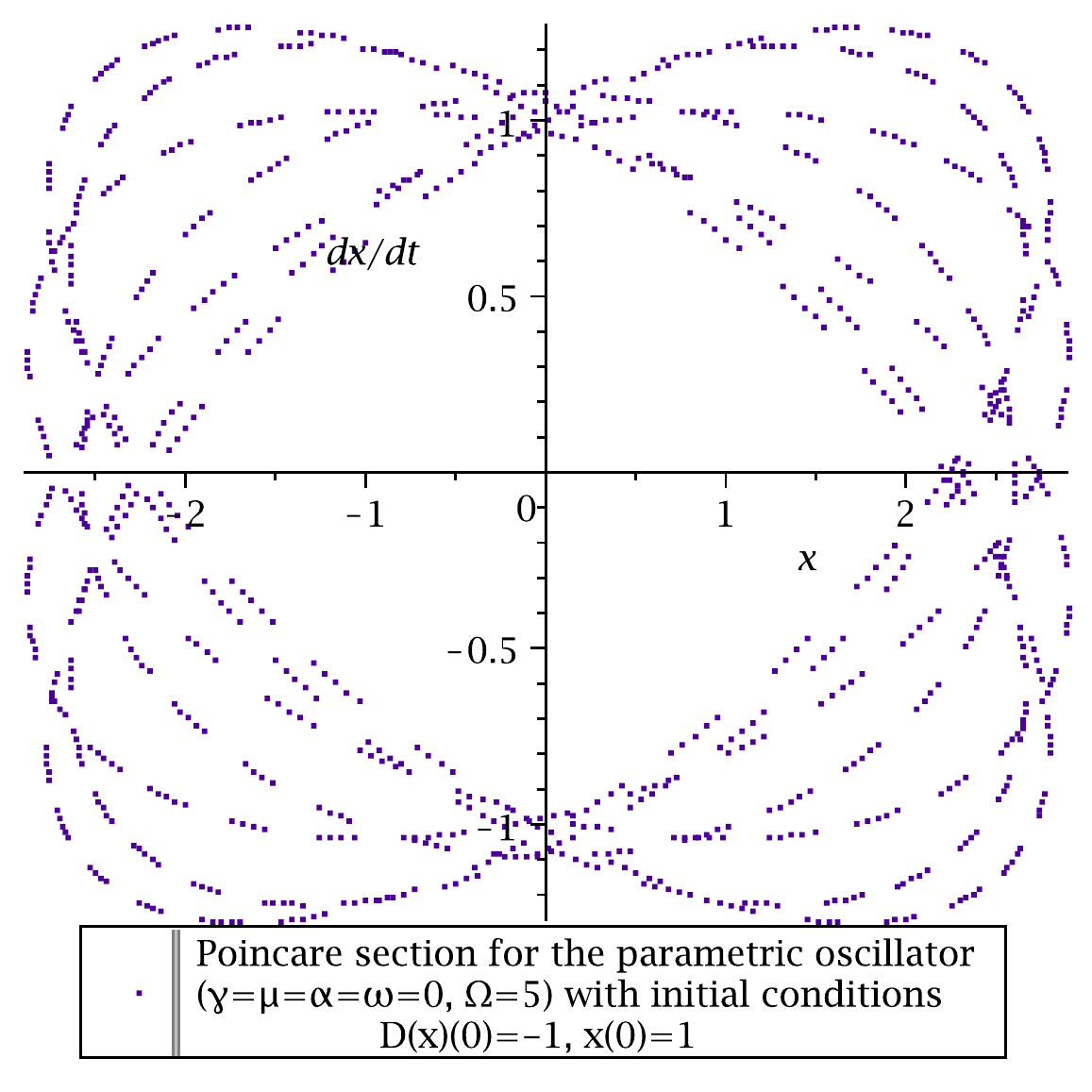}}
	\subcaptionbox{\label{sfig:i}}{\includegraphics[width=.3\textwidth]{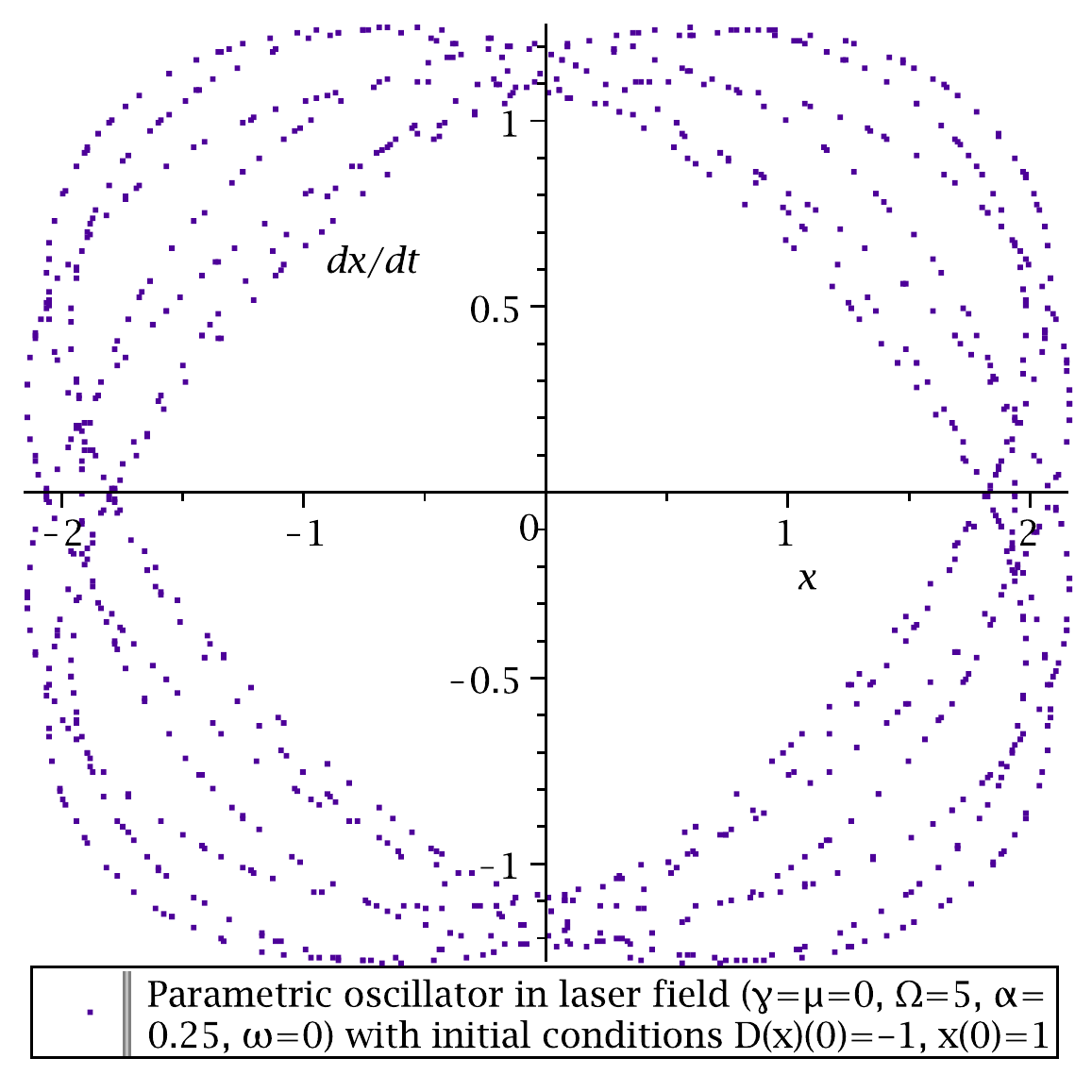}}
	\caption{Phase-space orbits and Poincar{\'e} sections for an ion confined in a nonlinear trap. Pictures reproduced from \cite{Mih10b} under permission of the authors. Copyright Institute of Physics.}
	\label{OrbitsDuff}
\end{figure}

A numerical integration of the equation of motion (\ref{cha1}) is performed by employing the fourth-order Runge–Kutta method \cite{Lynch17}. In order to illustrate the dynamics of the trapped particle (ion) we represent trajectories in the 2D phase space (phase portraits) \cite{Stee14, Lynch18} and in the extended phase space, as illustrated in Fig.~\ref{OrbitsDuff}, with an aim to emphasize regular and chaotic orbits.

By analyzing the associated phase portraits we observe that the cubic term in eq. (\ref{cha1}) $-\mu u^3$ provides a nonlinear restoring force for large values of $x$, while the linear term pushes away from the origin. The potential for this oscillator exhibits a double-well structure. For certain initial conditions there exists an unstable equilibrium point at $x = 0$, and given some damping the particle has to fall into one side of the well or the other if it approaches the equilibrium point with just enough energy to move over it. The homogeneous problem (non-driven oscillator) contains no surprises in it. Given an initial condition there is a unique phase-space trajectory that leads to the particle winding up at the bottom of one of the two wells, after the mechanical energy is converted into heat. When the oscillator is driven by a periodic force, the system can reach a limit cycle where as much mechanical energy is lost per cycle as it is dumped into the system by the crank. Chaos appears as a result of the two wells connected by an unstable equilibrium point. The phase portraits clearly reflect the existence of one (see Fig.~\ref{OrbitsDuff} c) or two attractors (see Fig.~\ref{OrbitsDuff} a), b), and d)), as well as fractal basin boundaries for the trapped particle (ion) assimilated with a periodically forced double-well oscillator. For some of the parameter values presented in Fig.~\ref{OrbitsDuff} the system clearly exhibits two periodic attractors corresponding to forced oscillations confined to the left or right well. Depending on the initial conditions the system can converge rapidly to one of the two attractors after an initial transient phase. The basins of attraction generally exhibit a complicated shape and the boundary between them is fractal \cite{Stro15}. As illustrated by the phase portraits there are particular cases when the dynamics is characterized by periodicity.

The Poincar{\'e} sections are also represented in Fig.~\ref{OrbitsDuff}. We emphasize the appearance of what we consider to be strange attractors in the ion dynamics. A strange attractor represents the limiting set of points to which the trajectory tends (after the initial transient) every period of the driving force. Fig.~\ref{OrbitsDuff}(f) represents a fractal set. This particular type of oscillator is also known as strange attractor, a clear indication that the system is chaotic \cite{Stro15, Cvita22}. Chaos prevails too in the other cases but the system exhibits periodic orbits (see Fig.~\ref{OrbitsDuff} g), h), and i).

\begin{figure}[!ht]
	\includegraphics[width=.5\textwidth]{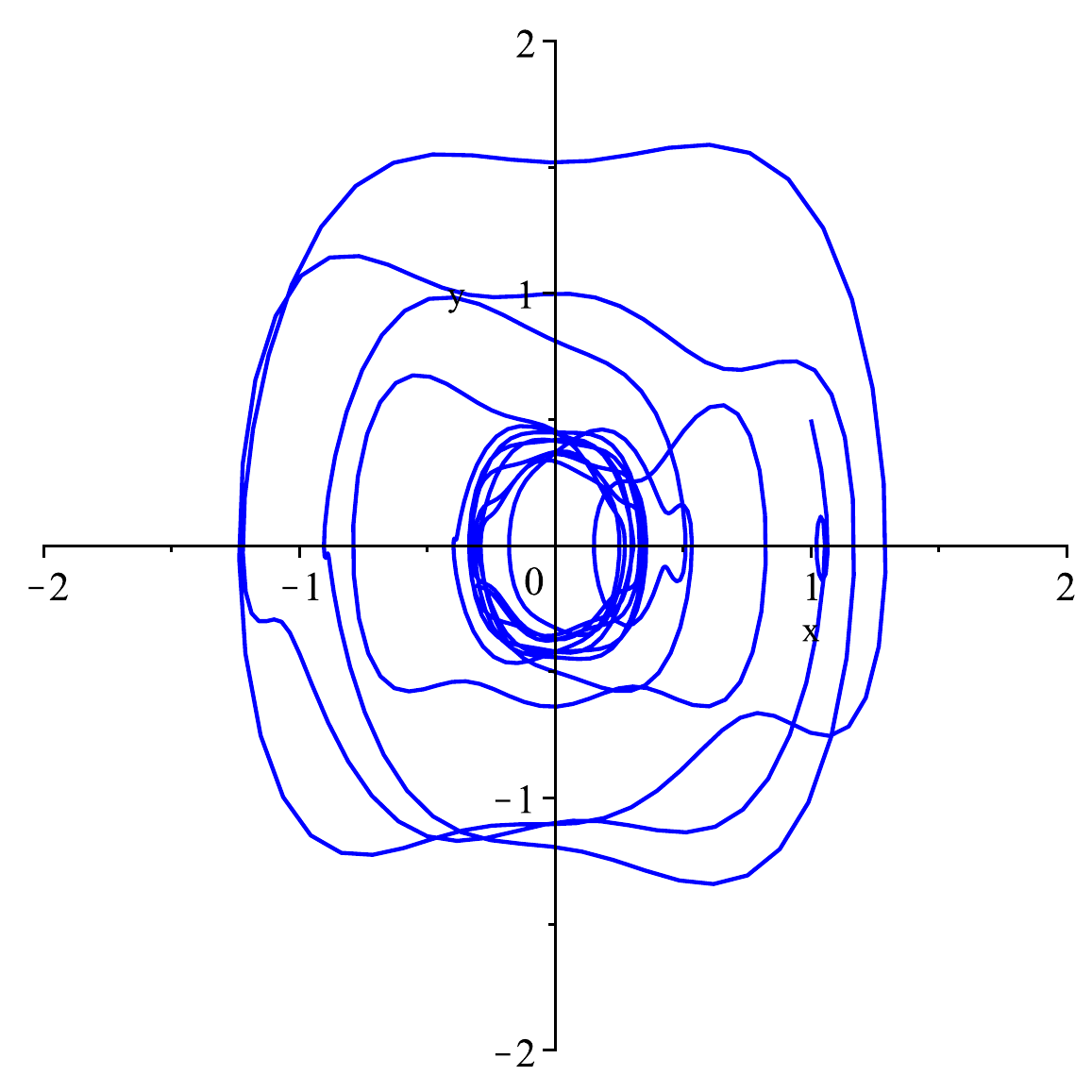}\hfill
	\includegraphics[width=.5\textwidth]{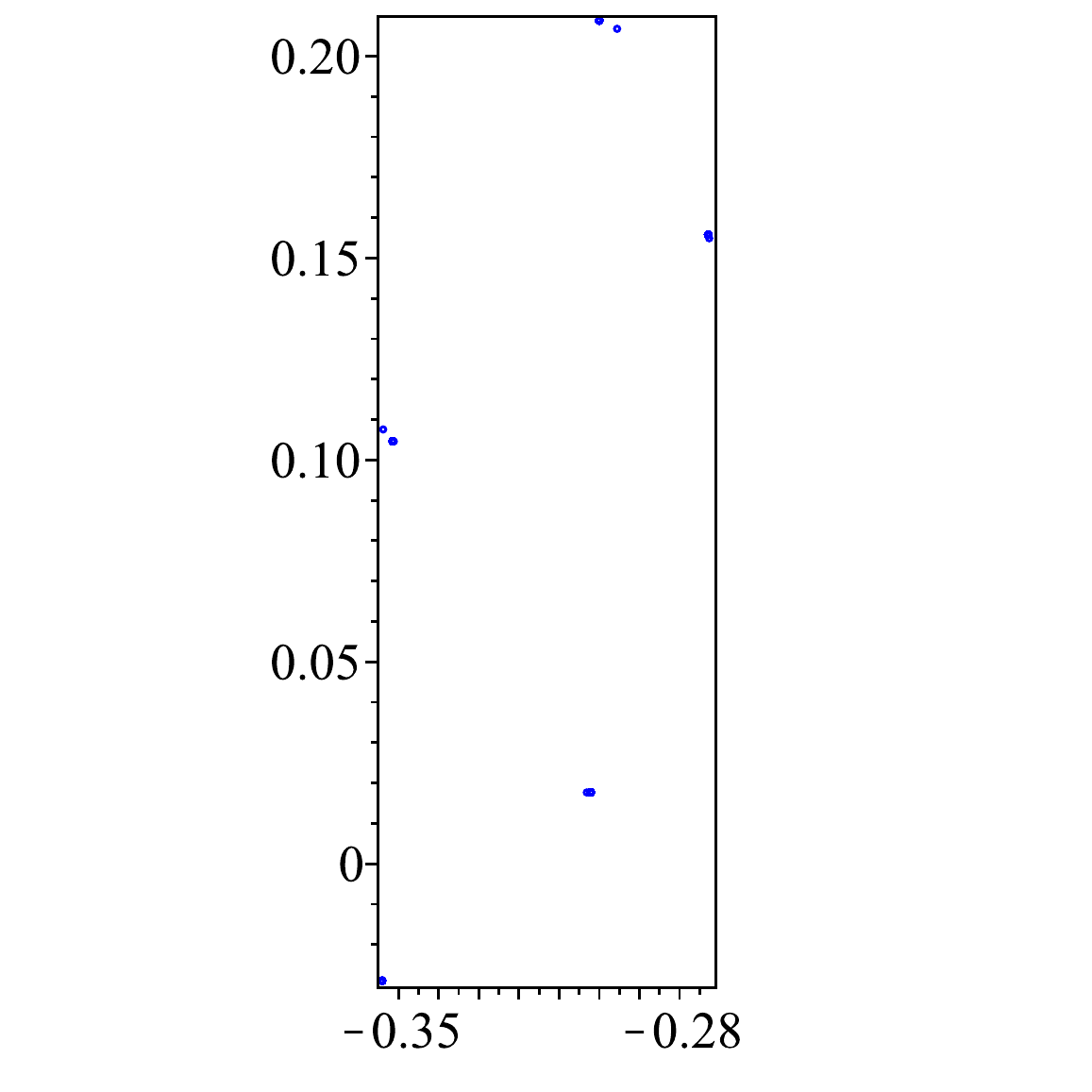}\hfill
	\\[\smallskipamount]
	\includegraphics[width=.5\textwidth]{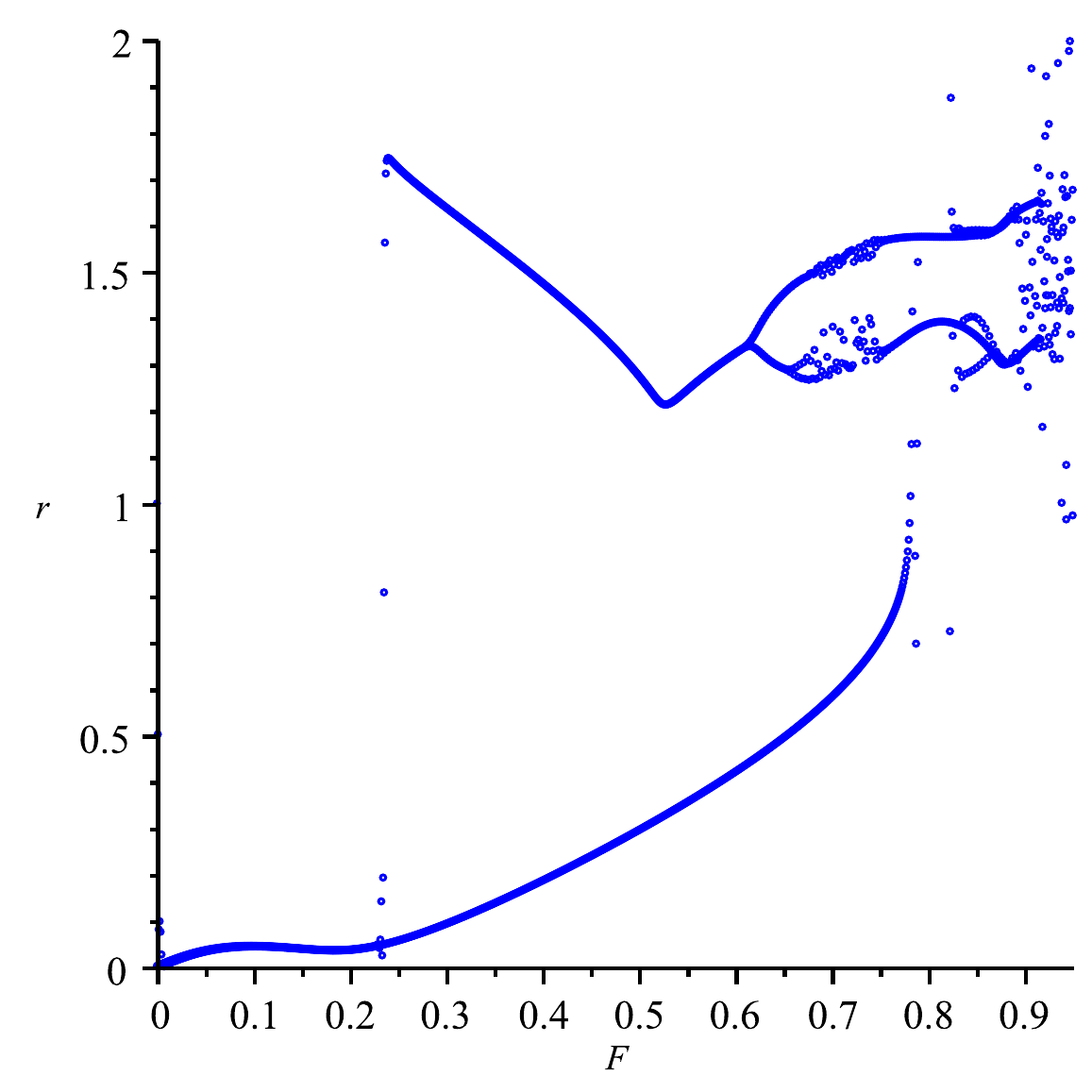}\hfill
	\includegraphics[width=.5\textwidth]{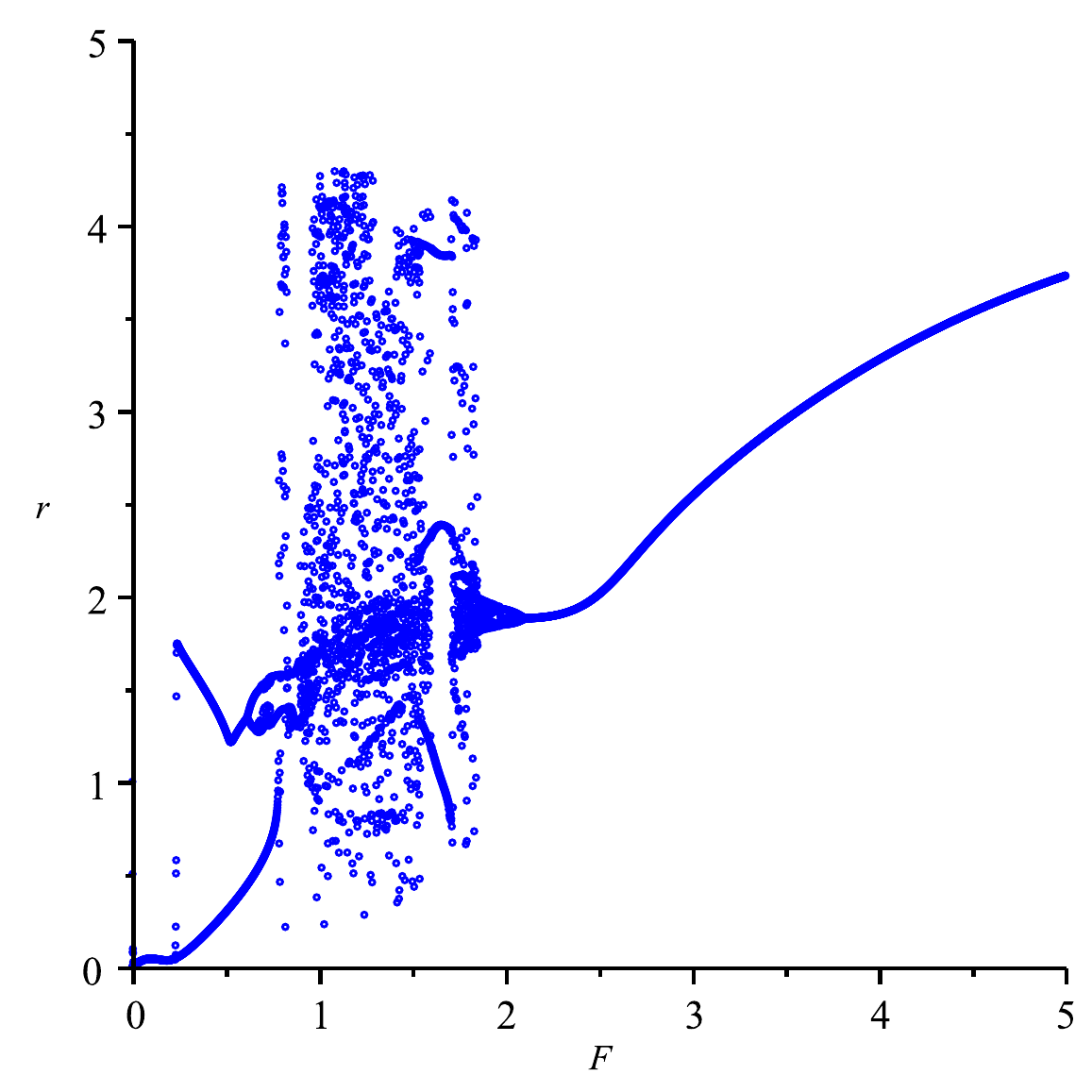}\hfill
	\caption{Phase portrait, Poincar{\'e} section and bifurcation diagrams for a particle (ion) confined in a nonlinear trap. The values of the control parameters are 	$a = 0.1$, $q=0.7$, $\gamma = 0.3$, $\omega = 1.25$, $\Omega = 4$, $F = 0.5$, $\mu = 1$, and $\alpha = 0$. The second bifurcation diagram corresponds to $0 < F < 5$. The first bifurcation diagram magnifies the area corresponding to $0 < F < 1$. Pictures reproduced from Ref. \cite{Mih10b} under permission of the authors. Copyright IOP Publishing.}
	\label{Duff2}
\end{figure}
\begin{figure}[!ht]
	\includegraphics[width=.5\textwidth]{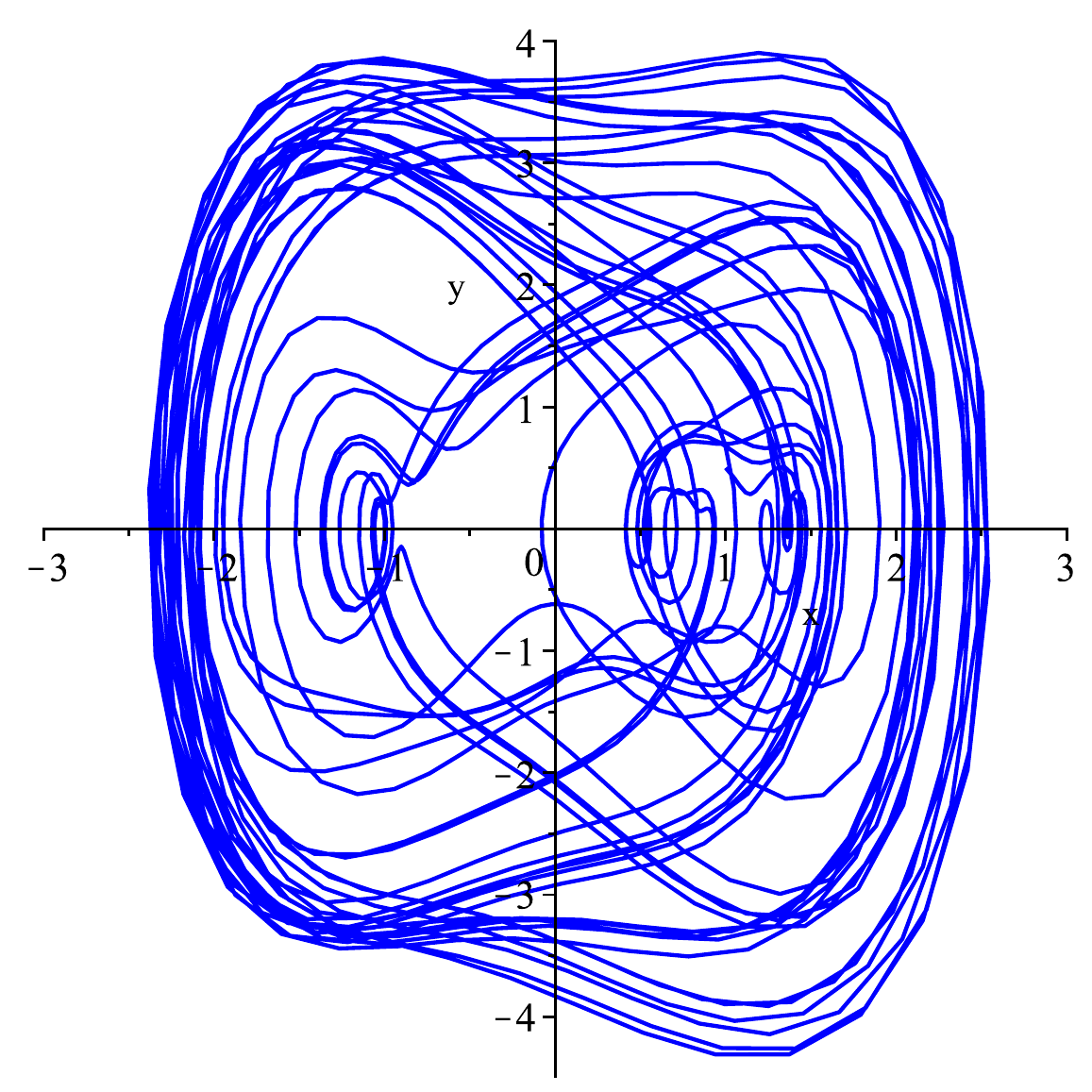}\hfill
	\includegraphics[width=.5\textwidth]{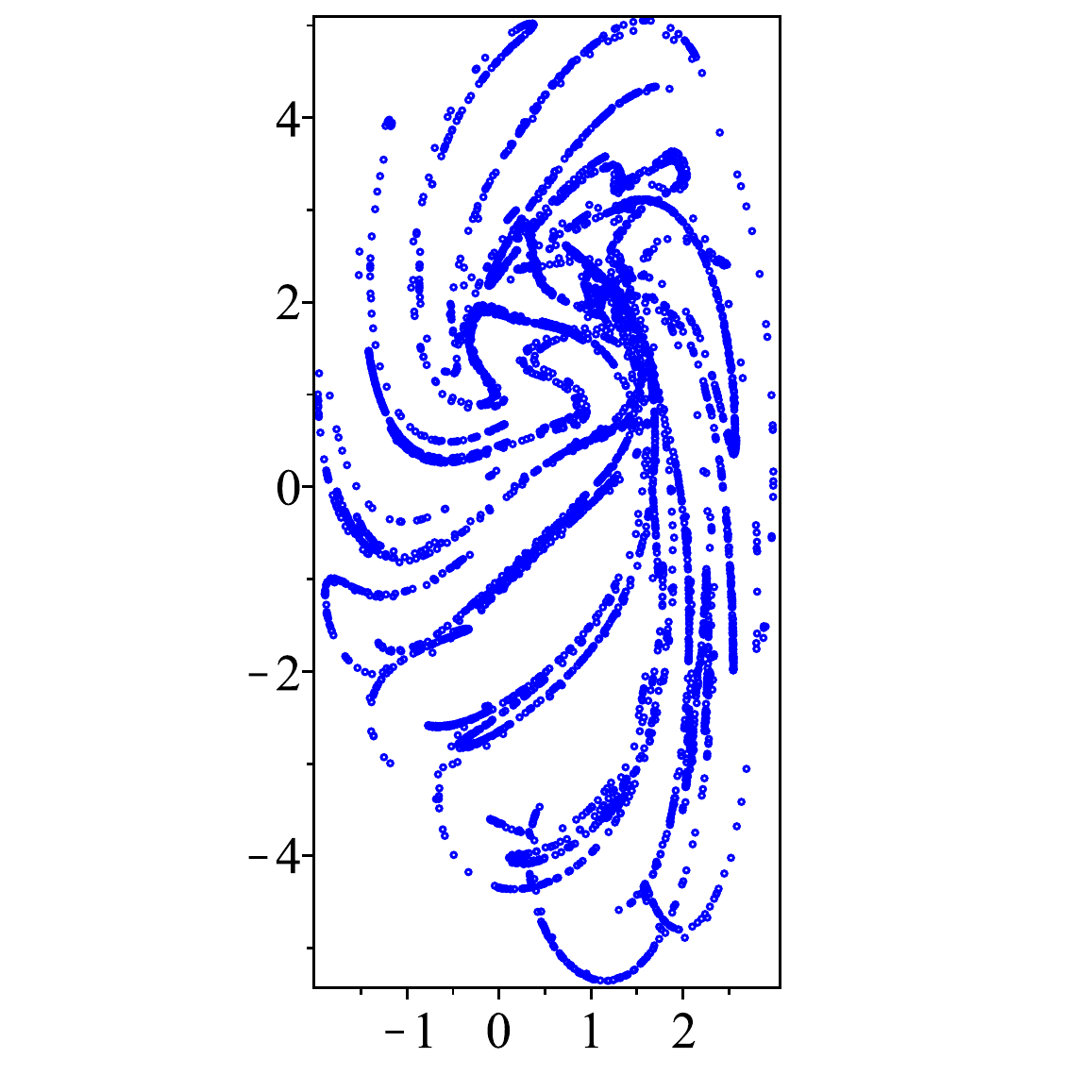}\hfill
	\\[\smallskipamount]
	\includegraphics[width=.6\textwidth]{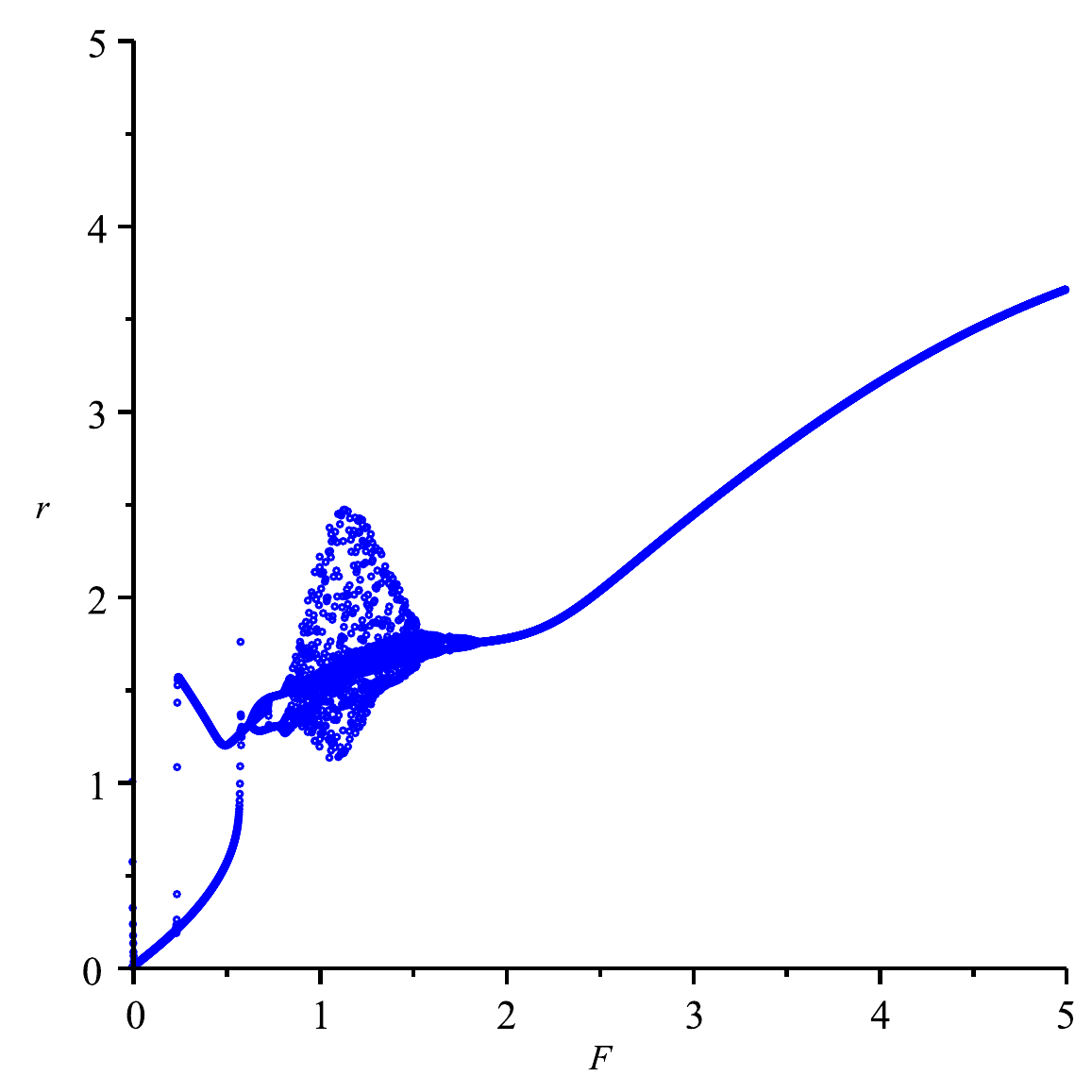}\hfill
	\caption{Phase portrait, Poincar{\'e} section and bifurcation diagram for an ion confined in a nonlinear Paul trap, in presence of laser radiation. The values of the parameters are $a = 0.1$, $q=0.7$, $\gamma = 0.3$, $\omega = 1.25$, $\Omega = 4$, $F = 2$, $\mu = 1$, and $\alpha = 0.3$. Pictures reproduced from Ref. \cite{Mih10b} under permission of the authors. Copyright IOP Publishing.}
	\label{Duff3}
\end{figure}

From calculus we can ascertain that the frontiers of the stability diagram are shifted towards negative regions of the $a$ axis in the plan of the control parameters $\left(a, q\right)$, as already reported by Hasegawa, Sevugarajan and Zhou \cite{Hase95, Sevu00, Sevu02, Zhou10}. In order to better characterize the phenomena involved we represent the phase portraits, the Poincar{\'e} sections and the bifurcation diagrams for an ion (particle) levitated in the trap, both in the absence and in presence of a laser field, as illustrated in Fig.~\ref{Duff2} and Fig.\ref{Duff3}. The damped Duffing oscillator generally exhibits an aperiodic appearance because the system is chaotic \cite{Kova11, Stro15}. $x\left(t\right)$ changes sign frequently which means that the particle crosses the hump repeatedly, as expected for strong forcing. The change of sign is in agreement with Ref. \cite{Sevu00} where the perturbed secular frequency of the ion is obtained by using a modified Lindstedt–Poincar{\'e} perturbation technique. Sevugarajan also shows this perturbation to be sign sensitive for octopole superposition and sign insensitive for hexapole superposition. 


A more detailed insight is supplied by the Poincar{\'e} section, which results by plotting $\left(x\left(t\right), y\left(t\right)\right)$ whenever $t$ is an integer multiple of $2 \pi$. Practically, we strobe the system at the same phase for each drive cycle. By looking at the Poincar{\'e} section we observe that the points fall on a fractal set which we interpret as a cross section of a strange attractor for equation (\ref{cha1}). The successive points $x\left(t\right), y\left(t\right)$ are found to hop erratically over the attractor while the system exhibits sensitive dependence on the initial conditions, which is the signature of chaos.

In case of a trapped ion (particle) and in presence of laser radiation, the phase portrait illustrates the existence of two attractors (as shown in Fig.~\ref{Duff3}) which seem to be periodic. We can discuss forced oscillations confined to the right or left well because two basins of attraction emerge. The points on the Poincar{\'e} section fall on a fractal set which represents the signature of chaos. Thus, laser radiation renders the motion chaotic. The bifurcation diagram shows a period-doubling bifurcation for $F \approx 0.7$ and a mixture of order and chaos for $0.85 \leqslant F \leqslant 1.9$. For larger values of the kicking term, ion dynamics is ordered. 

\subsubsection{Nonlinear Dynamics in a Paul trap. Conclusions}

We perform a qualitative investigation of the dynamical stability of an ion confined in a nonlinear quadrupole Paul trap, with anharmonicity resulting from the presence of higher order terms in the series expansion of the electric potential \cite{Mih10b}. The system exhibits a strongly nonlinear character. Regular and chaotic regions of motion are emphasized for the ion dynamics. System dynamics is chaotic when long-term behaviour is aperiodic. For particular initial conditions some of the solutions obtained present a certain degree of periodicity, although the dynamics is irregular. The damped parametric oscillator \cite{Zela20b} exhibits fractal properties and complex chaotic orbits. Chaotic (fractal) attractors are identified for particular solutions of the equation of motion. The motion on the strange attractor exhibits sensitive dependence on initial conditions. This means that two trajectories starting very close together will rapidly diverge from each other and will exhibit entirely different behaviour thereafter. Strange attractors are often fractal sets \cite{Stee14, Stro15}.

Global changes in the state of a physical system are generally described by means of dynamical maps. Illustrative examples span classical nonlinear systems undergoing transitions to chaos. Ref. \cite{Schi13} suggests a model that extends the concept of dynamical maps to a many-body system, where both coherent and dissipative elements are taken into consideration. The stroboscopic dynamics of a complex many-body spin model is analyzed by using a universal trapped ion quantum simulator. Experimental errors are discussed and novel experimental protocols are devised to mitigate against them.

Classical motion of a single damped ion confined in a Paul trap is usually described by a damped harmonic oscillator model such as the Duffing oscillator \cite{Kova11, Mih10b, Ake10}. A study of the quantum Hamiltonian for a boson confined in a nonlinear ion trap is performed in \cite{Mih10a} where the expected value of the quantum Hamiltonian on coherent states \cite{Dodon03, Cruz20, Phil14, Zela20c} is obtained. The resulting equation of motion is shown to be equivalent to the one describing a perturbed classical oscillator. Quantum damping motion of a trapped ion system is investigated in \cite{Qing17} by building a non-Hermitian Hamiltonian with dipole and quadrupole imaginary potential. Different real energy spectra and stable quantum states are obtained for both PT symmetry and asymmetry cases, along with the imaginary spectrum and decaying quantum state for the PT asymmetry case. The results apply to quantum dynamics of trapped ions. Ref. \cite{Rozh17} uses an original approach to explore nonlinear dynamics of a charged particle in a RF multipole ion trap, using the method of direct averaging over rapid field oscillations. The existence of localization regions for ion trap dynamics is identified. Illustrations of Poincar{\'e} sections demonstrate that ion dynamics is highly nonlinear, as also emphasized in \cite{Mih10b}. Ref. \cite{Mai19a} predicts a regime of anharmonic motion in which laser cooling becomes diffusive, while it can also turn into effective heating. This implies that a high-energy ion could be easily lost from the trap despite being laser cooled. This loss can be mitigated using a laser detuning much larger than the Doppler detuning. It is often more meaningful to characterize systems possessing complex dynamics through certain quantities involving asymptotic time averages of trajectories. Examples of such quantities are power spectra, generalized dimensions, Lyapunov exponents and Kolmogorov entropy \cite{Stee14, Cvita22, Lynch18}. Under particular conditions these quantities can be calculated in terms of averages of periodic orbits. The paper of Conangla {\emph {et al}} \cite{Cona20a} investigates the dynamics of the center of mass (CM) of a Brownian nanoparticle levitated in a Paul trap, focusing on the overdamped regime. An investigation of the nonlinear dynamics of charged microparticles levitated in a linear electrodynamic trap operated under viscous friction and STP conditions, is performed in Ref. \cite{Ryb20}. Based upon analytical modelling and the experimental results, the paper suggests an innovative method to infer the electric charge, mass and size of a certain microparticle confined in the linear electrodynamic trap. 

A novel approach to characterize nonstationary oscillators by means of the so called point transformations is demonstrated in \cite{Zela20a}. One of the main benefits of using the point transformation method lies in the fact that conserved quantities are preserved along with the structure of the inner product \cite{Stee07, Blum10}. In addition, these transformations can be constructed to be invertible which represents another remarkable feature. The authors demonstrate how point transformations enable one to solve the Schr{\"o}dinger equation for a wide diversity of nonstationary oscillators. Therefore, it is expected that the method suggested in \cite{Zela20a} can be applied to study the dynamics of particles confined in electromagnetic traps \cite{Major05, Bla06}.

We also mention an interesting paper of {\emph Flaj\v{s}manov\'{a} et al} \cite{Flaj20} that investigates stochastic (transient) dynamics of a nonlinear system, namely a weakly nonlinear Duffing
oscillator represented by an optically levitating nanoparticle (NP). The technique introduced enables one to ascertain the parameters of the levitated NP, as it can be also applied for lower temperatures and for recent quantum experiments.

\section{Non-neutral, complex plasmas. Coulomb systems. Examples and discussion}\label{Sec3}

\subsection{Complex plasmas and Coulomb systems}\label{ComplexCou}

Coulomb systems can be described as many-body systems consisting of identical particles that interact by means of electrostatic forces. When the potential energy associated to the Coulomb interaction is larger than the kinetic energy of the thermal (Brownian) motion \cite{Chav11}, the system is strongly coupled and it exhibits a strong spatial correlation between the electrically charged particles, similar to liquid or crystalline structures \cite{Boll84, Boll90a, Boni10c, Schwei98, Ander16}. Strongly coupled Coulomb systems encompass various many-body systems and physical conditions, such as dusty (complex) plasmas or non-neutral and ultracold plasmas \cite{Tsyto08, Fort06a, David01, Werth05a, Mendo13, Ott14}. 

\begin{figure}[bth]
	\begin{center}
		\includegraphics[scale=0.85]{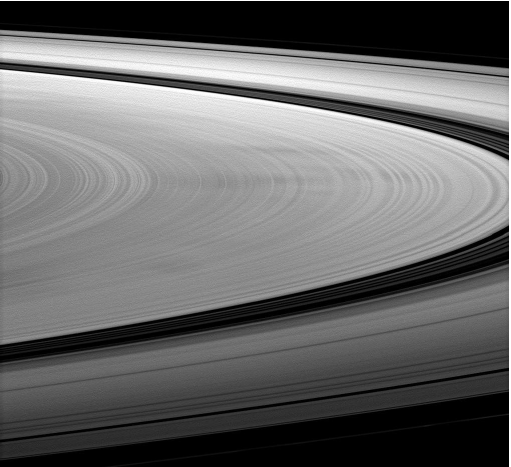} 
	\end{center}
	\caption[]{Cassini spacecraft images of dark spokes on Saturn's B ring. Photo reproduced from NASA https://solarsystem.nasa.gov/resources/14893/detailing-dark-spokes/. Copyright NASA.}
	\label{Saturn}
\end{figure}

Complex plasmas represent a unique type of low-temperature plasmas characterized by the presence of electrons, ions, neutral atoms and molecules, highly charged nano- or microparticles, by chemical reactions and by the interaction of plasmas with solid surfaces \cite{Fort05, Boni14, Shuk00, Khra04, Boni10a, Campa14}. The particles usually bear large electrical charges and exhibit long-range Coulomb interaction, which leads to the occurrence of strong coupling phenomena in the system \cite{Tsyto08} such as collective effects that result in the formation of plasma crystals \cite{Gebau03}. Complex plasmas are encountered in astrophysics as interstellar dust clouds, in comet tails or as spokes in the ring systems of giant gas planets \cite{Fort10a, Shuk02a}. They are also present in the mesosphere, troposphere, and magnetosphere of the Earth \cite{Shuk02b, Fort11, Pop11}, near artificial satellites and space stations, or in laboratory experiments \cite{Boni14, Morfi09}. Dust particle interaction occurs via shielded Coulomb forces \cite{Shuk00, Shuk02b}, the so-called Yukawa interaction \cite{Fort10a, Donko08}. Yukawa balls are reported in case of harmonically confined dusty plasmas \cite{Boni10b, Boni10a, Trevi06, Piel17}. Attention paid to this domain has witnessed a spectacular increase after the discovery of plasma crystals \cite{Schli96, Dub99, Piel17, Thom94, Tsyto07} and the detection of spokes in the rings of Saturn by the Voyager 2 mission in 1980 \cite{Fort10a, Shuk02b}, as illustrated in Fig.~\ref{Saturn}. 

Complex plasmas are intensively investigated in research laboratories as they are expected to shed new light on issues regarding fundamental physics, such as phase transitions \cite{Blu90}, self-organization, study of classical and quantum chaos or pattern formation and scaling \cite{Fort10a, Boni14, Boni10b, Morfi09}. Present interest is focused on strongly coupled Coulomb systems of finite dimensions \cite{Vlad05}. Particular examples of such systems would be electrons and excitons in quantum dots \cite{Boni14} or laser cooled ions confined in Paul or Penning type traps \cite{Major05, Vogel18, Werth09, David01}. 

Refs. \cite{Piel17, Piel02} represent comprehensive overviews on the dynamical processes that occur in complex (dusty) plasmas. Collective effects in dusty plasmas are presented in \cite{Melz05}. It is shown that the composite structure is liable for the remarkable properties of this particular type of plasma, among which the most prominent are the formation of liquid or solid phases under conditions of strong electrostatic coupling or the use of intensive nonuniform magnetic fields to levitate Coulomb clusters formed by charged diamagnetic particles \cite{Sav12}. Investigations begin with single particle effects, with an emphasis on levitation, on harmonic and nonlinear oscillations. Few particle systems and the Yukawa interaction are also examined. Many-body systems are described by means of low-frequency electrostatic waves and instabilities that occur in complex plasmas. Ref. \cite{Piel08} investigates some exotic features of complex plasmas such as charge accumulation in the ion wake, ion drag (under microgravity conditions) and spherical plasma crystals \cite{Piel17}. The characteristic features and intrinsic properties of complex plasmas are also explained in \cite{Khra09}. Inclusive reviews on complex plasmas are written by Fortov {\em et al} \cite{Fort05} and Morfill \cite{Morfi09}, while the reader interested in getting introduced into the domain should also refer to the work of Tsytovich, Fortov and Bonitz \cite{Fort10a, Boni14, Tsyto08, Boni10a}.    

A review on non-neutral plasmas levitated in ion traps and their characteristic properties can be found in \cite{Werth05a}. Ion traps represent versatile tools to investigate many-body Coulomb systems or dusty and non-neutral plasmas. The electrodynamic (Paul) trap uses a RF electric field which generates an oscillating saddle shaped potential that confines charged particles in a region where the electric field exhibits a minimum, under conditions of dynamical stability \cite{Ghosh95, Major05, Mih16a}. The dynamical time scales associated with trapped microparticles lie in the tens of milliseconds range, while microparticles can be individually observed using optical methods \cite{Davis02, Vis13, Vini15, Libb18}. As the background gas is dilute, particle dynamics exhibits strong coupling regimes characterized by collective motion \cite{Tsyto08, Fort06a, David01}. Dust particles may give birth to larger particles which might evolve into grain plasmas \cite{Tsyto08}. The mechanism of electrostatic coupling between the grains can vary widely, from the weak coupling (gaseous) regime to the pseudo-crystalline one \cite{Dub99, Werth05a, Lisin13}. Complex plasmas can be described as non-Hamiltonian systems of few or even many-body particles.

State-of-the-art results of particle-in-cell (PIC) simulations (considering ion-atom collisions) for dust particle charge, ion drag force, and interaction between grains in ultracold dusty plasmas, are presented in \cite{Sund20}. Both single grain and two grain systems are explored, by considering both streaming ions and equilibrium ions.

\subsection{Non-neutral plasmas. One component plasmas - OCPs}

First experimental observations of ordered structures consisting of charged iron and aluminium microparticles confined in a Paul trap, in vacuum, are reported in 1959 \cite{Wuerk59}. Phase transitions occur as an outcome of the dynamical equilibrium between the trapping potential and the inter-particle Coulomb repulsion \cite{Died87, Blu90}. The emergence of correlations is also reported for strongly coupled ion plasmas \cite{Boll90a, Ander16}. In 1991 an experiment reported on storage of macroscopic dust particles (anthracene) in a Paul trap operating under SATP conditions \cite{Wint91}. Electrodynamic traps and ion trapping techniques combined with laser cooling mechanisms \cite{Haro10, Werth09, Kno16} allow scientists to investigate the dynamics of small quantum systems and prepare them in well-controlled quantum states \cite{Quint14, Wine13, Bush13}. Trapped ions or particles represent one-component plasmas (OCP). The OCP model represents a reference that is used to study strongly coupled Coulomb systems \cite{David01, Dub99, Ott14, Tama99}. 

A one-component plasma (OCP) can be regarded in a very good approximation as the simplest statistical mechanical model of a Coulomb system. This feature has drawn vivid attention on OCPs in the last 40 years. For example, Ref. \cite{Tama99} employs an extension of the Debye-H{\"u}ckel (DH) theory to characterize OCPs. The approach enables one to perform analytic calculations of all the thermodynamic functions and of the structure factor, which represents a major progress. 

 A many-body system of charged particles interacting via a pairwise Yukawa potential, the so-called Yukawa one-component plasma (YOCP), represents a fair approximation for a class of physical systems \cite{Silve19}. Such systems are completely characterized by two parameters: the screening parameter and the nominal coupling strength (which we denote by $\Gamma$). In addition, it is already established that the collective spectrum of the YOCP is governed by a longitudinal acoustic mode, both in the weakly and strongly coupled regimes.

A non-neutral plasma can be described as a many-body collection of charged particles that does not embrace the overall charge neutrality condition. The various areas of application include: precision atomic clocks \cite{Van15, Koz18, Safro18}, trapping of antimatter plasma and antihydrogen production \cite{Quint14}, quantum computers \cite{Baut19, Bruz19}, nonlinear vortex dynamics and fundamental transport processes in trapped nonneutral plasmas, strongly-coupled one-component plasmas and Coulomb crystals \cite{Kel19, Boni08, Schne12b, Yos15, Zali19}, coherent radiation generation in free electron devices such as free electron lasers, magnetrons and cyclotron masers, as well as intense charged particle beam propagation in periodic focusing accelerators and transport systems \cite{David01}. 

A classical plasma which consists of ions {\em dressed} with electrons exhibits a well-defined thermodynamics. In the strong-coupling regime the {\em plasma} is a solid, then it becomes gradually a liquid when passing to the weak-coupling regime, after which it turns into a non-ideal gas and finally, in the weak-coupling limit, it behaves like an ideal classical gas \cite{Apost19}. 

New methods for ion trapping such as the Orbitrap, the digital ion trap (DIT), the rectilinear ion trap (RIT), the toroidal ion trap, development and application of the quadrupole ion trap (QIT) and of the quadrupole linear ion trap (LIT), along with the introduction of high-field asymmetric waveform ion mobility spectrometry (FAIMS) are reviewed in \cite{March17b}. 

\subsubsection{Electrodynamic traps as tools for confining OCPs}

If only a single component charged particle species is confined in a Paul trap, it represents a confined One Component Plasma (OCP). By modifying the trap parameters, the plasma can perform phase transitions between gas-like, liquid-like and crystal-like states. The crystal state is the most interesting one for experiments with trapped particles. The melting point of the crystal can be used to estimate a translational temperature of the trapped charged particles. The plasma state can be described by the coupling parameter \cite{Boll84, Wine88, Dub99, Boll03, Werth05a, Piel17} and it is essentially controlled by the trap parameters. Other examples of OCP systems are found in many other fields of physics as they have been investigated theoretically over time \cite{Ichi82, Ebel17, Boll90a}. 

In case of quadrupole traps the second-order Doppler effect is the result of space-charge Coulomb repulsion forces acting between trapped ions of like electrical charges. The Coulomb forces are balanced by the ponderomotive forces produced by ion motion in a highly non-uniform electric field. For large ion clouds most of the motional energy is found in the micromotion \cite{Major05, Berke98, Wang10}. Multipole ion trap geometries significantly reduce all ion number-dependent effects resulting through the second-order Doppler shift, as ions are weakly bound with confining fields that are effectively zero through the inner trap region and grow rapidly near the trap electrode walls. Owing to the specific map shape of trapping fields, charged ions (particles) spend relatively little time in the high RF electric field area. Hence, the RF heating phenomenon (micromotion) \cite{Nam17, Chen20, Rat20, Zhu21} is sensibly reduced. Multipole traps are also used intensively as tools in analytical chemistry to confine trapped molecular species that exhibit many degrees of freedom, and thus study cold collisions and low temperature processes \cite{Ger03, Trip06, Ger08a, Ger08b, West09, Robe18}. Space-charge effects are not negligible for such traps. Nevertheless, they represent extremely versatile tools to investigate the properties and dynamics of molecular ions or to simulate the properties of cold plasmas, such as astrophysical plasmas or the Earth atmosphere. 

The thermodynamic properties associated to a non-ideal classical Coulomb OCP rest on a single parameter, that is the coupling parameter $\Gamma$ which we introduce later in eq. \ref{gamma}. In the particular case of a Yukawa OCP when the pair interaction is screened by background charges, the thermodynamic state also depends on the range of the interaction via the screening parameter. An original approach used to define and measure the coupling strength in Coulomb and Yukawa OCPs is proposed in \cite{Ott14}, based on the radial pair distribution function (RPDF). 

\begin{figure*}[!ht]
	\begin{minipage}[h]{0.5\linewidth}
		\center{\includegraphics[width=\linewidth]{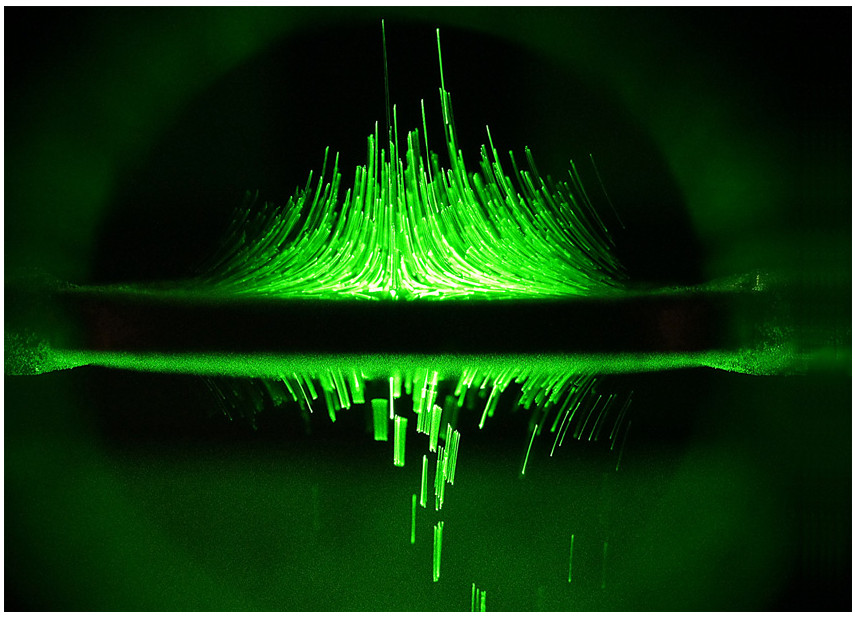}} \\$(a)$
	\end{minipage}
     \begin{minipage}[h]{0.53\linewidth}
    	\center{\includegraphics[width=\linewidth]{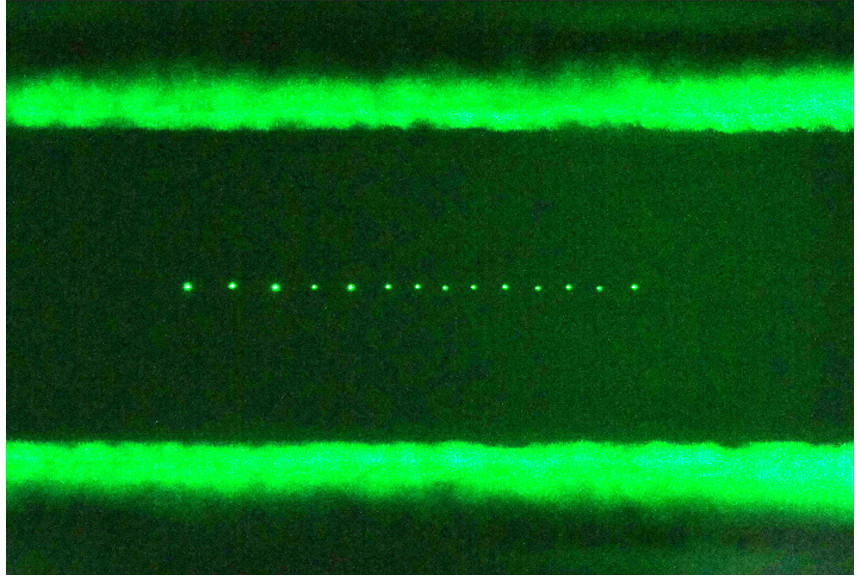}} \\$(b)$
     \end{minipage}
	\caption{Stable 2D and 3D structures consisting of $26 \ \mu$m-diameter particles trapped in microparticle electrodynamic ion traps (MEITs) with different geometries, operating under SATP conditions. Image (a) shows many hundreds of particles held in a ring trap, with a nearly horizontal ring plane. Image (b) illustrates levitation of 14 particles in a linear Coulomb crystal. Source: Images reproduced from \cite{Libb18} by courtesy of Prof. K. Libbrecht.}
	\label{Meit2}
\end{figure*}

Stable confinement of a single ion in the radio-frequency (RF) field of a Paul trap is well known as the Mathieu equations of motion can be analytically solved \cite{March05, Rand16, Li16}. This is no longer the case for high-order multipole fields where the equations of motion do not admit an analytical solution. In such case particle dynamics is quite complex and it is described by non-linear, coupled, non-autonomous equations of motion. The solutions for such system can only be found by performing a numerical integration \cite{Major05, Fort10a, Rieh04}. Fig.~\ref{Meit2} presents images of ordered 2D and 3D structures of microparticles levitated in ring and linear geometry microparticle electrodynamic ion traps (MEITs). 

Ref. \cite{Foot18} introduces an electrodynamic ion trap in which the electric quadrupole field oscillates at two different frequencies. By operating the trap as described the authors report simultaneous tight confinement of ions with extremely different charge-to-mass ratios, {\em e.g.} singly ionized atomic ions together with multiply charged nanoparticles (NPs). The stability conditions for two-frequency operation are inferred from the asymptotic properties of the Mathieu equation solutions \cite{Brou11, Rand16}, which emphasizes the effect of damping on parametric resonances. It is demonstrated that trap operation in a two-frequency mode (with frequency values as widely separated as possible) is most effective when the two species’ mass ratios and charge ratios are sufficiently large. The system can be assimilated with two {\em superimposed} Paul traps, each one of them operating close to a frequency optimised to tightly confine one of the species, which results in strong confinement for both particle species used. Such a hybrid trap and its associated operation mode grants an advantage with respect to single-frequency Paul traps, in which the more weakly confined species develop a sheath around a central core consisting of tightly confined ions \cite{Foot18}.  

Ref. \cite{Wies17} describes a novel cryogenic Penning trap geometry intended to perform high resolution spectroscopy experiments and enhanced resolution imaging of levitated particles. By employing of a non-neutral plasma method, a linear trapping model that accounts for large ion clouds is demonstrated, which will become the core of an atomic clock \cite{Pedro18}. Late experiments show that an equally spaced linear chain of ions can be used as a paradigm to study defect formation in a finite size 1D system \cite{Pedre20}. Moreover, investigation of the non-adiabatic transition between the chain and zig-zag topologies for such finite size systems enables one to perform tests of the underlying Homogeneous Kibble-Zurek model, using ion traps as tools \cite{Kel19, Pyka14}. Late experiments demonstrate the capabilities of two-dimensional Penning trap arrays as a scalable platform for quantum simulation and quantum computing with trapped atomic ions \cite{Jain20}. Moreover, state-of-the-art experimental results results establish radial 2D ion crystals as a robust experimental platform to implement a wide range of theoretical proposals in quantum computation and simulation \cite{Dono21}, while linear ion traps (LITs) enable scientists to perform studies with respect to heating of motional modes of a single ion and of extended ion crystals \cite{Kali21}. Other experimental investigations explore the possibility of constructing a large-scale storage-ring-type ion-trap system intended for storing, laser cooling, and controlling a large number of ions as a platform for scalable quantum computing (QC) and quantum simulations (QS) \cite{Shaf21}.  
   
Generally, investigations performed on micro- and nano-sized systems require a perturbation free, pristine environment, which can be achieved by means of trapping condensed matter in vacuum. Levitation techniques offer scientists the possibility to carry out quantum mechanics investigations of large systems, focusing on the transition between the quantum and classical worlds. Ref. \cite{Cona20b} reports "a novel hybrid levitation platform" based on a Paul trap and "a weak but tightly focused laser beam". The experimental setup achieves a deep trapping potential along with stable confinement and motion detection. In addition, the approach and setup look like a promising method to levitate and interrogate a large number of nanomaterials down to pressures of 1 mbar.

\subsection{Weakly-coupled plasmas}\label{weak}

Ref. \cite{Bryd99} is an extensive review of low density Coulomb systems, which presents results on the correlations of low density classical and quantum Coulomb systems under equilibrium conditions, in 3D space. The exponential decay of particle correlations in the classical Coulomb system, the Debye-H{\"u}ckel screening, is compared and confronted with the quantum case where strong arguments are offered to account for the absence of exponential screening. The paper also approaches experimental results and techniques for elaborate calculations that determine the asymptotic decay of correlations for quantum systems. 

Ref. \cite{Campo09} is an elaborate review paper that emphasizes recent advances on the statistical mechanics and out-of-equilibrium dynamics of solvable systems with long-range interactions. The review is focused on a comprehensive presentation of the concept of ensemble inequivalence, as exemplified by the exact solution in the microcanonical and canonical ensembles of mean-field type models. Long-range interacting systems display an extremely slow relaxation towards thermodynamic equilibrium and convergence towards quasi-stationary states, which represents an exotic feature. Such an unusual relaxation process can be explained by introducing an appropriate kinetic theory that relies on the Vlasov equation. In addition, a statistical approach based on a variational principle established by Lynden-Bell is demonstrated, which explains both qualitatively and quantitatively some features of quasi-stationary states.

By introducing a modified Thirring model and then comparing the equilibrium states of the unconstrained ensemble with respect to the canonical and grand-canonical ones, it is demonstrated that systems with long-range interactions can evolve into states of thermodynamic equilibrium in the unconstrained ensemble \cite{Late17}. Moreover, it also results that the parameter space that determines the possible stable configurations is expanded when the system is restrained by fixing the volume and particle number. These parameters oscillate in case of the unconstrained ensemble along with the system energy. First-order phase transitions are also reported for the model introduced. The control parameters for the unconstrained ensemble are temperature, pressure, and chemical potential, whilst the replica energy represents the free energy. In contrast, macroscopic systems with short-range interactions are unable to reach equilibrium states if the control parameters are temperature, pressure, and chemical potential.

\subsection{Strongly-coupled Coulomb systems. OCPs revisited.}\label{strong}

Since the 1980s, strongly coupled and non-neutral plasma systems confined in Paul and Penning traps are the subject of an intense scientific interest \cite{Ichi82, Boll84, Dub99, Wine88, Boll90a, Brew87, Totsu87, Ichi87, Dub88, Gil88, Boll90b, Dub90, Dub16a}, because of their intriguing properties and wide area of applications. 
 
A review on strongly coupled plasma physics and high-energy density matter is Ref. \cite{Muri04}, which focuses on OCPs and the physical mechanisms associated to these particular type of plasmas. Examples supplied such as white dwarfs, trapped laser-cooled ions, and dusty plasmas \cite{Piel17, Piel08}, better emphasize the wide area of applications and explain the huge interest towards investigating strongly coupled plasmas. As established in \cite{Dufty18} the kinetic theory for strongly coupled Coulomb systems is based on a density-functional theory (DFT) approach, namely the Kubo-Greenwood model.  

Ref. \cite{Brew87} reports on a strongly coupled non-neutral $^9$Be$^+$ ion plasma with a coupling parameter of approximately 100 or greater. The ions are confined in a Penning trap, then laser cooled. Measurements are performed with respect to the plasma shape, rotation frequency, density and temperature. Among the most remarkable works on strongly-correlated non-neutral plasmas are the papers of Dubin and O'Neil \cite{Dub99, Dub88, Dub90, Dub16a, Dub16b}. We also mention the work of Fortov {\em et al} \cite{Fort06a} which explores the physics of strongly coupled plasmas. A new frontier in the study of neutral plasmas are considered to be the ultracold neutral plasmas \cite{Kil05, Kil07, Rong09, Shuk10, Lyon15, Muri15}. 

Dust particles confined in an electrodynamic trap experience a large number of forces. The prevailing forces acting upon the micrometer sized particles of interest are the a.c. trapping field (whose outcome is the ponderomotive force caused by particle motion in a strongly nonlinear electric field) and gravity. The gravitational force can be expressed as \cite{Fort10a}

\begin{equation}
\mathbf{F}_g = m \mathbf{g} = \frac{4}{3} \pi a^3 \rho_d \mathbf{g} \ ,
\end{equation}  
where $\mathbf{g}$ stands for the gravitational acceleration, $m$ represents the particle mass, $a$ is the dust particle radius and $\rho_d$ corresponds to the dust particle density. The electric field yields a force 

\begin{equation}
\mathbf{F}_{el} = -Q \mathbf{e} = -Ze \mathbf{E} \ ,
\end{equation} 
where $Q = Ze$ is the electric charge of the dust particle and $e = 1.6 \cdot 10^{-19}$ C stands for the electric charge associated to the electron. When a temperature gradient arises in a neutral gas background, a force called thermophoretic drives the particles towards regions of lower gas temperature. The particle is also subject to the action of the ion drag force caused by a directed ion flow \cite{Tsyto08, Fort06a, Vasi13}. To summarize, the forces that act upon a particle (or an aerosol) include the radiation pressure force, the thermophoretic force \cite{Piel17}, the photophoretic force \cite{Liza18}, electric forces and possibly magnetic forces, to which we add aerodynamic drag and gravity \cite{Tsyto08}. 

A one-component plasma (OCP) consists of a single species of charge submerged in the neutralizing background field \cite{Fort06a, Dub99}. Single component non-neutral plasmas confined in Penning or Paul (RF) traps exhibit oscillations and instabilities owing to the occurrence of collective effects \cite{Werth05a}. The dimensionless coupling parameter that describes the correlation between individual particles in such type of plasma can be expressed as

\begin{equation}\label{gamma}
\Gamma = \frac{1}{4\pi \varepsilon_0 } \frac{q^2}{a_{WS} k_B T} \ ,
\end{equation}      
where $\varepsilon_0$ is the electric permittivity of free space, $q$ stands for the ion charge, $a_{WS}$ is the Wigner-Seitz radius, $k_B$ denotes the Boltzmann constant and $T$ represents the particle temperature. The Wigner-Seitz radius results from the relation

\begin{equation}
\frac{4}{3}\pi {a}^3_{WS} = \frac{1}{n} \ ,
\end{equation}  
where $n$ is the ion (particle) density. Although the Wigner-Seitz radius measures the average distance between individual particles, it does not coincide with the average inter-particle distance. The $\Gamma$ parameter represents the ratio between the potential energy of the nearest neighbour ions (particles) and the ion thermal energy, describing the thermodynamic state of an OCP. Low density OCPs can only exist at low temperatures. Any plasma characterized by a coupling factor $\Gamma > 1$ is called strongly coupled. For $\Gamma < 174$ the system exhibits a liquid-like structure, while larger values might indicate a liquid-solid phase transition into a pseudo-crystalline state (lattice) \cite{Wint91, Werth05a}. OCPs are supposed to exist in dense astrophysical objects \cite{Horn90}. An example of a high temperature system would be a quark-gluon plasma (QGP), used to characterize the early Universe and ultracompact matter found in neutron or quark stars \cite{Ebel17, Boni10b, Safro18, Fort11}.   

Colloidal suspensions of macroscopic particles and complex plasmas represent other examples of strongly coupled Coulomb systems (SCCS) \cite{Kal02}. Trapped micrometer sized particles interact strongly over long distances as they carry large electrical charges. Friction in air combined with large microparticle mass results in an efficient particle {\em cooling}, which leads to interesting strong coupling features. The signature of this phenomenon lies in the appearance of ordered structures, liquid or solid like, as phase transitions occur in case of such systems \cite{Major05, Kno14, Fort06a, David01, Dub99}. The presence of confining fields maintains the particles localized together in an OCP. Trapped and laser cooled ions offer a good, low temperature realization of a strongly coupled ultracold laboratory plasma. 

The equation of motion for a particle trapped in air can be expressed in a convenient form by introducing dimensionless variables defined as $Z = z/z_0$ and $\tau = \Omega t/2$, where $z_0$ represents the trap radial dimension (a geometrical constant) and $\Omega $ stands for the frequency of the a.c. trapping voltage. The nondimensional equation of motion can be cast into \cite{Kul11}

\begin{equation}\label{extforce}
\frac{d^2 Z}{dt^2} + \gamma \frac{dZ}{dt} + 2 \beta Z \cos(2 \tau) = \sigma \ ,
\end{equation}
where $\gamma$ stands for the drag parameter, $\beta$ is the a.c. field strength parameter and $\sigma$ represents a d.c. offset parameter, defined as \cite{Drew15}  

\begin{equation}
\gamma =\frac{6 \pi \mu d_p \kappa}{m\Omega} \ , \  \beta = \frac{4g}{b\Omega^2}\left(\frac{V_{ac}}{V^*_{dc}} \right) \;, \ \sigma = - \frac{4g}{z_0 \Omega^2}\left(\frac{V_{dc}}{V^*_{dc} } - 1\right) \ ,
\end{equation}
where $ V^*_{dc} $ satisfies
\begin{equation}
F_z - mg = q C_0\frac{V^*_{dc}}{z_0} .
\end{equation}
$C_0$ and $C_1$ are two geometrical constants ($ C_0 <1 $), while $ b = z_0C_0/C_1 $ stands for  the geometrical constant of the electrodynamic balance (EDB) \cite{Hart92, Singh18a, Davis90}. The three dimensional parameters $\left( \gamma, \beta, \sigma \right)$ control the levitated particle stability. For a negative charged particle the right hand term of eq. (\ref{extforce}) is a positive quantity. When the d.c. potential is adjusted to compensate the external vertical forces (gravity), then $V_{dc} = V^*_{dc}$, $\sigma = 0$ and the particle experiences stable confinement at the zero point, as long as th value of $\beta$ is not too large. Other relevant quantities are the gas viscosity coefficient $\mu$, the particle diameter $d_p$ and  mass $m$, while $\kappa$ represents a correction factor for Stoke's law for a non-spherical particle.

If a larger number of electrically charged particles are trapped together the Coulomb interaction will affect the individual particle motion and the space charge will alter the trap potentials. In case of particles with electric charges that bear the same sign, the Coulomb repulsion makes ideal trapping impossible. Coupling strongly depends on the trap parameters. A molecular dynamics simulation for hundreds of ions confined in a Paul trap is performed in \cite{Pre91}. The simulations account for the trapped particles micromotion and the inter-particle Coulomb interactions. A random walk in velocity has been implemented in these calculations with an aim to bring the secular motion to a temperature that is numerically measured. When coupling is large the ions build concentric shells that undergo oscillations at the RF frequency, while the ions within a shell create a 2D hexagonal crystal-like lattice \cite{Pre91}. Ref. \cite{Kal02} is a collection of works related to the domain of strongly coupled Coulomb systems, while Ref. \cite{Zhao08} focuses on molecular dynamics simulations with low energy ions levitated in a nanoscale Paul trap, in both vacuum and an aqueous environment. A concise introduction on the subject of Coulomb crystals that arise when trapped ions are cooled at very low temperatures is given in \cite{Drew15}.   

As already mentioned in Section \ref{ComplexCou}, complex plasmas comprise electrons, ions, neutrals, and macroscopic solid dust particles of nano- or micron size \cite{Fort10a, Ivlev12, Fort11}. The dust grains \cite{Dra03} may exhibit a positive or negative electric charge, depending on the charging mechanisms involved. Experiments performed show that dust grains acquire a negative charge typically of the order of $10^3 \div 10^5$ elementary charges, due to the electron and ion stream \cite{Asch12a}. As the dust particle component is strongly coupled with respect to other plasma components, occurrence of ordered structures is experimentally reported. These structures are also known as {\em plasma crystals} \cite{Thom94, Tsyto07, Chu94}. Ref. \cite{Asch12a} describes micron sized dust grains trapped in a low-temperature RF discharge, where the dynamic light scattering (DLS) technique is applied to the dust component of a complex plasma. The electric force that occurs in the plasma-wall sheath of the lower electrode compensates external forces such as gravity and ion drag. Intricate 3D crystal structures are experimentally observed. Relaxation into an equilibrium state of dust grain dynamics in plasma after interaction with a laser beam is experimentally and qualitatively investigated in \cite{Lisin13}, where the associated physical mechanisms are also discussed.  

Strong correlation effects that occur in classical and quantum plasmas are investigated in \cite{Ott14, Boni08}. Coulomb (Wigner) crystallization phenomena are analyzed, with an emphasis on one-component non-neutral plasmas in ion traps and on macroscopic two-component neutral plasmas. It is also explained how to achieve Coulomb crystallization in terms of the critical values of the coupling parameters, the distance fluctuations, and the phase diagram of Coulomb crystals. Ref. \cite{Fili12} focuses on numerical simulations of strongly coupled dust particles (carried out under SATP conditions), that take into account the influence of the buffer gas medium (viscosity) and the random forces exerted upon the particles. The simulations help in identifying optimum operating values for the electric field amplitude and frequency, that are required in order to achieve stable capture of dust particles in a linear Paul trap. The possibility to levitate Coulomb clusters that consist of charged diamagnetic particles in a nonuniform magnetic field is investigated both theoretically and experimentally in \cite{Sav12}. Numerical simulations performed illustrate the appearance of standing waves of the dust particle density, caused by the dynamic effects induced by the periodic external low frequency electric field. In addition, Ref. \cite{Stein17} investigates the neutral gas pressure effect on the crystallization of 3D cylindrical complex plasmas under laboratory conditions. 

The effect of correlations on a trapped plasma is also investigated in \cite{Ander16}. The properties of Coulomb crystals are explored along with the associated phase transitions. In addition, it is also demonstrated that correlations dramatically change the equilibrium and especially the dynamical properties of a trapped plasma. 

Calculation results of the inner pressure and energy of SCCS levitated in a Paul trap are also carried out by employing the statistical theory applied to the liquid state. Pair correlation functions of the Coulomb systems are thus inferred. The system total energy, the internal pressure and the Coulomb coupling parameter $\Gamma$ are computed using the inter-particle Coulomb potential \cite{Lapi18a}. Experimental and numerical investigations on wave‐like excitations in a system conisisting of a charged nylon filament and electrically charged microparticles levitated in a LIT are performed in \cite{Syr21a}. Numerical modelling shows the dynamic collective motion can be excited for high enough voltages supplied to the trap electrodes. 

Trapping of multiple Ba$^+$ ions in a single-beam optical dipole trap without RF or additional magnetic fields is investigated in \cite{Schmi18}. The approach used eliminates driven motion and combines features such as state-dependent confinement and nanoscale potentials, which are characteristic to optical trapping, with the desirable properties of trapped ions crystals, such as long-range interactions that exhibit collective motion. These results open new breakthroughs to investigate quantum dynamics in Coulomb crystals, such as the transition from linear strings to 2D zigzag structures, or the quantum superposition of such structures across the phase transition \cite{Schmi18}.

A simple nonisotope-selective method for ion trap loading, which relies on irradiating the trap electrodes (precoated with materials that exhibit a low work function) by using a 400 nm light-emitting diode (LED), is introduced in \cite{Zali19}. Trapping of single $^{24}$Mg$^+$ ions along with the occurrence of large ion crystals that comprise up to $10^3$ particles, are reported. It is  presumed that this method would be beneficial for applications such as portable ion frequency standards and commercial ion-based devices. Late progress and experimental results on polarization gradient cooling of Ca$^+$ multi-ion Coulomb crystals in a linear Paul trap is reported in Ref. \cite{Joshi20}. Coulomb crystals comprising two ion species simultaneously confined in RF traps are investigated in Ref. \cite{Ruiz20}, where a 3D model is introduced to simulate trapped bicrystals. 

The dynamics of a strongly coupled plasma in a trap is highly nonlinear, while computer simulations using molecular or Brownian dynamics (BD) \cite{Cona20a} should provide accurate quantitative data. Nonlinear effects produced by the viscosity of the surrounding medium are investigated in \cite{Rud20a} for a 3D toroidal Paul trap. An analytical description of the trap electric potential is suggested and quasi-equilibrium points located both inside and outside the trap are identified.     

Ref. \cite{Schmi20a} describes trapping of ultracold neutral Rb atoms and Ba$^+$ ions in a common optical potential, in absence of any RF fields. This approach overcomes the fundamental limitation associated with RF traps, which is due to micromotion-induced heating. It applies to a wide range of ion-atom species and it is assumed to open new perspectives in ultracold chemistry experiments and complex many-body dynamics. In case of conventional (RF) ion traps that confine ions with identical charge-to-mass ratio the trapping potential is approximately independent of the internal electronic state, while in the case of optical trapping state-dependent confinement is achieved \cite{Lamb17, Weck21}. This unique feature that conventional traps exhibit enables accurate control of the internal (electronic) and common external (motional) degrees of freedom, providing a starting point for quantum information processing \cite{Bruz19, Kau20, Rat20, Mehta20}, quantum simulation \cite{Baut19, Wang20} and quantum metrology \cite{Jord20, War20}. With respect to quantum metrology, remarkable progress on state-insensitive Paul traps has enabled state-of-the art atomic clocks \cite{Poli13, Lud15} with uncertainties down to $10^{-18} - 10^{-19}$ and performing investigations on the spatiotemporal variations of fundamental constants in physics at the cosmological scale \cite{Safro18, San19}. 

Collective dynamics of a collisionless nonneutral plasma, in the presence of stray electric field inside a Paul trap, is explored in \cite{Saxe21}. The initial distribution of levitated particles inside the trap is characterized by a Tsallis function. Presence of instabilities is indicated by a double humped time-averaged distribution function.

\subsection{Mesoscopic systems. Phase Transitions}

Experimental investigations of charged particles levitated in external potentials have recorded  significant progress that coincides with the advent of ion traps. The paper of Wuerker {\em et al} reports ordered structures obtained with aluminium microparticles that are cooled as an outcome of collisions with a background buffer gas \cite{Wuerk59}. The experiments describe crystallization into a regular array, followed by melting when both the RF trap potential and cooling force change. Other experiments also report similar phenomena \cite{Blu89}. The transition from an ion cloud to a crystal-like structure corresponds to a chaos-order transition. These intriguing phenomena are a convincing proof that a Paul trap represents a powerful tool to investigate the physics of few-body phase transitions and thus gain new insight on such mesoscopic systems \cite{Schli96, Walth95}.
 
The dynamics of a single stranded DNA (ssDNA) molecule can also be investigated by using nanoscale, linear 2D traps operating in vacuum \cite{Jose10}. By employing molecular dynamics simulations it is demonstrated that a line charge can be efficiently trapped for a well defined range of stability parameters. A 40 nm long ssDNA does not fold or curl in the Paul trap but can experience rotations around the CM. Rotations can be prevented by applying a stretching field in the axial direction, which results in enhanced confinement stability. 

A novel versatile method to achieve one-by-one coupling of single nano- and microparticles is demonstrated in Ref. \cite{Kuhl15}. The setup relies on a segmented linear Paul trap that levitates particles, in combination with an optical microscope, as this particular geometry enables fast particle characterization and alignment-free assembly of particles. Using fluorescent quantum dot clusters and dye-doped polystyrene beads, electromagnetic coupling is achieved by attaching them to spherical silica microresonators. Thus, coupled systems of levitated particles can be extensively investigated. The particles can also be deposited on the facets of optical fibers which is also experimentally demonstrated. An "optical fiber microfluidic control device based on photothermal effect induced convection" is demonstrated in \cite{Zhan19}, that exhibits the ability of allowing particle manipulation and sorting in  order of hundreds of microns. The setup design is simple as it uses three single-mode optical fibers regularly arranged, hence reducing the cost associated with optical tweezers.

Phase transitions represent an intriguing feature of complex plasmas and an important research focus in many-body physics over the last decades. The ordered structures collapse (sometimes the process is called {\em melting}) due to a variation of the discharge pressure or discharge power. This phenomenon is the outcome of the high thermal energies that dust particles acquire. Phase transitions of a 3D complex plasma can be investigated by means of the DLS technique \cite{Asch12b}. Anisotropy in the plasma-wall sheath plays an important role in this case \cite{Asch12a}. Physical mechanisms such as ion flow in the sheath yield to an ion two-stream instability and a phonon instability, both liable for the phase transitions and for the remarkably high kinetic energy of the dust component in the disordered phase state \cite{Schwei98}. Measurements on phase transitions phenomena are very delicate and very few experiments have been performed up to now, especially on systems with very few crystal layers \cite{Tsyto08}.

An area of vivid interest regarding nonneutral plasmas is connected to the phase transitions to liquid and crystal states when the coupling parameter $\Gamma$ \cite{Kul11, Ott14, Drew15} (see eq. \ref{gamma}) is sufficiently large, a condition that is satisfied for low temperatures. 

Phase transitions for trapped ions are investigated in Ref. \cite{Marci12}, where molecular dynamics simulations are employed to investigate a structural transition from a double ring to a single ring of ions. Qualitative investigations on the process of two-ion crystals splitting in segmented Paul traps and the challenges raised by the precise control of this process are presented in Ref. \cite{Kauf14}. Due to the strong Coulomb repulsion cold ions represent a remarkable tool to explore phase transitions under stable confinement conditions. Ref. \cite{Yan16} reports on ordered structures and phase transitions of up to sixteen laser-cooled $^{40}$Ca$^+$ ion crystals, levitated in a custom-made surface-electrode trap (SET). Experimental results are shown to exhibit very good agreement with numerical simulations. Preparation and coherent control of the angular momentum state of a two-ion crystal is demonstrated in \cite{Urban19}. Ref. \cite{Heaz19} is an overview of recent developments in cold ion chemistry, emphasizing on the use of Coulomb crystals for precision measurements, for the study of controlled ion-neutral reactions and to implement quantum information science. A novel technique based on universal control of two  $^{40}$Ca$^+$ ions confined in the same potential well, using a single gate laser beam, is explored in \cite{Seck20}. The method is expected to refine experimental protocols aimed at single-qubit addressing, while future progress would enable extending this technique to chains of four or more ions.

Furthermore, trapped ions act as sensitive detectors of weak forces and electric fields that excite ion motion. Late experiments report measurements of the center-of-mass (CM) motion of a 2D trapped-ion crystal consisting of around 100 ions \cite{Affo20}. The protocol implemented is free from noise associated with the CM mode. We also mention 2D ion crystals levitated in Penning traps, that enable one to perform investigations on quantum simulation or very accurate sensing experiments. For small amplitudes, the out-of-plane motion of such crystals can be characterized by a discrete set of normal modes called the drumhead modes, which enable the experimentalist to implement a range of quantum information protocols \cite{Shank20}. The authors demonstrate that in-plane thermal fluctuations in ion positions represent a major contributor to the broadening of the drumhead mode spectrum. In addition, the confining magnetic field causes unconventional in-plane normal modes, for which the average potential and kinetic energies are different. Thus, the experimental results encourage the pursuit towards designing improved techniques able to achieve cooling of the in-plane degrees of freedom. 

Spontaneous symmetry breaking is a fundamental concept in many areas of physics, including cosmology, particle physics and condensed matter. An example would consist in breaking of the spatial translational symmetry which underlies the formation of crystals and the phase transition from liquid to solid state. Using the analogy of crystals in space, breaking of the translational symmetry in time and the emergence of a {\em time crystal} was recently proposed but was later shown to be forbidden in thermal equilibrium. However, non-equilibrium Floquet systems which are subject to a periodic drive can exhibit persistent time correlations at an emergent subharmonic frequency. This new phase of matter has been dubbed a {\em discrete time crystal}. Experimental observation of a discrete time crystal in an interacting spin chain of trapped atomic ions is reported in Ref. \cite{Zhang18}. A periodic Hamiltonian is applied to the system under many-body localization conditions and a subharmonic temporal response that is robust to external perturbations is monitored. Observation of such a time crystal opens the door to the study of systems with long-range spatio-temporal correlations and novel phases of matter that emerge under intrinsically non-equilibrium conditions \cite{Zhang18}.

Ion crystals consisting of tens up to hundreds of ions levitated in a linear Paul trap are investigated in \cite{Oral19}, where a simulation code is used which computes the dynamics of such collections of ions in quasi-static RF fields characteristic for real linear quadrupole traps. Both micromotion and laser cooling are taken into account. 

Investigation of solid-liquid phase transitions in case of complex multi-body systems consisting of trapped ions (or electrically charged particles) represents a challenge both as an experiment or analytical modeling \cite{Kiet21}, due to the elaborate physical mechanisms involved. Late results indicate that a trapped-ion Wigner crystal represents a very promising candidate to perform controlled studies of such phase transitions under conditions of high atom resolution. Controlled melting of a pseudo-crystal of $^{174}$Yb$^+$ ions is achieved in Ref. \cite{Xu20}, where simultaneous coexistence of liquid and solid phases is reported.

\section{Strongly coupled plasmas confined in quadrupole and multipole electrodynamic traps}\label{Sec4}

\subsection{Stable 2D and 3D structures. Phase transitions}

As a general rule, the particle size and inter-particle distance allow one to use simple diagnostic means, {\em e.g.} optical measurements in the visible range. Results of such investigations can be used to characterize the properties of crystals, liquids and colloidal suspensions \cite{Ivlev12}. So far, plasma crystals have been obtained in plasmas of glow and high-frequency discharges at low pressures $\left( 1 \div 100 \ \textrm{Pa} \right)$ or by using a beam generated by a charged particle accelerator \cite{Fort05, Fort10a}. Investigations have been performed on crystal-liquid-gas phase transitions, on the temperature influence on dusty structures (starting from SATP conditions down to cryogenic ones) \cite{Fort02} or regarding the oscillations and waves that propagate in dusty plasma structures \cite{Fort05, Fort10a}. The response of dusty plasma structures with respect to external actions such as pulsed electric \cite{Vasi07, Vasi08, Pust09} and magnetic fields, laser radiation, thermophoretic force \cite{Vasi03a}, electron beam \cite{Vasi03b}, etc., has been extensively investigated. 3D structures consisting of a large number of particles are explored in experiments performed aboard the orbital space station “MIR” and the International Space Station (ISS) \cite{Fort10a, Myas17, Pust16, Lipa20}. Technological applications of dusty plasmas are considered in \cite{Bouf11}. 

As Coulomb crystals consisting of dust particles with electric charges of identical sign are unstable and particles are pushed apart by the Coulomb repulsion, a stable crystal can occur only in the presence of a trap potential \cite{Died87, Blu89, Marci12, Kauf14, Yan16}. These traps result for low pressure electrical discharges, in regions of strong non-uniform electrical field. The trap can develop in the near-electrode layer of a capacitive RF discharge or in a striation of the glow discharge positive column. The gravity force acting upon a charged particle is balanced by a longitudinal electrical field aligned in an axial direction, in case of a discharge that exhibits cylindrical symmetry. The radial electric field emerges due to the ambipolar diffusion of electrons and ions to the walls, which results in radial confinement of the charged particles. The electric charge of the dust particles can rise up to values that are several thousand times larger with respect to the electric charge of the electron, as it is determined by the electron temperature and the particle size \cite{Fort05, Fort10a}. The dimension of levitated dust particles generally lies in the $1 \div 50 \ \mu$m range, but not necessarily restricted to \cite{Wint91, Ghe98, Sto01, Sto08, Vis13}. The size range is determined by the electric charge of the particle and by the electrical discharge fields that accomplish levitation. The parameter fields for ordered dusty structures in low pressure electrical discharges are determined by the plasma parameters that limit the investigated properties of plasma crystals.

An additional restriction concerning the occurrence of Coulomb crystals in gas-discharge plasmas results from the fact that the electrical charges of the dust particles depend on the electron temperature, which in turn depends on the reduced electrical field intensity. If the electrical charges of dust particles and the trap that confines them are created independently, ordered structures of charged dust particles can be produced for various conditions. Ref. \cite{Fort06b} describes a method to generate dusty plasma crystals, that uses a proton beam with an average energy of about 2 MeV. The proton beam ionizes the gas and additional trap electric fields are generated by extra electrodes supplied with d.c. voltages. Once the pressure increases, the crystal is destroyed. Afterwards, levitation of charged particles no longer exists and particles are expelled out of the discharge as the electric fields that create the trap vanish. For example, the phenomenon occurs when the glow d.c. discharge striations fade and the longitudinal electrical field gradient that provides stability of the levitated particles (compensating the gravitational force) also vanishes. When the pressure increases, the radial electrical force drops off. Hence, in presence of an ion drag force and of the thermophoretic forces that develop due to a temperature gradient, confinement of particles in the radial direction becomes impossible \cite{Fort05, Fort10a}.

A demonstration on the occurrence and manipulation of a wide variety of ion Coulomb crystals (ICCs) consisting of small numbers of ions levitated in a 3D Penning trap, is performed in Ref. \cite{Mava13}. Various 2D and 3D structures are reported such as linear strings, intermediate geometries, planar structures, etc. In addition, a phase transition which starts from a linear string structure and ends into a zigzag geometry is observed for the first time in a Penning trap. Experimental results show good agreement with simulations, while the latter demonstrate that the rotation frequency of a small crystal is especially influenced by the laser parameters, independent of the ion number and of the axial confinement strength. The system described is shown to be of interest for applications which span areas such as quantum simulation \cite{Berm11}, QIP \cite{War20, Wan20}, or to perform tests of fundamental physics models starting from quantum field theory to cosmology \cite{Bla08, Vogel18, Bla10, Quint14}. Metastable ion Coulomb crystals (ICCs) in a polychromatic optical superlattice (OSL) created by rectified gradient forces are investigated in \cite{Kras19}.  

A late experiment reports on a simple monolithic Paul trap able to levitate up to two dozens of $^{171}$Yb ions in a stable 2D lattice \cite{Wang20}, which opens exciting possibilities to scale up the number of ion qubits within a single trap and to investigate 2D many-body quantum models \cite{Yos15}. Raman laser beams are then used to coherently manipulate these ion qubits.

\subsubsection{Microparticle electrodynamic traps-MEITs}

\begin{figure}[bth] 
	\centering
	\includegraphics[scale=0.97]{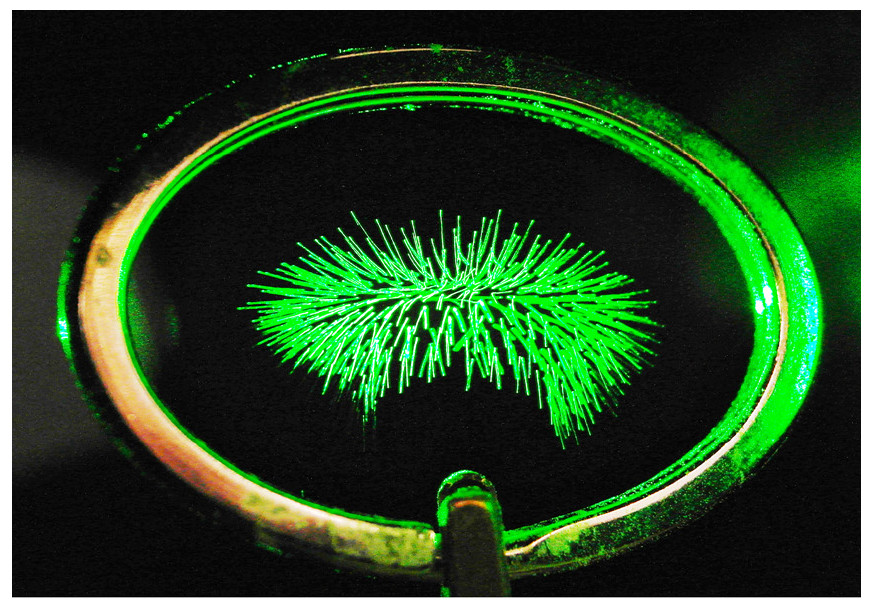}
	\includegraphics[scale=0.65]{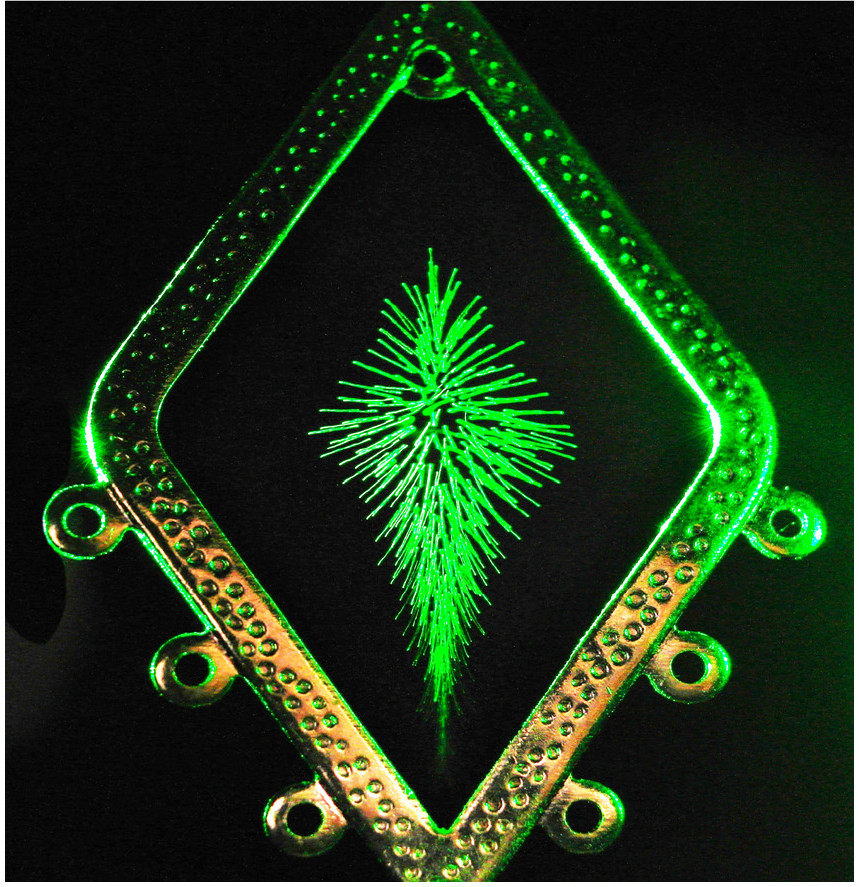}
	\caption{Stable 2D and 3D structures consisting of $26 \ \mu$m-diameter particles trapped under SATP conditions by ring-type microparticle electrodynamic ion traps (MEITs). The trapping voltage supplied to the planar ring electrode is 6 kV a.c. at 60 Hz frequency. The particles are negatively charged monodispersed Lycopodium club-moss spores illuminated with laser light. Source: images reproduced from \cite{Libb18} by courtesy of Prof. K. Libbrecht.} 
	\label{Meit}
\end{figure}

Further on we shortly present MEITs and the physics associated to them \cite{Libb18}. MEITs prove to be versatile instruments that are suited to characterize the physico-chemical properties of single charged particles with dimensions ranging between $100$ nm $\div 100 \ \mu$m, such as aerosols \cite{Kul11, Vogel13a, Kurt07}, liquid droplets \cite{Singh17, Singh18b, Lamb96}, solid particles \cite{Wuerk59, Wint91, Ghe95a, Sto16}, nanoparticles \cite{Sande14}, and even microorganisms \cite{Peng04} or DNA segments \cite{Jose10, Park12a}. Fig.~\ref{Meit} shows images of stable 2D and 3D structures consisting of $26 \ \mu$m diameter particles levitated under SATP conditions by ring-type microparticle electrodynamic ion traps (MEITs). The trapping voltage supplied to the planar ring electrode is 6 kV a.c. at 60 Hz frequency. The particles are negatively charged monodispersed {\em Lycopodium} club-moss spores illuminated with laser light \cite{Libb18}. A modified RF trap with a rotating toroidal electrode, used to levitate {\em Lycopodium Clavatum} vegetable spores is reported in \cite{Kost20}.

\subsubsection{Coulomb structure in a linear Paul trap affected by electrical pulses}

\begin{figure}[bth] 
	\centering
	\includegraphics[scale=0.43]{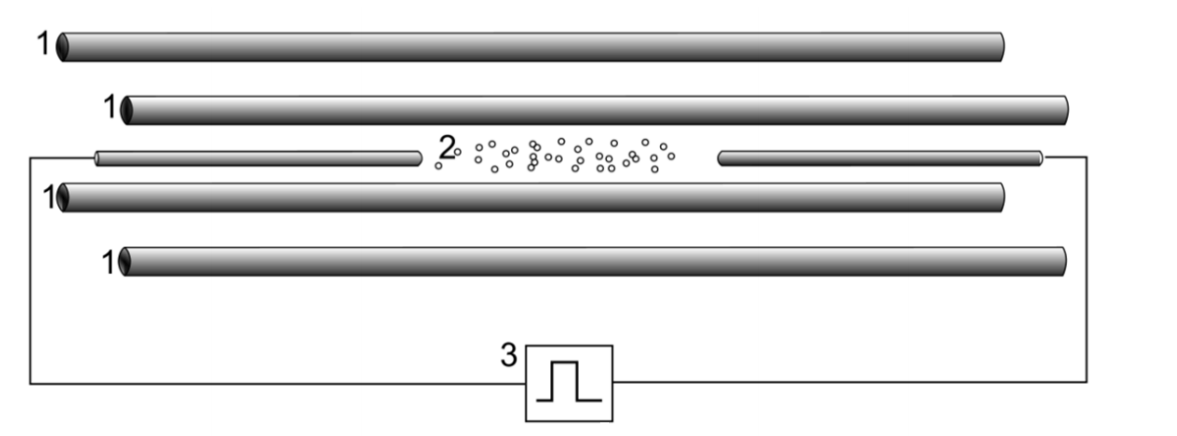}
	\caption{Scheme of a linear electrodynamic trap. Legend: 1 -– trap electrodes, 2 -– dust particles, 3 -– generator of rectangular pulses. Source: picture reproduced from \cite{Syr18} with kind permission of the authors.} 
	\label{setpuls}
\end{figure}

Fig.~\ref{setpuls} shows the schematic diagram of a power supply for an electrodynamic trap. Oscillations are excited by rectangular electric pulses applied to the additional electrodes located at the trap ends, separated by a distance of 5 cm. The pulse amplitude ranges between $10 \div 320$ V and the frequency lies between $1 \div 20$ Hz. The relative pulse duration can be adjusted in the interval $0.01 \div 0.99$. The electrodynamic trap presented in \cite{Syr18} consists of four cylindrical parallel, longitudinal electrodes ($10$ cm long, $4$ mm diameter), placed at the vertices of a square with a side of 2 cm. A sinusoidal voltage is applied, of opposite phase between adjacent electrodes. The a.c. voltage amplitude is around 4.5 kV at a frequency of 50 Hz.  

Polydisperse $Al_2O_3$ particles are used in the experiments described in Ref. \cite{Syr18}. Electrical charging of the particles is achieved by employing an induction method. The particles are placed on a flat electrode that is gradually shifted towards the electrodynamic trap. Then, a positive electric potential is applied to the electrode. When the potential is higher than 5 kV particles are attracted and then absorbed into the trap. The particle motion is recorded using a CCD camera HiSpec1 at a maximum resolution rate of $1240 \times 1024$ pixels. The frame rate can reach up to 524 frames per second at maximum resolution. Particle illumination is achieved by means of a flat laser beam with a diameter of 1 mm. The laser wavelength is 532 nm and the output power ranges up to 150 mW.

\begin{figure}[bth] 
	\centering
	\includegraphics[scale=0.5]{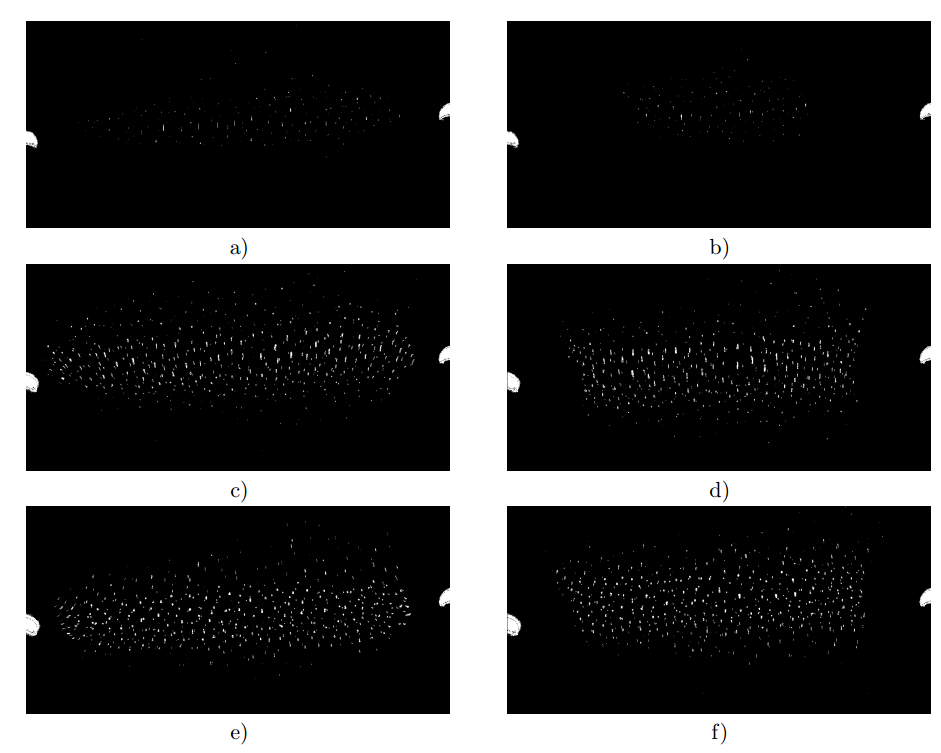}
	\caption{Coulomb structure in a linear electrodynamic trap at time instants of highest compression (right column) and stretching (left column): a), b) pulse frequency 1 Hz; c), d) pulse frequency 5 Hz; e), f) pulse frequency 9 Hz. Source: images reproduced from \cite{Syr18} with kind permission of the authors.}
	\label{pulse}
\end{figure} 

Fig.~\ref{pulse} illustrates oscillations of a Coulomb structure in a linear electrodynamic trap, for different values of the pulse frequency. The pulse amplitude is 320 V. The relative pulse duration is 0.5 while pulses are in phase. By increasing the frequency, the amplitude of the oscillations drops to almost undetectable at 20 Hz; the oscillations cease to affect the middle region of the structure. For a frequency of 1 Hz, the structure at the instant of time of maximum compression corresponds to the Coulomb structure in case of a constant potential supplied to the additional electrodes.

\subsection{Confinement regions in the voltage--frequency plane  }

\begin{figure}[bth] 
	\centering
	\includegraphics[scale=0.8]{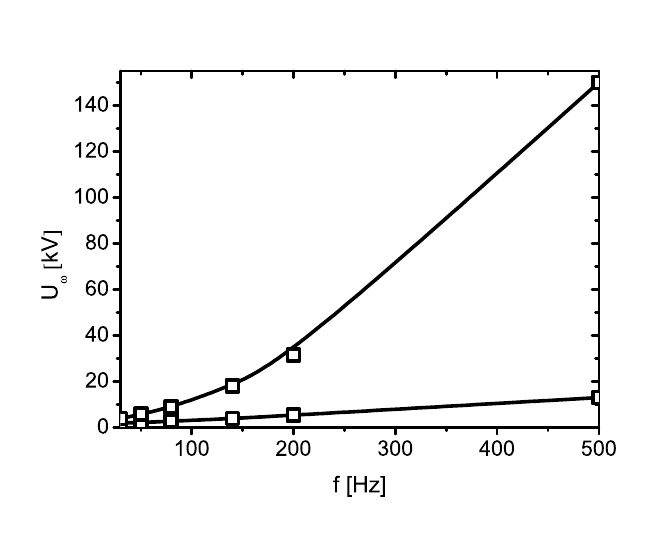}
	\caption{Boundaries for dust particle confinement in the $U_{\omega} - f$ plane. Experimental parameters: $f = 50 \div 500$ Hz, $U_{end} = 4000$ V, $\rho_{particle} = 0.38 \cdot 10^4$ kg/m$^3$, $r_{particle} =7 \ \mu$m, $Q_{particle} = 50000 e$, $\eta = 17 \cdot 10^{-6}$ Pa$\cdot$s - dynamic viscosity and $T = 300$ K. Source: picture reproduced from \cite{Lapi13} with kind permission of the authors.}
	\label{fig:5}
\end{figure}

Let us consider the confinement region in the $U_{\omega} - f$ plane, shown in Fig.~\ref{fig:5}. Regardless of the frequency value, the confinement region is bounded by lower and upper values of the a.c. trapping voltage. Besides this region the trap cannot confine dust particles. On the other hand, if the a.c. voltage is low the total restoring force that holds the particle inside the trap will dwindle. In such case the gravity force will prevail and particles will fall down. Otherwise, if the a.c. voltage value is too large, the trapping field might be large enough to push the particle out of the trap during a half-period of oscillation. 

\subsection{Mass dependent particle separation in electrodynamic traps} 

\begin{figure}[bth] 
	\centering
	\includegraphics[scale=0.6]{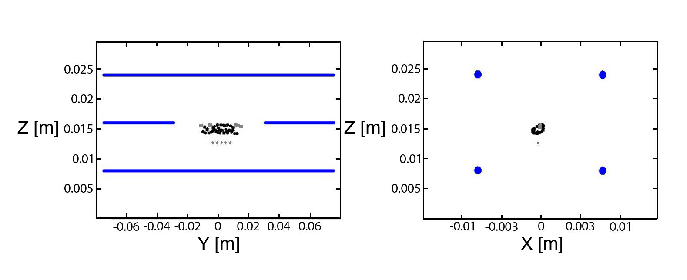}
	\caption{Typical dust particle levitation in a linear Paul trap. Left panel -- side view, right panel -- end view.  Experimental parameters: $f = 50$ Hz , $U_{end} = 900$ V, $U_{\omega} = 4400$ V, $r_{p} = 7\ \mu$m, $\eta = 17 \cdot 10^{-6}$ Pa$\cdot$s -- dynamic viscosity, $T = 300$ K.  Symbols: black circles -- particles of type A: 
	$\rho_{pA}=0.38 \cdot 10^4$ kg/m$^3$, $Q_{pA} = 9 \cdot 10^5 e$;
	grey squares -- particles of type B: $\rho_{pB}=0.76 \cdot 10^4$ kg/m$^3$, $Q_{pB} = 1.8 \cdot 10^6 e$; grey rhombus -- particles of type C: $\rho_{pC}=0.76 \cdot 10^4$ kg/m$^3$, $Q_{pC} = 9 \cdot 10^5 e$.}
	\label{fig:6}
\end{figure}

Fig.~\ref{fig:6} shows the results of simulations performed for dust particle confinement in a four wire (electrode) trap. We also observe levitation of trapped particles for three different values of the charge-to-mass ratio. Inside the trap there are 30 particles of type A (black circles) with ${Q/M}_A = 6.9 \cdot 10^{16}\ e$/kg and $m_{pA} =1.3 \cdot 10^{-12}$ kg, 5 particles of type B (grey squares) with ${Q/M}_B = 6.9 \cdot 10^{16}\ e$/kg and $m_{pB} = 2.6 \cdot 10^{-12}$ kg, along with 5 particles of type C (grey diamonds) with ${Q/M}_C = 3.4 \cdot 10^{16}\ e$/kg and $m_{pC} = 2.6 \cdot 10^{-12}$ kg.

Fig.~\ref{fig:6} illustrates how an electrodynamic trap can be used to perform dust particle separation. Type C particles are located below type A particles, as they have identical electric charge but are two times more heavier. Particles of type B and those of type A possess the same charge-to-mass ratio, but type B ones are two times more heavier and carry a double electrical charge in comparison with type A particles. Due to their double electric charge, type B particles are more strongly attracted and thus levitate closer to the trap axis. As illustrated in Fig.~\ref{fig:6}, type A particles establish an oscillating cylinder under the trap axis. The inter-particle interaction also contributes to the spatial separation and yields ordering of dust particles along the trap axis \cite{Lapi17}.

\subsection{Confinement of a maximum number of particles}

\begin{figure}[bth] 
	\centering
	\includegraphics[scale=0.2]{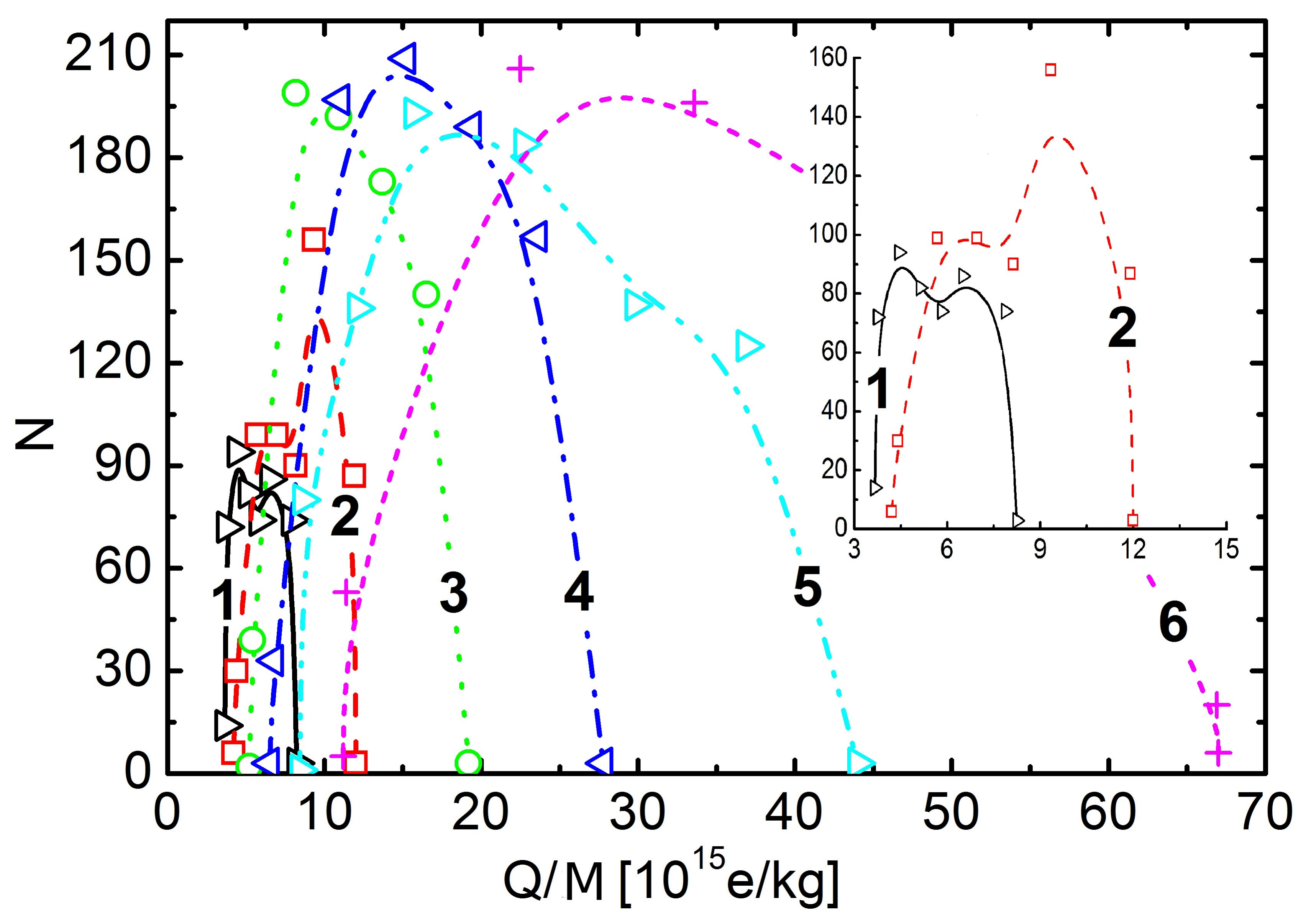} 
	\caption{Dependence of the maximum number of trapped particles on the frequency of the a.c. voltage and on the charge-to-mass ratio $Q/M$, under SATP conditions. The values of the a.c. voltage frequency are : $1-30$ Hz, $2-50$ Hz, $3-80$ Hz, $4-100$ Hz, $5-150$ Hz, $6-200$ Hz. Source: picture reproduced from \cite{Vasi13} with kind permission of the authors.}
	\label{capas}
\end{figure}

Fig.~\ref{capas} presents the regions of confinement under SATP conditions for a maximum number of particles (corresponding to the maximum trap capacity), for a particular value of the trapping voltage frequency and for different values of the charge-to-mass ratio $Q/M$. The region that confines a maximum number of particles lies within the curve corresponding to a specific value of the trapping voltage frequency. The range of $Q/M$ values modifies for a certain value of the maximum number of trapped particles, and it is determined by the boundary points of an intersection of the straight line drawn through a point (corresponding to a specific number of particles) with the curve that results out of the simulation. As the particle number in a dusty structure rises, the confinement region shrinks. Such a phenomenon is illustrated by the solid lines in Fig.~\ref{capas}.  

In order to establish the maximum trap capacity, a large number of particles are injected into the trap. As the frequency rises, the region that confines a fixed number of particles extends. The maximum number of particles corresponds to a maximum of the curve in Fig.~\ref{capas}. The highest number of particles confined in the trap is reached only for a single value of the charge-to-mass ratio $Q/M$.

\subsubsection{Experimental setup}
\begin{figure}[bth] 
	\centering
	\includegraphics[scale=0.45]{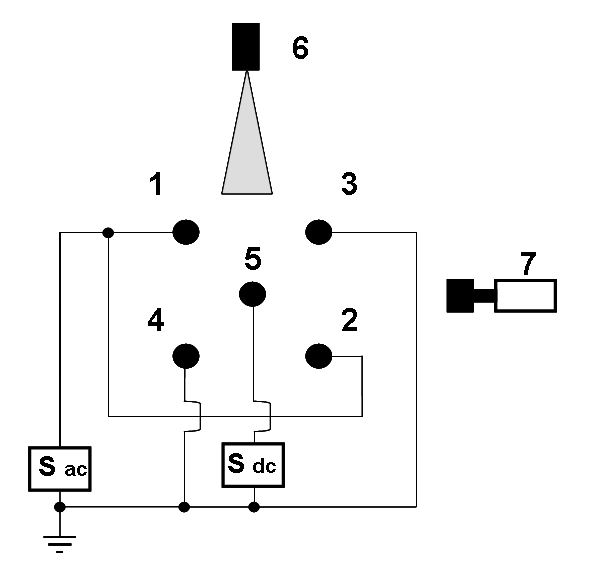} 
	\caption{Scheme of experimental setup used to study the behaviour of charged dust particles in a quadrupole dynamic trap. Legend: 1, 2, 3, 4, are the trap dynamic electrodes, 5 denotes the endcap electrodes, 6 is the laser diode, 7 stands for the high speed camera, $S_{ac}$ is a source of a.c. voltage and $S_{dc}$ is a source of d.c. voltage. Source: picture reproduced from \cite{Vasi13} with kind permission of the authors.}
	\label{setup}
\end{figure}

An experimental setup that uses a linear quadrupole trap to capture and confine electrically charged dust particles is presented in Fig.~\ref{setup}. The dynamic trap electrodes and the endcap electrodes are made of 12 cm long copper rods, with a diameter of 3 mm. The distance between the electrode axes is 1.3 cm. The endcap electrodes are placed along the trap axis at a distance of 65 mm apart. The cylindrical trap electrodes achieve radial confinement of the particles, while the endcap electrodes maintain the particles contained within the axial region of the trap. The a.c. voltage applied across the four cylindrical electrodes ranges within $0 \div 2000$ V and its frequency can be modified by means of a generator. A d.c. voltage of 900 V is applied between the endcap electrodes. To mitigate against the action of external electric fields, the trap is placed inside a grounded metal shield (chassis). The high voltage supply wires are placed at a safe distance, in order to prevent the electric field they generate from perturbing the trapping field. The wires are cabled along the metal surface of the shield. Particles are detected by means of a high speed video camera HiSpec at a maximum resolution rate of $1280 \times 1024$ pixels. To better observe the particles with dimensions ranging between $10 \div 150$ microns, a laser diode ($\lambda = 550$ nm, power $= 10 \div 100$ mW) is used or a laser sheet with a diameter of 150 microns. The trap is placed inside a transparent box with an aim to severely limit the influence of air flows that disturb particle dynamics \cite{Lapi18b}.

\subsubsection{Experimental details and characteristic parameters}
Polydisperse $Al_2O_3$ particles with dimensions ranging between $10 \div 80$ microns and hollow borosilicate glass spheres with dimensions within the $30 \div 100$ micron range are used in the experiments. As the numerical simulations illustrate, the quadrupole trap exhibits selectivity with respect to the mass and electric charge of the particles. Therefore, the trap will capture only those particles that specifically satisfy the required confinement conditions. In order to levitate the microparticles and build up a stable ordered structure, a special source of electrically charged particles has been devised and tested. Electrical charging of the microparticles is achieved by employing a streamer discharge in the electric field of a plane capacitor. The source allows one to positively charge particles up to $10^5 \div 10^6 e$ while in the same time it delivers them an initial velocity of $0.5 \div 2$ m/s.  

Selection of the particles with the desired charge-to-mass ratio is achieved in the electric field of a plane capacitor, towards which a flow of charged particles is directed. The capacitor plates are arranged in a horizontal plane and the distance between them is around 10 cm. In order to achieve stable confinement, a positive potential generated by a d.c. voltage source is applied to the lower capacitor plate. The voltage value is modified such as to balance the gravity force acting upon the charged particles. The particles that satisfy this condition move along the capacitor plates and they finally end up confined in the trap. Moreover, in order to provide stable levitation of the particles, a mandatory requirement consists in reducing the particle velocity. The neutral drag force efficiently brakes the particles by means of air friction (SATP conditions) when the capacitor electric field vanishes. The particles move across the quadrupole trap where low-velocity grains that possess an adequate charge-to-mass ratio are captured. We describe the method used to determine the particle electric charge. The trap supply voltage is initially switched off. As an outcome of gravity, particles fall and precipitate on a substrate. The dimensions of the precipitated particles are estimated using an optical microscope. The particle weight is evaluated by assuming it exhibits a spherical shape. The electric charge of the particle is then determined from the condition that the gravity force in the plane capacitor gap is balanced by the electric field in the area. The principle of the method in described in Section \ref{specharge}.

\subsection{Ordered Coulomb particle structures }
\begin{figure}[bth] 
	\centering
	\includegraphics[scale=0.53]{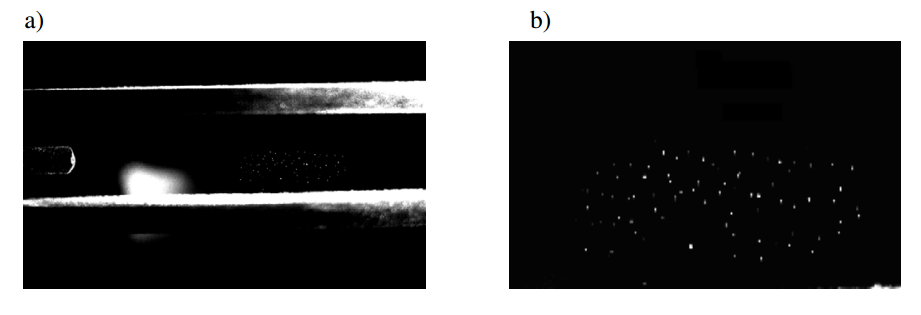} 
	\caption{a) Photo of the trap and ordered structure consisting of aluminium oxide particles with $10 \div 15 \ \mu$m diameter. $U_{ac} = 2000$ V, $f = 80$ Hz and $U_{dc} = 900$ V; b) Magnified image of an ordered Coulomb structure. The area of the view is $17 \times 9.5$ mm$^2$. The inter-particle distance between grains of similar size ranges between $400 \div 800 \ \mu$m, the distance between a big particle and a smaller one is around $1800 \div 2000 \ \mu$m. Source: images reproduced from \cite{Vasi13} with kind permission of the authors.}
	\label{fig5}
\end{figure}

Fig.~\ref{fig5} shows images of experimentally observed ordered Coulomb structures consisting of aluminium oxide particles with dimensions ranging between $10 \div 80 \ \mu$m, for a trapping voltage frequency value $f = 80$ Hz. The ordered structure usually comprises 50 up to 100 particles, as their number depends on the frequency and amplitude of the a.c. trapping voltage.

\begin{figure}[bth] 
	\centering
	\includegraphics[scale=0.65]{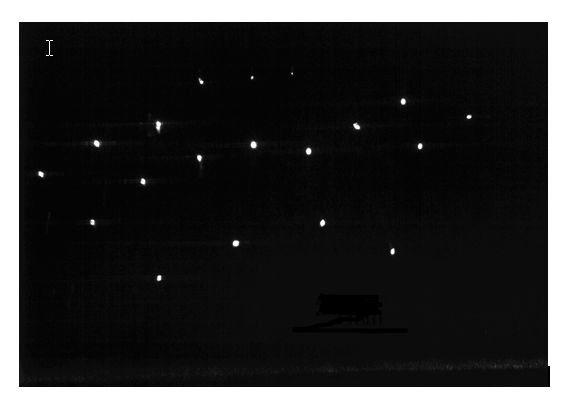} 
	\caption{Snapshot of a segment of an ordered Coulomb structure, for hollow glass spheres with diameter ranging between $30 \div 50 \ \mu$m. $U_{ac} = 2000$ V, $f = 80$ Hz, $U_{dc} = 900$ V. The area of the view field is $11\times 7.5$ mm$^2$. The inter-particle distance spans values from $1300 \div 1800 \ \mu$m. Source: image reproduced from \cite{Vasi13} with kind permission of the authors.}
	\label{fig6}
\end{figure}

Fig.~\ref{fig6} shows the structure established by hollow glass microspheres with diameter ranging between $30 \div 100 \ \mu$m. The maximum number of particles that create the ordered structure does not exceed 50. Owing to a large friction force, the chaotic movement of the particles is negligible as most of them oscillate at small amplitudes, in synchronism with the a.c. electric field. The particles that are closest to the trap axis exhibit low displacements, while only several particles located far from the trap axis oscillate with large amplitude. It seems likely that these particles are positioned near the border of the stability boundary. 

An ordered Coulomb structure was observed for about 12 hours. We show the Coulomb interaction is strong and the mean inter-particle distance remains quite unchanged. The electrical charge of the trapped particles is estimated using the method described above and we obtain values that span the interval $10^5 \div 10^6 \ e$. We emphasize that in contrast to dusty structures in a plasma, the trapped microparticles settle into a Coulomb system in absence of the neutralizing plasma background. Since the trap confines particles with equal charge-to-mass ratios, the ordered structures can incorporate particles of different sizes. An example of such a structure is shown in (Fig.~\ref{fig5} b) where in the lower part of the dusty structure one can distinguish a large particle. The inter-particle distance between equally sized grains varies between $400 \div 800 \ \mu$m, while the distance between the large particle and the smaller ones is around $1800 \div 2000 \ \mu$m. Considering that the trap confining force is proportional to a space shift and the repulsive force between the particles is of Coulomb type, it follows that the electric charge of the large particle is $6 \div 9$ times greater than the electric charge of smaller sized particles.

\begin{figure}[bth] 
	\centering
	\includegraphics[scale=0.55]{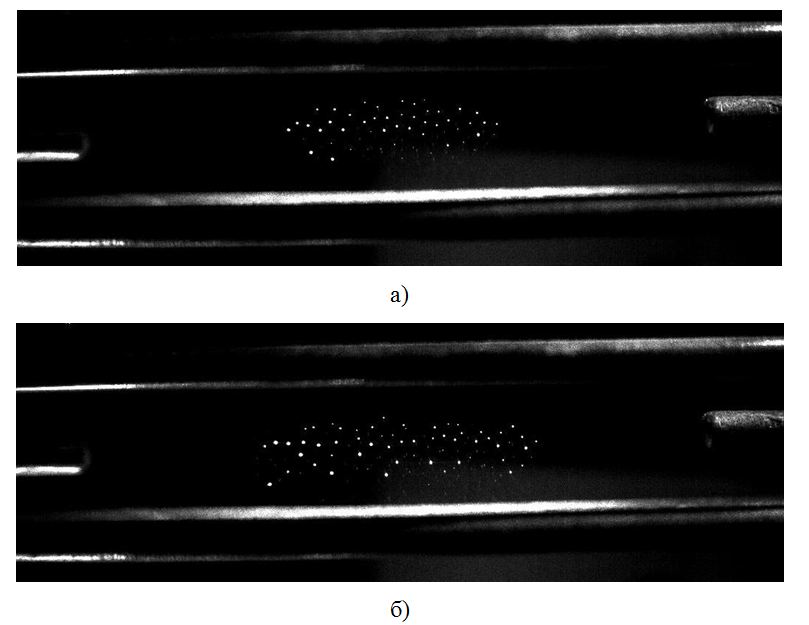} 
	\caption{Image of the trap and ordered Coulomb structure made of aluminium oxide particles with dimensions of about $10 \div 15$ microns. $U_{ac}= 2000$ V, $f = 80$ Hz, a) $U_{dc} = 900$ V; b) $U_{dc} = 750$ V. Source: images reproduced from \cite{Vasi13} with kind permission of the authors.}
	\label{fig7}
\end{figure}

By further lowering the value of the endcap voltage, particles reach the trap edges and leave by moving along the endcap electrodes. When the number of particles in the trap drops down to 10 or even less, microparticles align along the trap axis and set up a linear chain. Even if the length of the trap electrodes is equal to 12 cm, numerical simulations are performed considering a length of 15 cm for the cylindrical electrodes. Simulations show that a change in the trap electrodes length could result into a slight change in the electric field at a peripheral region, near the edge of the electrodes. As illustrated in Fig.~\ref{fig5} and Fig.~\ref{fig7}, particles are not present in this region. The same remark is valid with respect to the small difference between the $L_h$ values used in experiment and numerical simulation, respectively.

\begin{figure}[bth] 
	\centering
	\includegraphics[scale=1.1]{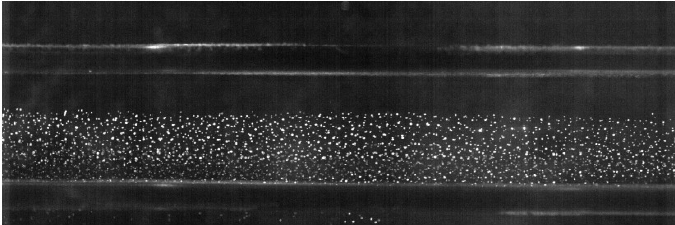} 
	\caption{Fragment of a large stable Coulomb structure containing about 2000 particles.}
	\label{large_stable}
\end{figure}

Ordered Coulomb structures that encompass a large number of particles are of interest for applications such as nuclear batteries \cite{Filip05}, where stable structures of radioactive particles are required to levitate in a high density gas media. These structures can be obtained using electrodynamic traps. Fig.~\ref{large_stable} shows a fragment of a large stable Coulomb structure that encloses around 2000 particles.

\subsubsection{Comparison of experimental and simulated results}
\begin{figure}[bth] 
	\centering
	\includegraphics[scale=0.47]{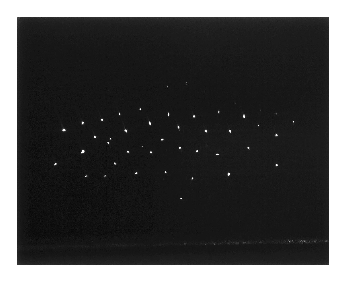}
	\includegraphics[scale=0.47]{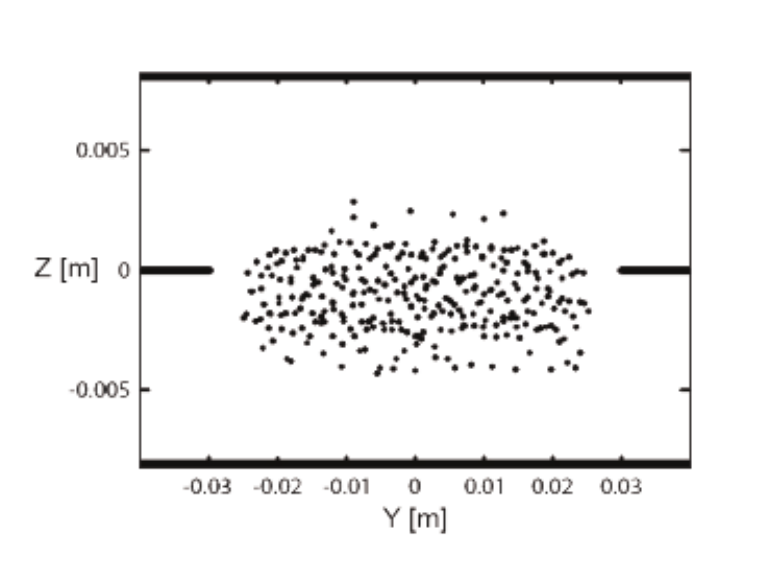}
	\caption{Comparison between experimental results and numerical simulations with respect to dust particle confinement in a linear Paul trap. Left panel -- typical experimental observation of the particle confinement phenomena, right panel -- numerical simulation of experiment. Experimental parameters: $f = 50$ Hz , $U_{end} = 900$ V, $U_{\omega} = 4400$ V, $r_p = 10 \div 15 \ \mu$m, $\rho_p = 0.38 \times 10^4$ kg/m$^3$ , $\eta = 17 \times 10^{-6}$ Pa$\cdot$s, $T =300$ K; Numerical simulation parameters: $r_{p}=10 \ \mu$m, $\rho_p = 0.38 \times 10^4$ kg/$m^3$, $Q_p = 4.5 \times 10^5 e$. Source: pictures reproduced from \cite{Lapi13} with kind permission of the authors.} 
	\label{fig:Experiment_Simulation}
\end{figure}

The left panel in Fig.~\ref{fig:Experiment_Simulation} presents an experiment of dust particle confinement in a slim Paul trap. Aluminium oxide Al$_{2}$O$_{3}$ ($\rho_p = 0.38 \times 10^4$ kg/m$^3$) dust particles with radius $r_p =1 0 \div 15 \ \mu$m are injected inside a two wire trap (2WT). The trap electrodes are supplied at an a.c. voltage $U_{\omega}\sin({\omega}t)$, where $U_{\omega} = 4400$ V. The estimated value of the electrical charge of trapped microparticles is $Q_{p} \sim 10^5 \div 5\times 10^5 \ e$. The experimental trap parameters are: wire (electrode) length -- $L_{m} = 15$ cm, distance between wires -- $L_{b} = 1.3$ cm, endcap electrode spacing -- $L_{h} = 6$ cm, endcap electrode length -- 4.5 cm. The d.c. voltage supplied to the endcap electrodes is $900$ V. 

A numerical simulation of this experiment was performed. The trap parameters' values  used in the simulation are: $r_p = 10 \ \mu$m, $\rho_p = 0.38 \times 10^4$ kg/m$^3$ and $Q_p =4.5 \times 10^5 e$. Simulation results on dust particle confinement are presented in the right panel of Fig.~\ref{fig:Experiment_Simulation}. Experimental and numerical simulation results are in good agreement, as they validate each other. 

\subsection{Hamiltonians for systems of $N$ particles. Ordered structures}

Collective dynamics for systems of ions confined in a 3D QIT that exhibits cylindrical symmetry is characterized in \cite{Major05, Ghe00}. We consider a system consisting of $N$ ions in a space with $d$ dimensions, ${\mathbb R}^d$. The coordinates in the space (manifold) of configurations ${\mathbb R}^d$ are denoted by $x_{\alpha j}\  , \; \alpha = 1, \ldots , N\ ,\; j=1, \ldots , d$. In case of linear, planar, and space models, we choose $d = 1$, $d = 2$ or $d = 3$, respectively. We introduce the kinetic energy $T$, the linear potential energy $U_1$, the quadrupole trap potential energy $U$ and the anharmonic trap potential $V$:

\begin{subequations}\label{stq7}
	\begin{equation}
	\label{stq7a}T = \sum_{\alpha = 1}^N \sum_{j=1}^d \frac 1{2m_\alpha }\ p_{\alpha
		j}^2 \  , \  \  U_1 = \frac 12 \sum_{\alpha =1}^N \sum_{j=1}^d \delta_j x_{\alpha j}\ 	,\;\
	\end{equation}
	\begin{equation}\label{stq7b}
	U = \frac 12 \sum_{\alpha = 1}^N \sum_{i, j=1}^d \kappa	_{ij} x_{\alpha j}^2 \  , \  \  V = \sum_{\alpha = 1}^N v\left( {\mathbf x}_\alpha , t \right) \ ,
	\end{equation}
\end{subequations}
where $m_\alpha $ is the mass of the ion denoted by $\alpha $, ${\mathbf x}_\alpha = \left(x_{\alpha 1}, \ldots , x_{\alpha d} \right) $, while $\delta_j$ and $\kappa_{ij} $ might eventually be time dependent functions. The Hamilton function associated to the system of $N$ ions can be cast as
$$
H = T + U_1 + U + V + W \ ,
$$
where $W$ represents the interaction potential between the ions. If the ions possess equal masses we introduce $d$ coordinates $x_j$ of the centre of mass (CM) of the system 

\begin{equation}
\label{mp9}x_j = \frac 1N \sum_{\alpha =1}^N x_{\alpha j} \  , \;
\end{equation}
and $d\left( N - 1\right) $ coordinates $y_{\alpha j}$ with respect to the relative motion of the ions

\begin{equation}
\label{mp9b}\ y_{\alpha j} = x_{\alpha j} - x_j \ , \;\  \sum_{\alpha = 1}^N y_{\alpha j}=0 \ .
\end{equation}

Further on we introduce $d$ collective coordinates $s_j$ and the collective coordinate $s$, defined as \cite{Mih21}
\begin{equation}
\label{mp10b}s_j = \sum_{\alpha =1}^N y_{\alpha j}^2 \  , \  \ \; s = \sum_{\alpha = 1}^N \sum_{j=1}^d y_{\alpha j}^2\ .
\end{equation}
From calculus we obtain 
\begin{equation}
\label{mp10}\sum_{\alpha =1}^N x_{\alpha j}^2 = N x_j^2 + \sum_{\alpha = 1}^N y_{\alpha j}^2 \ .
\end{equation}
\begin{equation}
\label{mp11}\ s_j = \frac 1{2N} \sum_{\alpha, \beta = 1}^N \left( x_{\alpha j} - x_{\beta j}\right)^2 \  , \;  \ s = \frac 1{2N} \sum_{\alpha , \beta = 1}^N \sum_{j=1}^d \left( x_{\alpha j} - x_{\beta  j}\right)^2 \  .
\end{equation}

From eq. (\ref{mp11}) it results that $s$ represents the square of the distance measured from the origin (fixed in the CM) to the point that corresponds to the system of $N$ ions from the space (manifold) of configurations. The relation $s = s_0$, with $s_0 > 0$ constant, determines a sphere of radius $\sqrt{s_0}$ with its centre located in the origin of the space (manifold) of configurations. In case of ordered structures of $N$ ions the trajectory is restricted within a neighbourhood $\left \|s-s_0\right\| < \varepsilon $ of this sphere, with $\varepsilon $ sufficiently small. On the other hand, the collective variable $s$ can be also interpreted as a dispersion:

\begin{equation}
\label{mp11b}s = \sum_{\alpha =1}^N \sum_{j=1}^d \left( x_{\alpha j}^2 - x_j^2 \right) \ .
\end{equation}
We now introduce the moments $p_{\alpha j}$ associated to the coordinates $x_{\alpha j}$. We also introduce $d$ impulses $p_j$ of the CM and $d \left( N - 1 \right) $ moments $\xi _{\alpha j}$ of the relative motion, characterized as

\begin{equation}
\label{mp11c}p_j = \frac 1N \sum_{\alpha =1}^N p_{\alpha j}\  ,\; \  \xi _{\alpha j}=p_{\alpha j} - p_j \  , \; \ \sum_{\alpha = 1}^N \xi _{\alpha j} = 0  \  ,
\end{equation}
with $p_{\alpha j} = - i\hbar \left( \partial / \partial x_{\alpha j} \right)$. We also introduce

\begin{equation}
D_j = \frac 1N \sum_{\alpha = 1}^N \frac{\partial}{\partial x_{\alpha j}} \ , \;\; D_{\alpha j} = \frac{\partial}{\partial x_{\alpha j}} - D_j \ , \;\; \sum_{\alpha = 1}^N D_{\alpha j} = 0 \ .
\end{equation}
In addition
\begin{equation}
\sum_{\alpha = 1}^N \frac{\partial^2}{\partial x_j^2} = ND_j^2 + \sum_{\alpha = 1}^N D_{\alpha j}^2 \ .
\end{equation}

When $d = 3$ we denote by $L_{\alpha 3}$ the projection of the angular momentum of the $\alpha $ particle on axis $3$. Then, the projections of the total angular momentum and of the angular momentum of the relative motion on the axis $3$, denoted by $L_3$ and $L_3^{\prime }$ respectively, satisfy  

\begin{subequations}\label{mp12} 
	\begin{equation}
	\label{mp12a}\sum_{\alpha = 1}^N L_{\alpha 3} = L_3 + L_3^{\prime }\; , \; \; L_{\alpha 3} = x_{\alpha 1} p_{\alpha 2} - x_{\alpha 2} p_{\alpha 1} \  ,
	\end{equation}
	\begin{equation}
	\label{mp12b}L_3 = p_1 D_2 - p_2 D_1 \  ,\;  L_3^{\prime } = \sum_{\alpha = 1}^N \left( 	y_{\alpha 1} \xi_{\alpha 2} - y_{\alpha 2} \xi_{\alpha 1} \right) \ .
	\end{equation}
\end{subequations}

In case of a quadrupole combined (Paul and Penning) trap that exhibits cylindrical symmetry, for a constant axial magnetic field $B_0$, the Hamilton function for a system of $N$ ions of mass $M$ and equal electric charge $Q$ can be cast into \cite{Major05, Mih18, Mih11}

\begin{equation}\label{mp13}
H = \sum_{\alpha = 1}^N \left[ \frac 1{2M}\sum_{j=1}^3 p_{\alpha j}^2 + \frac{K_r}2 \left( x_{\alpha 1}^2 + x_{\alpha 2}^2 \right) + \frac{K_a}2  x_{\alpha 3}^2 - \frac{\omega_c}2  L_{\alpha 3}\right] + W \ ,
\end{equation}
with
$$
K_r = \frac{M\omega_c^2}4 - 2 Q c_2 A \left( t\right) \ ,\; K_a = 4 Q c_2 A\left( t\right) \ ,\;\omega_c = \frac{Q B_0}M\ ,
$$
where the index $r$ corresponds to radial and index $a$ corresponds to axial, $\omega_c$ is the cyclotron frequency in the Penning trap, $c_2$ depends on the trap geometry and $A\left( t\right) $ represents a function that is periodic in time \cite{Mih18}. Then, $H$ can be expressed as the sum between the Hamiltonian of the CM of the system $H_{CM}$ and the Hamiltonian $H^{\prime }$ which characterizes the relative motion of the ions \cite{Mih21}:

\begin{subequations}\label{mp14}
	\begin{equation}\label{mp14a}
	H = H_{CM} + H^{\prime }\ ,
	\end{equation}
	\begin{equation}\label{mp14b}
	H_{CM} = \frac 1{2NM} \sum_{j=1}^3 p_j^2 + \frac{NK_r}2 \left( x_1^2 + x_2^2 \right) + \frac{NK_a}2 x_3^2 - \frac{\omega_c}2 L_3\ ,
	\end{equation}
	\begin{equation}\label{mp14c}
	H^{\prime} = \sum_{\alpha = 1}^N \left[ -\frac{\hbar ^2}{2M} \sum_{j=1}^3\xi_{\alpha j}^2 + \frac{K_r}2 \left( y_{\alpha 1}^2 + y_{\alpha 2}^2\right) +\frac{K_a}2y_{\alpha 3}^2\right] -\frac{\omega_c}2L_3^{\prime} + W \ .
	\end{equation}
\end{subequations}

The model we introduce here is based on Ref. \cite{Mih21}. It achieves a unitary approach aimed at generalizing the parameters for different types of 3D QIT and we apply this model to investigate the particular case of a combined Paul and Penning 3D QIT \cite{Li98}.         

We assume $W$ as an interaction potential that is invariant to translations (it only depends of $y_{\alpha j}$). The ion distribution within the trap can be illustrated by standard numerical programming \cite{Stee14, Aksa20, Sing10} using the Hamilton function given below   

\begin{multline}\label{mp15}
H_{sim} = \sum_{i=1}^n \frac 1{2M} \vv{p_{i}}^{2} + \sum_{i=1}^n \frac M2\left( \omega_1^2 x_i^2 + \omega_2^2 y_i^2 + \omega _3^2z_i^2\right) + \sum_{1 \leq i < j \leq n}\frac{Q^2}{4 \pi \varepsilon_0}\frac 1{\left| \vec r_i - \vec r_j \right| }\ ,
\end{multline}
where the second term describes the effective electric potential of the 3D QIT and the third term accounts for the Coulomb repulsion between the ions. 

The dynamics of systems of N ions confined in a 3D ion trap can be studied locally in the neighbourhood of the minimum configurations that describe ordered structures (Coulomb or ion crystals \cite{Kel19, Bahra19}). In order to perform a global description collective models with a small number of degrees of freedom can be introduced, or the Coulomb potential can be approximated with specific potentials for which the $N$-particle potentials are integrable. Small enough perturbations maintain the quantum stability although the classical system generally exhibits a chaotic behaviour. The system of $N$ ions that interacts via the Coulomb force also exhibits a continuous part of the energy spectrum, for which the classical counterpart consists of families of chaotic orbits. However, a weak correspondence can be established between the classical and quantum chaotic dynamics by using Husimi functions. Thus, it comes naturally to characterize quantum ion crystals \cite{Kel19, Yos15, Marci12} by means of the minimum points of the Husimi function \cite{Mih18, Mih11}. The equilibrium configurations can be determined by standard numerical programming \cite{Sing10}.

\subsection{Applications: The electrodynamic trap with corona discharge} 

Multiple experiments have been performed under SATP conditions, in an attempt to generate structures of electrically charged dust particles in the thermal plasma of a gas burner or in a corona discharge. In case of a thermal plasma Coulomb crystals do not occur \cite{Fort05} due to strong gas flows and high temperature gradients that result in fast destruction of the dust particles in the high temperature gas. Since many years the corona discharge was used to remove dust in electrostatic precipitators, to separate particles and to achieve colouring of powder. Therefore, the mechanisms used to electrically charge dust particles and their subsequent dynamics in the corona discharge are quite well known \cite{Cha95}. Nevertheless, stable dusty structures in the corona discharge have not been obtained yet. 

The possibility to use a corona discharge in a nuclear excited dusty plasma with an aim to achieve better stability of the plasma-dust structures and to accomplish a more efficient conversion of nuclear energy into radiation, is investigated in Ref. \cite{Fort10b}. The electric field distribution and other characteristics of the corona discharge in a nuclear excited dusty plasma are obtained in the special case of a cylindrical geometry, at pressures ranging between $1 \div 100$ atm. A mathematical model is developed to describe the behaviour of dust particles in a nuclear excited plasma. The model can characterize the following basic physical processes: (1) screening of Coulomb forces, (2) the energy exchange and the stochastic nature of the interaction of dust particles with a buffer gas and the surrounding plasma, and (3) the strong spatial inhomogeneity of a nuclear excited plasma. Calculations for the potential and nonpotential forces that act on dust particles have been carried out. In case of potential forces, stationary crystal-like dust structures are observed. In case of nonpotential forces, ordered vortices rotating towards each other are reported \cite{Ryk02a}. The production of well ordered plasma-dust structures out of a fissionable material is also shown to be possible under conditions of a corona discharge, in a nuclear excited plasma. Experimental studies of an argon-xenon gas mixture \cite{Ryk02a} demonstrate that micron sized dust particles can evolve into levitating stable and rotating ordered structures, whilst even the occurrence of dust crystals in an adequate trap is demonstrated.  

Ref. \cite{Lapi13} investigates the possibility and the physical conditions required to confine dust particles in a linear Paul trap operating under SATP conditions, in a corona discharge plasma or in a nuclear-excited plasma. The behaviour of dust particles can be simulated by means of the Brownian dynamics (BD) \cite{Chav11}. Numerical simulations are carried out for an adequate choice of the dust particle parameters and of the a.c. trapping voltage. Optimum values for the dust particle parameters and for the trapping voltage are established, with an aim to achieve stable confinement. The simulations performed allow one to identify the regions of stable confinement and the influence of the particle mass, electric charge, trapping voltage and frequency. The results of numerical simulations are in good agreement with the experimental results obtained in case of a linear Paul trap \cite{Lapi13}. 

Theoretical investigations have been performed on the virial equations that characterize a Coulomb system of charged particles levitated in a linear Paul trap, showing a clear dependence on particle and trap parameters \cite{Lapi19a, Syr19a}. The dynamics of pair correlation functions for a Coulomb system in an electrodynamic trap is thus obtained, as the system is highly nonideal and exhibits a coupling parameter of the order $\Gamma = 10^{10}$. In the framework of the statistical theory of liquids, the thermodynamic parameters of a strongly nonideal Coulomb microparticle structure confined in a linear Paul trap (operated under SATP conditions) are calculated using the BD method. The Coulomb potential of the inter-particle interaction and the calculated pair correlation functions of the Coulomb structure are used. The average inter-particle interaction parameter (the coupling parameter $\Gamma$), the internal energy of the Coulomb structure, and the pressure it imposes on the trap are inferred. It has been found that these parameters decrease with increasing size and charge of particles, caused by an increase in the average equilibrium inter-particle distance in the electrodynamic trap. As the system approaches a steady state the energy and pressure also drop off due to an increase in the average inter-particle distance, owing to a partial ordering of the Coulomb system of particles \cite{Lapi19b}. 

An electrodynamic trap that creates a corona discharge is investigated in \cite{Vlad18}. Neutral dust particles are injected into the trap where they are electrically charged in the plasma of a corona discharge. As a result the trap confines charged particles in presence of a plasma cloud. The electrodynamic trap consists of four horizontally positioned electrodes, made of wire. Each electrode is 10 cm long with a diameter of 300 $\mu$m. The distance between the electrodes is 10 mm. The Corona discharge between the electrodes is ignited using a voltage of 5 kV at a frequency of 50 Hz. The experiments use polydisperse Al$_2$O$_3$ powder. To observe the associated dynamics, particles are illuminated using a laser that operates at a wavelength of 532 nm and a maximum power of 150 mW. Particles are recorded by means of a HiSpec 1 digital camera.

\begin{figure}[bth] 
	\centering
	\includegraphics[scale=0.5]{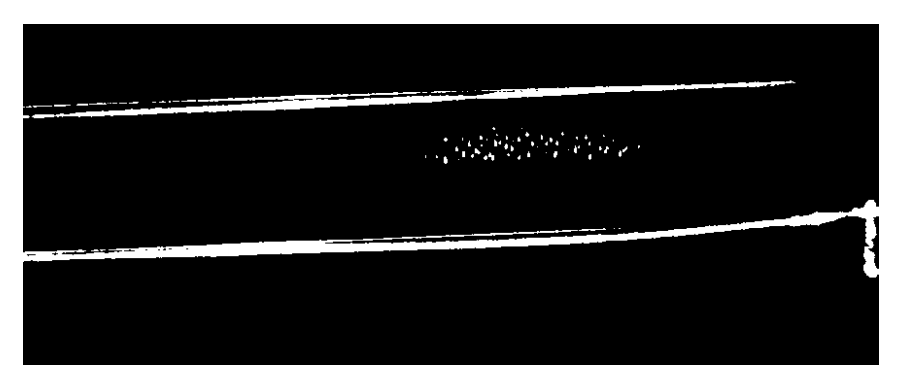} 
	\caption{Stable structure of aluminium oxide particles in a linear electrodynamic trap with corona electrodes. Source: image reproduced from \cite{Vlad18} with kind permission of the authors.}
	\label{coroneld}
\end{figure}

The electrically neutral particles are injected inside the trap from the upper side. The particles acquire electric charge in the field of a corona discharge during their downwards motion. Then, some of the particles are captured inside the trap. For particles with a diameter larger than 1 $\mu$m, the main charging mechanism is field charging. In this case, the charge of the particle can be estimated using the expression:
$Q = 3 \varepsilon_0 E d^2 \left( 1 + 2\frac{\varepsilon - 1} {\varepsilon + 2}\right) $ \cite{Pau32}, where $E$ is the electric field intensity, $\varepsilon_0$ represents the (absolute) vacuum permittivity, $d$ is the size of the particle and $\varepsilon$ stands for the relative permittivity. The electric field intensity essentially depends on the particle position inside the trap. For example, a 10 $\mu$m diameter particle located near an electrode can accumulate up to $4\times 10^4$ units of electron charge $e$. The captured particles create a stable structure near the trap axis. The presence of electric wind caused by direct movement of the ions complicates the particle confinement mechanism in a trap with corona discharge, with respect to the case of a classical electrodynamic trap. Fig.~\ref{coroneld} shows an image of a stable structure of aluminium oxide particles levitated in a linear electrodynamic trap with corona discharge. The amplitude of the a.c. voltage is 7 kV and the electric current value is around 80 $\mu$A. The trapped particle size lies between $10 \div 40 \ \mu$m \cite{Vlad18}. The dimensions of the ordered structure are $1.4 \times 0.2 \times 0.15$ cm, and it consists of approximately 55 particles. The average inter-particle distance is 0.8 mm.

\subsection{Applications: Charged microparticle confinement by oscillating electric fields in a gas flow}\label{gasflow}

Dust particles (microparticles) are often present in the atmosphere around nuclear power plants or in energy devices, such as etching and fusion installations. Removal of microparticles out of these devices is an important issue. One of the ways to improve the filtration efficiency lies in perturbing the microparticles by various physical factors with an aim to alter their physical properties. For example, in case of electrostatic filters, the microparticles build up electric charge in the corona discharge and then accumulate at the electrodes \cite{Peuk01}. Under certain conditions a corona discharge sensibly enhances the efficiency of electrostatic filters, thus providing a higher degree of cleaning \cite{Von67, Fink89}. However, electrostatic filters are not efficient in capturing microparticles with dimensions ranging between $0.6 \div 1.6 \ \mu$m. Unfortunately, the issues of selective particle removal cannot be solved by corona discharge precipitators.

Confinement of electrically charged microparticles in an electrodynamic trap is investigated in gas media under static, SATP conditions \cite{Vasi13, Lapi13}. The regions of particle capture are explored for a wide range of parameters such as: the microparticle electric charge and mass, the radius, the oscillating electric field strength and frequency. Different versions of improved linear electrodynamic trap geometries that achieve a more effective capture of microparticles are described in \cite{Ghe98, Sto01, Mih08, Lapi13, Ghe96a}. 

\begin{figure}[bth]
	\centering
	\includegraphics[scale=0.9]{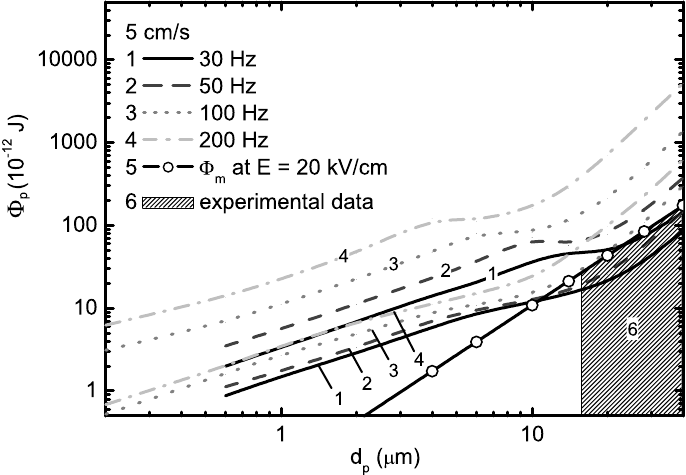} 
	\caption{Regions of confinement versus the microparticle diameter $d_p$. The regions are located between the upper and lower bounds designated by the corresponding lines, for different electric field frequencies: $1$ -- $f$ = 30 Hz, $2$ -- $f$ = 50 Hz, $3$ -- $f$ = 100 Hz, $4$ -- $f$ = 200 Hz. Line $5$ that describes $\Phi_m$ and the region denoted as 6 are related to the trap and particle parameters used in the experiments described in \cite{Lapi15c}. Source: picture reproduced (modified) from \cite{Lapi15c} under kind permission of the authors.}
	\label{Hz}
\end{figure}

Ref. \cite{Lapi15c} presents theoretical and experimental studies of charged microparticle capture in a linear Paul trap, in gas flow. The regions of microparticle and trap parameters required to achieve stable confinement are investigated under normal SATP conditions. The model used shows that the interaction force between the microparticle and the trap electrodes (2) depends on: 

\begin{equation}\label{intforce}
\Phi_p = \frac{U_{ac}Q_p}{2 \ln(R2/R1)} \  .
\end{equation}
Numerical simulations performed in \cite{Lapi15c} consider a microparticle characterized by a density $\rho_p = 3.99$ g/cm$^3$. We mention that $U_{ac}$ stands for the trapping voltage, while  $R_2$ and $R_1$ represent the radii of the grounded cylindrical shell surrounding the trap and trap electrode, respectively.

Fig.~\ref{Hz} presents the confinement regions for charged microparticles (areas $1$, $2$, $3$, $4$) in case of a gas flow velocity value around $5$ cm/sec. The lower and upper bounds of the particle confinement regions are denoted by the corresponding lines. Above the upper bound the electric field is strong enough to push microparticles out of the trap. Below the lower bound the particle cannot be captured as the trap field cannot compensate gravity forces. The black line (denoted as 5) corresponds to 

\begin{equation}
\Phi_m = \frac{U_{ac}Q_m}{2 \ln(R2/R1)}, \  Q_m = 3 \pi \varepsilon_0 d^2 E \left( 1 + 2\ \frac{\varepsilon - 1} {\varepsilon + 2}\right) \ ,
\end{equation}
where $Q_m$ is the maximum electric charge that a particle can accumulate in the corona discharge \cite{Pau32}. In the simulations performed the magnitude of the electric field strength in the corona discharge was chosen $E = 20$ kV/cm. The shaded region (denoted as 6) characterizes microparticles used in the experiments discussed below.

From Fig.~\ref{Hz} one can draw the following conclusions:
\begin{itemize}
	\item the regions of confinement for a particle with a diameter less than 10 $\mu$m are located above curve $5$, hence these particles cannot be trapped,
	\item confinement regions for frequencies ranging between $f \sim 30 \div 50$ Hz (areas $1$ and $2$) start for particle dimensions around $d \sim 0.6 \ \mu$m; in order to capture a particle of smaller dimensions, the electric field frequency has to reach a value of about $\sim 100$ Hz (areas $3$ and $4$), 
	\item as the electric field frequency rises, the voltage amplitude value required to achieve particle capture also increases. 
\end{itemize}

\begin{figure}[bth]
	\centering		
	\includegraphics[scale=0.35]{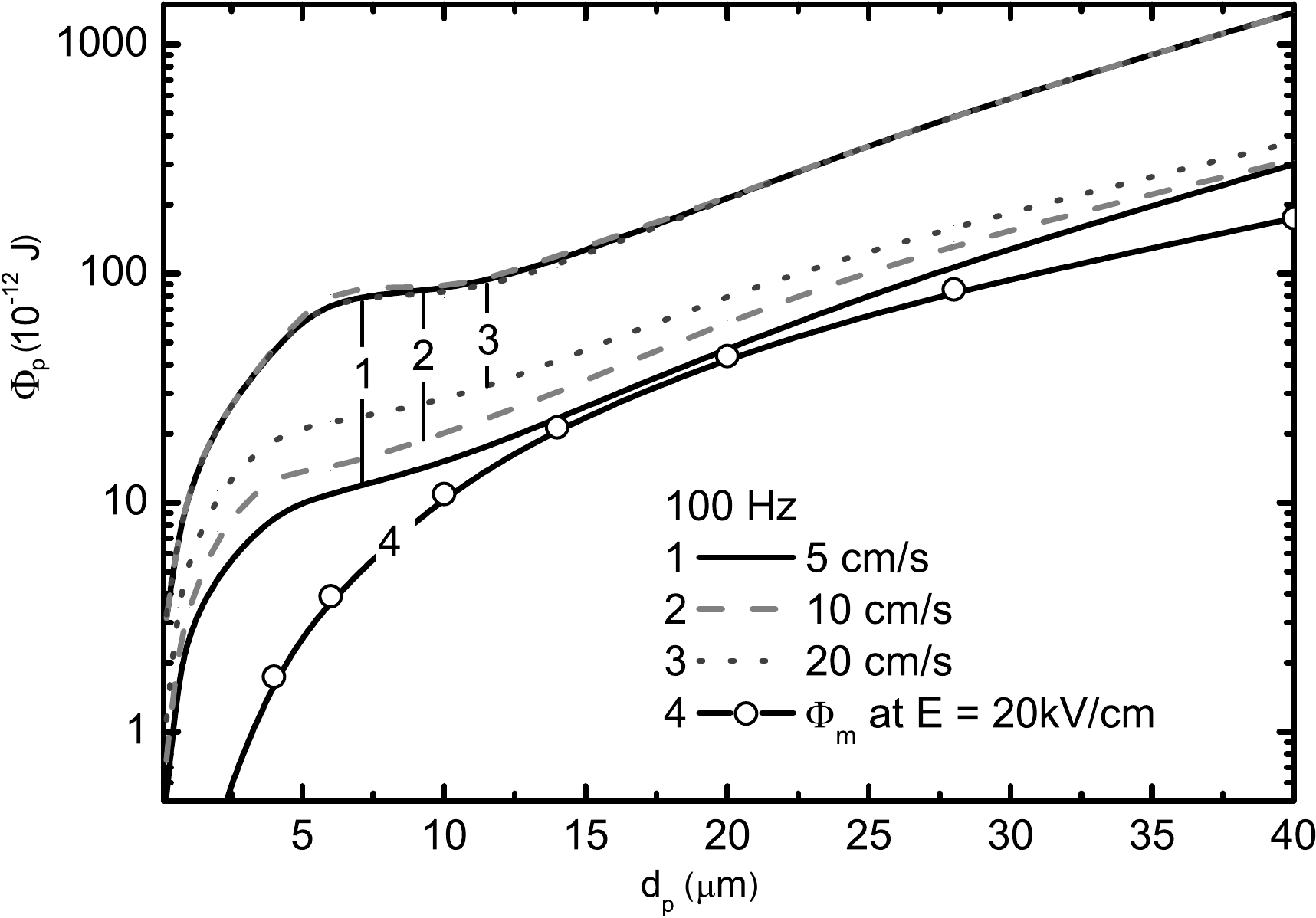}
	\caption{Regions of confinement (between corresponding lines) versus microparticle diameter $d_p$ for different gas flow velocities: $1$ -- $v_f = 5$ cm/s, $2$ -- $v_f = 10$ cm/s, $3$ -- $v_f = 20$ cm/s. Line $4$ corresponds to $\Phi_m$. Source: picture reproduced from \cite{Lapi15c} under kind permission of the authors.}
	\label{Hz2}
\end{figure}

Fig.~\ref{Hz2} shows the analogous regions of confinement versus microparticle diameter at frequency $f = 100$ Hz, for gas flow velocities equal to $5$ cm/sec, $10$ cm/sec and $20$ cm/sec, respectively. Fig.~\ref{Hz2} also illustrates that the confinement region becomes narrower by increasing the gas flow velocity, owing to the shift of the lower bound (border) upwards for virtually an unmodified position of the upper bound.

\begin{figure}[bth]
	\centering	
	\includegraphics[scale=0.8]{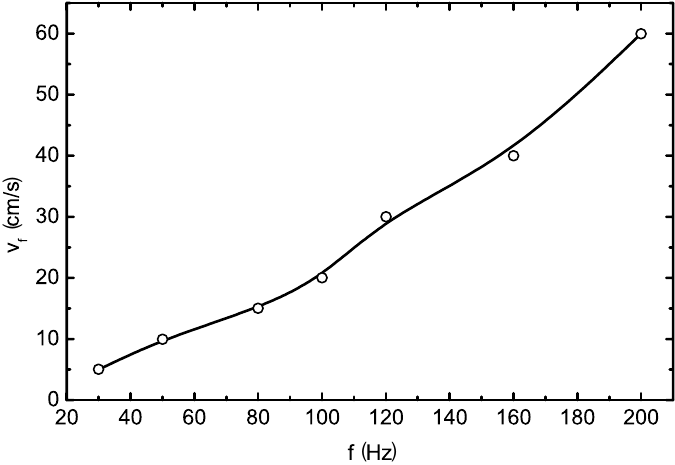}
	\caption{The upper bound of the particle capture region. $\Phi_p = 5 \times 10^{-13} \div 5.3 \times 10^{-9}$ J ($q_p = 4 \times 10^3 \div 4.2 \times 10^7 \ e$, $U_\omega = 8$ kV, $d_p = 2 \ \mu$m). Source: picture reproduced from \cite{Lapi15c} under kind permission of the authors.} 
	\label{Hz3}
\end{figure}

Fig.~\ref{Hz3} shows the upper bound of the particle capture region as a function of the gas flow velocity versus the electric field frequency (black line). The physical reason for the black line shift upwards lies in the fact that all dynamical time scales associated to the particle motion in the trap should be reduced by augmenting the gas flow velocity. 

\subsubsection{Experimental setup for the study of particle capture in gas flows}

\begin{figure}[bth]
	\centering		
	\includegraphics[scale=0.45]{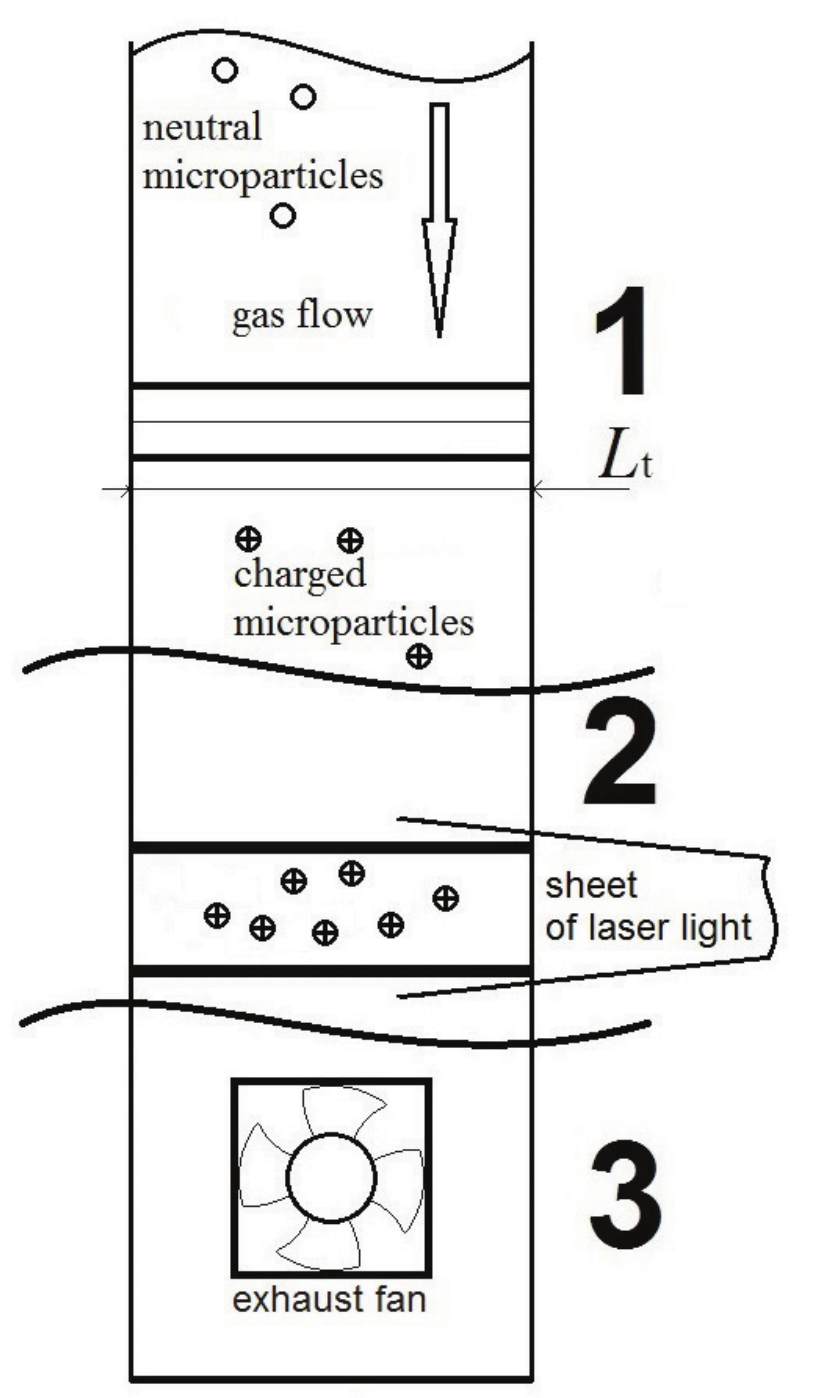}
	\caption{Sketch of an experimental setup used for charged microparticle capture in a linear trap, in gas flow. The gas channel consists of 3 modules: $1$ represents the corona discharge module, $2$ is the trap module and $3$ stands for the air-exhaust module. Source: image reproduced (modified) from \cite{Lapi15c} under kind permission of the authors.}
	\label{ExSheme}
\end{figure}

To study the physical mechanisms responsible for particle capture, the charged particles are required to drift as an outcome of the gas flow across the trap. An image of the experimental setup is presented in Fig.~\ref{ExSheme}. The setup consists of three separate modules located in a gas channel: $1$ represents the corona discharge module used to charge microparticles, $2$ is the trap module and $3$ stands for the air-exhauster module.

The corona discharge module consists of one row of discharge electrodes along with two rows of grounded electrodes, located above and below the discharge electrodes, at a distance of 12 mm apart. The discharge electrodes consist of wires with a diameter of $70 \ \mu$m, arranged at a distance of 1 cm apart with respect to each other. A d.c. voltage $U_c$ with a peak value up to $15$ kV is applied to the discharge electrodes. The grounded electrodes are made of metal rods with $d = 3$ mm diameter. This particular design generates an ion breeze (wind) in two opposite directions, thus compensating its influence on the column of air within the channel. 

Diagnostics and observation of microparticles is achieved using a laser light sheet that exhibits a 1 cm spot. The sheet height allows microparticle observation both in the trap and outside of it. The laser light sheet parameters are: wavelength $532$ nm and power up to $230$ mW. The microparticles are recorded using a HiSpec 1 Fastec Imaging camera located along the laser light sheet (Fig.~\ref{ExSheme}). 

The air-exhaust module is located at the end of the gas channel. The exhaust fan blows out (generates) a gas flow, with velocities that reach up to $v_f = 50 \pm 11$ cm/s. Polydisperse aluminium oxide Al$_2$O$_3$ powder is used in the experiments \cite{Lapi15c}. The microparticle density is $\rho_p = 3.99$ g/cm$^3$ for a typical size ranging between $4 \div 80 \ \mu$m. 

\begin{figure}[bth]
	\centering		
	\includegraphics[scale=2.2]{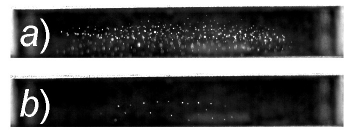}
	\caption{Structure of charged microparticles captured by an electrodynamic trap ($a$) in a static gas media ($v_f = 0$ cm/s) and ($b$) in a gas flow ($v_f = 50$ cm/s). Source: image reproduced from \cite{Lapi15c} under kind permission of the authors.} 
	\label{StrNoStr}
\end{figure}

The experiment starts by turning off the exhaust fan and by supplying a d.c. voltage $U_c = 15$ kV to the discharge electrodes. Then, particles are injected into the corona module. By travelling through the corona discharge area the particles acquire a positive charge $q$ and fall inside the trap module, where they are captured between the electrodes. Fig.~\ref{StrNoStr}(a) shows an example of a stable Coulomb structure of charged microparticles captured in the trap. The captured microparticles oscillate around equilibrium positions at a frequency of 50 Hz. Most of the particles are captured inside the trap, below the central axis, while others are trapped above it.

To study the gas flow effect on an ensemble of captured microparticles, the exhaust fan is turned on. Most of the microparticles are blown out of the trap and only a few particles remain trapped inside (see Fig.~\ref{StrNoStr}(b)). The particle structure shifts below the central axis of the trap if the inter-particle distances increase. 

\begin{figure}[bth]
		\centering		
		\includegraphics[scale=1]{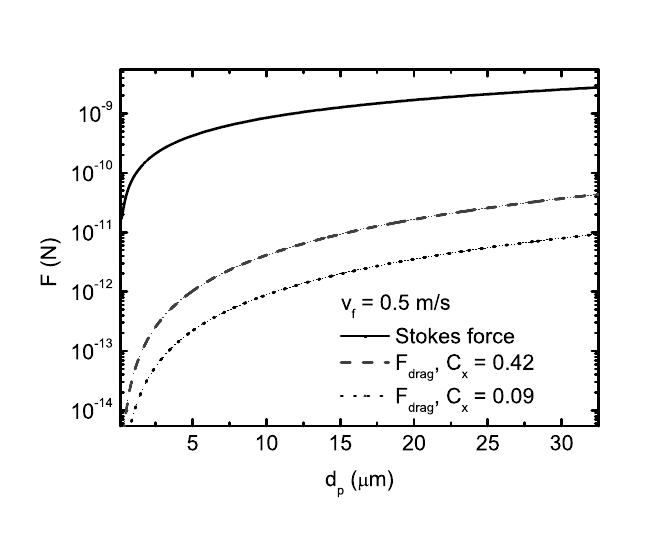}
		\caption{Dependence of Stokes and drag forces on microparticle diameter $d_p$. Source: picture reproduced (modified) from \cite{Lapi15c} under kind permission of the authors.}
		\label{Stok}	
\end{figure}

We emphasize that the regions of particle trapping observed in the experiments turn out to be larger with respect to those obtained from numerical simulations. Thus, from Fig.~\ref{Hz3} it follows that particle capture at a frequency $f = 50$ Hz is possible if the gas flow velocity is lower than $v_f$ = 10 cm/s, while in experiments the capture process is observed for $v_f$ = 50 cm/s. To explain this deviation let us note that in simulations the drag force of gas flow is estimated using the Stokes formula for spherical particles, while in the experiments performed the particle shape varies from spherical to slab. An alternative estimation of the drag force can be obtained from Newton's regime \cite{Soo67} for particles of different shapes. In this case the drag force is characterized by the expression $F_{\mathrm{drag}} = \rho_f {v_f}^2 S_p C_x/2$, where $\rho_f$ is the mass density of the gas, $v_f$ is the velocity of a microparticle relative to the velocity of the gas flow, $S_p$ is the reference area of the microparticle and $C_x$ is the drag coefficient that lies between $0.09 \div 1.15$. The dependence of the Stokes force and drag force $F_{\mathrm{drag}}$ versus the microparticle diameter is illustrated in Fig.~\ref{Stok}, for a value of the gas velocity $v_f = 50$ cm/s. In Fig.~\ref{Stok} the characteristic Stokes force is at least ten times larger than the drag force $F_{\mathrm{drag}}$, under identical conditions. Therefore, to achieve microparticle capture the gas velocity value could be higher while the value of $\Phi_p$ might be lower. The capture regions in Fig.~\ref{Hz} and Fig.~\ref{Hz2} are shifted downwards, while the capture region in Fig.~\ref{Hz3} is shifted upwards. Numerical estimations of the gas velocity for dust particle capture validate experimental results. 

\begin{figure}[bth]
	\centering		
	\includegraphics[scale=0.55]{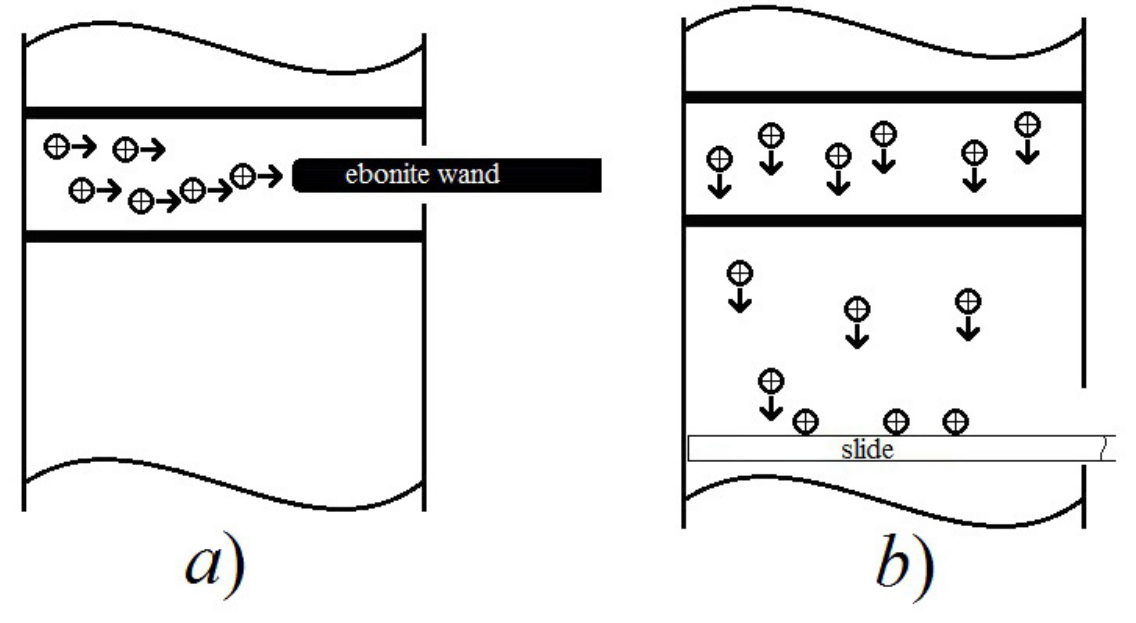}
	\caption{Picture of the mechanism of captured microparticles collection: ($a$) by using an electrically charged ebonite wand and ($b$) by means of a subject glass slide. Source: picture reproduced (modified) from \cite{Lapi15c} under kind permission of the authors.}
	\label{GetP}
\end{figure}

To measure the dimensions of microparticles captured in the electrodynamic trap we employ two methods that are explained in Fig.~\ref{GetP} (a) and (b). The first method (see Fig.~\ref{GetP} (a)) uses a hole in the wall of the gas channel between the trap electrodes. During the microparticle confinement experiment the hole is closed. When microparticles are captured the exhaust fan is turned off, the hole is opened and the clean ebonite wand that is initially negatively charged due to friction is inserted through the hole inside the gas channel. Positively charged microparticles are then captured by the ebonite wand, that is extracted from the gas channel afterwards. Then, the ebonite wand is discharged and microparticles are removed by shaking it, and precipitate on the subject glass. The second method (see Fig.~\ref{GetP} (b)) uses a slit in the wall of the gas channel under the trap. The slit is closed during the experiment. Following microparticle capture the exhaust fan is turned off and the slit is opened. The clean subject glass is inserted within the gas channel. The electrodynamic trap is then turned off and microparticles fall down on the subject glass. In case of both methods used the microparticle dimensions are measured by means of a microscope.

\begin{figure}[bth]
	\centering		
	\includegraphics[scale=0.85]{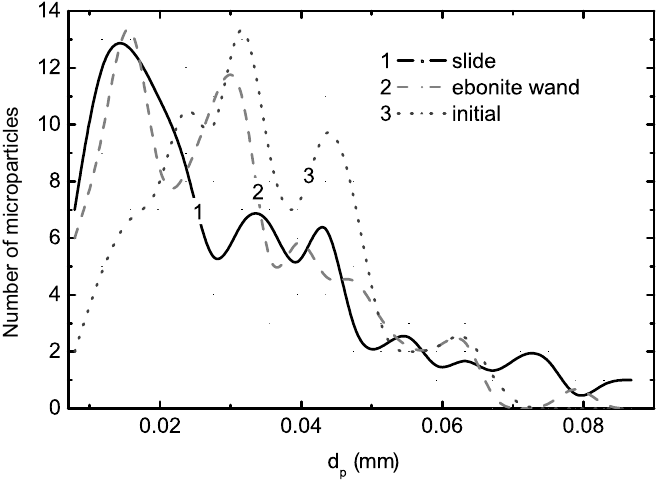}
	\caption{Initial and captured microparticle distributions: $1$ represents the distribution of particles on the subject glass; $2$ stands for the distribution of particles entrained by the ebonite wand; $3$ is the initial distribution. The average diameter of the microparticles in the experiment (curves $1$ and $2$) is $d_p = 32 \ \mu$m, with rms deviations $\sigma = 0.02$ and $\sigma = 0.01$, respectively. Source: picture reproduced (modified) from \cite{Lapi15c} under kind permission of the authors.}
	\label{Histo}
\end{figure}

In contrast to the simulations performed where spherical shaped particles are considered, the experiments use powder of polydisperse microparticles of complex shape. To define the effective particle size for each microparticle, the smallest particle size $d_{\mathrm{min}}$ and the largest one $d_{\mathrm{max}}$ are considered. The effective microparticle diameter is defined as $d_p = \frac{d_{\mathrm{min}} + d_{\mathrm{max}}}{2}$. The accuracy of the microparticle diameter measurement is around 4 $\mu$m. Fig.~\ref{Histo} illustrates three distributions of microparticles versus the effective particle size. The first one represents the initial distribution $P_i\left(d_p\right)$ of particles introduced in the flow. The second distribution refers to microparticle dimensions obtained by using the ebonite wand $P_e\left(d_p\right)$, while the third distribution is obtained by using the subject glass slide $P_g\left(d_p\right)$.

\begin{figure}[bth]
	\centering
	\includegraphics[scale=0.95]{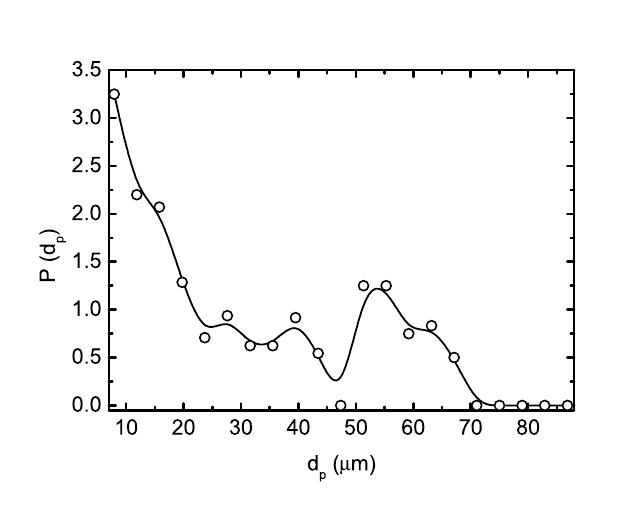} 
	\caption{Relative distribution of the particles versus their effective size. Source: picture reproduced (modified) from \cite{Lapi15c} under kind permission of the authors.}
	\label{Ratio}
\end{figure}

Fig.~\ref{Ratio} presents the function $P\left(d_p\right)$, described by the expression 

\begin{equation}
P\left(d_p\right) = \frac{P_e\left(d_p\right) + P_g\left(d_p\right)}{2P_i\left(d_p\right)}\ .
\end{equation}
 
The function $P\left(d_p\right)$ represents in fact the relative distribution of particles, depending on their effective dimensions. In Fig.~\ref{Ratio} the maximum of the captured microparticles distribution is shifted towards small sized particles with respect to the initial powder. The result indicates that the trap is selective, as the capture process is more efficient for smaller sized particles.

\subsection{Applications: Cleaning dusty surfaces using electrodynamic traps}

\begin{figure}[bth]
	\centering
	\includegraphics[scale=0.5]{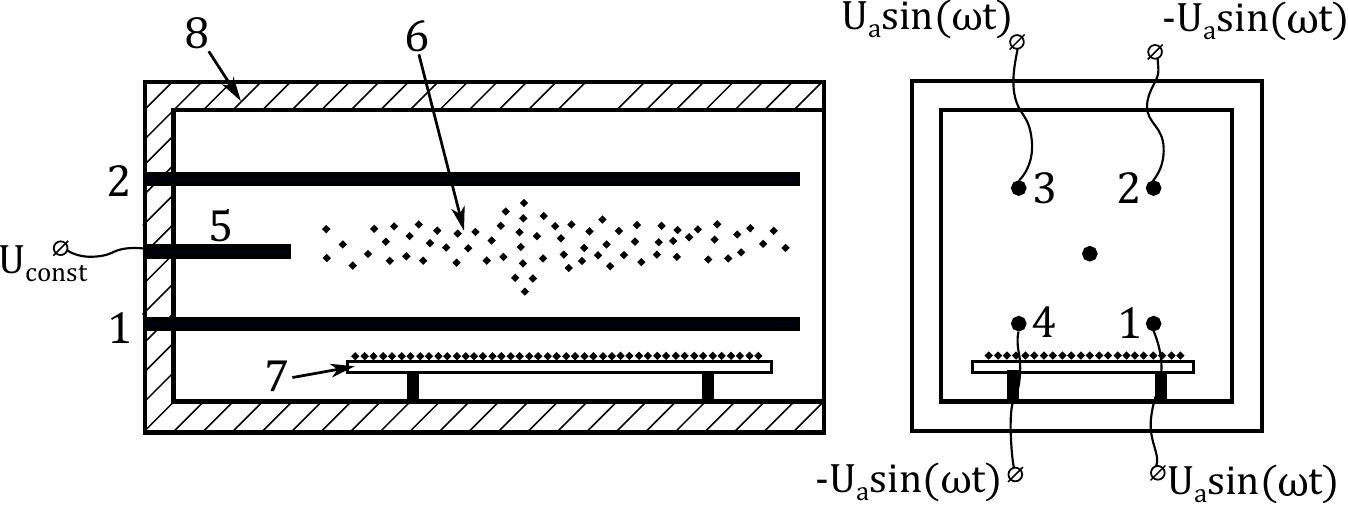} 
	\caption[]{Trap design, front and side views. Notations: 1$ \div$ 4 -- trap electrodes, 5 -- endcap electrode, 6 -- levitated particles, 7 -- glass substrate with dust particles, 8 -- optically transparent plastic box. Source: picture reproduced (modified) from \cite{Deput19} under kind permission of the authors.}
	\label{trap}
\end{figure}

An interesting application of linear electrodynamic traps is presented in Refs. \cite{Deput19, Syr21b}, more precisely a new method to achieve cleaning of dusty surfaces. A linear quadrupole electrodynamic trap (shown in Fig.~\ref{trap}) was designed and tested, consisting of four parallel cylindrical steel electrodes, each one with a diameter of 3 mm and a total length of 30 cm. The electrodes are located on the vertices of a 2 cm side square. The a.c. voltage $U_a\sin(\omega t)$ at frequency $f = \omega /2\pi = 50$ Hz is applied to the trap electrodes (where opposite electrodes are in phase), while the electric potential phase shift between neighbouring electrodes is equal to $\pi$. The electrodynamic trap is placed inside an optically transparent plastic box to guard against instabilities induced by air flows \cite{Lapi18b}. To forbid particle escape at the left edge of the trap (see Fig.~\ref{trap}) an additional electrode is mounted, supplied with a d.c. electric potential $U = 1$ kV. We remark that trapped particles are restricted to escape even when there is no endcap electrode present as a result of the repulsive potential that occurs, created by charged particles that accumulate at the left wall of the plastic box. Electrically neutral polydisperse Al$_2$O$_3$ particles are placed on a glass substrate, located below the trap. Video recording of the particles that are illuminated by means of a laser beam is achieved using a HiSpec 1 video camera. All experiments are carried out under SATP conditions. 

\paragraph{Experimental results. Capture and retention of particles} 

\begin{figure}[bth]
	\centering
	\includegraphics[scale=0.75]{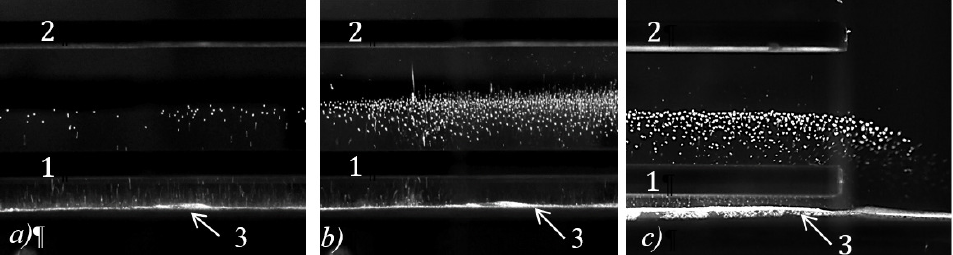} 
	\caption[]{Images of Al$_2$O$_3$ particles dragged from the glass substrate, located at a distance of 0.5 cm with respect to the trap electrodes, for different trapping voltage amplitudes: $a) U_a = 3.5$ kV, $b)$ and $c)$ $U_a = 5$ kV; 1 and 2 indicate the trap electrodes, and 3 denotes the glass substrate that holds the particles. Source: images reproduced (modified) from \cite{Deput19} under kind permission of the authors.}
	\label{fig9.2}
\end{figure}

As the amplitude $U_a$ of the a.c. voltage increases to a critical value (for example $2.5$ kV, in case of a $0.5$ cm distance between the substrate and the trap electrodes), the electric field begins to attract and then capture dust particles. Fig.~\ref{fig9.2} shows images of  Al$_2$O$_3$ particles captured in the trap. These particles are attracted from the glass substrate, located at a distance of $0.5$ cm apart with respect to the trap electrodes. By observing the space delimited by the substrate (3) and the electrode denoted by (1), one can notice how particles are dragged from the substrate towards the trap. Particles captured in the trap that slowly move towards the free end are also visible. When the $U_a$ voltage value rises from $3.5$ kV (Fig.~\ref{fig9.2}$a$) to $5$ kV (Fig.~\ref{fig9.2}$b$), the number of confined particles significantly increases. Fig.~\ref{fig9.2}$c$ shows the open end of the trap. We notice how trapped particles shift towards the open end, then leave the trap and fall down. This effect can be used to clean very dusty surfaces and then collect the trapped particles into a special container located at the open end.

\begin{figure}[bth]
	\centering
	\includegraphics[scale=0.9]{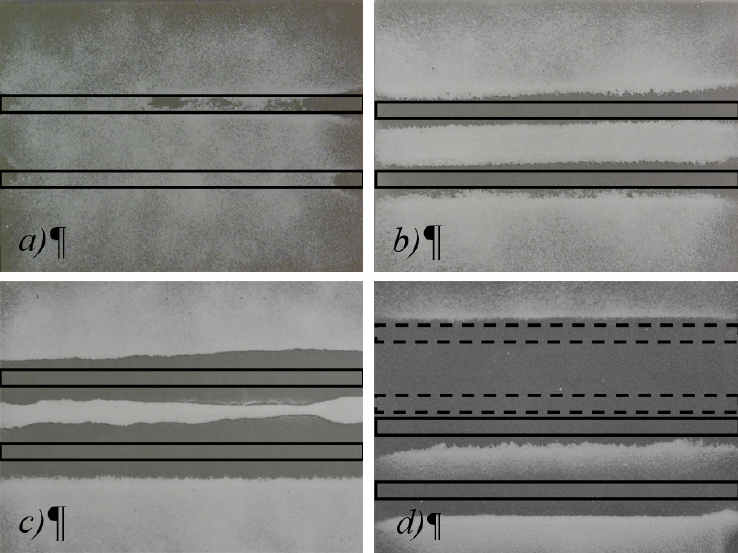} 
	\caption[]{Images of a cleaned substrate, placed at different distances $l$ apart with respect to the electrodes of the electrodynamic trap, for $U_a = 5$ kV: $a) \ l = 1$ cm, $b)$ and $d) \ l = 0.5$ cm, $c) \ l = 0.15$ cm. Projections of the electrodes are represented by solid lines. In image $d)$ the initial and final projections of the electrodes are recorded by dashed and solid lines, respectively. Source: images reproduced (modified) from \cite{Deput19} under kind permission of the authors.}
	\label{fig9.3}
\end{figure}

Fig.~\ref{fig9.3} shows images of a glass substrate located at $1$ cm distance (\ref{fig9.3}a), $0.5$ cm distance (\ref{fig9.3}b) and $0.15$ cm distance (\ref{fig9.3}c), with respect to the trap electrodes supplied at $U_a = 5$ kV. Dark areas in the images indicate substrate regions that are cleaned of particles. These areas are located near the projection of electrodes (indicated by solid lines) on the substrate. When the distance between the substrate and the electrodes decreases, the surface area rendered free of particles increases. Fig.~\ref{fig9.3}d shows an image of the substrate obtained after it was shifted normal to the axis of the electrodes, parallel to its own surface. The wide dust-free areas are clearly visible. Dashed lines denote the initial projections of the trap electrodes on the substrate while solid lines mark their final position.

\begin{figure}[bth]
	\centering
	\includegraphics[scale=0.9]{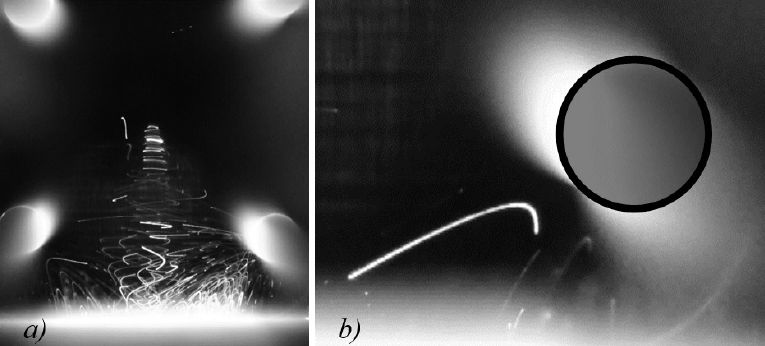} 
	\caption{Particle trajectories in an electrodynamic trap (observed from the edge of the trap). Voltage amplitudes: a) $U_a = 4$ kV, b) $U_a = 3$ kV. Source: image reproduced (modified) from \cite{Deput19} under kind permission of the authors.}
	\label{fig9.4}
\end{figure}

Fig.~\ref{fig9.4} shows an image of particle trajectories (observed from the edge of the trap) for an exposure time equal to one period $T = 2\pi/\omega$ of the a.c. voltage. As illustrated, some of the particles are dragged inside the trap. A certain number of particles exhibit curved trajectories near the electrodes, an indication of the phenomenon of particle reflection followed by a shift (movement) oriented downwards. Fig.~\ref{fig9.4}b shows an enlarged image of a single particle trajectory near the bottom right electrode (the electrode is marked by a black circle, while the white spot represents a glare). The particle is reflected when the electrode changes its polarity. An analysis of Fig.~\ref{fig9.4} allows us to advance an interesting assumption about the mechanism of particle drag into the trap. Electrical charging of dust particles might be caused by polarization and triboelectric physical mechanisms. A neutral dust particle located on a substrate is polarized by the electric field. Then, the particle begins to move and accumulates electric charge as an outcome of friction with the substrate and with other particles \cite{Low80}.

\begin{figure}[bth]
	\centering
	\includegraphics[scale=0.55]{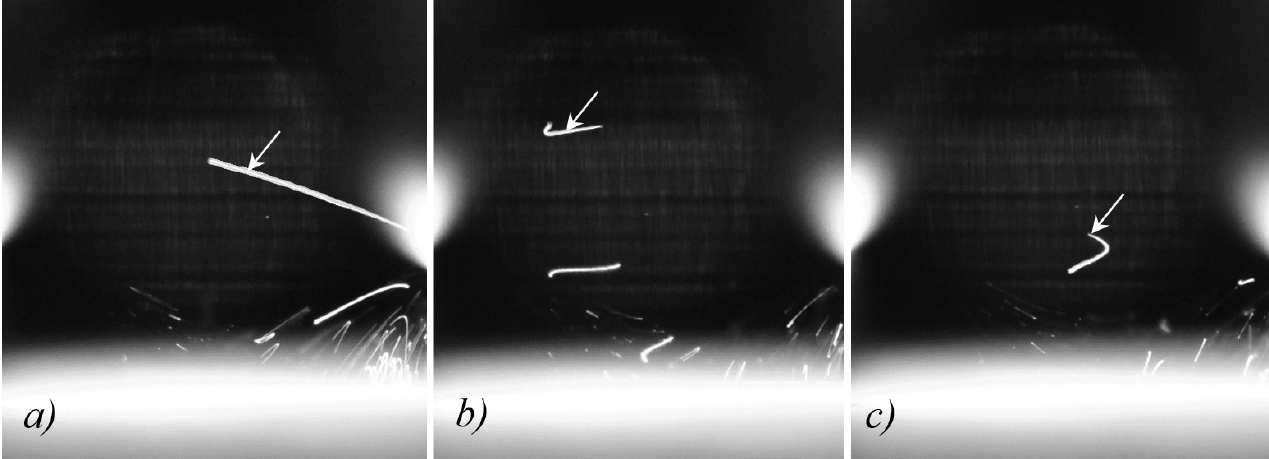} 
	\caption{Particle trajectory (marked by an arrow) for an a.c. voltage amplitude $U_a = 2.5$ kV, a value that is insufficient to achieve trapping. The time interval between two successive images is $3T$ while the exposure time is equal to $T$ (the period of the a.c. voltage). Source: images reproduced (modified) from \cite{Deput19} under kind permission of the authors.}
	\label{fig9.5}
\end{figure}

Charged particles whose velocity is directed towards the trap (for an appropriate polarity of the electrode) can be captured. However, even in this case the trap not will hold all particles. Fig.~\ref{fig9.5} shows several successive images of the trajectory of a particle whose velocity is directed towards the trap volume, after reflection from the electrode (Fig.~\ref{fig9.5}a). Since the voltage amplitude value $U_a = 2.5$ kV is too low for confinement, the particle falls down (Fig.~\ref{fig9.5}b, \ref{fig9.5}c).

\begin{figure}[bth]
	\centering
	\includegraphics[scale=0.48]{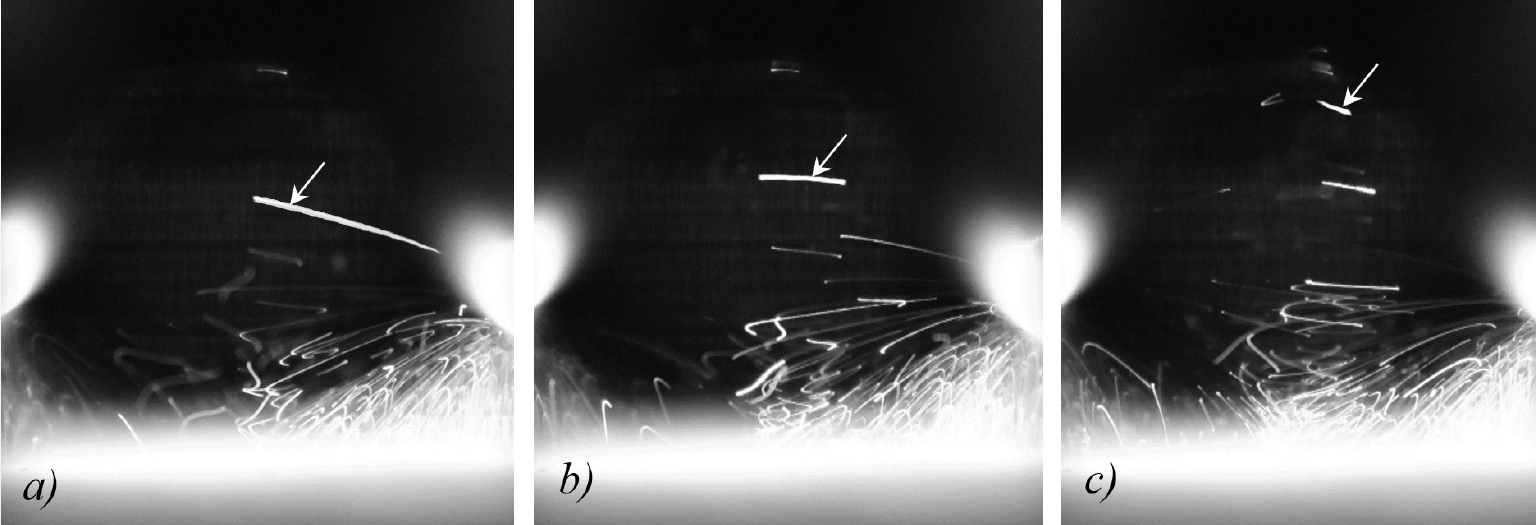} 
	\caption{Trajectory of a trapped particle (marked by an arrow) for an a.c. voltage amplitude $U_a = 3.5$ kV. The particle shifts towards the trap centre. The time interval between successive images is $3T$ while the exposure time is equal to $T$. Source: images reproduced (modified) from \cite{Deput19} under kind permission of the authors.} 
	\label{fig9.6}
\end{figure}

When the a.c. voltage amplitude increases, the conditions for particle uplift and capture are enhanced. Fig.~\ref{fig9.6} shows three images of the trajectory of a particle reflected from the electrode towards the trap volume, for a voltage amplitude value of $3.5$ kV. In this case the confining forces generated by the oscillating electric field intensify and the particle ascends towards the trap centre. Other captured particles behave in a similar manner.

\subsection{Numerical simulation of dust particles behaviour in electrodynamic traps}\label{s:model}

Experimental observations discussed above agree with the results of numerical simulations performed with respect to charged particle confinement. Two independent numerical approaches are used, namely the methods of Brownian and molecular dynamics. By using methods specific to Brownian dynamics (BD) \cite{Vasi13, Syr16a} we perform numerical simulations of the motion of charged particles for different linear trap setups. The trapping conditions for both a single particle and an ensemble of particles are thus obtained.

\begin{figure}[bth]
	\centering
	\includegraphics[scale=0.85]{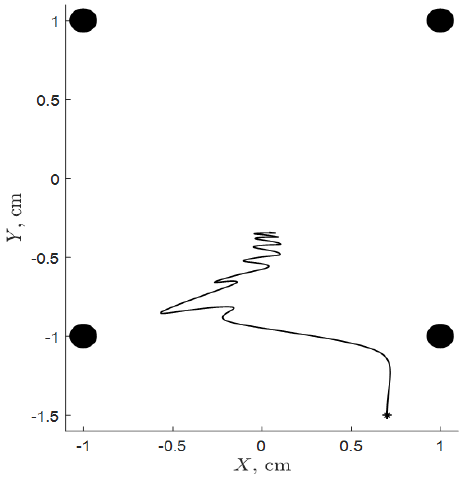} 
	\caption{Simulated trajectory of a particle whose velocity is directed upwards. The horizontal and vertical coordinates of the particle relative to the trap axis are plotted. Source: picture reproduced (modified) from \cite{Deput19} under kind permission of the authors.}
	\label{fig9.7}
\end{figure}

The idea behind simulations is to find out which conditions trigger the uplift of charged particles from the substrate, followed by their capture inside the trap. It is quite natural that for low trapping voltages the electric field is not able to lift the charged particle. Experimental results show that this issue occurs for trapping voltage amplitudes that are below 2.5 kV. As we move towards higher voltage amplitudes $U_a$, if the particle velocity at an initial moment of time is oriented towards the trap then the capture probability is high. Fig.~\ref{fig9.7} shows the simulated particle trajectory for $U_a = 3.7$ kV, by using the molecular dynamics method. At an initial moment, the particle with a radius of 20 $\mu$m and an electric charge of around $10^5e$ is placed below the bottom of the trap. The particle velocity value is about $0.5$ m/s, directed upwards. In Fig.~\ref{fig9.6} the value of the voltage amplitude under simulation is close to the experiment value of $3.5$ kV. The particle is captured inside the trap and its trajectory is similar to the one observed experimentally, as illustrated.

\begin{figure}[bth]
	\centering
	\includegraphics[scale=0.8]{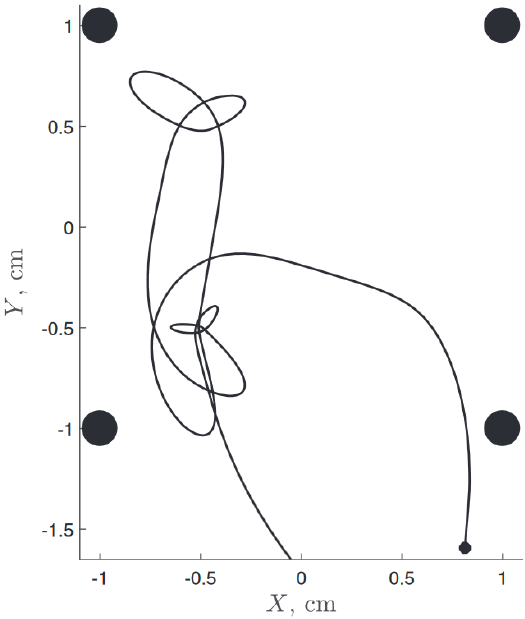} 
	\caption{Numerical simulation of the dynamics of a particle with a radius of $15 \ \mu$m and an electric charge of $2.3 \times 10^6 e$, captured inside the trap, for an a.c. voltage amplitude of $8$ kV. The small dot close to the $X$ axis denotes the initial particle position. The horizontal and vertical coordinates of the particle relative to the trap axes are plotted. Source: picture reproduced (modified) from \cite{Deput19} under kind permission of the authors.} 
	\label{fig9.8}
\end{figure}

By using numerical simulations we have also tried to establish the conditions under which the charged particle is lifted from the surface of the substrate, then enters the trap and leaves it afterwards, as experimentally observed. Fig.~\ref{fig9.8} shows a trajectory simulation for a charged particle initially located under the trap (asterisk), when $U_a = 8$ kV; the particle radius is $r = 15 \ \mu$m and the electric charge value $q = 2.3 \times 10^6 e$. Numerical simulations are performed using the BD method. The particle is lifted inside the trap where it follows a complex motion between the electrodes before quickly falling down in the end, as an outcome of an agreement in the direction of gravity and electric forces. Simulations performed allow us to identify the conditions that enable lifting and capture of charged particles inside the trap, for sufficiently large values of the electric charge and trapping voltage. Both the experiment and simulations performed demonstrate that in order to achieve particle confinement, the electric charge should be about $3 \times 10^5 e$ while the voltage amplitude range lies between $3 \div 5$ kV. For higher values of these parameters ($10^6e$, $8$ kV) the efficiency of particle capture deteriorates significantly. This conclusion is important in practice, especially when using linear traps to clean dusty surfaces, since it indicates that an increase of the oscillating voltage amplitude does not always result into enhanced efficiency for particle capture mechanisms.

\begin{figure}[bth] 
\centering
\includegraphics[scale=0.6]{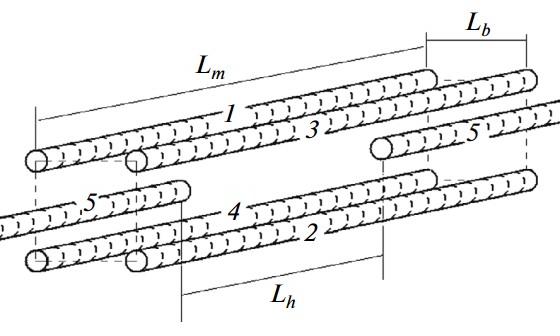}	
\caption{Sketch of a linear quadrupole trap. 1, 2, 3, and 4 denote the trap dynamic electrodes, while 5 stands for the endcap electrodes. Source: picture reproduced (modified) from \cite{Vasi13} with kind permission of the authors.}
\label{fig:trap}
\end{figure}

In order to establish the regions of stable particle confinement, mathematical (numerical) simulations on the dynamics of a single particle or of an ensemble of charged particles confined in a quadrupole Paul trap are carried out in \cite{Vasi13}, for different air pressure values. A sketch of the quadrupole trap geometry used is shown in Fig.~\ref{fig:trap}. The trap consists of four cylindrical electrodes (with diameter $d = 3$ mm and length $L_m = 15$ cm) supplied by means of an a.c. voltage and two 4.5 cm long endcap electrodes supplied with a d.c. voltage. The endcap electrodes are located at each edge of the geometrical trap axis, separated by a distance of $L_h = 6$ cm. The distance between the axes of the dynamical electrodes is $L_b = 1.3$ cm. 

The dust particle motion is described by means of BD \cite{Lapi13}, taking into account the viscosity of the buffer gas, as well as the stochastic forces that act on dust particles owing to neutral and plasma particles. The inter-particle interaction and the dust particle interaction are taken into account by considering Coulomb forces. The dynamics of dust particles is assumed to be described by a Langevin equation:

\begin{equation}\label{Lang}
m_d\frac{d^2 r_i}{dt^2} = F_{tr}(r_i) + F_{int}\left( r_i \right) - 6 \pi \eta R_d \left( \frac{d r_i}{d t} - v_f \right) + F_{Br}\left( r_i \right) + F_{mg} \ ,
\end{equation}
where $m_d$ is the dust particle mass, $r_j$ is a radius-vector for the j-th particle; j = 1, 2, ..., $N$; $N$ represents the number of dust particles; $F_{tr}(r_j) = -\nabla_i \overline{U}$ is an external potential force of the trap; $F_{int}(r_i) = -\nabla_i U$ is the force that describes pair interaction between particles, while $U$ stands for the potential energy of dusty particle interaction;  $F_{mg}$ is the gravity force; $F_{Br}$ is the Langevin delta correlated source of forces which models random forces acting on a dust particle, owing to buffer gas particles; $R_d$ is the radius of a dust particle, $\eta \approx 0.02$ mPa$\cdot$s represents the dynamical viscosity of air at SATP conditions and $v_f$ is the velocity of the air flow. Dust particle dynamics is simulated by the numerical integrator of the Langevin equation.  

The physical factors that affect dust particle dynamics can be estimated by means of the typical values of the forces that act upon the particles. The typical physical parameters values used in the numerical simulations are: inter-particle distance $r \sim 10^{-4}$ m; typical electric charge of the particle $Q \sim  10^5e$; trapping voltage $U \sim  10^3$ V. Thus, we obtain the following estimations for the forces in the Langevin equation in case of particles with a diameter of several microns: 

\begin{align}
F_{int}\left( r_{s},r_{i} \right) = \frac{k \cdot q_{s} \cdot q_{i}}{r^2} \sim 10^{-10} N, \ 
F_{tr}\left( r_{i} \right) = U  q_i/r \sim 10^{-8} N, \ 
\\ F_{mg}^{i} = m_{i}g \sim 10^{-10} \div 10^{-13} N. 
\end{align}

When approaching a neutral molecule an ion polarizes it and hence interacts with the induced dipole. The force that describes the interaction is 

\begin{equation}
F_{Br}\left( r_{i}\right) = \frac{2 \tilde{\alpha} q^2}{r^5} \sim 10^{-10} ,
\end{equation}
where $\tilde{\alpha}$ stands for the polarizability of the molecule and 
$\tilde{\alpha} \sim r_{neutral}^2 / 8$  \cite{Rai91}. The Stokes force $F_{Stokes} = 6 \pi \eta R_{d} dr_{i}/dt$ takes values ranging between $ 10^{-11} \div 10^{-13} N$, for particle velocities $dr_{i}/{dt} \approx 10^{-2} \div 10^{-4}$ m/s.

\subsubsection{Electromagnetic field of the trap}

To simulate the electromagnetic field in the trap we assume the wires are surrounded by a grounded cylindrical electrode that creates a cylindrical capacitor, with a much larger diameter with respect to the distance between the wires $D >> L_b$. Under this assumption each electrode and the surrounding cylinder can be considered as a cylindrical capacitor. By assigning a voltage difference between an electrode and an external cylinder, one can calculate the charge accumulated on the internal electrode. To allow for finite length of the electrodes and calculate the forces acting on a dust particle, each electrode is mathematically divided into small sections and it is assumed that each of these sections can be regarded as a point charge. The resulting Coulomb force acting on a charged dust particle is estimated as a sum of the contribution of each point charge located at the electrodes. By dividing the wires on a large number $\tilde{N}$ of uniformly charged points, the total force acting upon any particle from a wire can be calculated as a sum of Coulomb forces:

\begin{equation}
F_{int}(r_{i}) = \sum_{k}\frac{LUQ_{i}}{2\tilde{N} \ln \left(\frac{R_2}{R_1}\right)\left(r_i - r_k\right)^2} \ ,
\end{equation}
where $L$ is the electrode length, $U$ represents the amplitude of the applied a.c. voltage, $\omega =2\pi f$ is the a.c. voltage frequency, $Q_i$ stands for the electric charge of particle $i$, $r_i$ and $r_k$ are the radius -- vectors of the dust particle and of the point on the wire, respectively, $R_2$ is the radius of a grounded cylinder and $R_1$ denotes the radius of a wire.

\begin{figure}[bth] 
	\centering
	\includegraphics[scale=0.55]{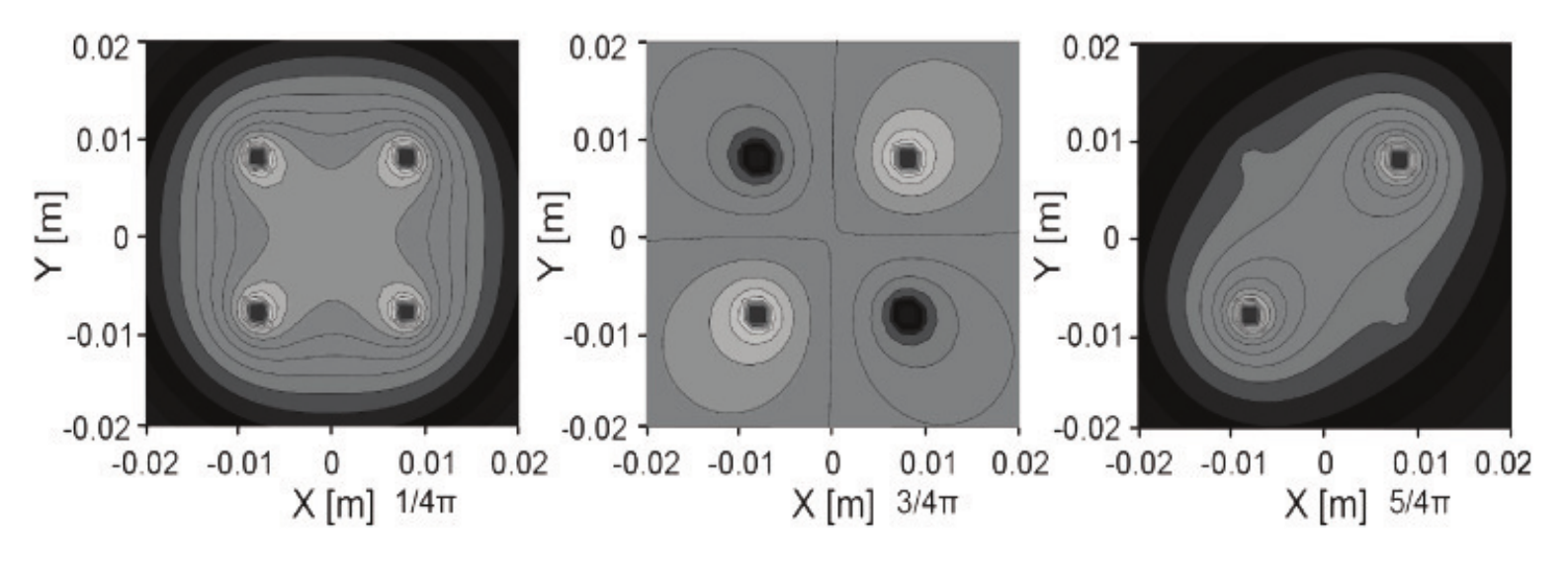}
	\caption{Contour plots of the time evolution of the electric field. Two diagonally connected electrodes are supplied at $\tilde{U} = U_{\omega}cos({\omega}t)$ while the other two electrodes are supplied at $U = U_{\omega}sin({\omega}t)$. The darker regions in the figure correspond to potential wells that attract the particles, while the lighter ones describe potential hills that repel the grains. In the left picture the inner region of the trap corresponds to a potential hill, while in the right picture it corresponds to a potential well. The potential fields are located at the trap centre, normal to the trap axis, so that the contribution of the electrostatic field generated by the endcap electrodes can be disregarded. Source: image reproduced from \cite{Lapi13} with kind permission of the authors.}
	\label{fig:surface}
\end{figure} 

A uniform charge distribution along the trap wires can occur in case of a more or less uniform dust particle distribution along the trap axis. In case of endcap electrodes with fixed potential such a situation does not appear. To estimate the electric field generated by the endcap electrode it is necessary to solve a cylindrical Laplace equation in orthogonal elliptic - hyperbolic coordinates ($\sigma - \tau$).

\begin{equation}
\Delta U = \frac{1}{\alpha^2(\sigma^2 - \tau^2)} \left[ \frac{\partial}{\partial\sigma}
\left[(\sigma^2 - 1)\frac{\partial U}{\partial \sigma}\right] 
+ \frac{\partial}{\partial\tau}\left[(1 - \tau^2) \frac{\partial U}{\partial \tau}\right] \right] \ ,
\end{equation}
where $2 \alpha = L_h$ is the distance between the endcap electrodes. Hence, in case of a dust particle, the electrostatic intensity at a certain point in the space of coordinates ($\sigma$, $\tau$) can be expressed as: 

\begin{eqnarray}
E\left(\tau, \sigma\right) \propto \exp \left(- \frac{1}{1 - \alpha^2}\ln \left(\tau^2 + \frac{\alpha^2 \sigma^2 - 1}{1 - \alpha^2}\right) + \frac{1}{1 - \alpha^2}\ln \left(\frac{1}{\alpha^2 - 1}\right) -1 \right) \ ,\\ 
\nonumber
\tau = \sqrt{\left(z + \alpha \right)^2 + x^2 + y^2} - \sqrt{\left( z - \alpha \right)^2 + x^2 + y^2} \ ,\\
\nonumber
\sigma = \sqrt{\left(z + \alpha \right)^2 + x^2 + y^2} + \sqrt{\left( z - \alpha \right)^2 + x^2 + y^2} \ .
\end{eqnarray}

The a.c. voltage $U_{\omega}$ is applied to the trap electrodes as follows: two diagonal electrodes are supplied with an a.c. voltage $U = U_{\omega}\cos({\omega}t)$ while the other two electrodes are supplied at $\tilde{U} = U_{\omega}\sin({\omega}t)$. The trapping voltage frequency is ${\omega}= 2 \cdot \pi f$ (as shown in Fig.~\ref{fig:trap}). The setup generates an oscillating electric field inside the trap, whose equipotential surfaces for three different instances of time : $t = \frac 14 \pi$, $\frac 34 \pi$, $\frac 54 \pi$, are illustrated in Fig.~\ref{fig:surface}. In order to prevent dust particle escape, a d.c. voltage $U_{end}$ is supplied between the endcap electrodes that repels particles towards the inner volume of the trap.

\subsubsection{Regions of single particle confinement}

\begin{figure}[bth] 
	\centering
	\includegraphics[scale=0.53]{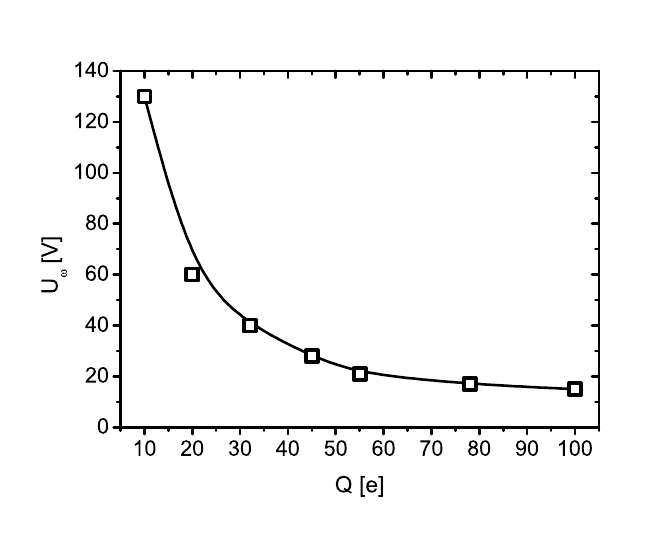}
	\includegraphics[scale=0.53]{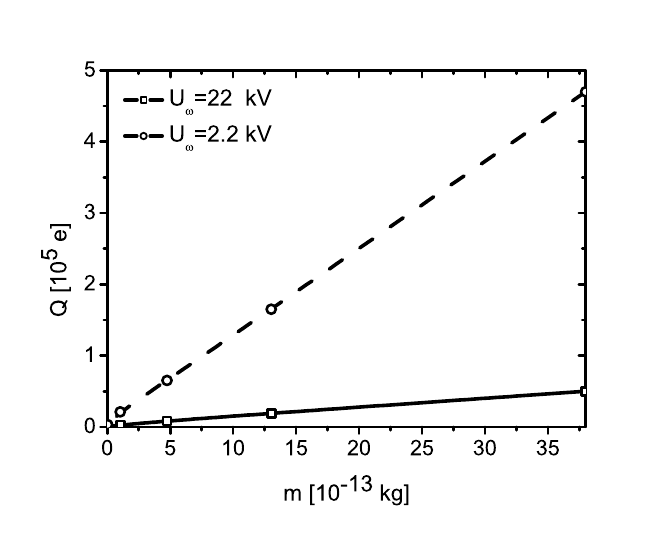}
	\caption{The lower boundary of the regions of dust particle confinement. Left panel: $U_\omega$ versus particle charge; Right panel -- $Q$ versus particle mass for two different a.c. voltages. The area above the line corresponds to particle confinement. Left panel parameters: $Q_{particle} = 10 e$, $U_{\omega} = 135$ V, $f = 50$ Hz, $U_{end} = 135$ V, $\rho_{particle} = 1,5 \cdot 10^4$ kg/m$^3$ , $r_{particle} = 1 \ \mu$m, $\eta = 17 \cdot 10^{-6}$ Pa$\cdot$s - dynamic viscosity and $T = 300$ K. Right panel parameters: $f = 100$ Hz, $U_{end} = 700$ V, $U_{\omega} = 2,2 \div 22$ kV, $\rho_{particle} = 0.38 \cdot 10^4 $ kg/m$^3$, $r_{particle} = 1 \ \mu$m, $\eta = 17 \cdot 10^{-6}$ Pa$\cdot$s - dynamic viscosity and $T = 300$ K. Source: picture reproduced from \cite{Lapi13} with kind permission from the authors.}
	\label{fig:3}
\end{figure}

We revert to the dynamic equation (\ref{Lang}). The trapping force exerted upon a dust particle depends on the product between the electric charge and the a.c. trapping voltage. Hence, for the lower boundary of the dust particle confinement region, an inverse dependence between the a.c. voltage and the electric charge is expected. The left panel in Fig.~\ref{fig:3} confirms this dependence that is predicted by calculations.

The right panel in Fig.~\ref{fig:3} also shows the analogous lower boundary of dust particle confinement for two different a.c. voltages, namely $U_{\omega} = 2200$ V and $U_{\omega} = 22000$ V. The lower boundary is characterized by the dependence of the dust electric charge versus its mass. These results support the prominent role of the charge-to-mass ratio $Q/m$, as it results from equation \ref{Lang} and from the above estimations of the r.h.s. forces in eq. \ref{Lang}, for the experimental conditions considered. The applied a.c. voltages are lower than the breakdown voltage for air.  

\begin{figure}[bth] 
	\centering
	\includegraphics[scale=0.45]{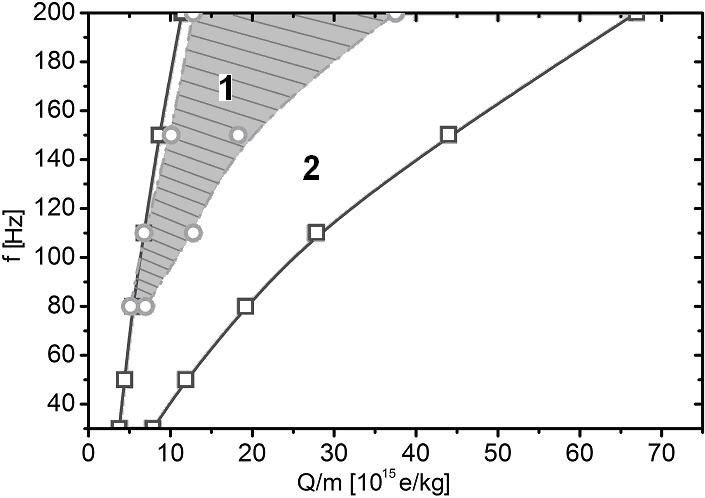} 
	\caption{Regions of stable confinement of a single dust particle in dependence of frequency $f$ and charge-to-mass ratio $Q/m$, for different dynamical viscosities $\eta =1.7 \ \mu$Pa$\cdot$s (shaded region 1) and $\eta =17 \ \mu$Pa$\cdot$s (unshaded region 2). The particle and trap parameters are: $Q = 20500 \div 685000 e$, $U = 4400$ V, d.c. voltage = $900$ V, particle density $\rho = 0.76 \times 10^{-4}$ kg/m$^{3}$ and particle radius $9 \ \mu$m. Source: picture reproduced from \cite{Vasi13} with kind permission of the authors.}
	\label{sumnjp}
\end{figure}

Let us consider the regions of stable confinement for a single particle and investigate their dependence on the a.c. voltage frequency. No matter the frequency value, the confinement region is limited by upper and lower values of the charge-to-mass ratio $Q/m$ (Fig.~\ref{sumnjp}). In a similar manner, the confinement region is limited by lower and upper frequencies, regardless of the value of the charge-to-mass ratio $Q/m$. Beyond this region the trap cannot confine particles. By increasing the dynamical viscosity of the medium, the confinement region becomes wider. As an example, Fig.~\ref{sumnjp} illustrates the results of simulations for two different values of the viscosity: $\eta = 1.7 \ \mu$Pa$\cdot$s (squares) and $\eta = 17 \ \mu$Pa$\cdot$s (circles). Dissipation of the dust particles kinetic energy in case of a low value dynamical viscosity of the medium is lower, therefore the velocity and kinetic energy of a particle are larger due to the electric field. For higher particle velocities the boundary frequency value should be higher, in order to prevent particle escape out of the trap at half--cycle of the a.c. voltage. The results of calculations presented in Fig.~\ref{sumnjp} demonstrate that by using quadrupole dynamic traps it is possible to confine particles with larger mass and dimensions, compared to the situation of low pressure plasmas of RF or d.c. glow discharges.

\section{Waves in Plasmas}\label{waves}
\subsection{Solitary waves in long charged particle structures confined in the linear Paul trap}\label{SolWave}

Charged dust particles embedded into plasmas do not only change the electron -- ion composition and thus affect conventional wave modes (e.g., ion -- acoustic waves), but they also: (a) introduce new low-frequency modes associated with microparticle motion, (b) alter the dissipation rates, and (c) give rise to instabilities \cite{Piel17}, etc. Investigations on dust acoustic solitary waves in plasmas have been carried out over more than several decades \cite{Pecs20}. The particle electric charges vary in time and space, which leads to important qualitative differences between dusty plasmas \cite{Piel17} and usual multicomponent plasmas \cite{Fort05, Fort10a, Boni14, Chop14, Khra04, Vlad05, Shuk92, Chat12, Ghai17, He18, Set18}, because in addition to positive ions and electrons other components are also present, such as negative ions or dust particles \cite{Aro20, Gore20}. Dust acoustic waves in a charged dust component have been investigated in a complex plasma at low-pressure, a model for which the analytical and numerical fluid models and kinetic approaches are treated extensively in \cite{Fort05, Fort10a, Khra04}. The main difficulty in solving this issue lies in the fact that a complex plasma represents a nonlinear medium where the waves of finite amplitude cannot be considered independently. Nonlinear phenomena in complex plasmas are very diverse due to a large number of different wave modes that can be sustained. The wave amplitude can reach a nonlinear level owing to different physical processes and mechanisms. This is not necessarily an external forcing or an outcome of the wave instabilities, it can also represent a regular collective process of nonlinear wave steepening. In absence of dissipation (or for low dissipation) nonlinear steepening can be balanced by wave dispersion, which in turn can result in the formation of solitons \cite{Lan13}. When dissipation in the system is large it can overcome the role of dispersion, case when the balance between nonlinearity and dissipation can generate shock waves \cite{Fort05, Fort10a, Khra04}. A very effective method to study these phenomena is the particle-in-cell (PIC) simulation \cite{Ludw12, Zhang14, Gao16, Med18a}. 

The essential features of the shock and solitary waves, which are associated with positive ion dynamics and dust charge fluctuation, have been investigated by employing the reductive perturbation method. It was demonstrated that dust charge fluctuation represents a source of dissipation, responsible for the development of dust-ion-acoustic shock structures \cite{Mamun10}. The pseudopotential method can be used to investigate the occurrence of solitary structures, along with the characteristics of solitary waves and double layers, which are associated with positive ion dynamics and pressures of degenerate electrons and positrons.   Propagation of nonlinear waves \cite{Rud13} in dusty plasmas is investigated in Ref. \cite{Pak09}, where the Kadomtsev–Petviashivili (KP) equation is inferred for an unmagnetized dusty plasma with variable dust charge and two temperature ions. Moreover, the system is better characterized by means of a Sagdeev potential which enables one to explore the stability conditions and the existence of solitonic solutions. A travelling rarefaction soliton propagates in most cases. The amplitude of solitary waves of the KP equation diverges at critical values of the plasma parameters. Solitonic solutions of the modified KP equation with finite amplitude are derived for such case \cite{Pak09}. A study of the characteristics of freak waves in a dusty plasma containing two temperatures ions is presented in Ref. \cite{Set18}, by modulating the KP equation to infer the nonlinear Schrödinger equation (NLSE). Nonlinear propagation of dust-acoustic (DA) solitary waves \cite{Ghai17, Denra18} in a three-component unmagnetized dusty plasma consisting of Maxwellian electrons, vortex-like (trapped) ions and arbitrarily charged cold mobile dust rain, is investigated in \cite{Rah14}. It is demonstrated that the dynamics of small but finite amplitude DA waves is regulated by a nonlinear equation of modified Korteweg-de Vries (mK-dV) type, also known as the cylindrical Kadomtsev-Petviashvili equation \cite{Mann20}. The basic characteristics and propagation of dust-acoustic (DA) shock waves (DASHWs) \cite{Chop14} in self-gravitating dusty plasmas containing massive dust of opposite polarity, trapped ions and Boltzmann electrons, are explored in \cite{Ema18, Sumi19}. The reductive perturbation technique is applied to infer the standard modified Burgers equation (mBE). Collision of ion-acoustic solitary waves in a collisionless plasma \cite{Vlad11} with cold ions and Boltzmann electrons is investigated by means of numerical simulations \cite{Med18a, Med18b}. Ref. \cite{Li21} investigates propagation of a shock wave in a strongly coupled complex plasma, which is demonstrated to be characterized by a Korteweg-de Vries–Burgers equation. Recent experimental realization of an ultracold dusty plasma and observation of self-organization phenomena \cite{Bolt19} has yielded to particle-in-cell (PIC) simulation with Monte-Carlo-collisions for the wake behind a dust particle, as an outcome of ion focusing at superfluid helium temperature (2 K) \cite{Sund20}. All these nonlinear phenomena have important applications in space and astrophysical environments. 

In contrast to the studies mentioned above, we firstly discuss experimental generation of density waves in a strongly coupled one component Coulomb system consisting of micron sized particles (interacting via a non screened Coulomb potential), under SATP conditions (in air). Microparticles are levitated in a long, linear quadrupole electrodynamic Paul trap \cite{Syr19b}. In particular, the possibility of generating density waves in the form of individual solitary humps is also demonstrated. The physical possibility of generating waves that are equivalent to the solitary density waves is discussed by Arnold \cite{Arno92}, as it is caused by the nonuniform velocity distribution of moving particles. 

The analysis of experimentally observed solitary density waves \cite{Syr19b} is based on some assertions of the catastrophe theory \cite{Arno92} that allow one to identify these waves as {\em caustics}. The conclusion is based on the fact that the density profile of experimentally obtained solitary waves can be described under a good accuracy by a theoretical dependence predicted by the caustic theory \cite{Arno92} ($const/\sqrt{\epsilon}$, where $\epsilon$ is the distance with respect to the singularity of the caustic). Strictly speaking this classification can be applied to collisionless systems of particles, when small deviations from constancy in the initial velocity distribution lead to particle accumulations for sufficiently long periods of time. According to Arnold's remarks \cite{Arno92} {\em 'this conclusion still holds when one goes from a one dimensional medium to a medium filling a space of any dimension, and when one allows for the effects on the motion of particles of an external force field or a field originating from the medium, and also when the effects of relativity and the expansion of the universe are accounted for \dots . 
	Ya. B. Zel'dovich called such caustics pancakes ('bliny' in Russian; at first pancakes were interpreted as galaxies, later as clusters of galaxies) \dots    
	The predictions of the theory of singularities for the caustics geometry, wave fronts and their metamorphoses, have been completely confirmed in experiments \dots '}.  

As demonstrated in Refs. \cite{Syr19b, Syr19c} solitary caustics can be considered as new experimental support of the general versatility of the caustic theory in describing different physical phenomena, not only in collisionless systems of particles but also when the inter-particle interaction and the interaction with external fields in viscous media are strong. Further, we show that the generation of solitary density waves in strongly coupled one component Coulomb systems of particles is possible under SATP conditions, when the energy losses caused by air viscosity can be compensated by the energy contribution of the a.c. electric field in the trap.   

Ref. \cite{Lin19} reports on propagation of compressional shocks in two-dimensional (2D) dusty plasmas. Molecular dynamics (MD) simulations \cite{Hos15} performed under various conditions are used to explain the physical phenomena involved.

\subsection{Multiparticle trapping in a linear Paul trap} 

\begin{figure*}[bth]
	\centering
	\includegraphics[scale=0.5]{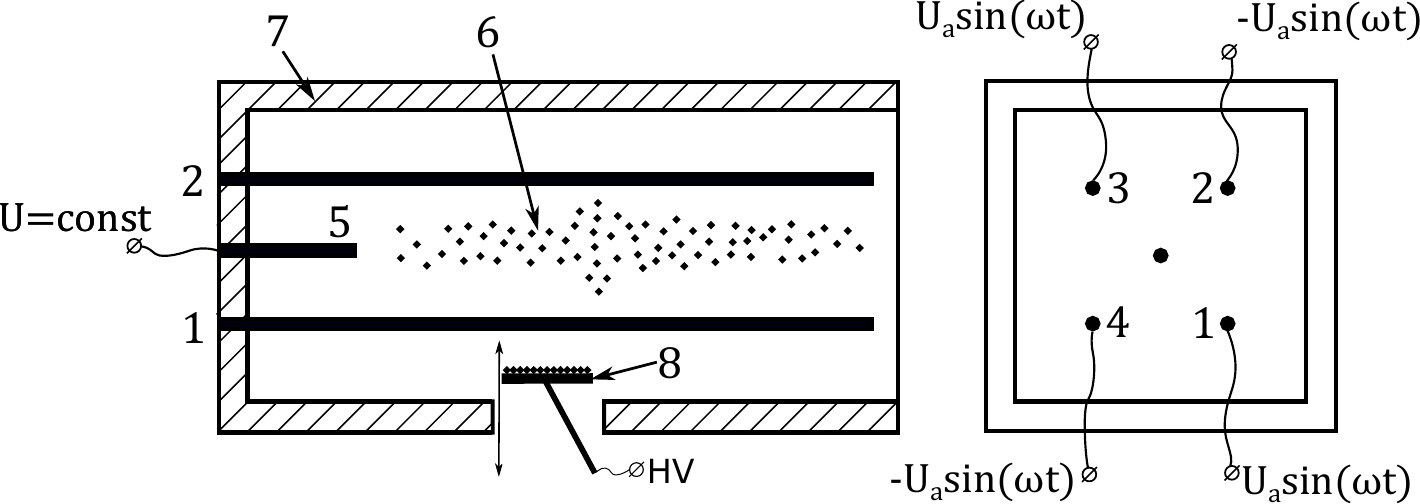}	
	\caption{Image of the trap with front and side views. Legend: $1 \div 4$ -- trap electrodes, 5 -- end electrode, 6 -- confined particles, 7-- optically transparent plastic box, 8 -- flat charging electrode.}
	\label{Trap} 
\end{figure*}

Experimental investigations on the excitation and development of dust particle density disturbances in charged particle structures are performed using a linear quadrupole electrodynamic trap (Fig.~\ref{Trap}) \cite{Syr19b}. The trap consists of four parallel steel cylindrical electrodes, each one of them 30 cm long, with a diameter of 3 mm. An a.c. electric potential $U_a\sin (2\pi ft)$ with frequency $f = 50$ Hz is applied between the four trap electrodes, placed at the corners of a square of 2 cm side. The phase shift of the electric potential between adjacent electrodes is equal to $\pi$. At the left edge of the trap (see Fig.~\ref{Trap}) an additional electrode is mounted, supplied at a constant electric potential $U = 1$ kV, that prevents particle escape along the trap axis. The right edge of the trap is left open but end electrode effects \cite{Lapi15b} also forbid particle escape.

\begin{figure*}[bth]
	\centering
	\includegraphics[scale=0.35]{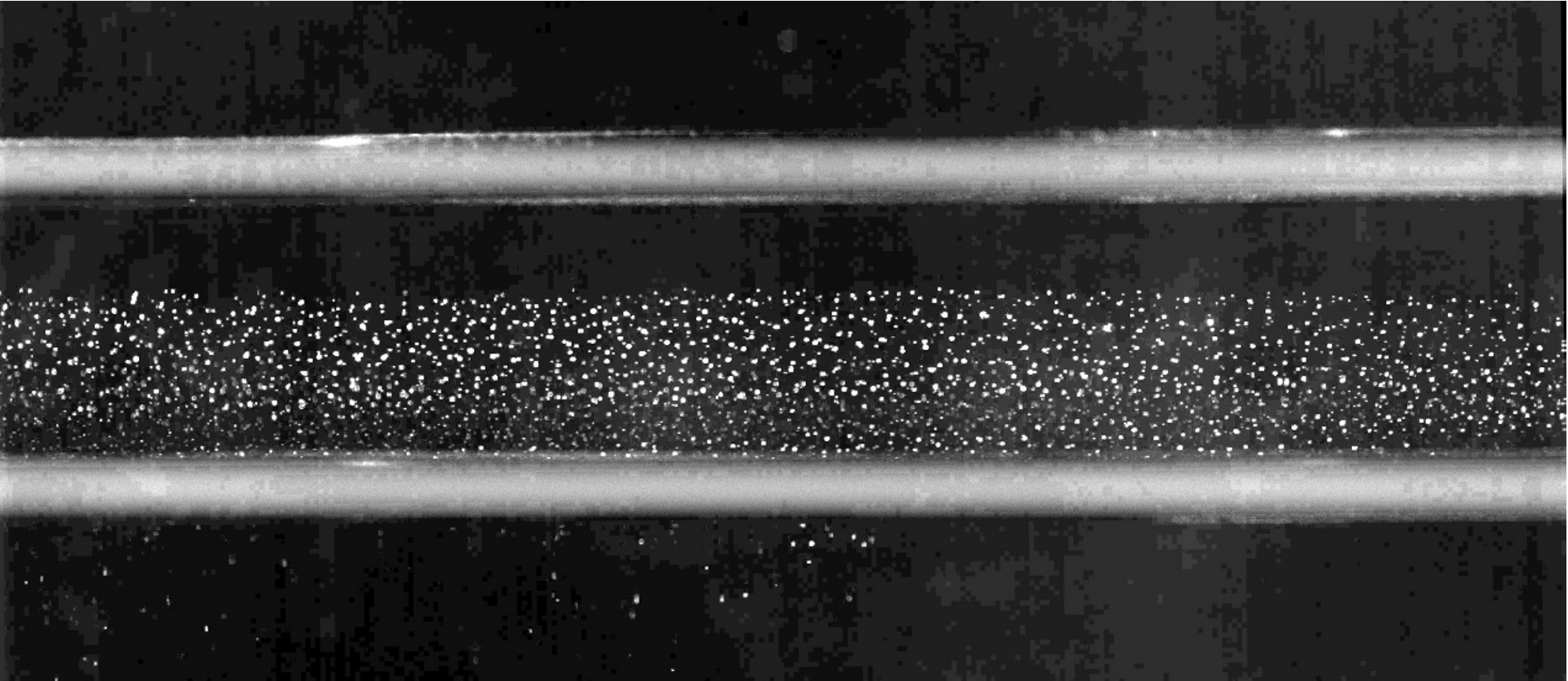}
	\caption{Stable structure of charged dust particles in an electrodynamic trap.}
	\label{Stable_structure}
\end{figure*}

The electrodynamic trap is placed inside an optically transparent plastic box to mitigate against air flows \cite{Lapi18b}. Polydisperse Al$_2$O$_3$ particles are trapped, with diameter ranging between $10 \div 40 \ \mu$m. The particles are positively charged on the surface of a flat electrode (labelled as 8) supplied at 10 kV, by using an induction method (see Fig.~\ref{Trap}). The electrode is brought inside the trap from below, through a hole drilled in the plastic box. An electric field produced between the charging electrode and the trap electrodes accelerates the particles, which are then attracted in the trap. Afterwards, the flat electrode is removed. Video recording of the particles (illuminated using a laser beam) is performed by means of a HiSpec 1 video camera. As demonstrated in Ref. \cite{Vasi13} the trap captures particles for a distinct range of charge-to-mass ratios. 

Fig.~\ref{Stable_structure} shows stable structures obtained for charged dust particles, when the a.c. voltage $U_a = 3.6$ kV. Although the observed structure slightly oscillates with the frequency of the a.c. supply voltage, the inter-particle distances remain approximately constant. In order to characterize particle correlations, the electric charge of the particles and the average inter-particle distance are estimated. The electric charge of the particles is not measured in \cite{Vasi13}. The average electric charge value is estimated using some of the results previously obtained in \cite{Syr16b, Vasi18}, and the average value found is about $5 \times 10^4 e$. The average inter-particle distance is around 1 mm, estimated from typical images of particle configurations using a specially written computer code. Thus, the Coulomb interaction at the average inter-particle distance is approximately equal to $10^5$ in units of atmospheric gas temperature at $300$ K.

\subsection{Pair correlation functions} 

\begin{figure*}[bth]
	\centering
	\includegraphics[scale=0.4]{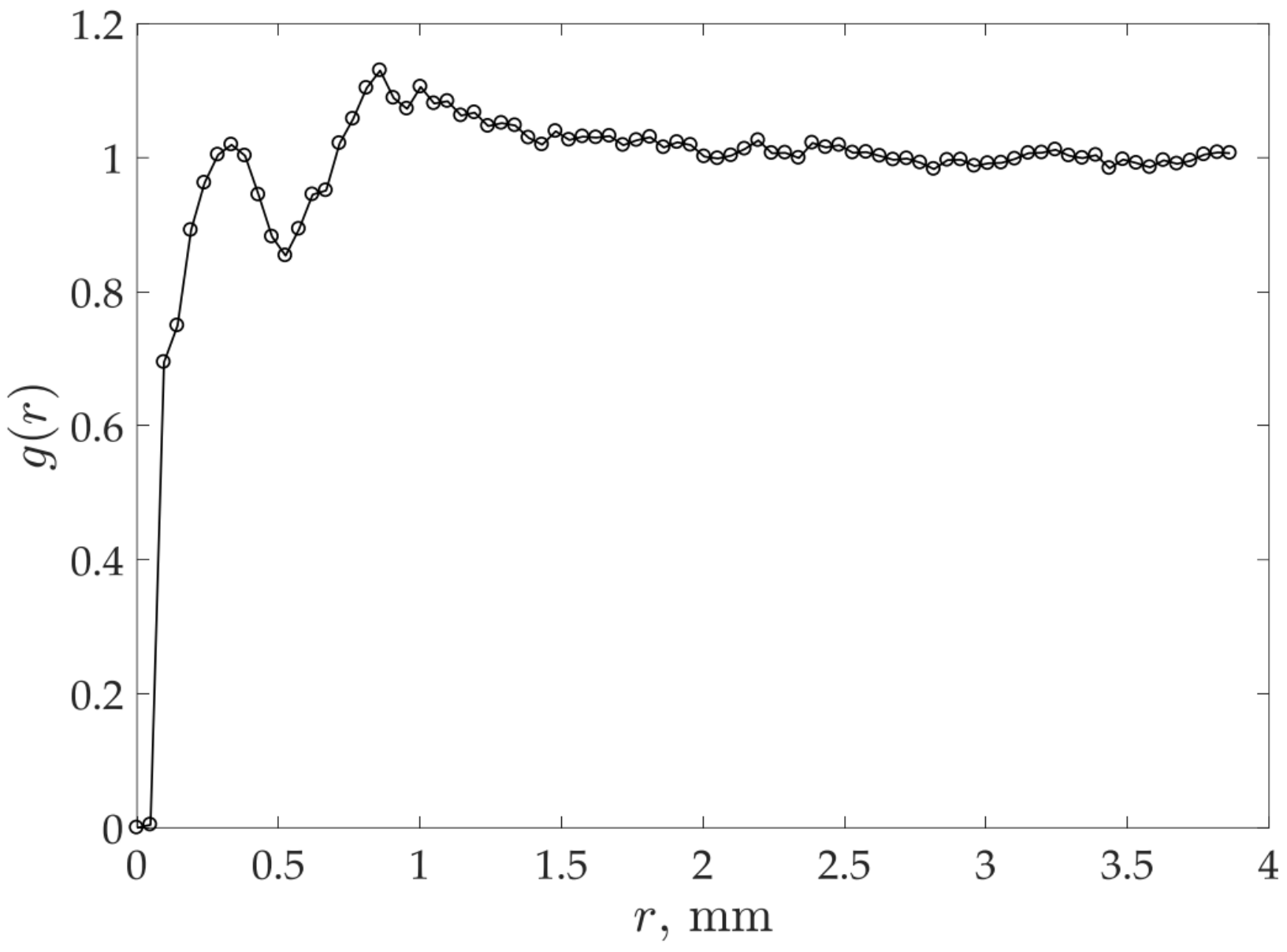}
	\caption{Pair correlation function for a Coulomb structure.}
	\label{cor}
\end{figure*}

The average pair correlation function \cite{Syr19a} is obtained for structures analogous to those presented in Fig.~\ref{Stable_structure}. By video recording the particle motion at a frame rate of two hundred frames per second, the pair correlation function is estimated using a special computer code for each frame. The average pair correlation function is shown in Fig.~\ref{cor}. To obtain the correct correlation function, the diameter of the laser beam should not exceed the mean inter-particle distance. Two cylindrical lenses are used to create a flat laser beam with a diameter of roughly 0.25 mm. The lenses are arranged so that the constriction region is achieved at the centre of the Coulomb structure. 

The pair correlation function presented in Fig.~\ref{cor} exhibits two peaks, a feature that can be explained due to various dimensions and electric charge values of the particles that compose the Coulomb structure. The main peak occurs for an inter-particle distance of roughly 1 mm. The suggested pair correlation function makes it possible to state the existence of short -- range order structures and the absence of long -- range ones, which is typical for liquid -- like particle ordering. Hence, the Coulomb structure consists of dust particles that are strongly correlated for short distances, while the pair correlation function points out to the liquid -- like ordering with a correlation radius of roughly two average inter-particle distances. From the physical point of view, such a behaviour of the pair correlation functions can be explained by chaotic overlapping of the long -- range Coulomb {\em tails} of the inter-particle interaction.

\subsection{Occurrence and evolution of solitary density waves in a linear trap} 

\begin{figure*}[bth]
	\begin{minipage}[h]{1\linewidth}
		\center{\includegraphics[width=0.7\linewidth]{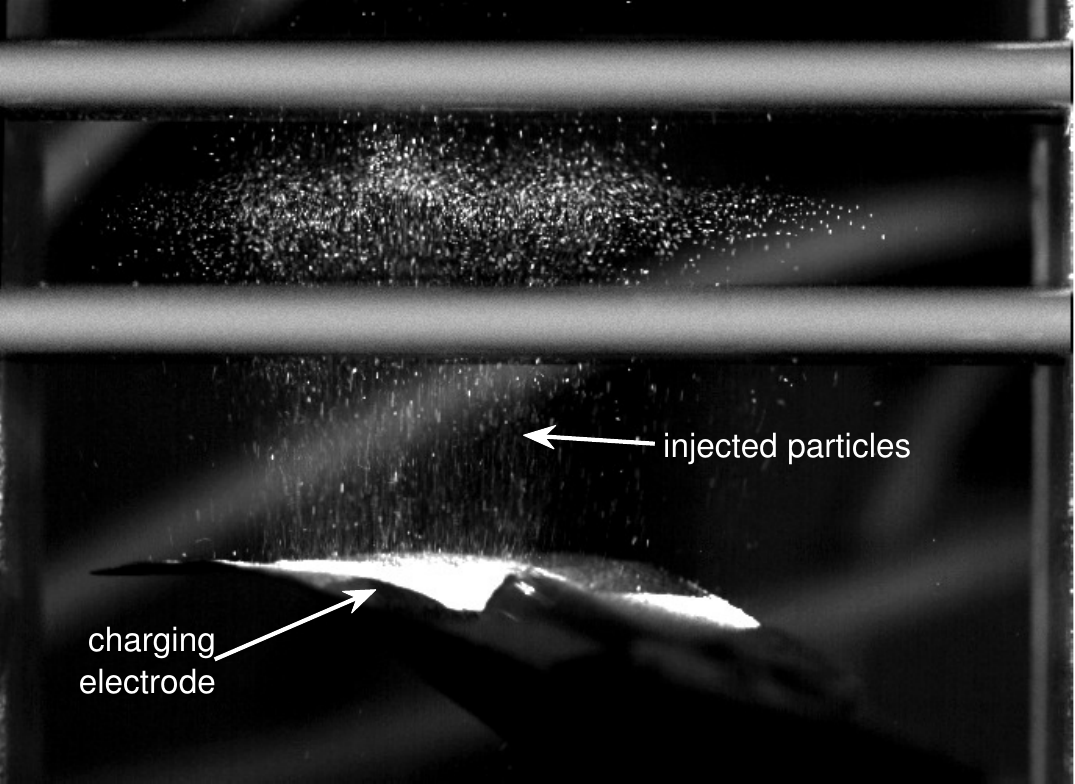}} \\$(a)$
	\end{minipage}
	\begin{minipage}[h]{1\linewidth}
		\center{\includegraphics[width=0.9\linewidth]{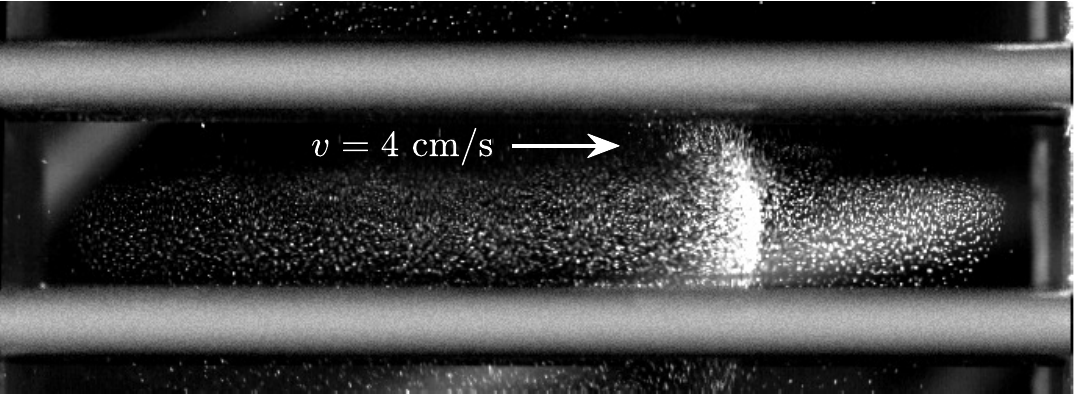}} \\$(b)$
	\end{minipage}
	\caption{Occurrence of hump density waves in a Coulomb structure after injection of additional particles.}
	\label{injection}
\end{figure*}

Density waves can arise in such type of Coulomb structures after the injection of additional particles (Fig.~\ref{injection}). Additional particles are injected into the trap, originating from the surface of the flat electrode (Fig.~\ref{injection}$(a)$), after which a hump density wave occurs in the Coulomb structure that propagates towards the right edge of the trap at a velocity of around 4 cm/s (Fig.~\ref{injection}$(b)$).

\begin{figure*}[bth]
	\begin{minipage}[h]{0.9\linewidth}
		\center{\includegraphics[width=0.71\linewidth]{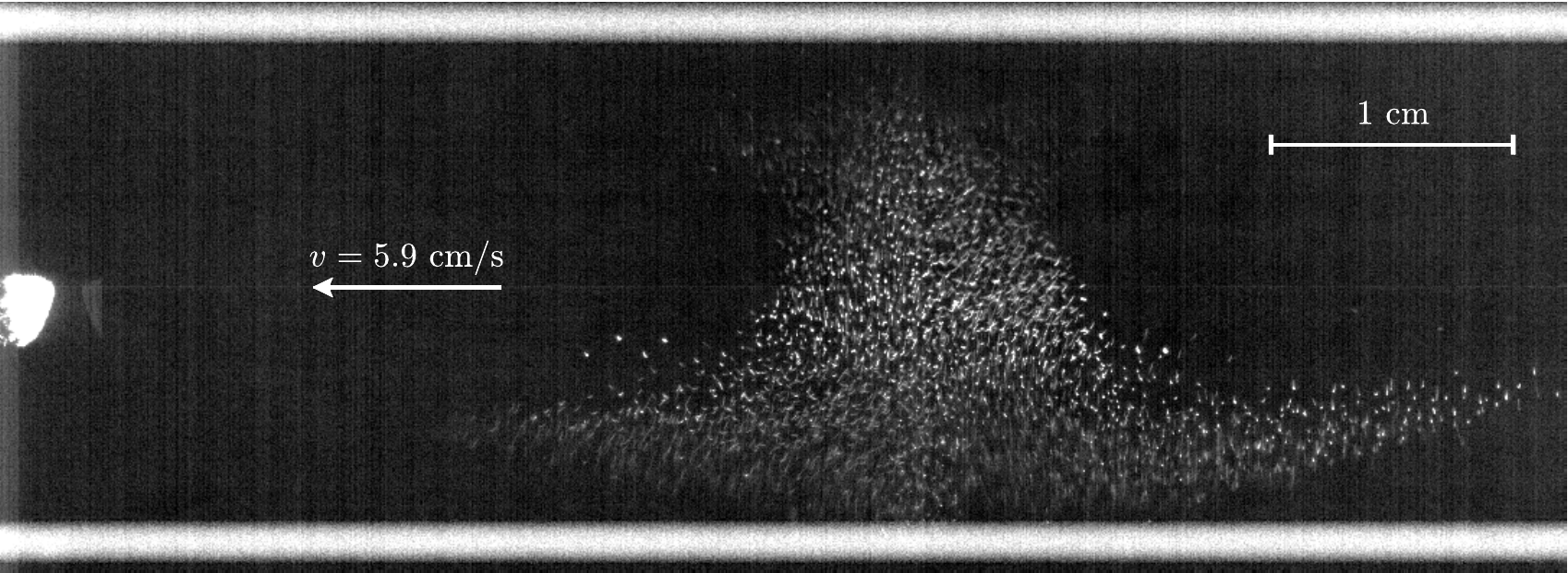}} \\$(a)$
	\end{minipage}
	\begin{minipage}[h]{0.9\linewidth}
		\center{\includegraphics[width=0.71\linewidth]{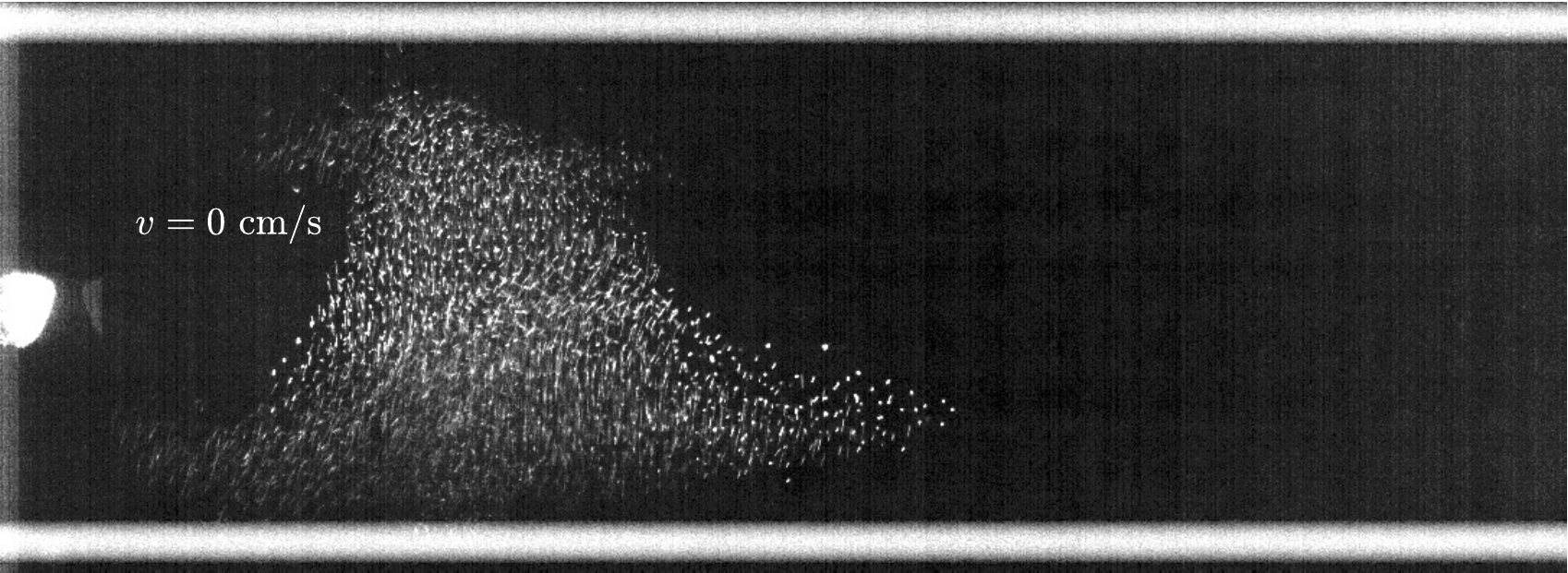}} \\$(b)$
	\end{minipage}
	\begin{minipage}[h]{0.9\linewidth}
		\center{\includegraphics[width=0.71\linewidth]{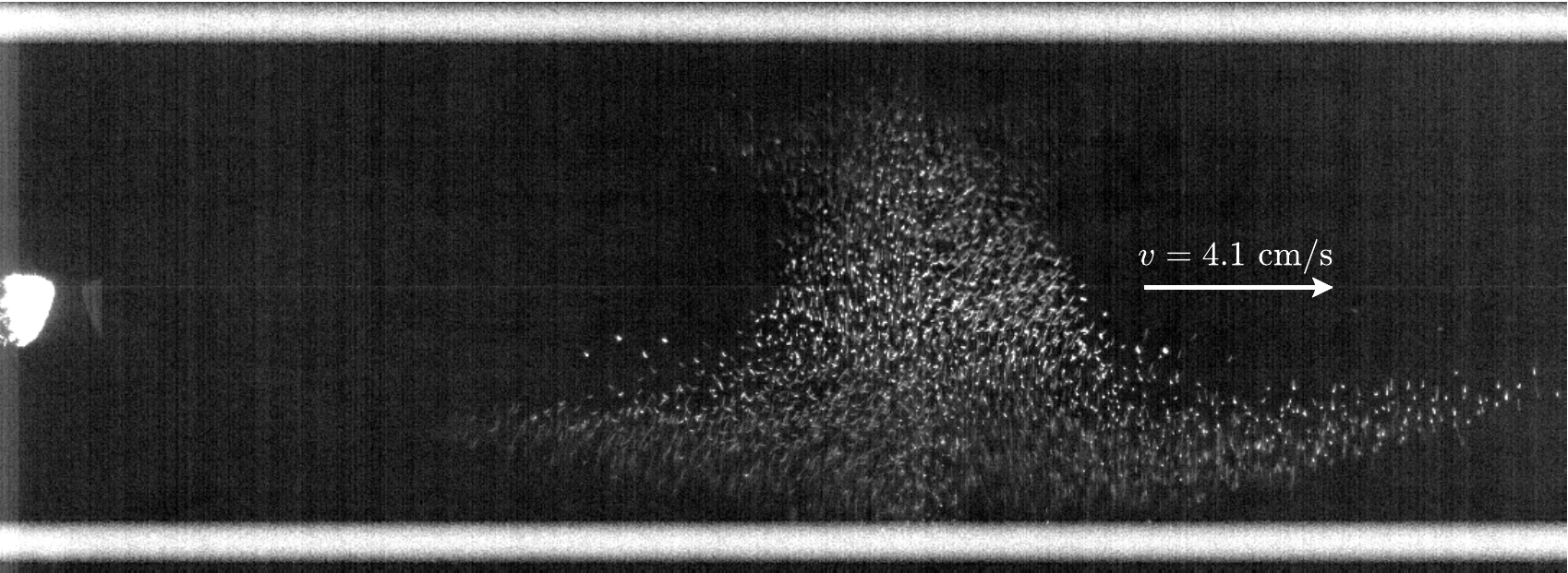}} \\$(c)$
	\end{minipage}
	\caption{Reflection of the wave from the left edge electrode (bright spot). $U_A = 5.1$ kV and $f = 50$ Hz. $(a)$ The wave propagates towards the locking electrode; $t = 0$ s; $(b)$ The wave stops at the end of the locking electrode at $0.3$ s; $t = 0.43$ s; $(c)$ The wave propagates in the opposite direction; $t = 1.01$ s.}
	\label{reflection}
\end{figure*}

Density waves can also occur in a stable Coulomb structure when the operating parameters of the electrodynamic trap are modified, for example when the a.c. trapping voltage rises up to a value of $5.1$ kV. In this case dust density waves appear in the long range dust particle structure (see Fig.~\ref{Stable_structure}). During propagation the wave shape changes very little. Propagation of a density wave in the vicinity of an end electrode is illustrated in Fig.~\ref{reflection}. The wave propagates towards the end electrode with a velocity of $5.9$ cm/s (Fig.~\ref{reflection}$(a)$). After approaching the end electrode within a minimum distance of $1$ cm (Fig.~\ref{reflection}$(b)$), the wave stops for approximately $0.5$ s then starts propagating in the reverse (backward) direction with a velocity of $4.1$ cm/s (Fig.~\ref{reflection}$(c)$). We emphasize that, the characteristic time of all dust particle dynamic processes in the trap is related to a frequency value that is typical of the order of several Hz or tens of Hz. An analogous reflection occurs when the hump wave approaches the right edge of the trap. The reflection is caused by the longitudinal electric field that is present at the trap edge, as an outcome of the electric field lines curvature \cite{Lapi15b}.

\begin{figure*}[bth]
	\begin{minipage}[h]{1\linewidth}
		\center{\includegraphics[width=0.9\linewidth]{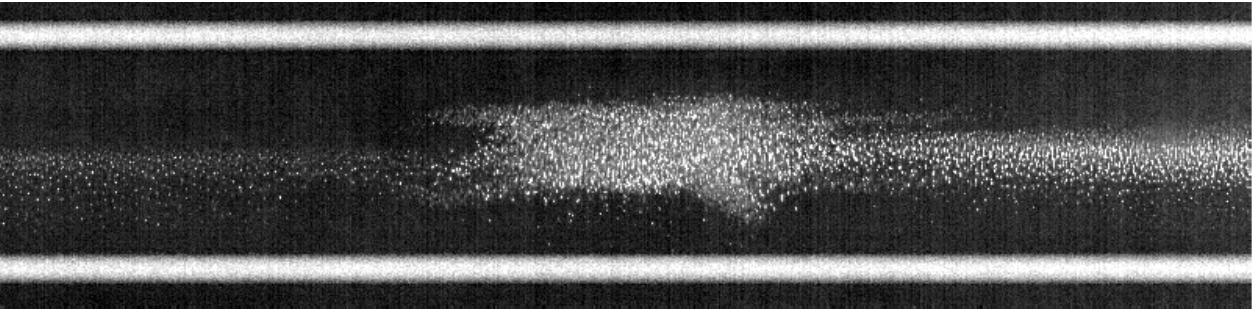}} \\$(a)$
	\end{minipage}
	\begin{minipage}[h]{1\linewidth}
		\center{\includegraphics[width=0.9\linewidth]{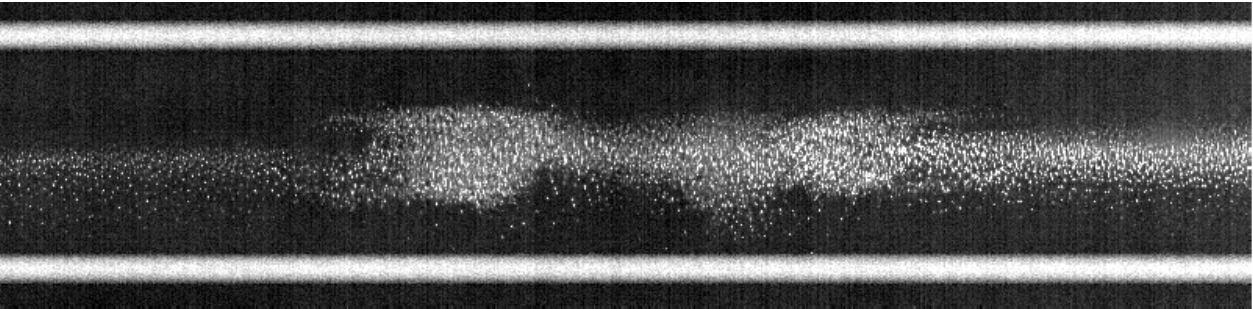}} \\$(b)$
	\end{minipage}
	\begin{minipage}[h]{1\linewidth}
		\center{\includegraphics[width=0.9\linewidth]{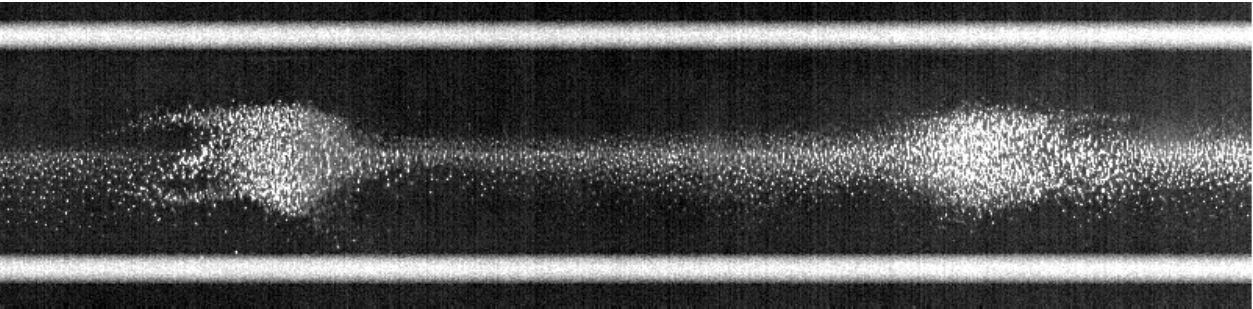}} \\$(c)$
	\end{minipage}
	\caption{Time evolution of a hump density wave. $U_A = 5.1$ kV, $f = 50$ Hz. (a) $ t = 0 $ s;  (b) $ t = 0.42 $ s; (c) $  t = 1.1 $ s.}
	\label{separation}
\end{figure*}

Fig.~\ref{separation} illustrates the time evolution of the density hump wave that appears in the central region of the trap. One second later after its ocurrence the hump wave divides into two separate humps (Fig.~\ref{separation}$(b)$) that propagate in opposite directions, with a relative speed of $7.3$ cm/s.

\begin{figure*}[bth]
	\centering
	\includegraphics[scale=0.35]{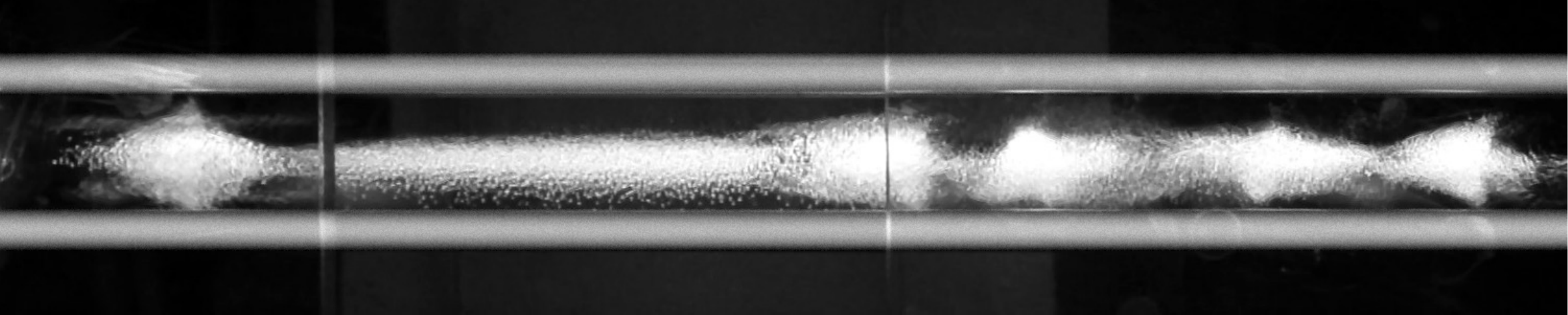} \\	
	\caption{Hump density waves chaotically propagating in different directions for $U_A = 5.1$ kV and $f = 50$ Hz}
	\label{five_caustics}
\end{figure*}

More interesting experimental observations are shown in Fig.~\ref{five_caustics} which illustrates the simultaneous existence of five hump density waves, for $U_a = 5$ kV and $f = 50$ Hz. For these particular values of the trap parameters, density waves chaotically propagate along the trap in different directions while they can also merge and split. 

\subsection{One-dimensional particle-in-cell simulations}  

In order to describe the dynamics of a system of electrically charged particles levitated in a Paul trap a fairly complex theoretical model is required, that takes into account both the time-varying electric field created by the trap electrodes and the space-charge field generated by the particles. Up to now, there exists no adequate theoretical model in the literature to describe such phenomenon. This is the main reason why a theoretical description of such phenomenon starts from considering a one component plasma (OCP) fluid model, an approach that can at least qualitatively clarify certain physical features of the system \cite{Syr19b}. 

Simple simulations of density waves are performed using the one-dimensional kinetic approximation under which a system of identical charged particles is considered. Each particle has an associated mass $m$ and an electric charge $q$. It is assumed that the particles are {\em cold} and their dynamics can be described in a collisionless approximation. The corresponding system of Vlasov equations can be expressed as: 

\begin{equation}
\label{eq1}
\frac{\partial f}{\partial t} + v \frac{\partial f}{\partial x} - \frac{\partial \varphi}{\partial x} \frac{\partial f}{\partial v} = 0 , \quad   
\frac{\partial^2\varphi}{\partial x^2} = - n \,, \quad   
n = \int\limits_{-\infty}^{\infty}\!f(x,v,t)\,dv\,,
\end{equation}
where $f$ is the particle distribution function, $n$ denotes the particle density, $\varphi$ represents the electric field potential, while $ x $, $ t $ and $ v $ represent the coordinate, time and velocity, respectively. The equations are given in normalized form with 

\begin{center}
	\begin{tabular}{ c c c }
		$n_{0}$, $L$,     & $[m/4\pi q^2n_{0}]^{1/2}$, & $L [4\pi q^2n_{0}/m]^{1/2}$, \\ 
		$4\pi qn_{0}L^2$, & $4\pi qn_{0}L$, & $(mn_0)^{(1/2)}/\sqrt{4\pi}qL$ \\  
	\end{tabular}
\end{center}
\begin{equation}
\label{eq2}
\end{equation}
as units of density, length, time, velocity, potential, electric field and particle distribution function, respectively. $n_0$ stands for the maximum value of the particle density at $t = 0$, while $L$ is the half-length of the particle structure under consideration.

\subsubsection{Initial and boundary conditions} 

The system of equations (\ref{eq1}) should be built up using additional initial and boundary conditions that characterize a particular problem. Several possible initial particle distributions are taken into account. However, it should be noted that the experimental particle density is nonuniform due to the end effects that crop up in the endcap electrodes area \cite{Syr19b}. Hence, to study the occurrence of density waves and perform numerical simulations it is necessary to use a nonuniform density distribution as an initial condition. We emphasize on the fact that density waves start to develop precisely in the end regions, but only for a.c. supply voltages higher than $5.1$ kV. As an example we present the evolution of a system of identical positively charged particles, characterized by the following density distribution at an initial moment of time $t = 0$: 

\begin{equation}
\label{eq3}
n =  
\left \{
\begin{tabular}{ccc}
0 & \mbox{at} & $-1.1 \leq x \leq -1.0$ \\
$\cos[\pi (x + 0.5)]$ & \mbox{at} & $-1.0 \leq x \leq 0.0$ \\
$\cos[\pi (x - 0.5)]$ & \mbox{at} & $1.0 \leq x \leq 1.1$ \\ 
0 & \mbox{at} & $1.0 \leq x \leq 1.1$ \ .
\end{tabular}
\right \} 	
\end{equation}

We assume the particles are {\it cold} and their flow velocity is zero at an initial moment of time $t = 0$. The constant value of the electric field, namely $E = E_0$, is defined at the left boundary of the region ($x = -1.1$), while $ E = - E_0 $ defines the right boundary ($ x = 1.1 $). Calculations are performed by choosing a value $ E_0 = 0.1 $. Thus, it is assumed that the electrodes are located at the edges of the system and the electric fields in the area prevent electrically charged particles to escape out of the region $ [- 1.1, 1.1] $. 

Time evolution of a system of charged particles is studied using the particle-in-cell (PIC) method, developed and widely used to investigate the dynamics of a collisionless plasma \cite{Ludw12, Zhang14, Gao16, Med18a, Hicks19b}. By using the PIC method, the material medium is represented by a set of particles that move under the influence of an electromagnetic field, which in turn is determined by the particle distributions and their associated electric currents.  

Generally, the particle motion obeys certain boundary region conditions. In this case, for particles located at the boundaries $ x = \pm 1.1 $, there are no additional conditions other than the influence of the given electric field. Particles that are sufficiently fast can leave the system at one of the boundaries. The motion of each charged particle depends on its initial position. It turns out that for a convenient representation of the solutions obtained, it makes sense to distinguish particles located at $t = 0$ in the region $x < 0$ (particles 1) from particles whose initial coordinates lie in the region $x > 0 $ (particles 2). In calculus we consider an identical value $8 \cdot 10^5$ for the number of particles in each {\em sort}. To represent the dynamics in the phase plane (space) only data for each 200th particle have been used.

\subsection{Caustic solitary density waves in collisionless approximation}

\begin{figure}[bth] 
	\centering
	\includegraphics[viewport = 192 420 433 796,scale=0.82]{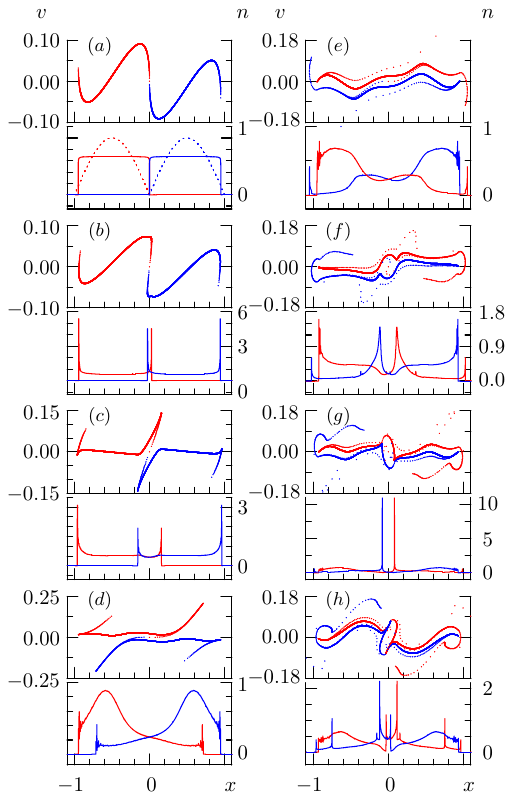}
	\caption{The phase plane for particles 1 (red line) and particles 2 (blue line) top in each double figure. The particle density for particle species 1 (red line) and the particle density for particle species 2 (blue line) are presented in the bottom panel of each double figure. The particle density is represented by the right axis of coordinates. The figures refer to different moments of time $t$: (a) 1.9, (b) 2.7, (c) 4.0, (d) 7.0, (e) 13.5, (f) 19.0, (g) 22.1, (h) 24.0. The dashed line in Fig. (a) shows the initial density distribution.}
	\label{figC}
\end{figure}

We consider the results of the numerical simulation of the problem under investigation. Fig.~\ref{figC} shows the evolution of the particle system state at different moments of time. For each instant of time, the phase plane and the density distribution are shown in a double figure. It can be seen that for $t = 1.9$ (Fig.~\ref{figC}(a)) the initial cosine distribution of the density for each {\em sort} of particles (shown by the dotted curves) changes into a homogeneous one.

At the left and right boundaries of the region of uniformly distributed particles, as well as in the point of contact ($ x = 0 $) of particles 1 and particles 2, a sharp change in the particle velocities is reported. Basically, it is exactly at these points that the derivatives of the particle velocities with respect to the coordinate tend to infinity and the flow begins to acquire a two-stream nature. More clearly, these phenomena are observed later at $t = 2.7$ (Fig.~\ref{figC}(b)). At the points where the two streams merge, density peaks appear for each {\em sort} of particles. The peak density of particles is several times larger than the initial density amplitude. The location of particle stream mergers can be defined as a caustic by analogy with an optical caustic, that represents the place where the rays merge \cite{Arno92}. At $t = 2.7$ the formation of four caustics is clearly visible. Two caustics develop near the boundaries of the region due to particle reflection at the electrodes. In the trap centre caustics arise, because the initial non-uniform particle distribution evolves with the emergence of particle flows that move towards each other. The phase curve $v\left(x\right)$ of all particles (namely, particles 1 and particles 2) is three-valued, corresponding to a three-stream flow.  

\newpage

At subsequent moments of time corresponding to $ t = 4 $ (Fig.~\ref{figC}(c)) and $ t = 7 $ (Fig.~\ref{figC}(d)) the peak values of the density in caustics drop, while the phase space curves illustrate well the fact that density peaks are located at the points where the streams merge. We emphasize that the two extreme caustics developed near the electrodes remain practically still for a long time. Simultaneously, the two central caustics diverge from the centre and give rise to caustic waves. We point out that caustic motion is observed in the study of nonlinear dynamics of one-dimensional flows for a self-gravitating cold non-dissipative substance \cite{Gure93}. In this case particles are attracted to each other and particle countermotion arises naturally. Development of caustics in a system consisting of several kind of charged particles is also possible. Such a phenomenon is observed in a simulation of the expansion of a plasma with negative ions, in vacuum \cite{Med10}.

At subsequent instants of time $ t \ge 9$, the central caustics goes across almost immobile extreme caustics and vanishes at the electrodes. In such case the particles leave the region located between the electrodes. It is interesting to note that only particles 2 (initially located in the region $ x > 0 $) exit (leave) to the left, while particles 1 (initially located in the region $ x < 0 $) exit to the right. In this situation with two near-electrode caustics, escape of particles continues till $t = 13$. Then, particles cease to cross the boundaries and begin to stem near-electrode caustics, which now consists of particles of another {\em sort} (Fig.~\ref{figC}(e), $ t = 13.5 $). These new caustics slowly approach the previously developed near-electrode caustics and together with them new particle streams arise.

Stream interaction leads to the formation of ring-like structures in the phase plane (space), around which there are large enough {\em humps} of the particle density (Fig.~\ref {figC}(f), $t = 19$). Gradually, these {\em humps} of density contract and then turn into peaks of increasing height. The maximum density is achieved at $t = 22.1$  (Fig.~\ref {figC}(g)). Simultaneously, closed ring-shaped structures in the phase plane begin to gradually rotate and remodel (Fig.~\ref{figC}(h), $t = 24$). Moreover, the number of streams increases and new caustics are generated both in the centre and at the peripheral region. In time, the number of caustics will increase and the distance between them will shrink. 

Thus, a system of particles with identical mass and electric charge, confined between the reflecting electrodes, evolves such as to create two near-electrode caustics and two central caustics which eventually diverge from the centre and reach the end electrodes. Experimentally obtained central caustics are illustrated in Fig.~\ref{separation}. Particles can leave the interelectrode region. As time goes by, development of new caustics is possible both near the electrode area and in the central region. New central caustics also diverge from the centre. Eventually, the flow will acquire a multi-caustic and multi-stream character.

\subsection{Caustic solitary density waves in the strongly coupled Coulomb charges in Paul trap}   

\begin{figure}[htb]
	\begin{minipage}[h]{1\linewidth}
		\center{\includegraphics[width=0.8\linewidth]{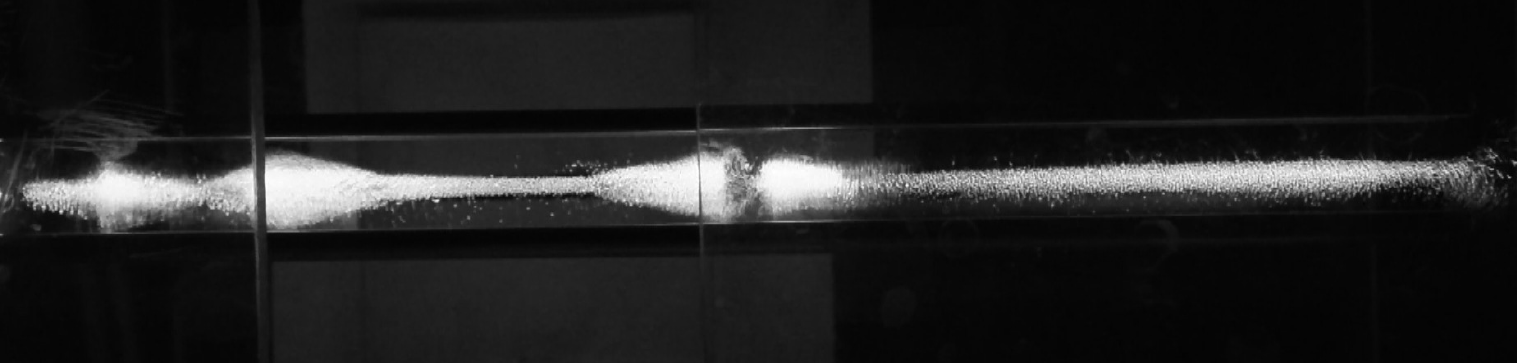}} \\$(a)$ \\ 
	\end{minipage}
	\begin{minipage}[h]{1\linewidth}
		\center{\includegraphics[width=0.8\linewidth]{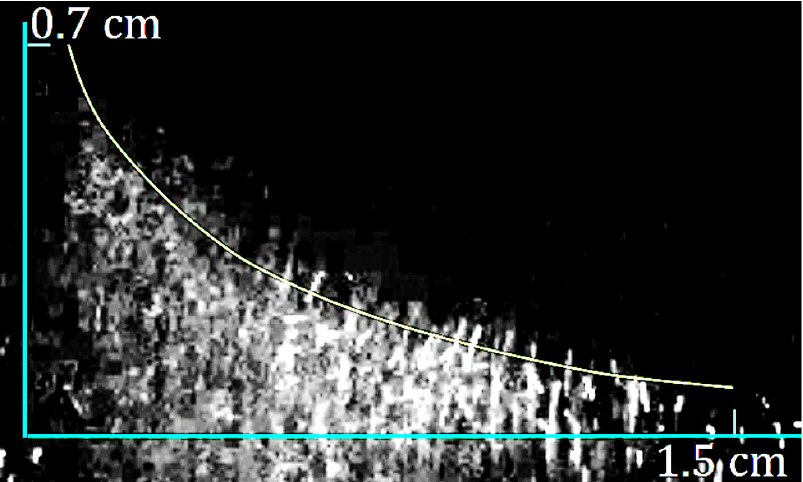}} \\$(b)$
	\end{minipage}
	\caption{(a) Solitary density waves chaotically propagating in different directions for $U_a = 5$ kV and $f = 50$ Hz. (b) Typical example of the upper part of the axillary symmetric density hump with line approximating its external bound.}
	\label{Stable_structure1}
\end{figure}

Solitary density waves appear in the long range structure of charged dust particles (see Fig.~\ref{Stable_structure} (a) and \ref{Stable_structure1} (a)), when the a.c. voltage supplied to the trap electrodes rises up to $5.1$ kV \cite{Syr19b}. In such case density waves chaotically propagate along the trap axis in different directions. 

Fig.~\ref{Stable_structure1} (b) presents the upper part of the axillary symmetric solitary density hump (wave) with its external upper bound approximated by a smooth line. Similar accumulations of particles are investigated in \cite{Arno92}. Following the basic ideas presented in \cite{Arno92}, we explain the physical reason which leads to the occurrence of density waves in experiments and simulations \cite{Syr19b}. To achieve such step we briefly remind the results of density wave simulation. As mentioned before, at $t = 1.9$ (Fig.~\ref{figC}(a)) the initial cosine distribution of the particle density of each {\em sort} of particles changes into a space homogeneous one, but nonuniform in the velocity distribution. The field $v\left( x \right)$ at time $t = 1.9$ can be considered as an initial velocity distribution of particles that are uniformly distributed along the trap axis. 
\begin{figure}[htb]  
	\center{\includegraphics[width=1\linewidth]{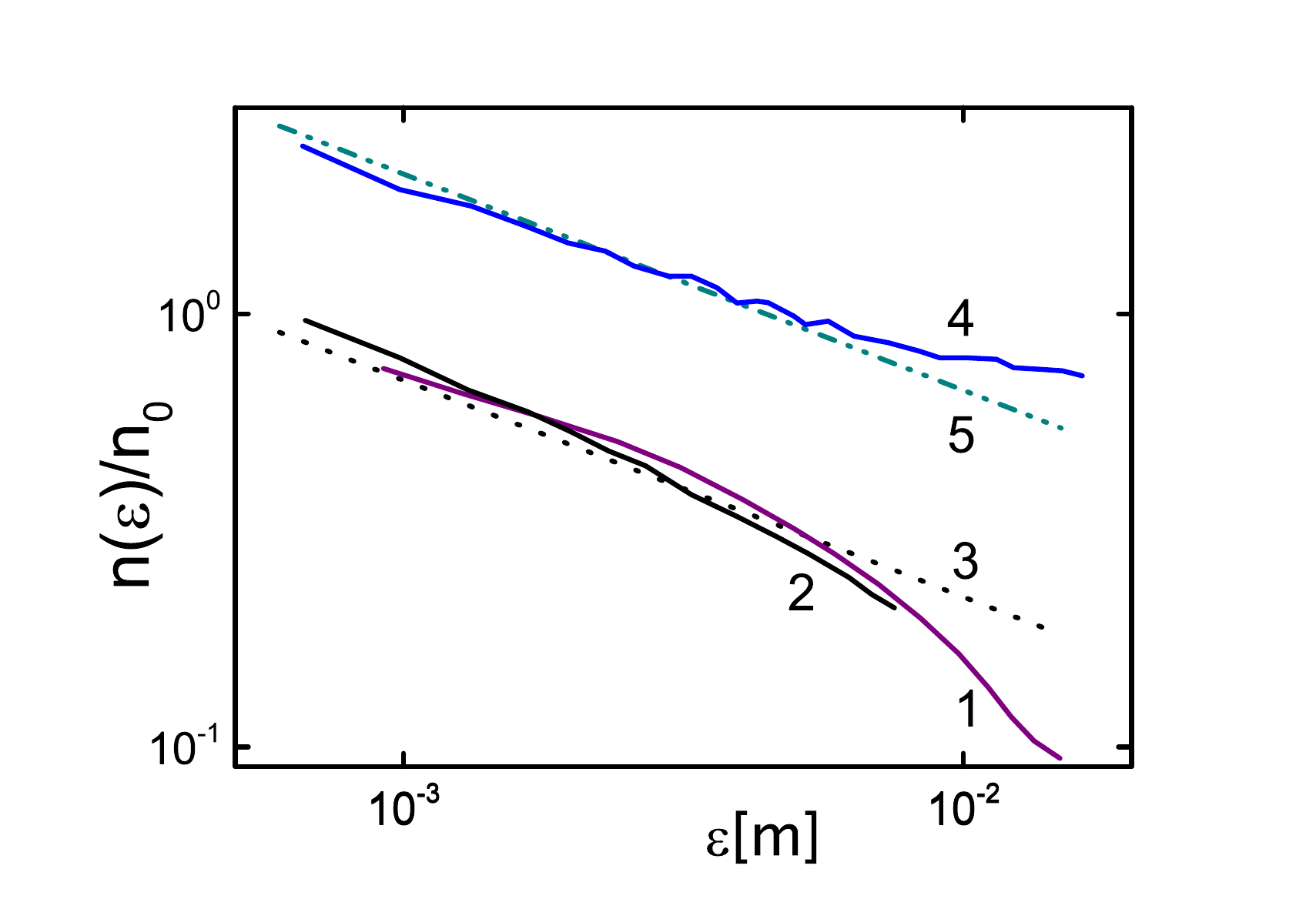}} \\		
	\caption{Experimental and simulated space density distribution of typical density humps of generated waves. Legend: 1 - experimental data; 2 - left caustic at $t = 4.0$ in Fig.~\ref{figC} (c) ; 3 - $const_1/\sqrt\epsilon$; 4 - centre caustic at $t = 2.7$ in Fig.~\ref{figC} (b); 5 - $const_2/\sqrt\epsilon$.} 
	\label{abc}
\end{figure}  

When the particles begin to move, starting from a certain moment in time faster particles begin to leave slower ones behind. Thus, the initially uniform space distribution (density) of particles changes with time. So, deviations from stability in the initial velocity distribution lead to accumulation of particles, as illustrated in Fig.~\ref {figC} (b) and (c) (density peaks are analogues of traffic congestion). If the density profile in these accumulations can be described by a dependence proportional to $\epsilon^{-1/2}$ ($\epsilon$ is the distance from the top of the density hump), we can identify the observed density humps as {\em caustics} \cite{Arno92}.  

The bottom panel in Fig.~\ref {abc} presents density profiles of experimental and simulated particle accumulations along with their dependence $const/\sqrt{\epsilon}$ versus distance $\epsilon$, with fit constants ($const_1$ and $const_2$) at different moments of time. Experimental and simulated density profiles of wave humps are in a good agreement with each other, and depend on $const/\sqrt{\epsilon}$. Therefore, we can identify particle accumulations as caustic density waves.

\subsection{Energy income and energy loss for density waves in a Paul trap}  

Let us consider the energy that is characteristic for particle motion in a classical linear Paul trap \cite{Paul90}. If the $z$ axis is directed along the trap symmetry axis, the Mathieu equation that describes dynamics of a single particle can be expressed as  
\begin{subequations}
	\begin{eqnarray}\label{dimensional}
	\frac{d^2x}{d\tau^2} + (a_x + 2 q_x \cos{2 \tau})x + b\frac{dx}{d\tau} = 0,\\
	\frac{d^2y}{d\tau^2} + (a_y + 2q_y \cos{2 \tau})y + b\frac{dy}{d\tau} = 0,
	\end{eqnarray}
\end{subequations}
where $$a_x = -a_y = \frac{4QU}{mr_0^2\omega^2}, \ q_x = - q_y = \frac{2QV}{mr_0^2\omega^2} \ ,$$ are the adimensional characteristic constants for the trap geometry, $V$ is the amplitude of the a.c. voltage, $U$ stands for the d.c. voltage ($V = 0$ in our case), $\tau = {\Omega t}/2$ represents the dimensionless time, $b = {2\gamma}/{m\omega}$ is the damping parameter, $Q$ denotes the particle charge, $\gamma = 6\pi \eta  R$ stands for the linear damping coefficient, $\eta$ is the dynamic air viscosity under SATP conditions, while $R$ represents the radius of a spherical particle. 

\begin{figure}[htb]
	\centering	
	\includegraphics[scale=0.21]{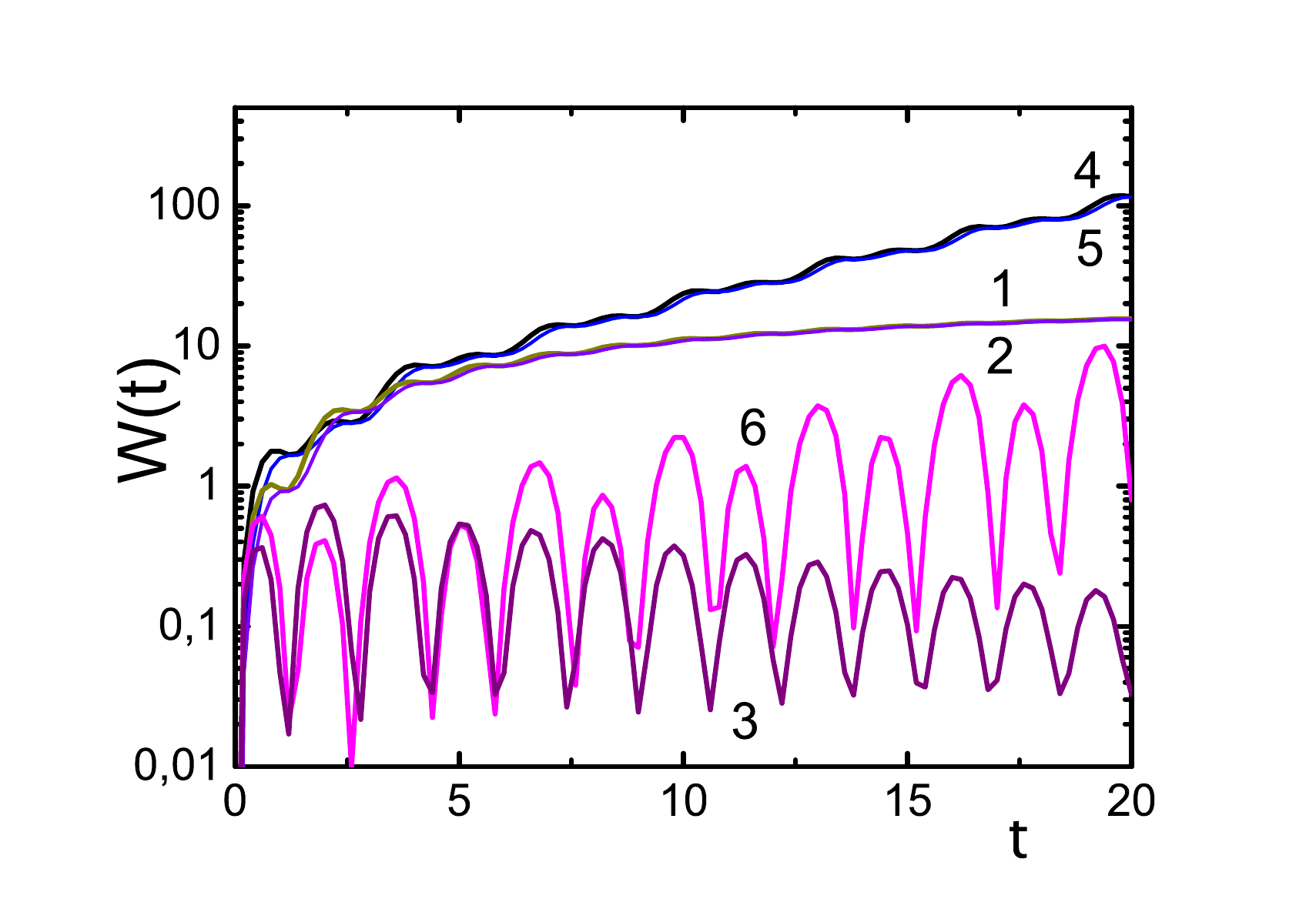}	
	\includegraphics[scale=0.21]{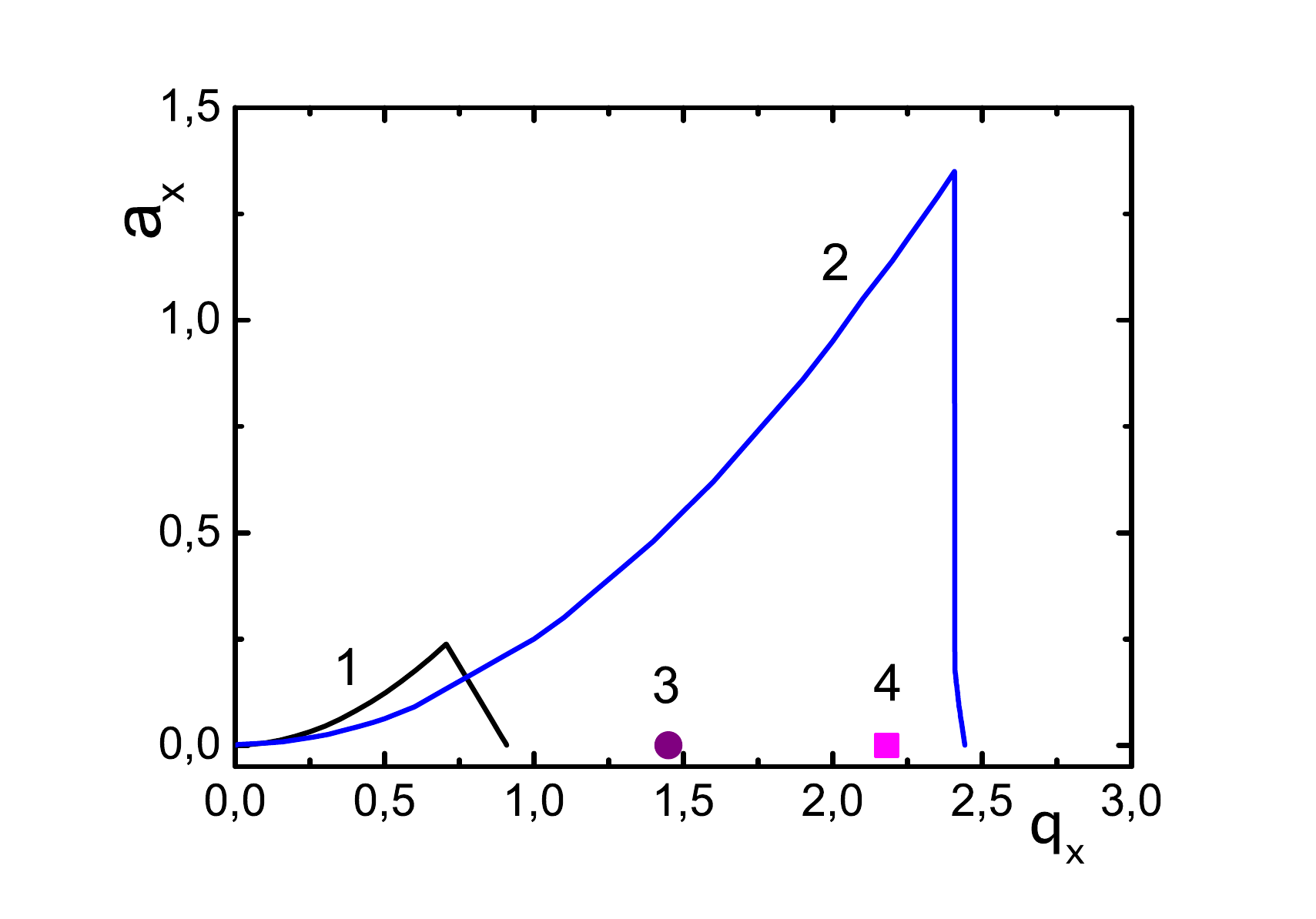}	
	\caption{(Left panel) Energy income and energy loss for a single particle confined in the trap. Stable particle structure (Fig.~\ref{Trap} (b)): lines 1 - $W_{E}$, 2 - $ W_{fr}$, 3 - ($W_{E}-W_{fr}$). Density wave generation (Fig.~\ref{Stable_structure1} (a)) : lines 4 - $W_{E}$, 5 -$ W_{fr}$, 6 - ($W_{E}-W_{fr})$. Mechanical work $W$ is given in conditional units. (Right panel) Stability diagram of a single particle levitated in a quadrupole Paul trap, in the ($a_x - q_x$) plane. Above the stability lines $1$ and $2$ the particle is ejected from the trap (more details can be found in \cite{Nas01, Hase95, Vini15}). For line 1 air viscosity is zero; in case of line 2 air viscosity is considered for SATP (normal) conditions. Experimental results: 3 -- stable particle structure (Fig.~\ref{Trap} (b)),  4 -- dust wave generation (Fig.~\ref{Stable_structure1} (a)).}
	\label{work}
\end{figure}

Stable behaviour of a dust particle structure or the occurrence of density waves depend on the energy income and energy loss for a particle levitated in a Paul trap. We consider the mechanical work per particle of the electrical field in the trap $W_{E}$ and the mechanical work of the viscosity forces $W_{fr}$. The work of the electric field can either increase or decrease the charged particle energy, while particle energy losses are defined by the work performed by the viscosity forces. The corresponding mechanical works are described by the scalar product between forces and particle displacements: 

\begin{eqnarray}\label{Wtr}
W_{E}(t)= \int_0^t d\tau  \left( \mathbf{F}^{E}(\tau) | \frac{d\mathbf{r}}{d\tau}\right), \\
W_{fr}(t)= \int_0^t d\tau  \left( \mathbf{F}^{fr}(\tau) | \frac{d\mathbf{r}}{d\tau}\right),
\end{eqnarray}
where brackets denote the scalar product of vectors $\mathbf{F}^{E}(\tau)$ and $\frac{d\mathbf{r}}{d\tau}$. Particle motion in the trap can occur along the classical trajectories $\mathbf{r}(\tau)$ with velocity ${d\mathbf{r}}/{d\tau}$, where $\mathbf{F}^{E}(\tau)$ is the electrical trap force acting upon a particle and $\mathbf{F}^{fr}(\tau) = -6 \pi \eta  R \frac{d\mathbf{r}}{d\tau}$ represents the damping force. Fig.~\ref{work} illustrates the time evolution of the energy income $W_{E}(t)$ and energy loss $W_{fr}(t)$ for a single levitated particle.  

Shifts in oscillations of lines observed in Fig.~\ref{work} illustrate that changes in particle velocity are lagging with respect to changes in the electrical trap forces, that are proportional to the particle acceleration. This is the physical reason behind the energy exchange between particles and the electric field within the trap. The decay of the difference ($W_{E}-W_{fr}$), indicated by line $3$, shows the transition to a stable regime in Fig.~\ref{Trap} (b). Line $6$ indicates that higher a.c. voltages supplied to the trap electrodes result in an increase of the energy income, which represents the physical reason for density waves that develop in many particle structures (Fig.~\ref{Stable_structure} (a)).  

The right panel in Fig.~\ref{work} shows the stability diagram for $30 \ \mu$m particles (which is the most probable particle size according to \cite{Syr16b}) in the $a_x - q_x$ plane, when $\gamma = 0$ (line 1) and $\gamma = 6 \pi \eta R$ (line 2). Point 3 in Fig.~\ref{work} corresponds to experimental parameters that characterize a stable Coulomb structure in Fig.~\ref{Trap} (b). Point 4 located near the boundary of the stability region corresponds to density wave generation (Fig.~\ref{Stable_structure1} (a)). These points are obtained as numerical solutions of the equations (\ref{dimensional}). Further, we obtain the experimental diagram of exciting density waves in strongly coupled Coulomb systems of particles confined in air, under SATP (normal) conditions.   

The physical mechanism responsible for particle motion along the trap axis in a viscous gas is not yet completely explained. The mechanisms responsible for energy transfer from the transverse oscillations of the trap electric field into the longitudinal direction are still unknown. This change is likely possible due to the strong inter-particle interaction that results in fast {\em thermalisation}, in case of strongly coupled nonequilibrium systems of particles. Other possibilities are shown in \cite{Bark10} where dielectric nanoparticles levitated in a dynamic trap are considered and both degrees of freedom are coupled to the cavity field. The longitudinal electric field can also occur as a result of the accumulating volume charge that appears after the build up of charge density accumulations. Contrary to the case of the glow gas-discharge plasma, physical mechanisms such as energy transfer in all particle degrees of freedom related to the dependence of the electric charge on space position or random fluctuations \cite{Bel99, Vaul99, Fili05} do not occur in our experiments, due to electric charge conservation for periods of time of up to several hours \cite{Syr16b}. A detailed treatment of the problem of energy redistribution in the particle motion along the trap axis represents an objective of further investigations.

\subsubsection{Discussion}

Ref. \cite{Syr19b} represents the first pioneering work on density wave propagation in a linear structure of strongly coupled charged particles. The solitary density wave is demonstrated to occur as a single hump in case of a quadrupole electrodynamic trap operating under SATP (normal) conditions, when the energy losses due to air viscosity can be compensated by the energy increase owing to the a.c. trapping field. Generation of density waves is demonstrated by adding charged particles to the electrodynamic trap or by performing a smooth increase of the a.c. voltage amplitude. The physical possibility of analogous hump density waves (caustics) is discussed by V. I. Arnold \cite{Arno92}, as it is caused by the nonuniform velocity distribution of the dust particles.	

Density waves that crop up can be identified as caustics according to Arnold's definition \cite{Arno92}, as the experimental wave density profile can be described by theoretical dependencies predicted by the caustic theory ($const/\sqrt{\epsilon}$, where $\epsilon$ is the distance from the caustic singularity). The results of the experimental investigations can be considered as new proofs that confirm the versatility of the caustic theory in describing different physical phenomena, not only for collisionless systems of particles but also when the inter-particle interaction and the interaction with external fields in viscous media are strong. Generation of a density wave in a quadrupole electrodynamic trap has to be demonstrated, when energy losses due to the drag force can be compensated by the energy contribution of the a.c. trapping field. Therefore, these waves can be regarded as an outcome of some sort of self-sustaining waves in a non-equilibrium environment and may be identified as autowaves. The mathematical apparatus required to describe the wave phenomenon needs to be further developed. In opposition, the energy of solitons has to be preserved as they are required to satisfy a certain set of additional physical criteria \cite{Lan13}.

\subsection{Studies on the interaction between an acoustic wave and levitated microparticles in a linear Paul trap}\label{AcoustWave}

We study the effect that a low frequency acoustic wave inflicts upon a cloud of particles levitated in a linear Paul trap, operating under SATP conditions. An acoustic wave represents a type of longitudinal wave that consists of a sequence of pressure pulses or elastic displacements of the material, whether gas, liquid, or solid, in which the wave propagates. The speed of an acoustic wave is determined by the temperature, pressure, and elastic properties of the medium it crosses. In case of an electrodynamic trap, application of an acoustic wave creates an additional force field that superposes over the trapping field. The combined effect of the two force fields may result in thermalisation and stabilisation of the particle. Thus, improved manipulation of levitated particles is achieved by implementing better control over the position in space, as well as a more precise selection of the specific charge ratio $Q/M$. In case of a 3D Paul trap particle dynamics strongly depends on the specific charge ratio. As the effect of an acoustic wave is essentially mechanic, it is possible to decouple the mass $M$ and the electric charge $Q$ in the equation of motion. The parameters that characterize an acoustic wave are the intensity and the associated frequency. Both parameters span a wide range of values so that the mechanical effect of an acoustic wave can be determined in a precise manner. The experiments performed are presented in detail in Refs. \cite{Sto11, Sto08}. 

Generally, a two-dimensional (2D) quadrupole field is generated by applying a trapping RF-voltage, $V_0 \cos \Omega t$, between diagonally connected pairs of cylinders, mounted parallel and equidistantly spaced around the trap central axis. Particle dynamics in the $x - y$ (radial) plane, normal to the trap longitudinal axis $z$, is governed by the same Mathieu type of equation as is the case for a 3D cylindrical Paul trap \cite{Major05, Stock16}. The setup used in \cite{Sto11} consists of 4 equidistant, radially spaced electrodes (E1 $ \div$ E4), and two endcap electrodes (E5 and E6), as illustrated in Fig.~\ref{ac1}. The a.c trapping voltage $V_{ac} = V_0 \cos \left( \Omega t\right)$ is applied between the E2 and E3 electrodes, where $\Omega = 2 \pi \nu_0$, $\nu_0 = 50$ Hz. A d.c. voltage $U_x$ is applied to the lower electrode E4 with an aim to perform particle diagnosis and compensate gravity, while the upper trap electrode E1 is electrically connected to the ground potential. An axial confinement voltage, denoted by $U_0$, is applied between the endcap electrodes (denoted as E5 and E6) in order to prevent particle escape along the trap axis. Under the assumption that the trap length $L$ is sensibly larger than the trap radius $R$, the electric potential near the axis region (afar from the ends) is \cite{Major05}

\begin{equation}\label{acpot}
\Phi \left(x, y, t \right) = \frac {x^2 - y^2}{2R^2} \left( U_0 + V_0 \cos \Omega t \right) \ , \ \ x, y \ll R
\end{equation}    

The trap geometry used in \cite{Sto11} is variable ($L = 3 \div 7$ cm), a common feature that is shared with other previously designed linear traps \cite{Ghe98, Sto01}. During the experiments the value of $L$ is kept constant at $3.5$ cm. According to \cite{Sto11} the equations of motion in the $x - y$ (radial) plane for a particle levitated in a Paul trap and located in a region close to the trap axis, that undergoes action of an acoustic wave, can be expressed as

\begin{eqnarray}\label{Acou1}
M \frac{d^2 x}{dt^2} = - \frac{Q}{R^2}\left( U_0 + V_0 \cos \Omega t \right) x - k \frac{dx}{dt} + F_{ac x} \left( t \right)  \ , \\
M \frac{d^2 y}{dt^2} = \frac{Q}{R^2}\left( U_0 + V_0 \cos \Omega t \right) y - k \frac{dy}{dt} + F_{ac y} \left( t \right) \ ,
\end{eqnarray}            
where $F_{ac}$ is the force exerted upon the particle by the acoustic wave, while the second term in each equation describes the drag force experienced by an object that moves through a fluid, either liquid or air. The drag force is contrary to the direction of the incoming flow velocity. For example, the drag force acting upon an object that moves in a fluid is

\begin{equation}
D = \frac 12 C \rho A v^2  \ ,
\end{equation} 
where $C$ stands for the drag coefficient that varies with the speed of the body in motion, $\rho$ is the density of the fluid, $A$ represents the cross-section area of the body that is normal to the flow direction and $v$ denotes the speed of the object with respect to the fluid. Considering spherical particles, the force of viscosity (or the drag force) acting on a small sphere moving through a viscous fluid is given by the Stokes law

\begin{equation}
F_d = 6 \pi \eta R_p v \ ,
\end{equation}
where $\eta $ is the dynamic viscosity (in case of air, under SATP conditions, $\eta = 18.5 \ \mu$Pa$\cdot $s), $R_p$ stands for the particle radius and $v$ denotes the flow velocity relative to the particle. The microparticle weight is considered negligible. Further, we introduce the adimensional time denoted as $\xi = \Omega t/2$ \cite{Major05}, which modifies the system of equations (\ref{Acou1}) into

\begin{equation}\label{Acou2}
\frac{d^2u}{d\xi^2} + \delta \frac{du}{d\xi} + \left( a_u + 2 q_u \cos \left( 2\xi \right)\right) - f_{ac\ u} = 0 ,\ u = x, y \ ,
\end{equation} 
with
\begin{equation}\label{Acou3}
a_x = - a_y = \frac{4QU_0}{M\Omega^2 R^2} , q_x = -q_y = \frac{2QV_0}{M\Omega^2 R^2} , \ \delta = \frac {12 \pi \eta R_p v}{M\Omega} , \ f_{ac\ u}\left( t\right) = \frac{4 F_{ac\ u}}{M\Omega^2} \ .
\end{equation} 

\begin{figure*}[!ht]
	\centering
	\includegraphics[scale=1.1]{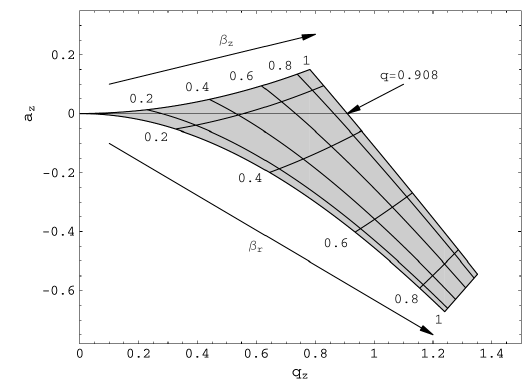}	
	\caption{First stability domain of the Paul trap including lines of constant values for $\beta_r$ and $\beta_z$. Picture reproduced from \cite{Major05} by courtesy of Prof. G. Werth.}
	\label{PaulStab} 
\end{figure*}

The values of the adimensional parameters determine the frontiers of the stability domains for solutions of the Mathieu equation \cite{Major05, March05}. The first stability domain of the Mathieu equation is illustrated in Fig.~\ref{PaulStab}. If the adiabatic approximation is valid  $\left( |a_x|, |a_y| \ll 1 \ \textrm{and} \ |q_x|, |q_y| \ll 1 \right)$ , the solutions of the system of eqs. (\ref{Acou2}) are bound and may be expressed as \cite{Major05}

\begin{eqnarray}
x\left( t\right) = x_0 \cos\left(\omega_x t + \varphi_x\right) \left( 1 + \frac{q_x}{2}\cos\Omega t \right) , \\
y\left( t\right) = y_0 \cos\left(\omega_y t + \varphi_y\right) \left( 1 + \frac{q_y}{2}\cos\Omega t \right) \ ,
\end{eqnarray}  
where $x_0, y_0, \varphi_x, \textrm{and} \ \varphi_y$ are determined from the initial conditions, and
\begin{equation}
\omega_x = \frac{\Omega}{2}\sqrt{\frac{q_x^2}{2} + a_x} , \ \omega_y = \frac{\Omega}{2}\sqrt{\frac{q_y^2}{2} + a_y} \ .
\end{equation}

Thus, the dynamics of a single particle (ion) in the radial ($x-y$) plane is a harmonic oscillation (called {\em secular} motion) at two characteristic frequencies, $\omega_x$ and $\omega_y$, which is amplitude modulated at the drive frequency $\Omega$ ({\em micromotion}) \cite{Libb18, Wang10}. If one disregards the micromotion, ion dynamics in the radial plane $x-y$ can be assimilated with the motion of a particle in a harmonic potential. 

\subsubsection{Experimental setup}

\begin{figure}[bth]
	\centering
	\includegraphics[scale=1]{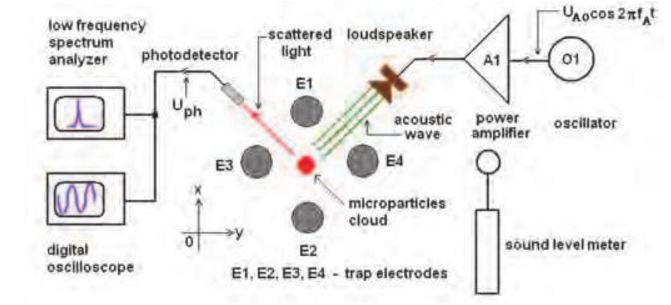}	
	\caption{Experimental setup for acoustic wave excitation. Measurement chain. Picture reproduced from \cite{Sto11} with kind permission of O. Stoican.}
	\label{ac2}
\end{figure}

The experimental setup (shown in Fig.~\ref{ac1}) relies on the method described in \cite{Schle01}. The beam generated by a cw laser diode (650 nm, 5 mW) is directed along the longitudinal axis of the trap. The electric potential along the trap $z$ axis is well characterized by eq. (\ref{acpot}). A photodetector (PD) placed outside the trap, normal to the beam direction, is used to collect a fraction of the radiation scattered by the trapped microparticles. Background light effects are mitigated using a dedicated electronic circuit. Acoustic excitation of trapped microparticles is performed by means of a loudspeaker that generates a monochromatic acoustic wave with frequency $\nu_A$. The scattered radiation intensity is amplitude modulated by the particle motion. The spectrum of frequency components or more precisely the frequency-domain representation of the signal, is visualized by means of a spectrum analyzer. The experimental setup (shown in Fig.~\ref{ac2}) is an enhanced version of the previous setup used in \cite{Sto08}.  

\begin{figure}[htb]
	\centering
	\includegraphics[scale=1]{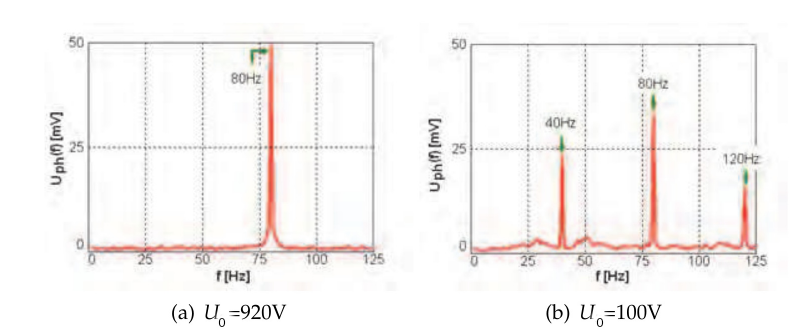}	
	\caption{Spectra of the photodetector output voltage in absence of acoustic excitation. Experimental parameters: $\nu_0 = 80 $Hz, $V_0 = 3.3$ kV, and $U_x = 0$ V. Picture reproduced from \cite{Sto11} with kind permission from O. Stoican.}
	\label{ac3}
\end{figure}

\begin{figure}[htb]
	\centering
	\includegraphics[scale=1.3]{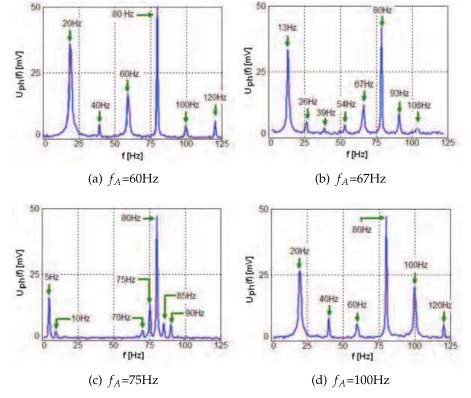}	
	\caption{Spectra of the photodetector output voltage in case of acoustic wave excitation of the microparticles. Experimental parameters: $\nu_0 = 80$ Hz, $V_0 = 3.3$ kV, $U_x = 0$ V, $U_0 = 920$ V, and sound level $\approx$ 85 dB. $f_A$ in the picture corresponds to $\nu_A$ in our case. Picture reproduced from \cite{Sto11} with kind permission from O. Stoican.}
	\label{ac4}
\end{figure}

Polydisperse Al$_2$O$_3$ microparticles with dimensions ranging between $60 \div 200 \ \mu$m are used. When the operating point in the stability diagram of the Mathieu equation reaches the boundary ({\em springpoint}, according to \cite{Davis90}), large amplitude oscillations of the microparticle occur which end up in particle loss. Experimental data, acquired in absence and in presence of the acoustic excitation, are reproduced in Fig.~\ref{ac3} and Fig.~\ref{ac4}. 

Fig.~\ref{ac4} clearly illustrates how an acoustic wave induces additional lines in the motional (dynamic) spectrum of trapped microparticles. If the value of the acoustic wave frequency $\nu_A$ is close to the value of the a.c. voltage frequency $\nu_0$, an amplitude modulation of the photodetector voltage is reported. The phenomenon is identical with the beat frequency between two wavelengths. A numerical analysis is also performed in \cite{Sto11} while the results show good agreement with the experiment. Based on the experimental results it is supposed that the operating point in the stability diagram, for the linear trap under test, corresponds to $q_x \approx 0.908$ (namely $Q/M \approx 5.4 \times 10^{-4}$ C/kg, according to eq. \ref{Acou3}). If the the microparticle mass and the specific charge ratio $Q/M$ are known, then the electric charge value, $Q$, can be estimated.

\section{Multipole linear electrodynamic traps. Discussion and area of applications}\label{Sec6}

We review some of the most important milestones that describe experiments with microparticles in quadrupole and especially multipole traps, in an attempt to describe state of the art results in the field. Among the first papers aimed at investigating the confinement of ion clouds in multipole traps, we can mention Refs. \cite{Walz94} and \cite{Mih99}. Particle (ion) dynamics was investigated by means of numerical integration of the equations of motion and ion stability was shown to be phase and position dependent. The trajectories envelope and the phase-dependence of ion location in quadrupole and hexapole RF traps, have been explored.

An electrodynamic trap used for investigating charging processes of a single grain under controlled laboratory conditions is proposed in \cite{Bera10}. A linear cylindrical quadrupole trap is used and every electrode is split in half, in order to achieve a harmonic trapping potential and thus perform precise measurements on the specific charge-to-mass ratio, by determining the secular frequency of the grain. This approach relies on a method previously applied in case of microparticles \cite{Sto08}. In Ref. \cite{Vasi13} mathematical simulations are used to investigate a dust particle’s behaviour in a MEIT with quadrupole geometry. Regions of stable confinement of a single particle are reported in dependence of the frequency and charge-to-mass ratio. An increase of the medium’s dynamical viscosity has the effect of extending the confinement region for charged particles. Ordered Coulomb structures of charged dust particles that occur in a quadrupole trap operated under SATP conditions, are also reported by the group from JIHT Moscow \cite{Lapi16b}. Very recent papers focus on measuring the charge of a single dust particle \cite{Deput15a, Lapi16a}, on the effective forces that act on a microparticle levitated in a Paul trap \cite{Lapi15b, Lapi16c}, or investigate confinement of microparticles in a gas flow \cite{Lapi15a, Deput15b}. Other papers report on the dynamics of micrometer sized particles levitated in a linear electrodynamic trap that operates under SATP conditions, where time variations of the light intensity scattered by microparticles are recorded and analyzed in the frequency domain \cite{Sto11, Vis13}. Ref. \cite{Marmi13} presents a RF/high voltage pulse generator that provides suitable waveforms, required to operate a planar multipole ion trap/time-of-flight mass spectrometer.

MEITs and the physics associated to them are investigated in Ref. \cite{Libb18}. Ring electrode geometry and quadrupole linear trap geometry centimetre sized traps are examined, used as tools for physics teaching labs and lecture demonstrations. A viscous damping force is introduced in order to characterize the motion of particles confined in MEITs operating under SATP conditions \cite{Vini15}. It is demonstrated that Stokes damping describes well the mechanism of particle damping in air with respect to the microparticle species used. The secular force in the one-dimensional case is inferred. Nonlinear dynamics is also observed \cite{Nay04} as such mesoscopic systems \cite{Porr08} represent versatile tools aimed at performing integrability studies and investigations on quantum chaos \cite{Mih10b}. 

Investigations on higher pole traps intended to be used for ion trapped based frequency standards are firstly reported in Ref. \cite{Prest99}. A 12-pole trap is tested and ions are shuttled into it originating from a linear quadrupole trap. The outcome of the experiments is a clear demonstration that a multipole trap holds larger ion clouds with respect to a quadrupole trap. The paper also emphasizes on the issue of space charge interactions that are non-negligible in a multipole trap. The Boltzmann equation that characterizes large ion clouds in the general multipole trap of arbitrary order is solved. The authors explain that fluctuations in the numbers of trapped ions influence the clock frequency much less severely than in the quadrupole case, which motivates present and future quest towards developing and testing multipole traps for high-precision frequency standards \cite{Van15}. Ref. \cite{Burt06} reports on the JPL multipole linear ion trap standard (LITS) which demonstrates excellent frequency stability and improved immunity from two of its remaining systematic effects: the second-order Doppler shift and the second-order Zeeman shift. The authors report developments that reduce the residual systematic effects to less than $ 6 \times 10^{-17} $ and the highest ratio of atomic transition frequency to frequency width (atomic line quality factor) ever demonstrated until then for a microwave atomic standard operated at room temperature. 

An Electron Spectrometer MultiPole Trap (ES-MPT) setup was devised by Jusko and co-workers \cite{Jusko12}. A radio-frequency (RF) ion trap and an electron spectrometer are used in the experiment with an aim to estimate the energy distribution of the electrons produced. Results of simulations and first experimental tests with monoenergetic electrons from laser photodetachment of $O^{-}$ are presented. Other papers of the group from the Charles Univ. of Prague approach the study of associative photodetachment of $H^{-} + H$ using a 22-pole trap combined with an electron energy filter \cite{Rou09}, study of the capture and cooling of $OH^{-}$ ions in multipole traps \cite{Trip06, Otto09}, or investigations of  $H^{-}$ ions in RF octopole traps with superimposed magnetic field \cite{Rou10}. 

Numerical simulations of the kinetic temperature of ions levitated in a cryogenic linear multipole RF trap that are subject to elastic collisions with a buffer gas, are presented in \cite{Asva09}. Ion temperature dependence on the trapping parameters is analyzed in detail. A disadvantageous ion-to-neutral mass ratio combined with high trapping voltages might result in ion heating, at temperatures much higher compared to the trap temperature. Transversal probability distributions are obtained for a 22-pole ion trap and various trapping voltages. As the trapping voltage increases the ions are pushed closer to the RF electrodes. A new design of a 22-pole cryogenic electrodynamic trap is discussed in \cite{Asva10} where buffer gas-cooled H$_2$D$^+$ ions are confined at a nominal trap temperature of 14 K. One of the experimental issues in case of cryogenic traps lies in the realization of a RF resonator that should dissipate very little power \cite{Gando12}. A linear octopole ion trap that is suitable for collisional cryogenic cooling and spectroscopy of large ions is investigated in \cite{Boya14}. The strong radial confinement achieved by the trap is superior to the performance of a 22-pole trap as it enhances the dissociation yield of stored ions to 30 \%, which recommends it for laser spectroscopy studies \cite{Dem15}. A 22 pole ion trapping setup intended for spectroscopy is tested in \cite{Asva14}, using CH$_5^+$ and H$_3$O$^+$ ions that are buffer gas cooled by using He at a temperature of 3.8 K. A late experiment reports on a linear cryogenic 16-pole wire ion trap developed for cryogenic ion spectroscopy at temperatures below 4 K \cite{Geis21}, demonstrating low ion temperatures and promising perspectives for high sensitivity spectroscopy. 

Multipole ion trap designs based on a set of planar, annular or concentric electrodes, are presented in \cite{Clark13}. Such millimeter (mm) scale traps are shown to exhibit trap depths as high as tens of electron volts. Several example traps are investigated, as well as scaling of the intrinsic trap characteristics with voltage, frequency, and trap size. Stability and dynamics of ion rings in linear multipole traps as a function of the pole number is discussed in \cite{Carta13}. Multipole traps present a flatter potential in their centre and therefore a modified density distribution compared to quadrupole traps. Crystallization processes in multipole traps are investigated in \cite{Champ13} where the dynamics and thermodynamics of large ion clouds in traps of different geometry is explored. Applications of QIT and multipole ion traps span areas such as Quantum Information Processing (QIP) \cite{Bruz19, War20, Rat20, Wan20}, metrology of frequencies and fundamental constants \cite{Quint14, Orsz16, Kno15}, production of cold molecules and the study of chemical dynamics at ultralow temperatures (cold ion–atom collisions) \cite{Asva12, Gian19}. 

A recent paper reports on a microfabricated planar ion trap featuring 21 d.c. electrodes \cite{Bahra19}, intended to study interactions of atomic ions with ultracold neutral atoms. The design also integrates a compact mirror magneto-optical chip trap (mMOT), used for cooling and confining neutral $^{87}$Rb atoms that are transferred into an integrated chip-based Ioffe-Pritchard trap potential. Ref. \cite{Hicks19b} shows that RF charged particle traps, such as the Paul trap or higher order multipole traps, represent excellent tools to levitate quasi-neutral plasmas and uses PIC simulation to characterize them. Moreover, the presence of positive and negative plasma species mitigates the ejection of particles that occurs as an outcome of space charge repulsion. 

The temperature dependent evolution of an ion ensemble in a 3D QIT is explored in \cite{Aksa20} for different RF voltage waveforms, taking into account octopole field contributions. An in-phase four electrode bar trap is introduced in \cite{Rud20b} for which the potential distribution is similar to an octupole trap. The trap is intended for microparticle investigation, while it can also generate an ideal octupole field. In addition, RF linear multipole traps exhibit sensitivity with respect to electrode misalignment or geometric imperfections in their design, which eventually leads to symmetry breaking and disrupts trap operation \cite{Marche21}.

\subsection{Improved  stability in multipole ion traps. Optimizing the signal-to-noise ratio}

In case of quadrupole traps the second-order Doppler effect is the result of space-charge Coulomb repulsion forces acting between trapped ions of like electrical charges. The Coulomb forces are balanced by the ponderomotive forces generated due to ion motion in a highly non-uniform electric field. For large ion clouds most of the motional energy is found in the micromotion. Multipole ion trap geometries significantly reduce all ion number-dependent effects that result through the second-order Doppler shift, as ions are weakly bound with confining fields that are effectively zero through the inner trap region and grow rapidly close to the trap electrode walls. Because of the particular shape of the trapping fields charged particles spend a small amount of time in the high RF electric field area, which helps in mitigating the RF heating phenomenon. Multipole traps are also used as tools in analytical chemistry with an aim to confine trapped molecular species that exhibit many degrees of freedom \cite{Trip06, Ger08a, Ger08b, West09}. Space-charge effects are not negligible for such traps. Nevertheless, they represent extremely versatile tools to investigate the properties and dynamics of molecular ions or to simulate the properties of cold plasmas, such as astrophysical plasmas or the Earth atmosphere. Stable confinement of a single ion in the RF field of a Paul trap is well known as the Mathieu equations that describe particle motion can be analytically solved \cite{Ghosh95, Major05, Baril74, March05}.

New Structures for Lossless Ion Manipulations (SLIM) module were recently developed to explore ion trapping at a pressure of 4 Torr \cite{Zhang15}. Based on such setup ions can be trapped and accumulated with up to 100 \% efficiency, contained for minimum 5 hours with insignificant losses and then rapidly ejected outside the SLIM trap. Such features open new pathways for enhanced precision ion manipulation.

\subsection{Regions of stability in multipole traps. Discussion}

A linear quadrupole Paul trap uses a combination of time varying and static electric potentials in order to create a trapping configuration that confines charged particles such as ions, electrons, positrons, and micro or nanoparticles (NPs) \cite{Orsz16, Kno15, Hart17}. A radiofrequency (RF) or a.c. voltage is used in order to generate an oscillating quadrupole potential in the $x-y$ plane which achieves radial confinement of the trapped particles. Axial confinement of positively charged ions (particles) is achieved by means of a static potential applied between two endcap electrodes located at the trap ends, along the trap axis ($z$ plane). The RF or a.c. field induces an effective potential which harmonically confines particles in a region where the trapping field exhibits a minimum, under conditions of dynamic stability \cite{Major05, March05, Vini15, Libb18}. As it is almost impossible to achieve quadratic a.c. and d.c. endcap potentials for the whole trap volume, it can be assumed that near the trap axis the electric potential can be regarded as harmonic which is a sufficiently accurate approximation.

As already pointed out, a single ion can experience stable confinement in the RF field of a quadrupole Paul trap. This is no longer the case for high-order multipole fields where the equations of motion do not admit an analytical solution. In the latter case particle dynamics is quite complex, as it is described by non-linear, coupled, non-autonomous equations of motion. The solutions for such system can only be found by performing numerical integrations \cite{Major05, Fort10a, March05, Rieh04}. The assumption of an effective trapping potential still holds and ion dynamics can be separated into a slow drift (the {\em secular motion}) and the rapid oscillating micromotion \cite{Otto09, Champ09}. The effective potential in case of an ideal cylindrical multipole trap can be expressed as

\begin{equation}
V^{*}(r) = \frac{1}{4} \frac{n^2(qV_{ac})^2}{m \Omega^2 {r_0}^2}{\left(\frac{r}{r_0}\right)}^{(2n - 2)} \ ,
\end{equation}         
where $V_{ac}$ is the amplitude of the RF field and $n$ represents the number of poles (for example, $n = 2$ for a quadrupole trap, $n = 4$ for an octopole trap and $n = 6$ for a 12-pole trap). The larger the value of $n$, the flatter the potential created in the trap centre and the steeper the potential close to the trap electrodes. 

Recent experiments show that in case of linear multipole RF (Paul) traps, single ring structures occur as an outcome of the radial potential geometry. By investigating the stability of these tubular structures as a function of the number of trap poles, it is demonstrated that ions arrange themselves as a triangular lattice folded onto a cylinder \cite{Carta13}. Numerical simulations performed for different lattice constants supply the normal mode spectrum for these ion structures.   

Nonlinear dynamics of a charged particle in a RF multipole ion trap is investigated in \cite{Rozh17} and the expression of the 2D electrical potential is supplied. The existence of localization regions for ion dynamics in the trap is demonstrated. The associated Poincar{\'e} sections illustrate ion dynamics to be highly nonlinear. Ref. \cite{Rud19} proposes a new method to analyze the potential field in multipole ion traps, which exhibits a good agreement with both numerical calculations and experimental data. Results show that in case of multipole traps, the associated dynamics exhibits multiple stable quasi-equilibrium points. The high-order Laplace fields that appear in multielectrode trap systems yield quasi-periodic or chaotic nonlinear ion motion, which is demonstrated as an Ince-Strutt-like stability diagram. A numerical demonstration is performed for a 22-pole trap \cite{Rud19}.

\subsubsection{Multipole trap geometries. Experimental setups}

\begin{figure}[bth]
	\centering
	\includegraphics[scale=0.83]{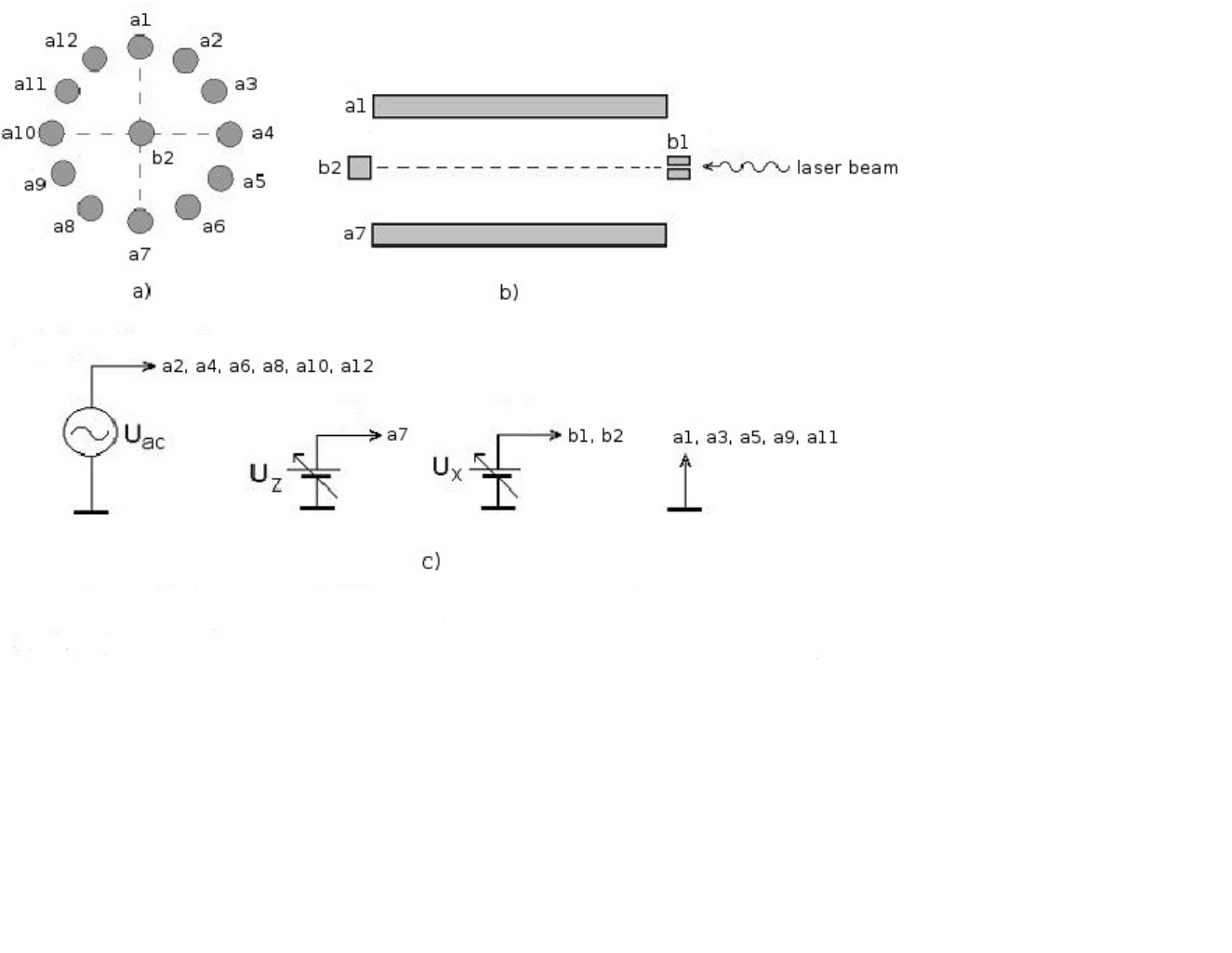} 
	\vspace{-6cm}
	\caption{A sketch of a 12-electrode linear Paul trap geometry. a) cross-section; 
	\newline b) longitudinal section; c) electrode wiring. Picture reproduced from \cite{Mih16a} with kind permission of the authors. Copyright AIP.}
	\label{12pole}
\end{figure}

We report three setups of Multipole Microparticle Electrodynamic Ion Traps (MMEITs). The first geometry consists of eight brass electrodes of 6 mm diameter equidistantly spaced around a 20 mm radius, and two endcap electrodes located at the trap ends, at 65 mm apart from each other. The second trap geometry consists of twelve equidistantly spaced brass electrodes (a$_1 \div$ a$_{12}$) and two endcap electrodes ($b_1$ and $b_2$), as illustrated in Fig.~\ref{12pole}. The electrode diameter is 4 mm, the trap radius is 20 mm, and the electrode length does not exceed 85 mm. 

\begin{figure}[bth]
	\centering
	\includegraphics[scale=0.65]{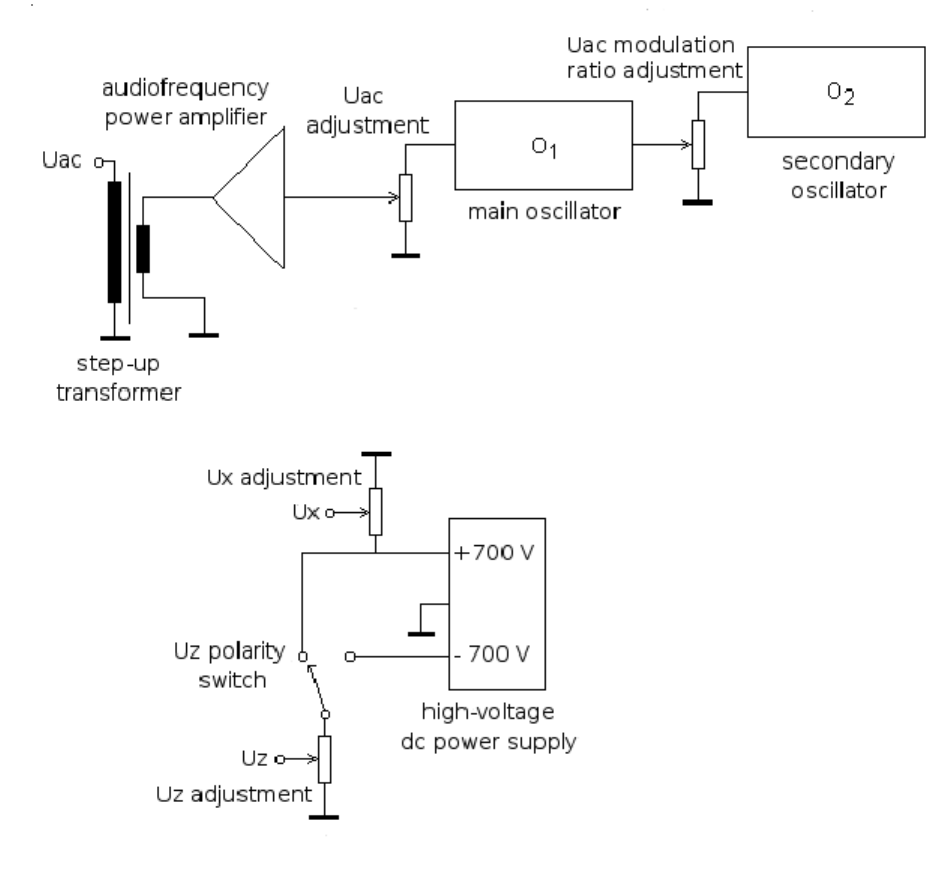}
	\caption{Block scheme of the electronic supply unit. Picture reproduced from \cite{Mih16a} with kind permission of the authors. Copyright AIP.}
	\label{supply}  
\end{figure}

Fig.~\ref{12pole} shows a sketch of a 12-electrode trap geometry designed and tested. Both setups are intended to study the occurrence of stable and ordered patterns for different microparticle species. Alumina microparticles (with dimensions ranging between $60 \div 200$ microns) are used to illustrate the trapping phenomenon, but other species can be considered. Specific charge measurements on different trapped microparticle species can also be performed by refining the setup. The 12-electrode trap is characterized by a variable geometry \cite{Sto11, Ghe98, Sto01}, as its length can vary between $10 \div 75$ mm. The endcap electrode, denoted as $b_2$, is located on a piston which slides along the $z$ axis of the trap. An electronic supply system delivers an a.c. voltage $V_{ac}$ with an amplitude ranging between $0 \div 4$ kV and a variable frequency that lies in the $40 \div 800$ Hz range, required in order to achieve radial trapping of charged particles. A high voltage step-up transformer, driven by a low frequency oscillator (main oscillator) $O_1$, delivers the $V_{ac}$ voltage, as shown in Fig.~\ref{supply}. An auxiliary oscillator, denoted as $O_2$, is used to modulate the amplitude of the $V_{ac}$ voltage. The modulation ratio can reach a peak value of $100\%$ while the modulation frequency lies in the $10 \div 30$ Hz range. 

\begin{figure}[bth]
	\centering  
	\includegraphics[scale=0.16]{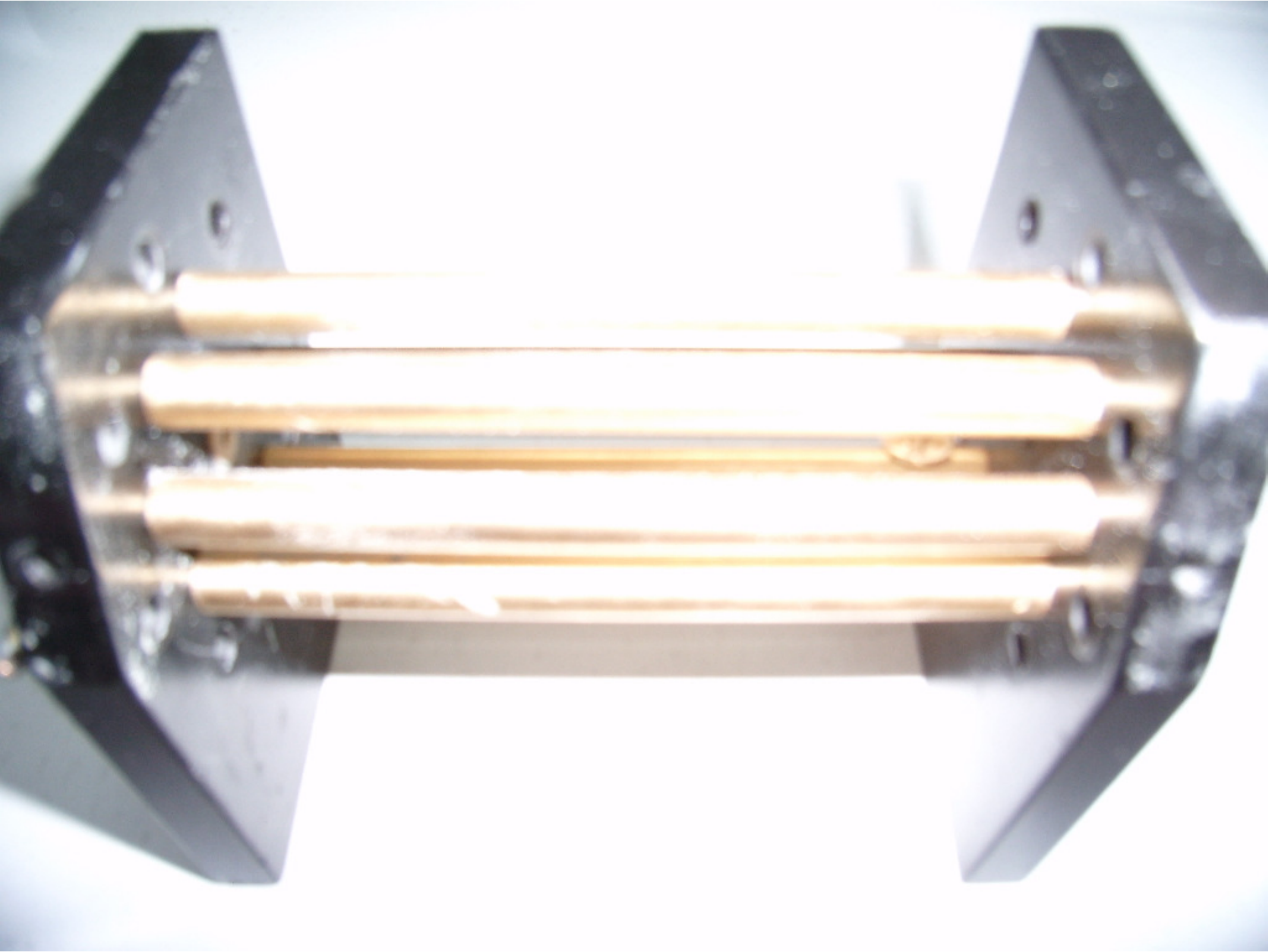}
	\includegraphics[scale=0.16]{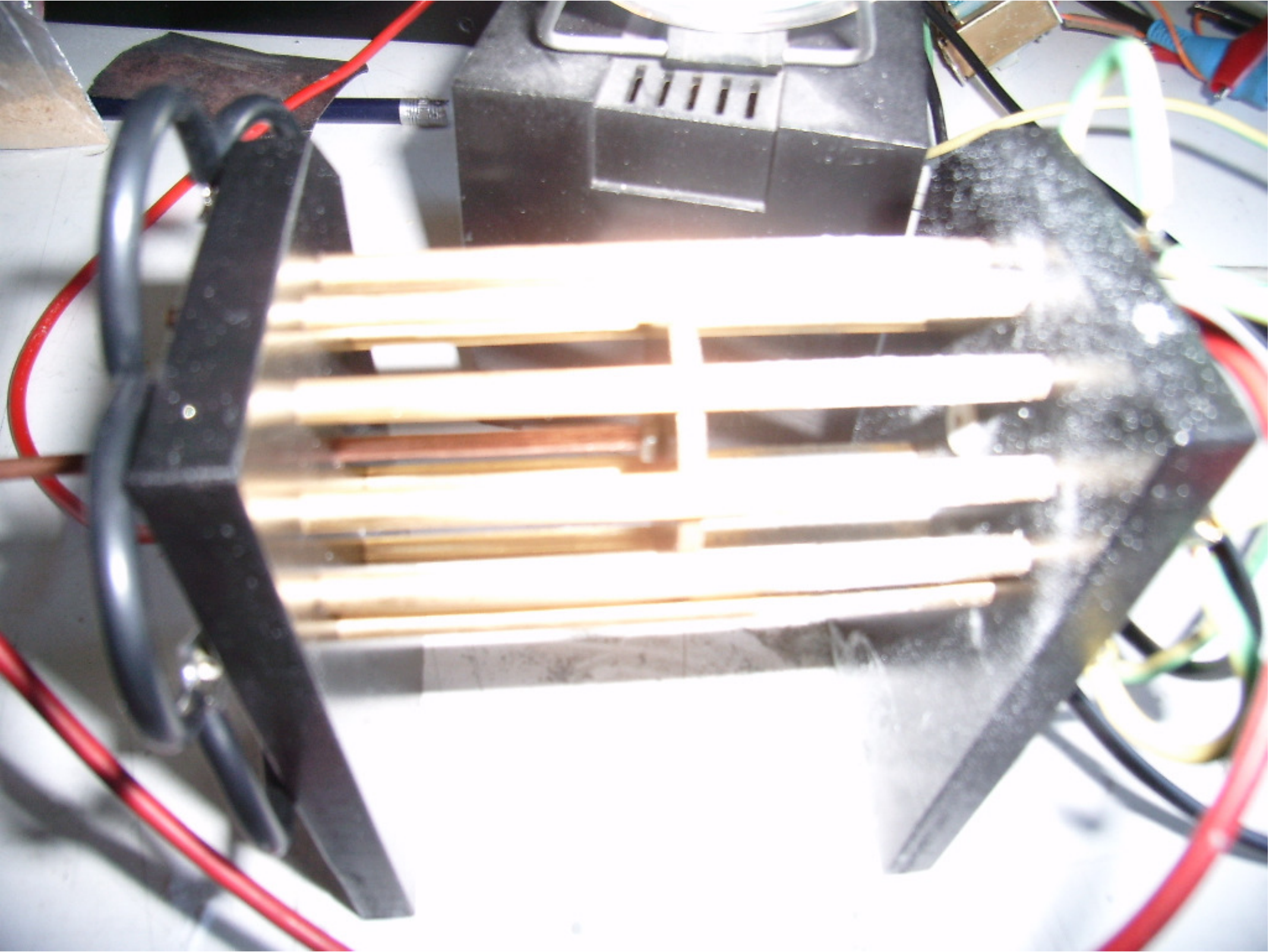} 
	\caption{Photos of an 8-electrode and of a 12-electrode linear Paul trap geometries, designed and tested in INFLPR. Images reproduced from \cite{Mih16a} with kind permission of the authors. Copyright AIP.} 
	\label{traps}
\end{figure}

Images of both 8-electrode and 12-electrode traps are presented in Fig.~\ref{traps}. The electronic system also supplies a variable d.c voltage $U_x$ (called diagnose voltage) applied between the upper and lower multipole trap electrodes, used to compensate the gravitational field and shift the particle position along the $x$ axis. Ions confined in linear Paul traps under ultrahigh vacuum conditions do arrange themselves along the longitudinal $z$ axis and within a large region around it where the trapping potential is very weak. The situation is sensibly different in case of electrically charged microparticles which we explain in Section \ref{model}. An extra d.c. variable voltage $U_z$ is applied between the trap endcap electrodes in order to achieve axial confinement and prevent particle loss at the trap ends. The $U_x$ and $U_z$ voltages range between $100 \div 700$ V, and their polarity is reversible. Both the $U_x$ and $U_z$ voltages are obtained using a common d.c. double power supply that delivers a voltage of $\pm \ 700$ V at the output. Because the absorbed d.c. current is very low, the d.c. voltages are supplied to the electrodes by means of potentiometer voltage dividers, as shown in Fig.~\ref{supply}. 

\begin{figure}[!ht]
	\begin{minipage}[h]{0.47\linewidth}
		\center{\includegraphics[width=\linewidth]{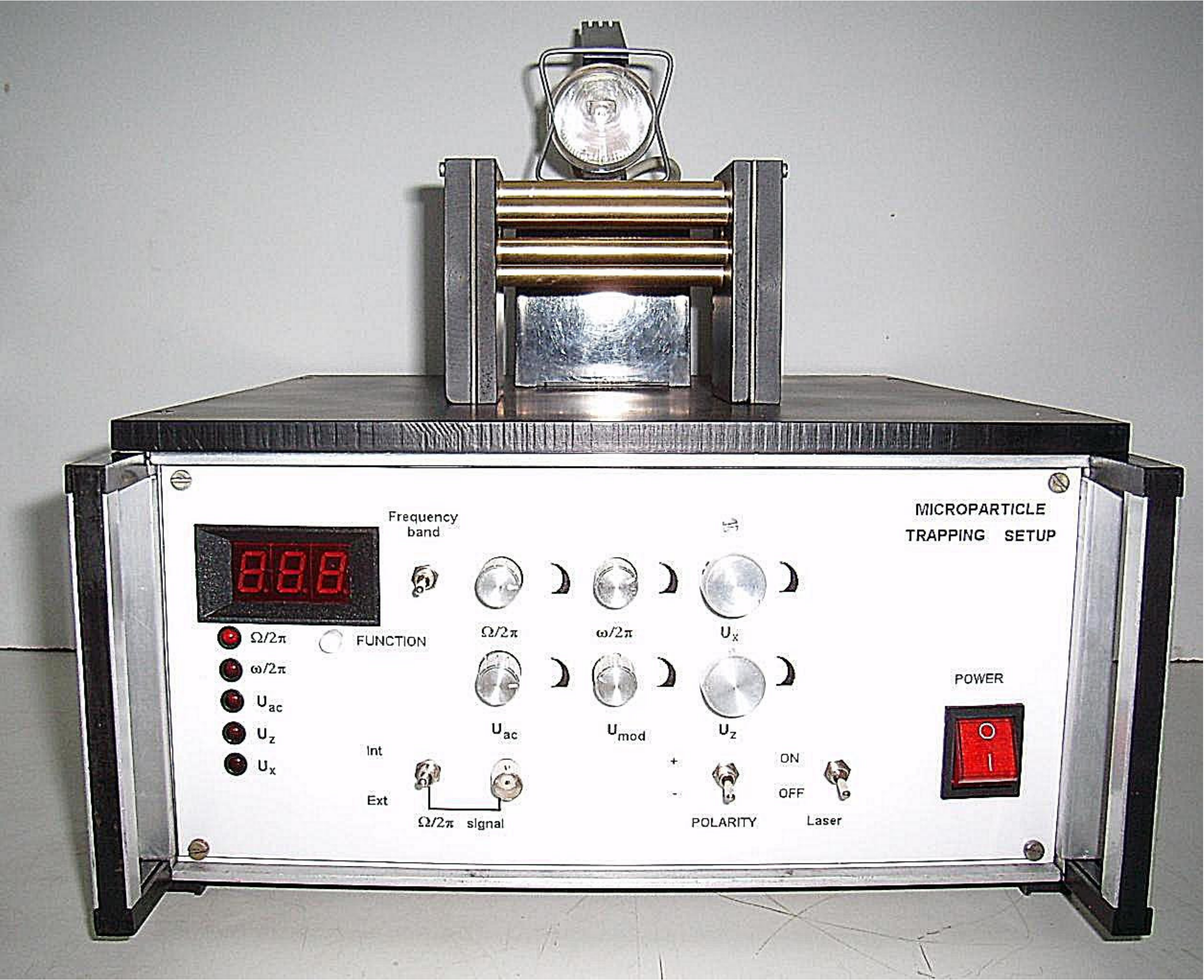}\\a)}
	\end{minipage}
	\begin{minipage}[h]{0.52\linewidth}
		\center{\includegraphics[width=\linewidth]{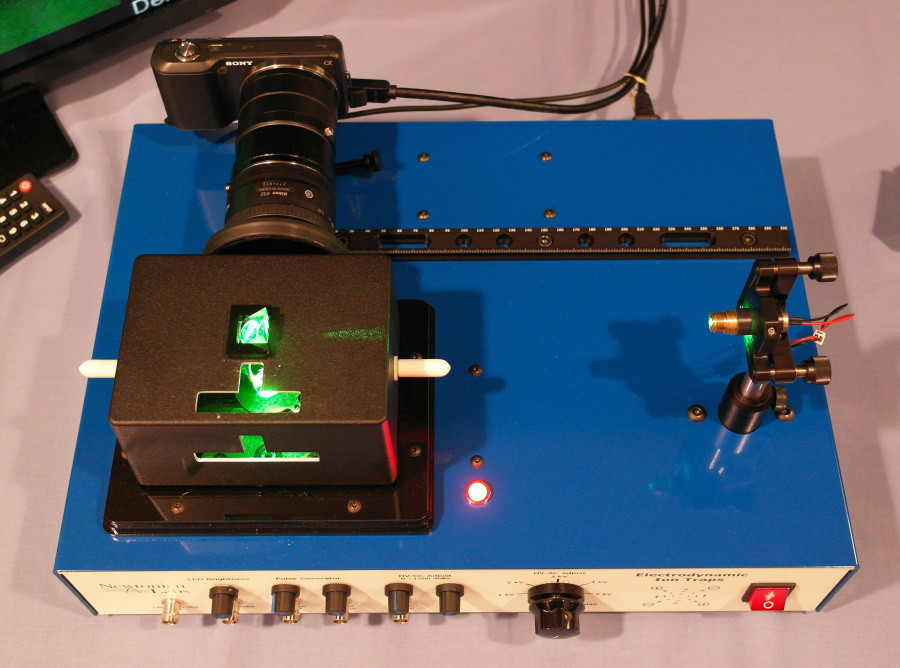}\\b)}
	\end{minipage}
	\caption{(a) Electronic supply unit used for the $8$-pole, $12$-pole and $16$-pole traps designed in INFLPR, versus (b) EIT setup with Ring Trap geometry included, designed and tested by Newtonian Labs. Image (a) reproduced from \cite{Mih16a} under kind permission of the authors, and image (b) reproduced from  http://newtonianlabs.com/eit/BirdsEyeView1.html, by courtesy of Prof. K. Libbrecht} 
	\label{ACDCsupply}
\end{figure}

The supply system is a single unit which delivers all the required trapping voltages, as illustrated in Fig.~\ref{ACDCsupply}. A microcontroller based measurement circuit allows separate monitoring and displays both the supply voltage amplitudes and frequencies (for the a.c. voltage and modulation voltage). In order to visualize and diagnose the trapped particles, the experimental setups are equipped with two different illumination systems. The first system is based on a halogen lamp whose beam is directed normal to the trap axis. The second system consists of a laser diode securely attached to one of the endcap electrodes, whose output beam is oriented parallel to the trap axis.  

The trap geometries investigated are fitted to study complex Coulomb systems (microplasmas) levitated in multipole dynamic traps, operating at SATP conditions. Experimental results clearly demonstrate that the stability region for multipole traps is sensibly larger with respect to the case of a quadrupole trap (EDB configuration) \cite{Hart92, Libb18, Vasi13, Prest99}. 

\begin{figure}[bth]
	\begin{center}  
		\includegraphics[scale=0.4]{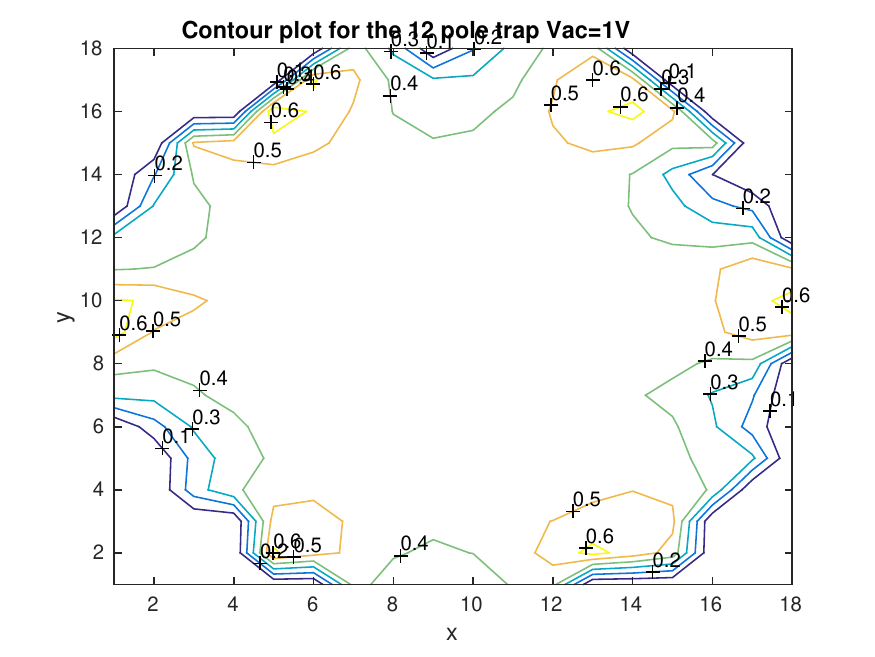}
		\includegraphics[scale=0.4]{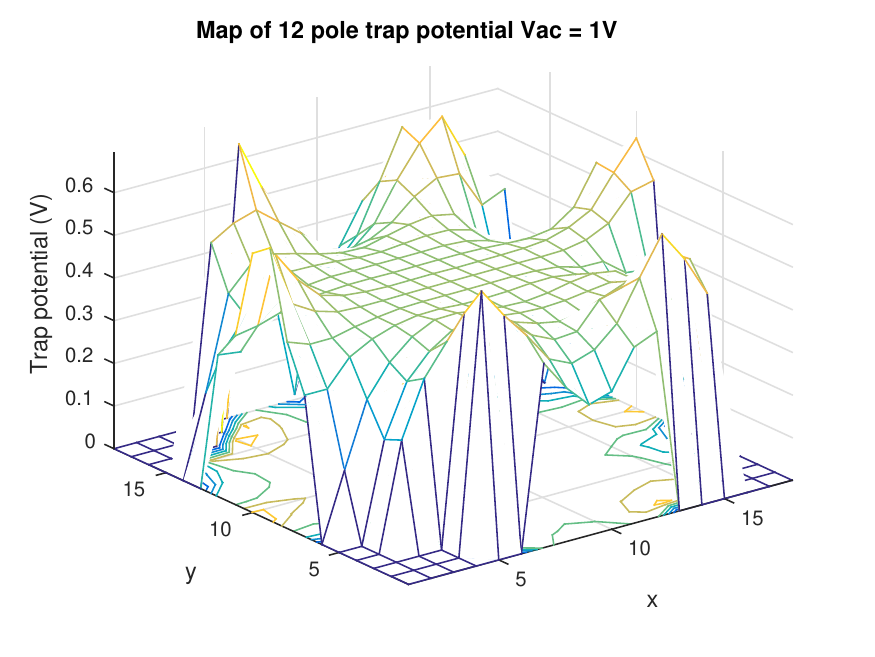}
		\includegraphics[scale=0.45]{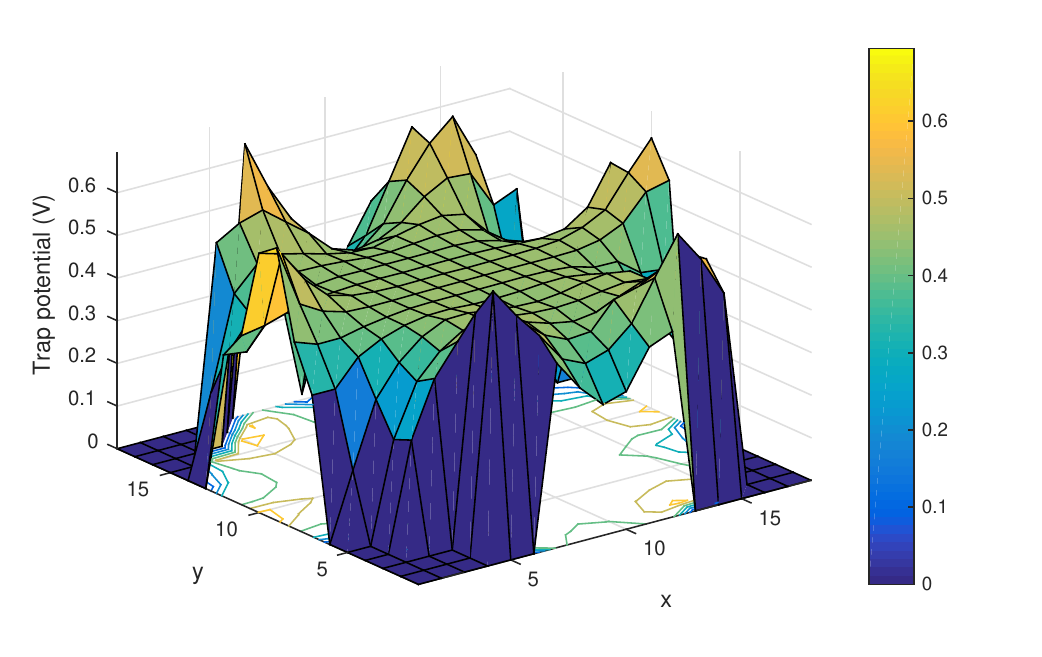}
	\end{center}
	\caption{Contour and 3D plots for a 12-electrode Paul trap potential when $V_{ac} = 1$ V. Picture reproduced from \cite{Mih16a} with kind permission of the authors. Copyright AIP.} 
	\label{12pole1V}
\end{figure}

\begin{figure}[bth]
	\centering  
	\includegraphics[scale=0.4]{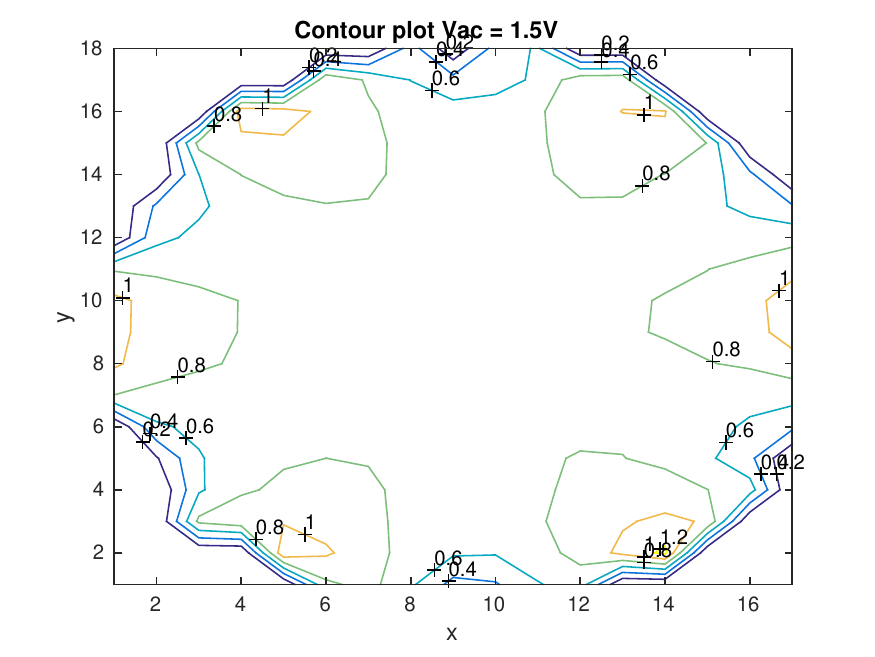}
	\includegraphics[scale=0.4]{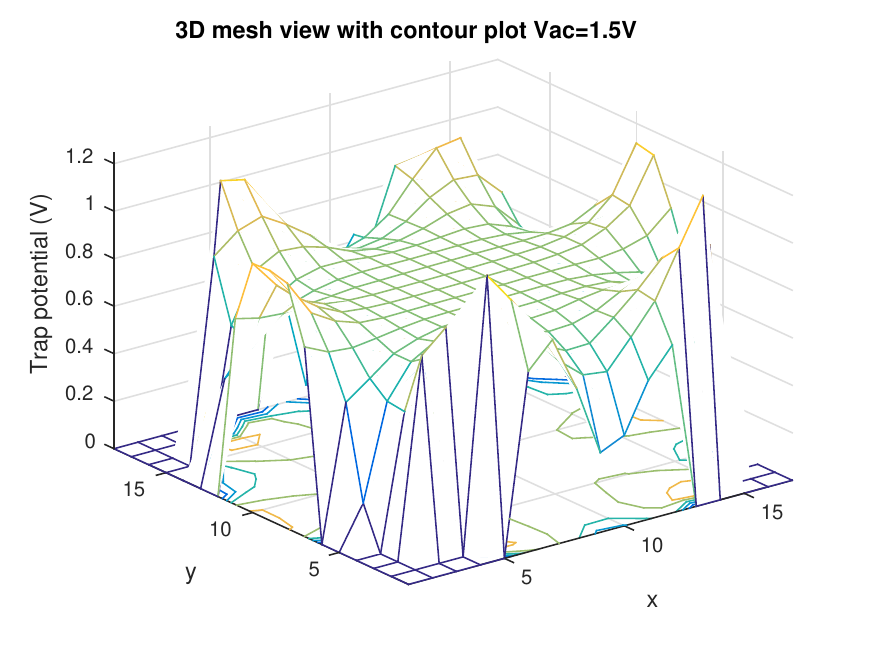}
	\vspace{.5cm}
	\includegraphics[scale=0.43]{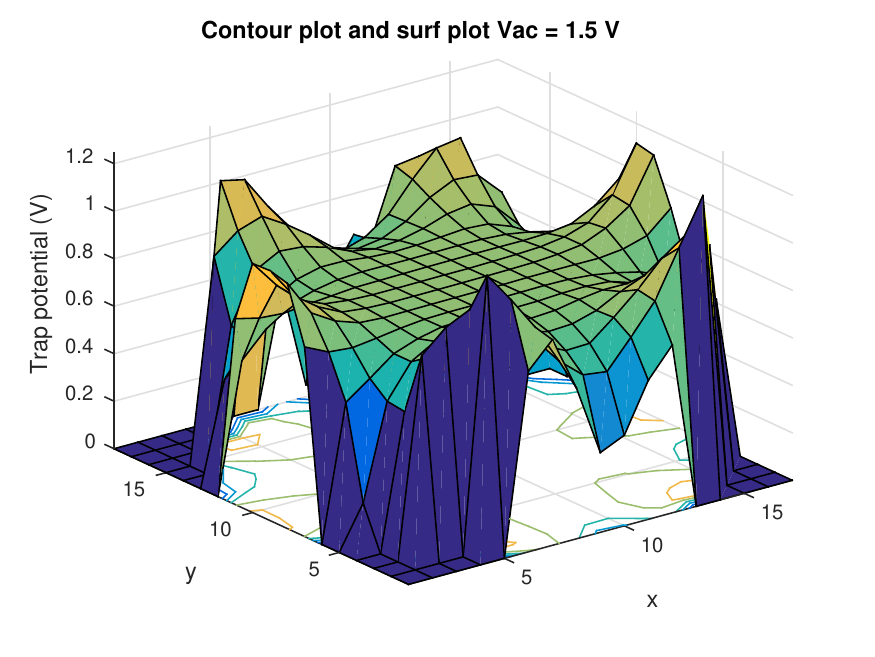}
	\caption{Contour and 3D plots for a 12-electrode Paul trap potential when $V_{ac} = 1.5$ V. Picture reproduced from \cite{Mih16a} with kind permission of the authors. Copyright AIP.} 
	\label{12pole15V}
\end{figure}

We use the electrolytic tank method to map the RF field potential in the $12$-electrode and $16$-electrode traps. A precision mechanical setup is used to chart the electric field within the traps. Dedurised water is used as an electrolyte solution. Two a.c. supply voltage values are used: $1$ V and $1.5$ V, respectively, which represent the measured amplitude values. Measurements are performed for an a.c. frequency value of $\Omega = 2\pi \times 10$ Hz. Figures \ref{12pole1V} and \ref{12pole15V} show experimentally obtained contour and 3D maps of the trap potential (rms values). 

\subsubsection{Results}\label{results}

\begin{figure}[!ht] 
	\begin{center}
		\includegraphics[scale=1.5]{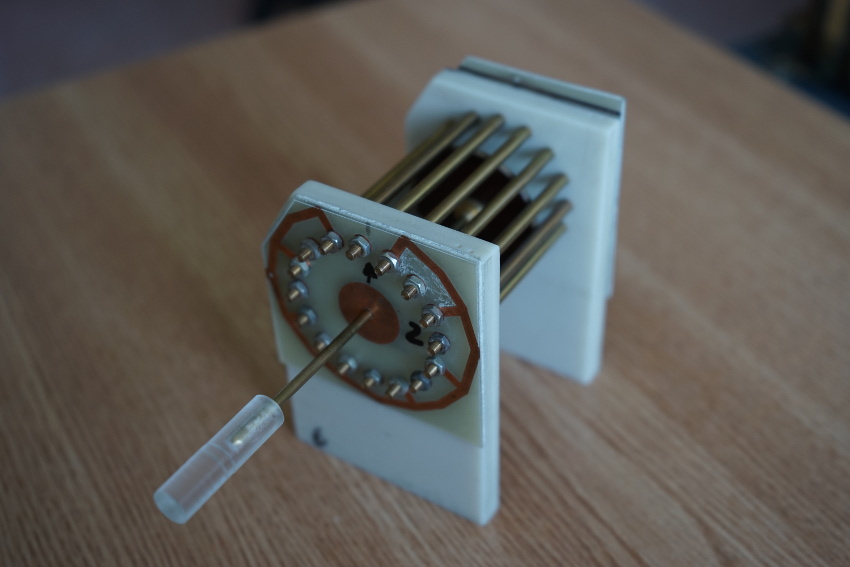}
		\caption{Image of a 16 electrode trap with variable geometry, designed and tested in INFLPR, intended for levitating microparticles under SATP conditions. The trap radius is $23$ mm. Image reproduced from \cite{Mih16b} with kind permission of the publisher.}
		\label{16pole}
	\end{center}
\end{figure}

Different linear multipole electrodynamic trap geometries operating under SATP conditions have been investigated. Work was focused on an 8-electrode, a 12-electrode and a 16-electrode trap geometry, respectively, with an aim to study conditions for stable microparticle confinement and to illustrate the emergence of planar and volume structures for these microplasmas. Microparticles are radially confined due to an a.c. trapping voltage $V_{ac}$ which is set at $2.5$ kV. Levitated microparticles can be shifted both axially and vertically using two d.c. voltages: $U_z$ and $U_x$. The geometry of the 8-electrode trap has proven to be very critical, different radii have been used and the trap has gone through intensive tests with an aim to optimize it. It presents a sensibly higher degree of instability with respect to a 12-electrode and a 16-electrode geometry. A 16 electrode trap designed in INFLPR is shown in Fig.~\ref{16pole}. The trap geometry consists of 16 brass electrodes of $60$ mm length and $4$ mm diameter, equidistantly spaced on a $46$ mm diameter. If the AC potential is not too large $\left(V_{ac} < 3.5 \ \text{kV}\right)$ stable oscillations will occur, until the d.c. potential is adjusted to balance vertical forces such as gravity. When this condition is fulfilled the oscillation amplitude becomes vanishingly small and the particle experiences stable trapping \cite{Kul11, Sein16}.

The 12-electrode (pole) trap has been studied more intensively as particle dynamics is far more stable. The traps are loaded with microparticles using a miniature isolated screwdriver. The peak of the screwdriver is inserted into the alumina powder. When setting the screwdriver to one of the trap electrodes potential the particles are instantly charged and a small part of them are confined, depending on their energy and the phase of the a.c. trapping field. A trapped particle microplasma results (very similar to a dusty plasma, which is of great interest for astrophysics), consisting of tens up to hundreds of particles. Such a setup is suited in order to study and illustrate particle dynamics in electromagnetic fields, along with the appearance of ordered structures, crystal like formations \cite{Mih16a, Ghe98, Sto01, Vini15, Libb18}. 

\begin{figure}[!ht] 
	\begin{center}
		\includegraphics[scale=0.25]{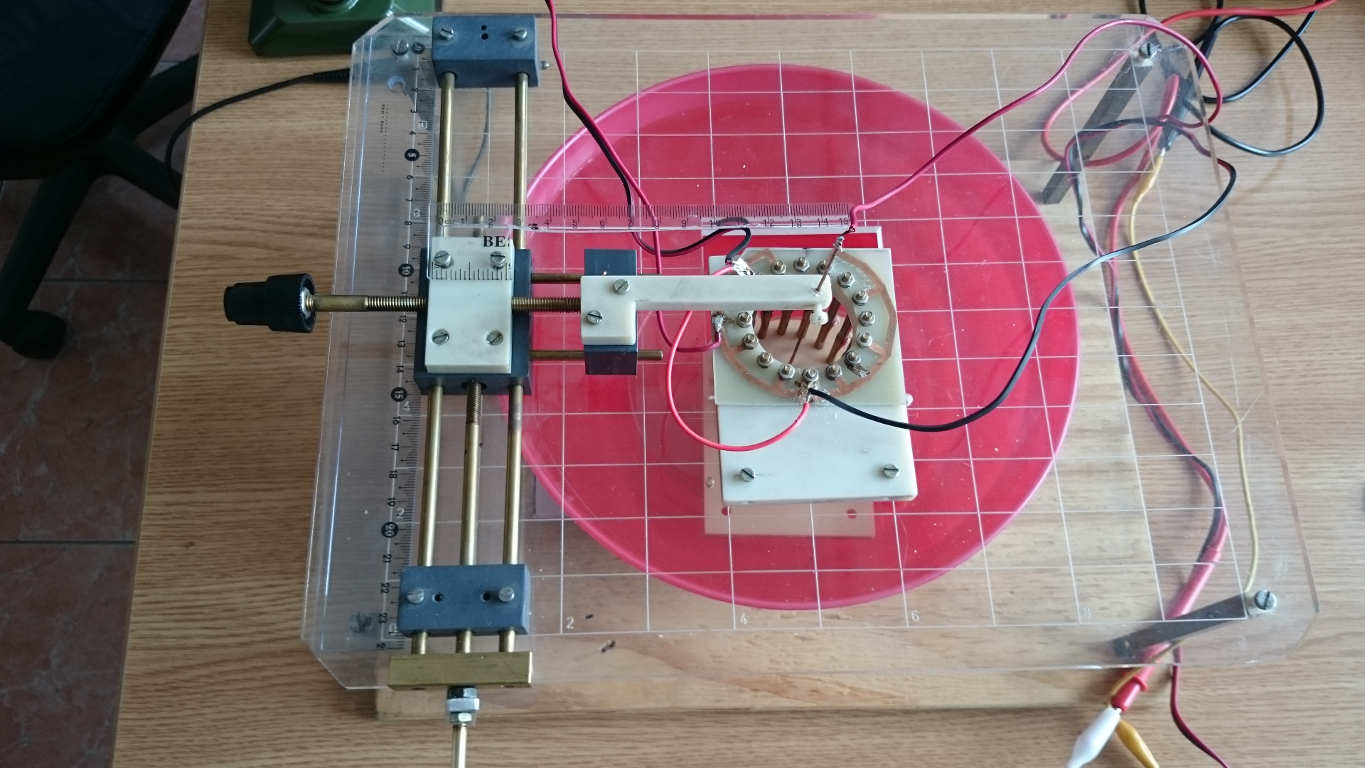}
		\caption{Use of the electrolytic tank method to chart the electric field map within the 16 electrode trap. Image reproduced from \cite{Mih16b} with kind permission of the publisher.}
		\label{tank}
	\end{center}
\end{figure}

\begin{figure}[h!tb]
	\centering
	\includegraphics[width=0.45\linewidth]{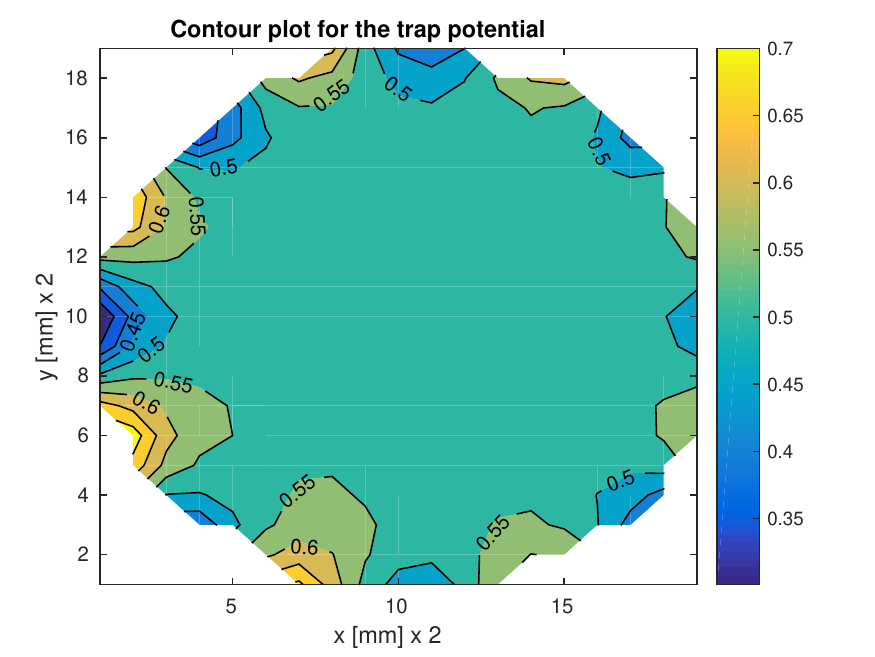}
	\includegraphics[width=0.5\textwidth]{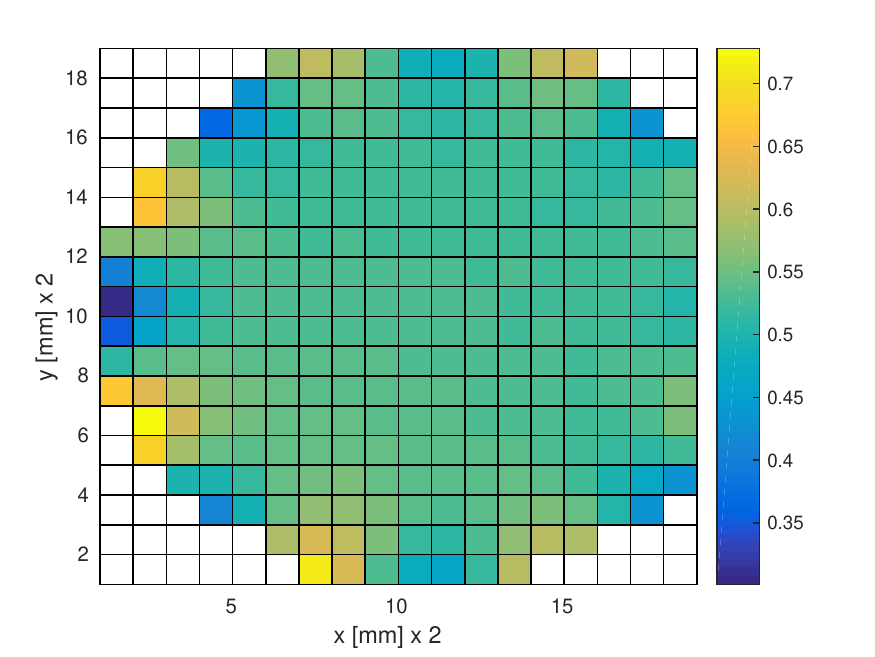}
	\includegraphics[width=0.48\textwidth]{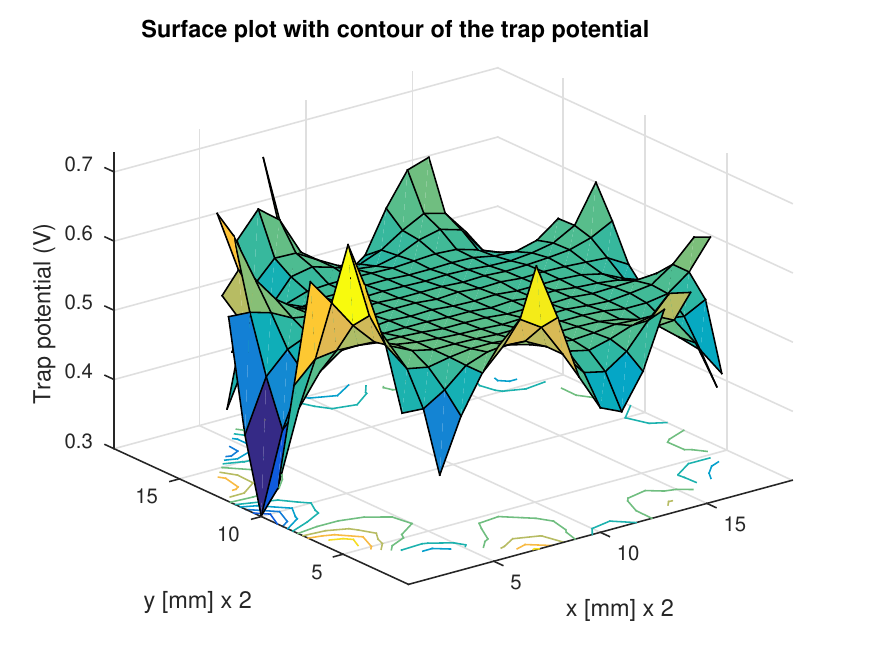}
	\includegraphics[width=0.48\textwidth]{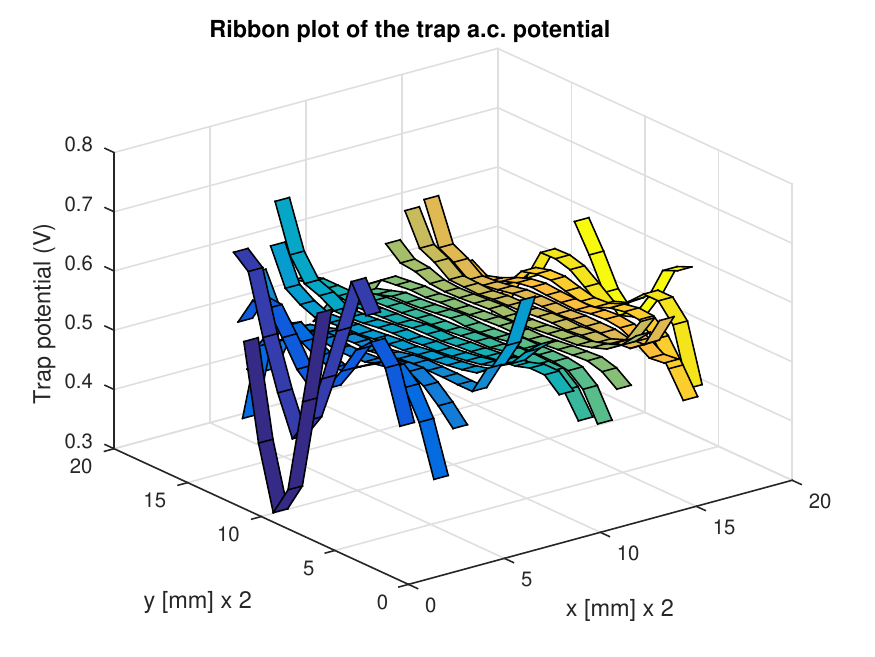}
	\caption{Maps of the 16-electrode trap potential: contour plot, pseudocolor plot, surface plot and ribbon plot, for an input sine wave of $1.5$ V amplitude ($1.037$ V rms). Picture reproduced from \cite{Mih16b} with kind permission of the publisher.}
	\label{mapchart15V}
\end{figure}

The a.c. potential in a $12$ electrode and in a $16$ electrode trap is mapped using an electrolytic tank filled with dedurised water. Further, we describe the method used in case of a $16$-pole trap. The trap is immersed in water, for around $38$ mm out of its total length ($60$ mm). A needle electrode is used to measure the trap voltage, immersed at $16$ mm below water level. The needle electrode can be displaced $36$ mm both horizontally and vertically, using a precision mechanism. The trap potential is mapped by means of a transversal (radial) section located $22$ mm apart with respect to the immersed end, using $2$ mm step displacements for both the vertical and horizontal position. The experimental setup, including the trap and electrolytic tank, is shown in Fig. \ref{tank}. The Paul trap is supplied with a sine wave delivered by a function generator at $46.1$ Hz frequency. An oscilloscope is used to monitor the sine wave. The rms value of the a.c. voltage is measured using a precision voltmeter. Maps of the trap potential are obtained for a $0.5$ V amplitude sine wave supplied to the trap electrodes. The sine wave is applied between even and respectively uneven electrodes, that are electrically supplied as shown in Fig.~\ref{12pole}. In Fig.~\ref{mapchart15V} we show maps of the trap potential for a supply voltage of $1.5$ V amplitude ($1.037$ V rms value) sine wave.

Practically, stable confinement is achieved especially in a 12-electrode and in a 16-electrode trap.  We report thread-like formations (strings), and particularly 2D (some of them zig-zag) and 3D structures of microplasmas. The stable structures observed tend to align themselves with respect to the $x$ component of the radial field. The ordered formations are not located along the trap axis but rather in the proximity of the electrodes. The laser diode is shifted away from the initial position (along the trap axis) towards outer regions of the trap. We report stable confinement for many hours and even days. Laminar air flows that cross the trap volume break an equilibrium which might be described as somehow fragile, while some of the particles can get neutralized at the electrodes. Nevertheless, the trap was also tested under conditions of intense air flows and experimental observations demonstrate that most of the particles remain trapped, even if they rearrange themselves after being subject to intense and repeated perturbations. When a transparent plastic box is used to shield the trap, a sensible increase in the dynamical stability of particle motion is achieved, along with the emergence of larger clouds. Experimental evidence demonstrates that a 16 electrode trap design is characterized by an extended region of lower field, compared to the 8-electrode and 12-electrode geometries we have tested. Moreover, the 16-electrode trap design is suitable for various applications where larger signal-to-noise ratios are required.

We illustrate below maps of the trap potential, for a supply voltage of $1.5$ V amplitude ($1.037$ V rms value) sine wave. 

\begin{figure}[bth]
	\centering 
	\includegraphics[scale=0.205]{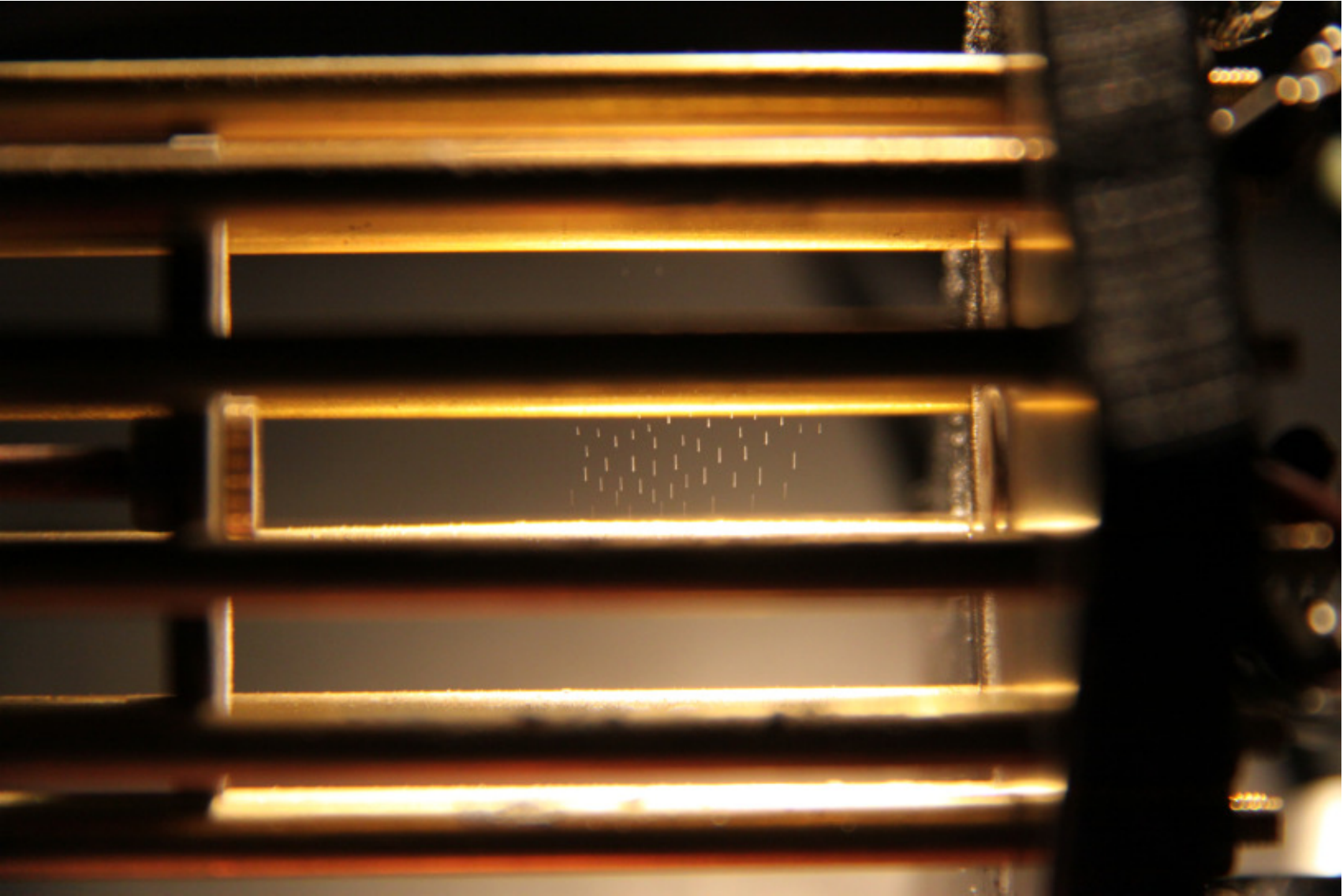}
	\includegraphics[scale=0.153]{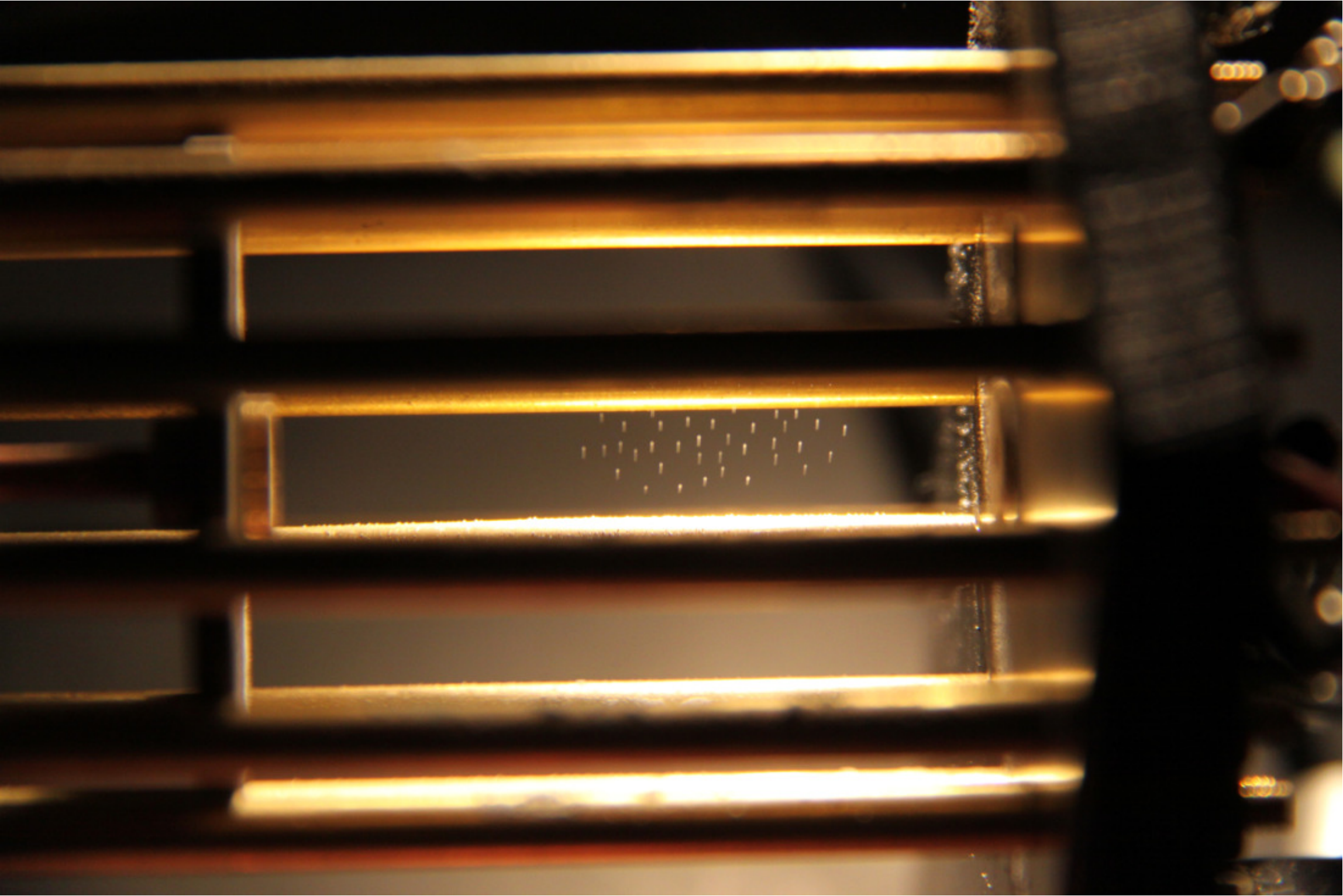}
	\vspace{1cm}
	\includegraphics[scale=0.15]{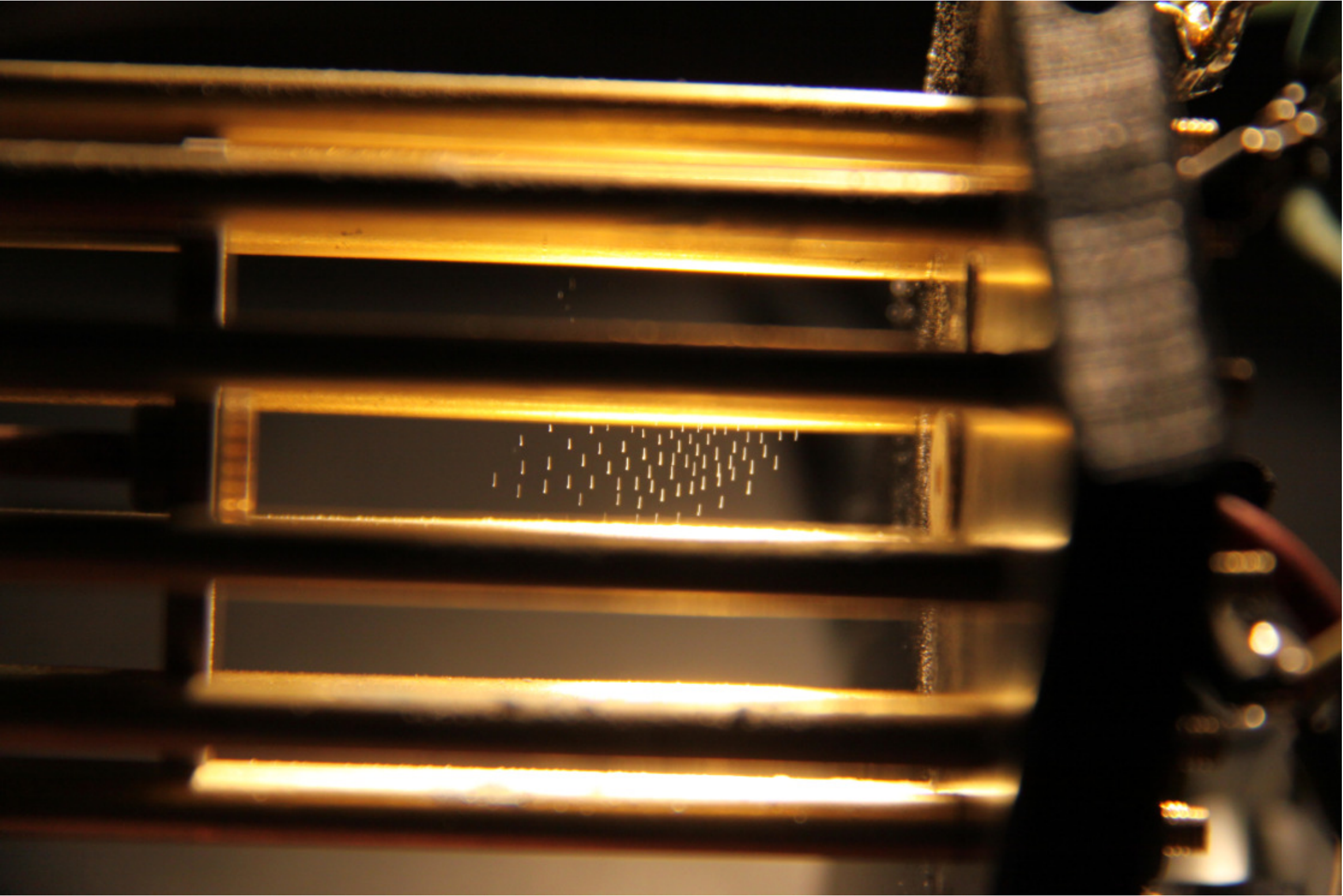}
	\includegraphics[scale=0.152]{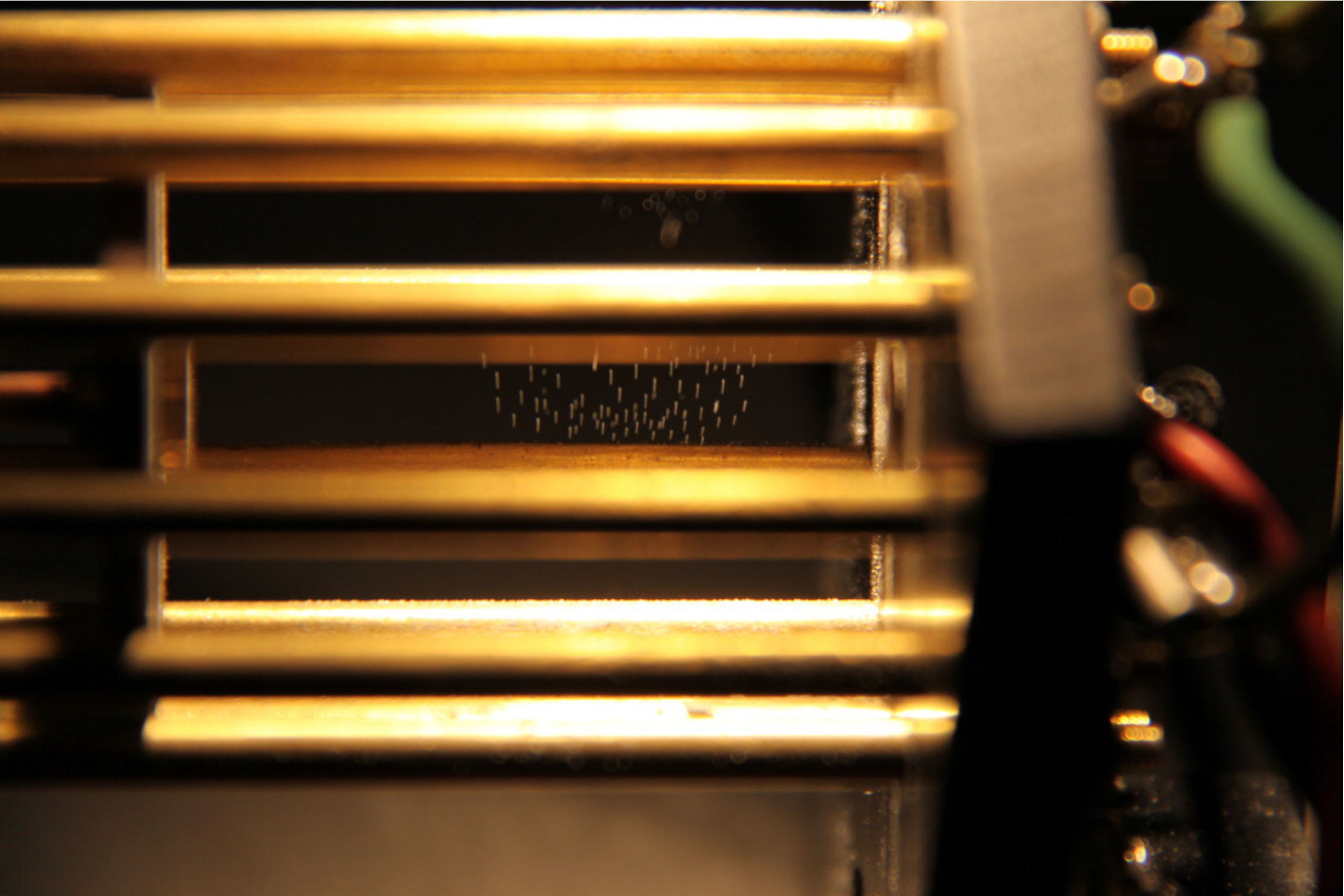}
	\caption{Photos of ordered structures observed in a 12-electrode Paul trap. Images reproduced from \cite{Mih16a} with kind permission of the authors. Copyright AIP.} 
	\label{microplasmas}
\end{figure}

Fig.~\ref{microplasmas} shows images that illustrate stable structures observed and captured in case of a 12-pole trap. All images are taken with a high sensitivity digital camera, using a halogen lamp that illuminates the trap. Pictures taken using the laser diode are less clear due to light reflection by the trap electrodes. We report filiform structures consisting of large number of microparticles far from the trap centre, in regions where the trapping potential is extremely weak. This leads to the conclusion that the microparticle weight is not balanced by the trapping field in the region located near the trap centre. Moreover, most of the ordered structures observed are generally located very close to the trap electrodes, at distances about $2 \div 10$ mm apart from them. The a.c. frequency range is swept between $\Omega = 2\pi \times 50 \div 2 \pi \times 100$ Hz. 

The trap geometries investigated are characterized by multiple regions of stable trapping, some of them located near the trap electrodes. Most of the images taken illustrate such phenomenon along with the numerical simulations performed. At least three or four regions of stable trapping are identified, but due to camera limitations focusing is achieved for a limited region of space. When looking with bare eyes and for an adequate angle of observation spatial structures are observed, located at different regions within the trap volume. We can ascertain that the multipolar trap geometries investigated, and especially the linear 12- and 16-electrode Paul traps, exhibit an extended region where the trapping field almost vanishes. The amplitude of the field rises abruptly when approaching the trap electrodes area which leads to stable trapping, along with the occurrence of planar and volume structures in a layer of about a few millimetres thick which practically spans the inner electrode space. We report particle oscillations around equilibrium positions where gravity is balanced by the trapping potential. Close to the trap centre particles are almost {\em frozen}, which means they can be considered motionless due to their very low oscillation amplitude. The amplitude increases as one moves away from the trap centre. We also report regions of dynamical stability for trapped charged microparticles located far away from the trap centre, as illustrated in Fig.~\ref{microplasmas}. The values of the specific charge ratios for  alumina microparticles range between $5.4 \times 10^{-4} \div 0.13 \times 10^{-3}$ C/kg \cite{Sto01, Sto08}. 

To summarize, we report trapping of microparticles in multipole linear ion (Paul) traps (MLIT) operating in air, under SATP conditions. We suggest such traps can be used to levitate and study different microscopic particles, aerosols, and other constituents or polluting agents present in the atmosphere. The investigations performed are based on previous results and experience \cite{Ghe98, Mih08, Sto08, Vis13, Fili12, Lapi15b}. Numerical analysis is used to characterize microparticle dynamics.

\subsection{Analytical and numerical modelling. Instabilities. Mitigation}\label{model}

In addition to the experimental work, numerical simulation of charged particle dynamics is carried out under conditions close to the experiment. Brownian dynamics (BD) is used to study charged microparticle motion and thus identify regions of stable trapping. Numerical simulations take into account the stochastic forces of random collisions with neutral particles, the viscosity of the gas medium, the regular forces produced by the a.c. trapping voltage and gravity. Thus, microparticle dynamics is characterized by a stochastic Langevin differential equation \cite{Vasi13, Fili12}: 

\begin{equation}\label{eq.1}
m_p \frac{d^2 r}{dt^2} = F_t(r)-6 \pi \eta r_p \frac{dr}{dt} + F_b + F_g
\end{equation}
where $m$ and $r_p$ represent the microparticle mass and radius vector, respectively,  $\eta = 18.2 \:\mu$Pa$\cdot$s is the dynamic viscosity of the gas medium and $F_t(r)$ is the ponderomotive force. The $F_b$ term stands for stochastic delta-correlated forces accounting for stochastic collisions with neutral particles, while $F_g$ is the gravitational force. We consider a microparticle mass density value $\rho_p = 3700$ kg/m$^3$ \cite{Sto08, Vis13}, valid for all simulations performed \cite{Mih16a, Lapi16d}. In order to solve the stochastic differential equation (\ref{eq.1}), a numerical method developed in \cite{Skeel02} is used. The average Coulomb force that acts on a microparticle (as an outcome of the contribution of each trap electrode) can be expressed as the vector sum of the forces of point-like charges uniformly distributed along the electrodes, as demonstrated in \cite{Vasi13, Lapi15b}:

\begin{equation}
\label{eq.2}
|F_t(r)| = \sum \limits_s \frac{L U q}{2 N \ln{\left(\frac{R_2}{R_1}\right)}(r_s - r)^2},
\end{equation}
where $L$ is the length of the trap electrodes, $U$ is the trapping voltage: $V_{ac} \sin(\Omega t)$ or $V_{ac} \sin(\Omega t+\pi)$, $q$ is the microparticle charge, $N$ is the number of point-like charges for each trap electrode, $R_2$ and $R_1$ represent the radii of the grounded cylindrical shell surrounding the trap and trap electrode, respectively, while $r$ and $r_s$ denote  vectors for microparticle and point-like charge positions, respectively. Numerical simulations  run using the following trap parameters: length of electrodes $L = 6.5$ cm, $V_{ac} = 2$ kV, $R_2 = 25$ cm, $R_1 = 3$ mm and a trap radius value $r_t = 2$ cm. For this model the computation results depend on $\Phi_p$, defined in Sec. \ref{gasflow} by eq. \ref{intforce}.

\begin{figure}[bth]
	\centering
	\includegraphics[scale=0.22]{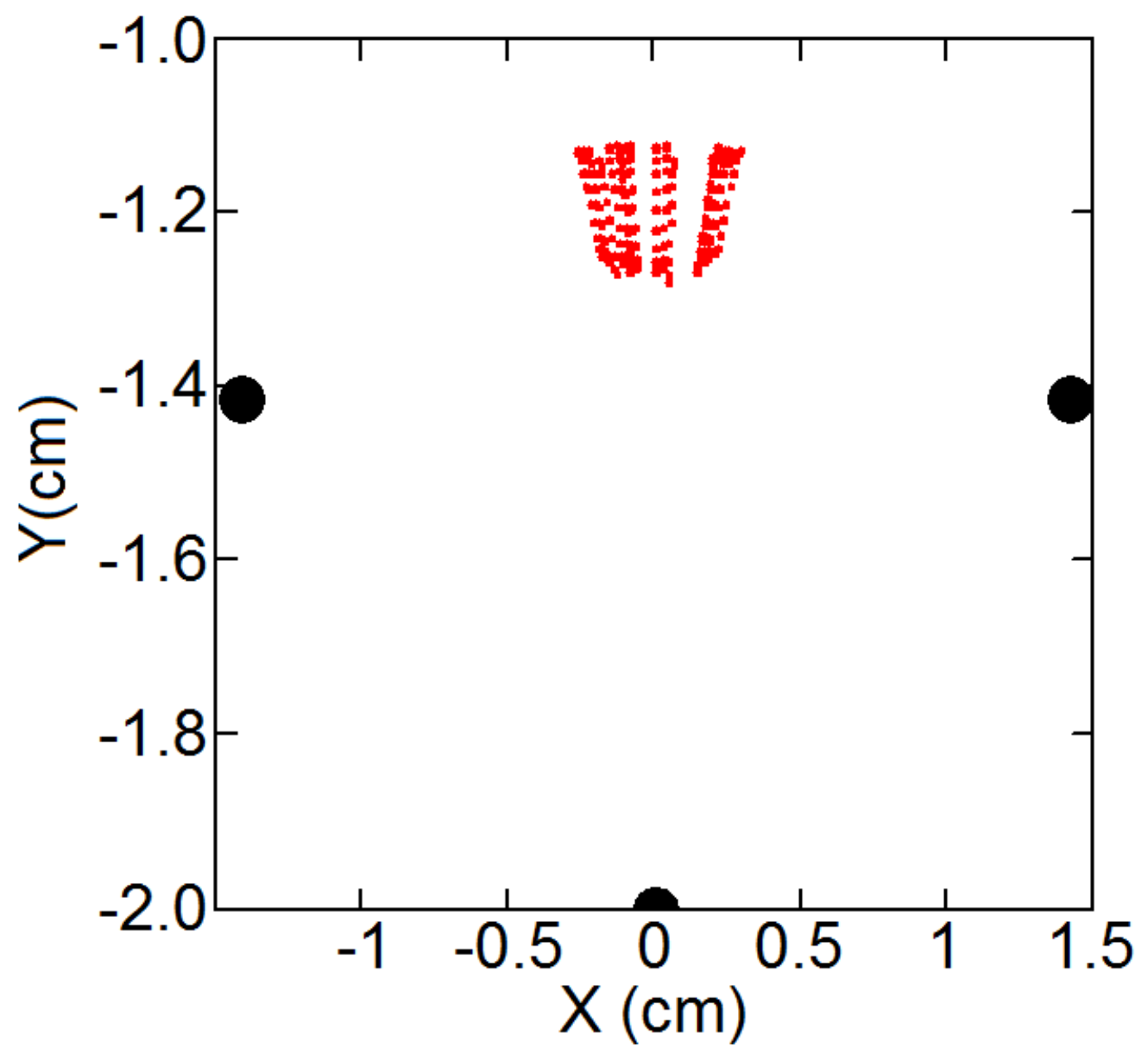}
	\includegraphics[scale=0.22]{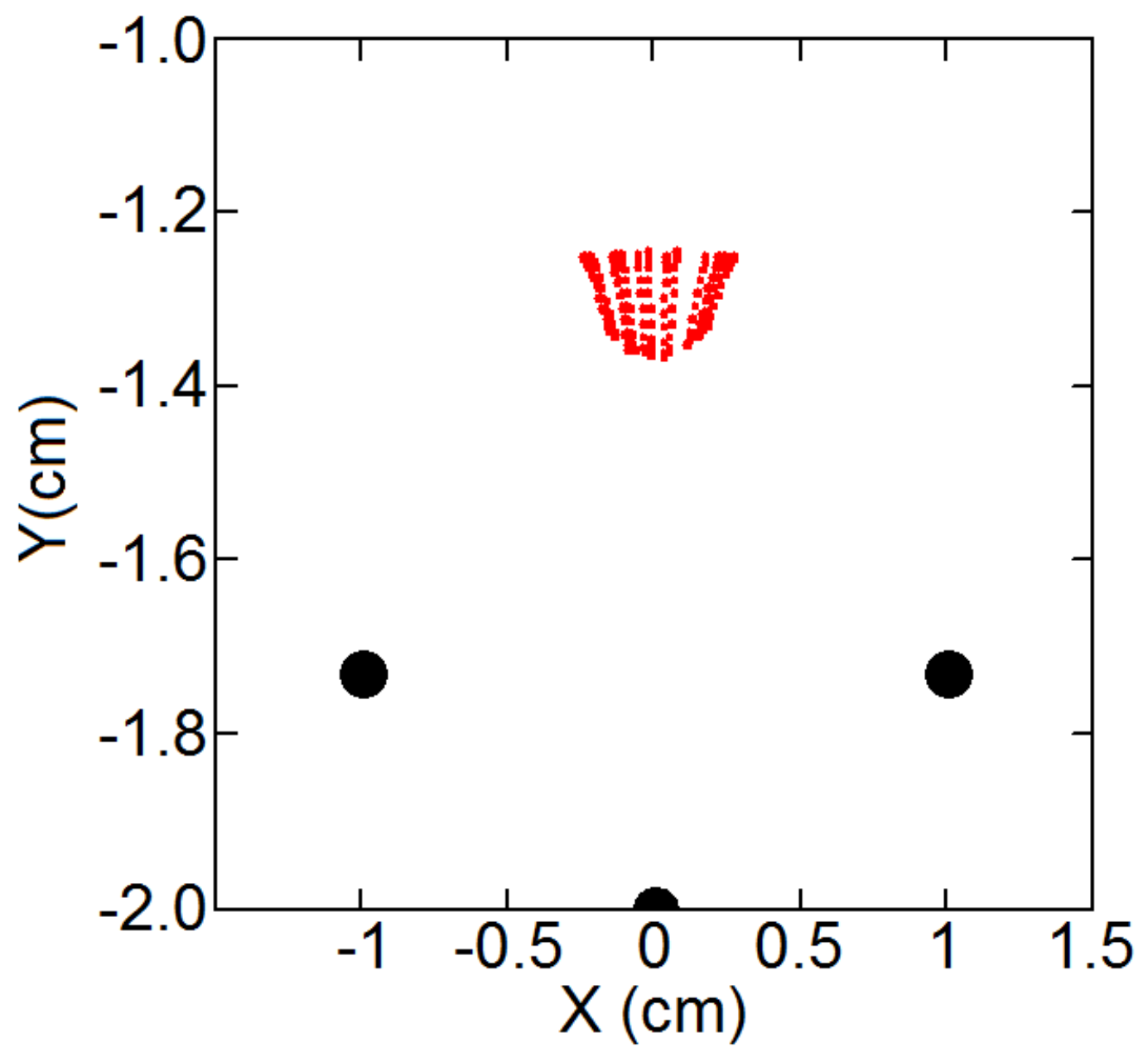}
	\includegraphics[scale=0.22]{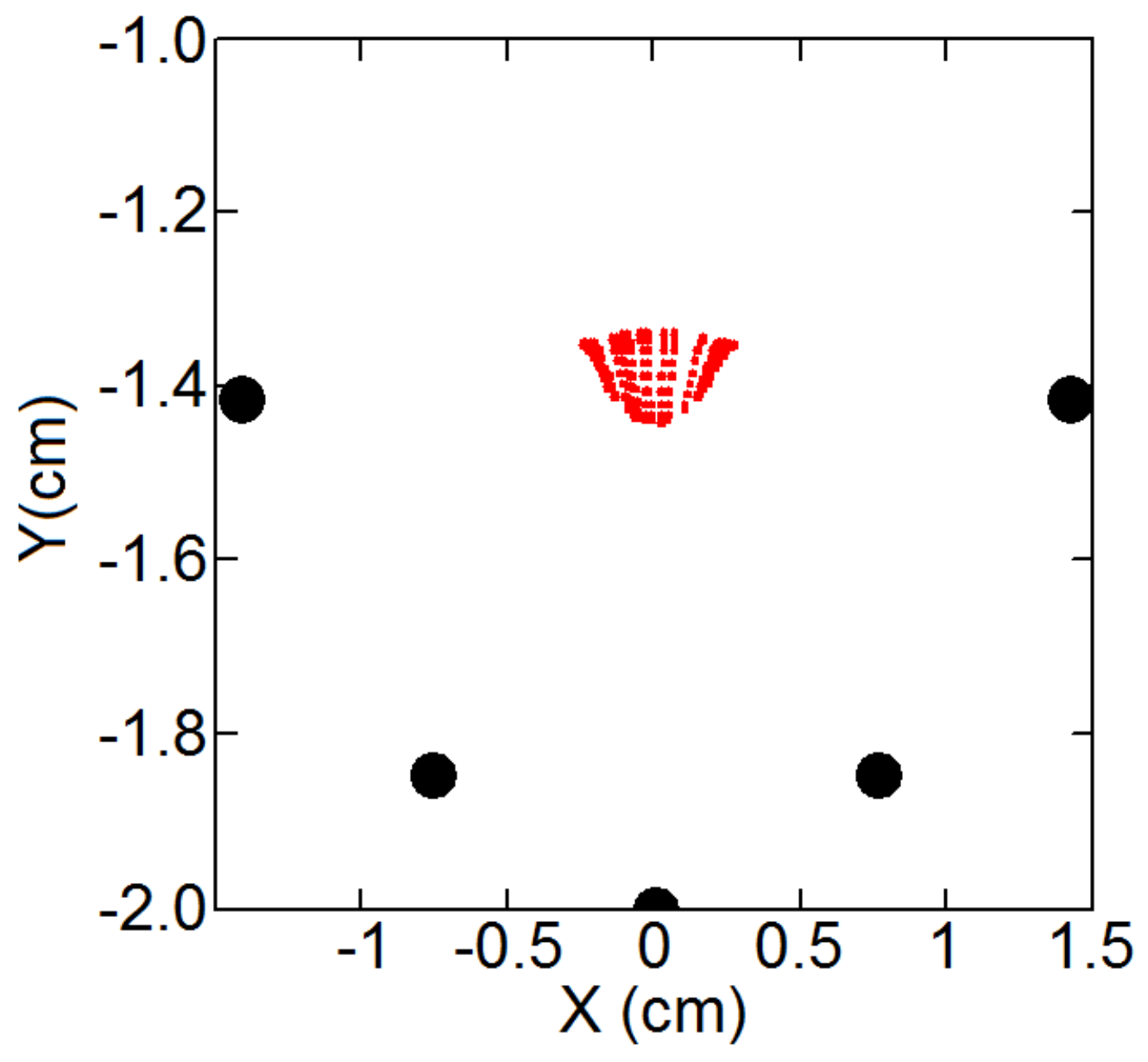}
	\caption{End views of microparticle tracks in (a) 8-electrode, (b) 12-electrode, and (c) 16 electrode traps when $f = 60$ Hz. Big black dots correspond to trap electrodes (not in scale). The microparticle electric charge value is chosen $q = 8 \cdot 10^4 e$. The d.c. voltage is disregarded in simulations. Images reproduced from \cite{Mih16a} with kind permission of the authors (AIP Copyright) and from \cite{Mih16b} with kind permission from the publisher.}
	\label{tracks8-16}
\end{figure}

To investigate the influence of the trap electrode number on the stability of alumina (dust) particles, the average amplitude of particle oscillations is explored. We study the associated dynamics for a number of 20 microparticles confined in the trap and average the amplitudes of particle oscillations. To achieve that we choose a period of time long enough (around ten periods of the a.c. trapping voltage) so as to obtain stable particle oscillations. Trajectories for 20 particles confined in 8-electrode, 12-electrode and 16-electrode traps, respectively, are shown in Fig.~\ref{tracks8-16}. As the number of trap electrodes increases, particle trajectories are shifted downwards due to a smaller gradient of the a.c. electric field, as illustrated in Fig~\ref{tracks8-16}.

In order to achieve dust particle confinement, besides the a.c. trapping voltage amplitude there are several other parameters that have to be considered: the a.c. voltage frequency, the influence of the trap geometry (by means of the trap geometric parameters), the particle dimension and its geometrical shape, or the inter-electrode distance value. Refs. \cite{Lapi15a, Lapi16d} demonstrate the connection between stable dust particle confinement and the a.c. field frequency or specific charge ratio $Q/M$, in case of the trap geometries considered and for two different values of the viscosity. The first trap geometry consists of four wires (see Fig.~\ref{fig:trap}) while the second geometry comprises only two wires, supplied at an a.c. voltage $U_{\omega}\sin({\omega}t)$. 

\begin{figure}[bth] 
	\centering
	\includegraphics[scale=0.44]{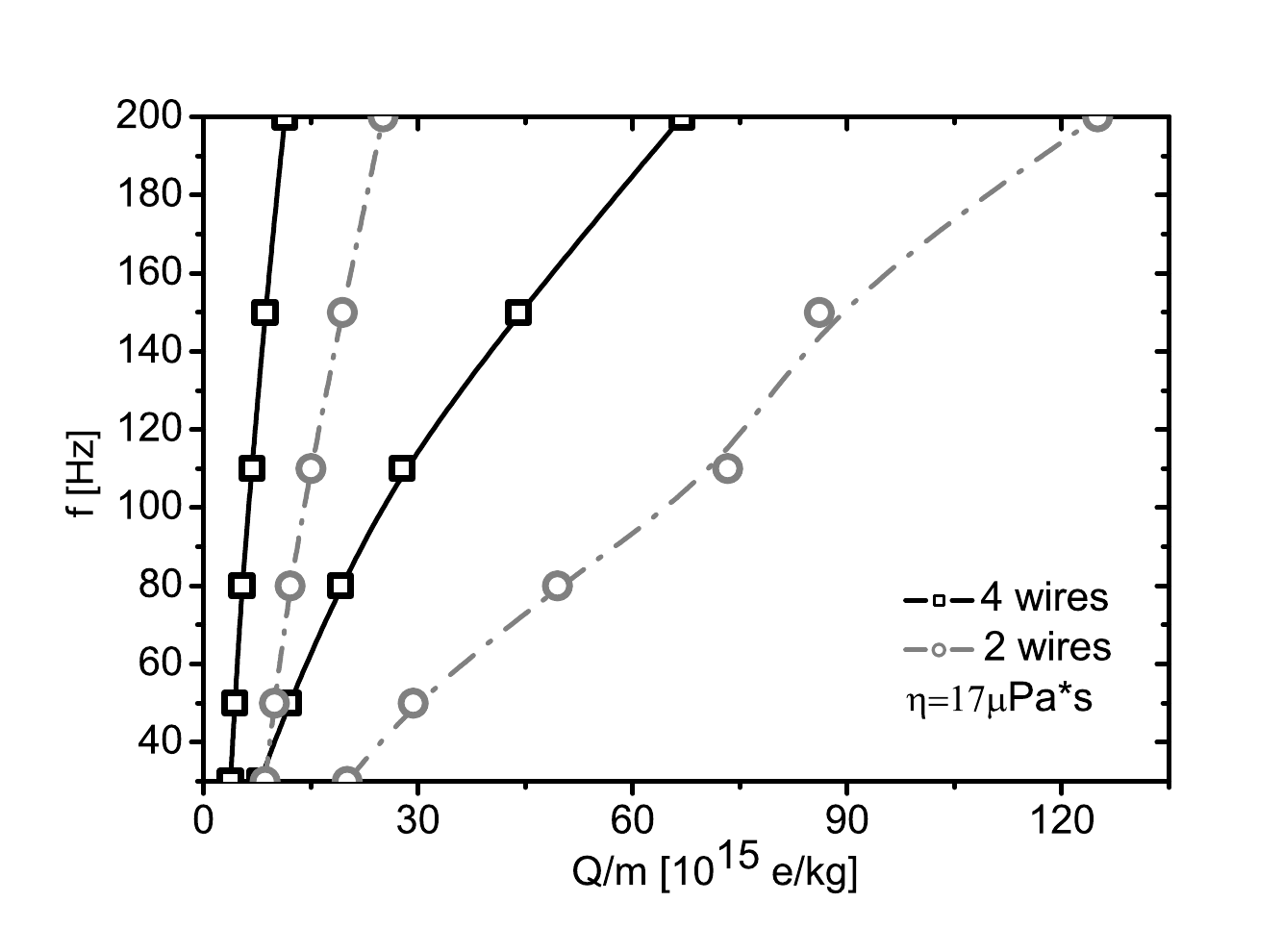}
	\includegraphics[scale=0.44]{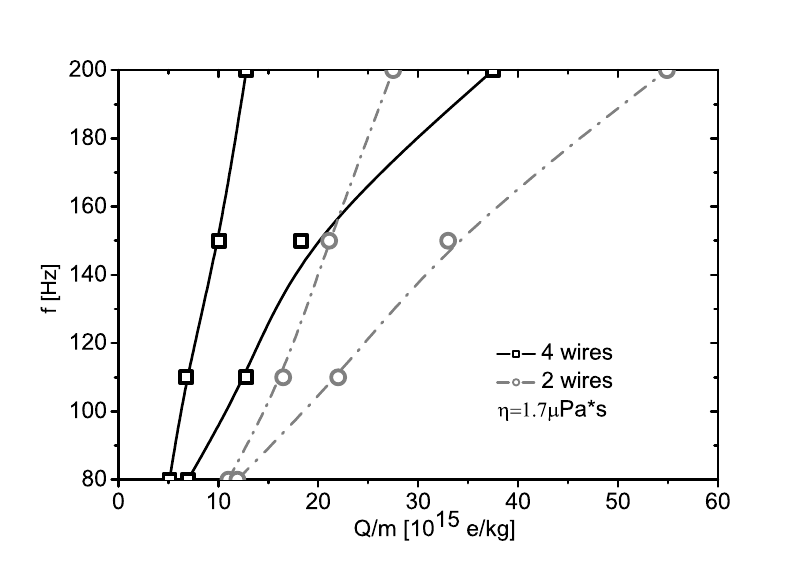}
	\includegraphics[scale=0.44]{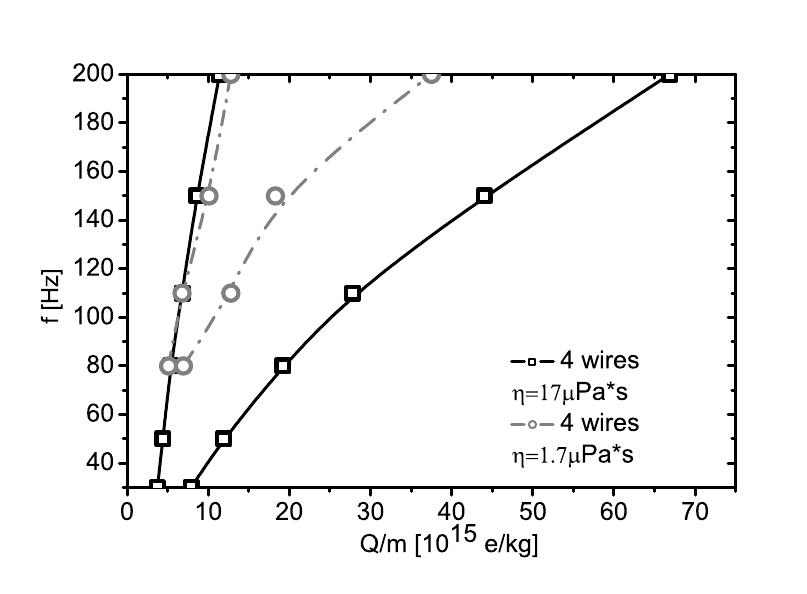}
	\caption{Regions of dust particle confinement in the $f$ -- $Q/M$ plane for two different trap geometries and two different values of the viscosity. The parameter values are: $f = 30 \div 200$ Hz, $U_{\mathrm{end}} = 900$ V, $U_{\omega} = 4400$ V, $\rho_{p} = 0.38 \times 10^4$ kg/m$^3$, $r_{p} = 7 \ \mu$m, $Q_{p} = 2 \times 10^4 e \div 6.8 \times 10^5 e$, $\eta = 17 \times 10^{-6}$ Pa$\cdot$s -- dynamic viscosity, $T = 300$K. Source: picture reproduced (modified) from \cite{Lapi15a} with kind permission of the authors.}
	\label{fig:Regions_geometries}  
\end{figure}

As it follows from Fig.~\ref{fig:Regions_geometries}, the regions of confinement are bounded by lower and upper values of the specific charge ratio, regardless of a.c. voltage frequency value. Beyond these regions the trap cannot confine dust particles. In the left panel of Fig.~\ref{fig:Regions_geometries} the confinement region characteristic to the two wire trap (2WT) is shifted, while in the same time it is wider with respect to the confinement region in case of a four wire trap (4WT). This result is an outcome of the smaller boundary gradients for the potential field inside a 2WT trap with respect to the case of a 4WT. Thus, in order to maintain particles levitated inside a 2WT trap a higher value of the charge-to-mass ratio is required. 

A comparison between the left and central panels in Fig.~\ref{fig:Regions_geometries} shows how the confinement region becomes wider for both trap geometries as the viscosity increases. Simulations are carried out for two different values of the dynamic viscosity: $\eta_1 = 1.7 \times 10^{-6}$ Pa$\cdot$s and $\eta_2 = 17 \times 10^{-6}$ Pa$\cdot$s. The confinement regions also depend on the complicated interplay between the trap confining forces and viscosity, suppressing dust particle oscillation and possible resonance. A lower viscosity value results in smaller dissipation of dust particle energy, thus increasing the velocity and implicitly the energy that a particle acquires from the trap field. Experimental results suggest that a 4WT geometry is better, as it implies a lower value of the specific charge ratio $Q/M$ or of the a.c. voltage $U_{\omega}$ required to confine particles.   

In order to solve the stochastic differential equation (\ref{eq.1}), a numerical method developed in \cite{Skeel02} is used. Numerical simulation is performed considering the following trap parameter values: electrode length $L = 6.5$ cm, $R_2 = 25$ cm, $R_1 = 3$ mm, trap radius $r_t = 4$ cm, and a.c. voltage amplitude $U_{\omega} = 2$ kV.

\begin{figure}[bth] 
	\centering
	\includegraphics[scale=0.15]{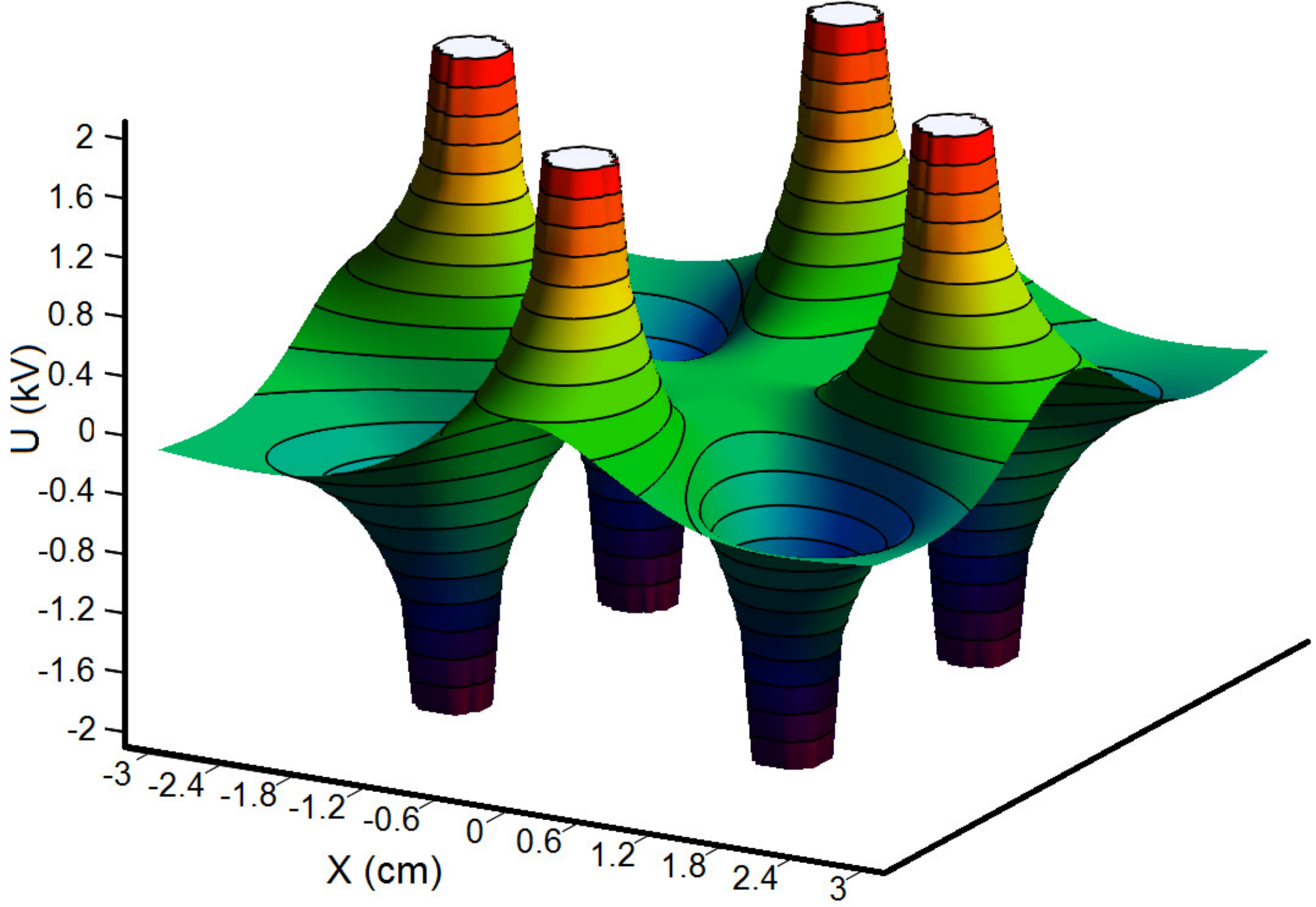}
	\includegraphics[scale=0.15]{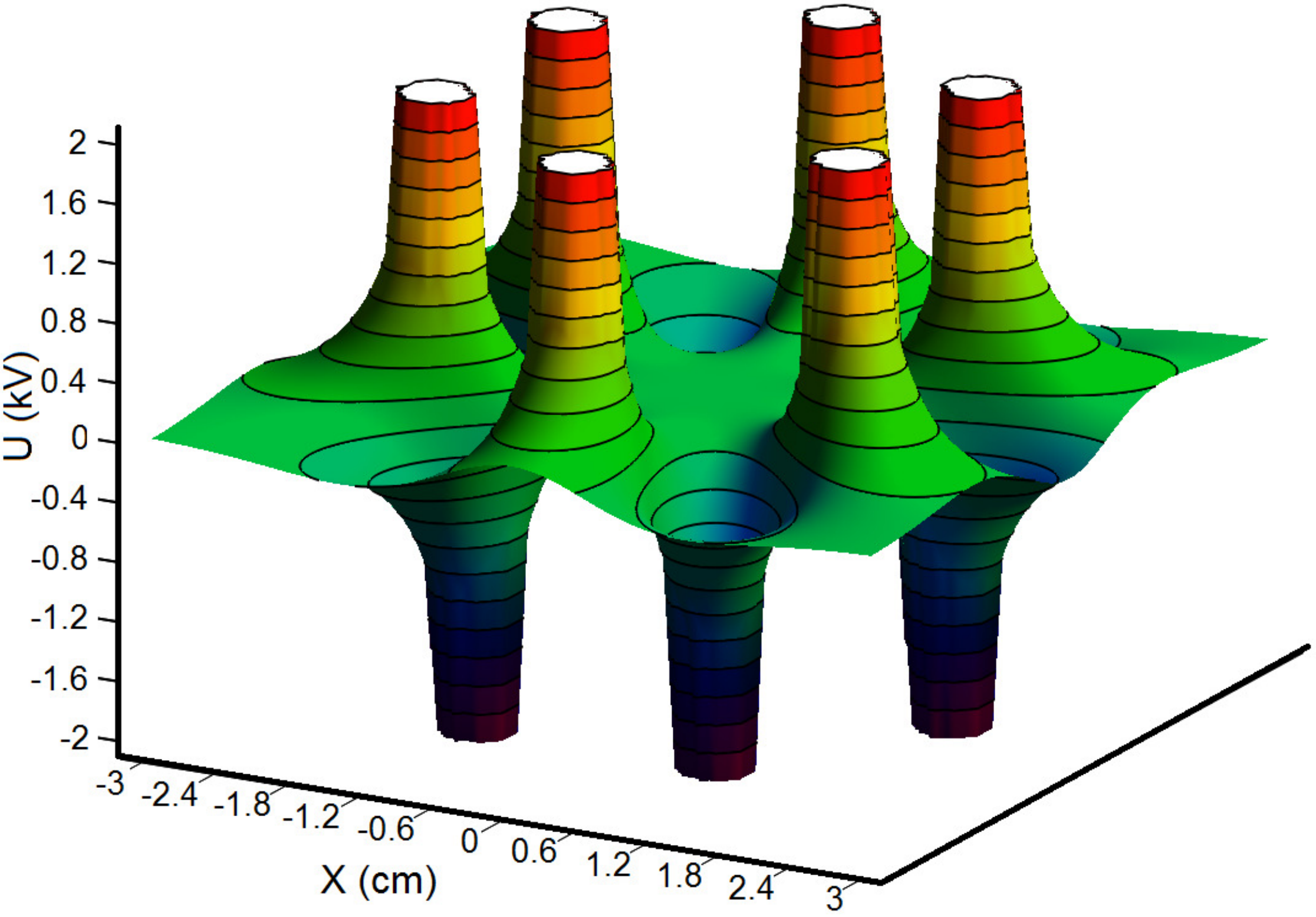}
	\includegraphics[scale=0.2]{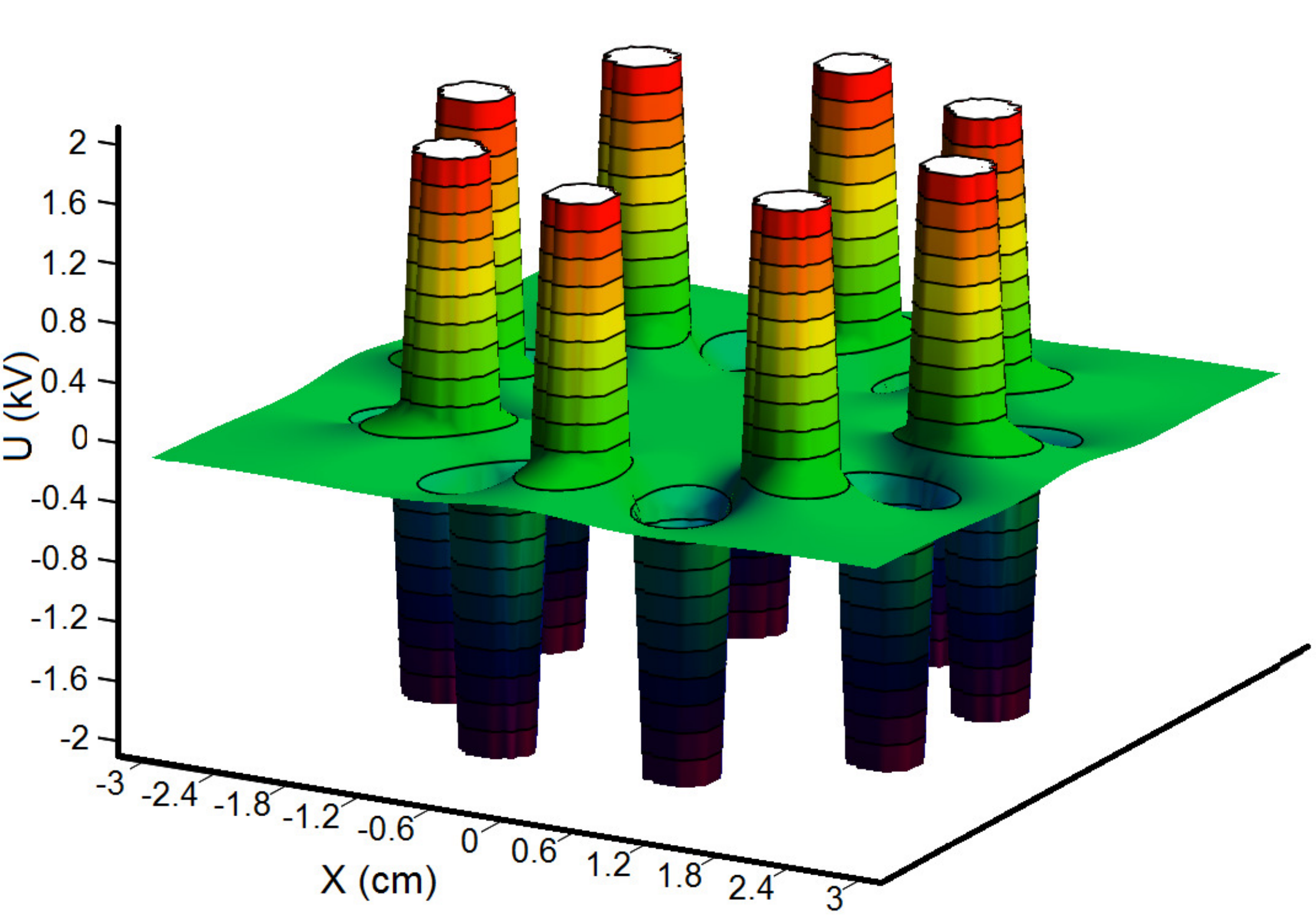}
	\caption{3D plots for a (a) 8 electrode trap, (b) 12 electrode trap, and (c) 16 electrode trap. Source: pictures reproduced from \cite{Mih16a, Lapi16d} with kind permission from the authors. Copyright AIP.}
	\label{468}
\end{figure}

\begin{figure}[bth] 
	\centering
	\includegraphics[scale=0.45]{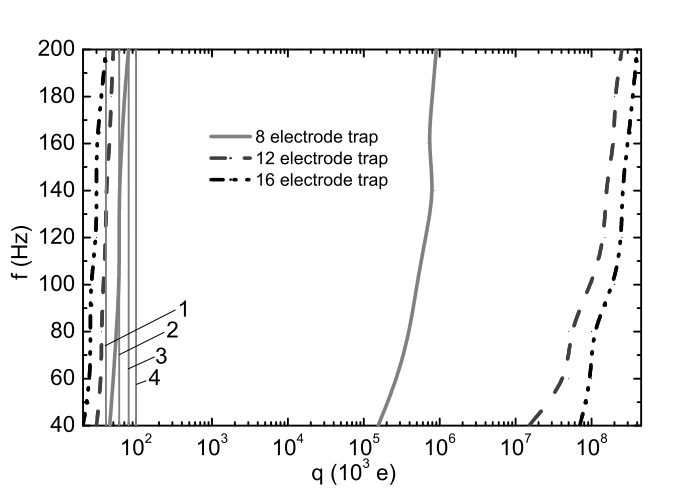}
	\caption{Regions of single particle confinement, depending on the frequency $f$ of the a.c. trapping voltage and on the particle charge $q_p$. The parameter values for the microparticle species used in calculus are: radius $r_p = 5 \ \mu$m, electric charge $q_p = 3 \times 10^4 e \div 5 \times 10^{11} e$. The vertical lines 1 -- 4 correspond to electric charge values of $q_p = 4, 6, 8, 10 \times 10^4 e$, considered in order to estimate the oscillation amplitude in the trap. Source: graphic reproduced from \cite{Mih16b, Lapi16d} with kind permission of the authors. Copyright AIP.}
	\label{468Stab}
\end{figure}

In Fig.~\ref{468} hills correspond to potential barriers and pits correspond to potential wells that attract microparticles. White holes inside the hills correspond to the trap electrodes. Every half-cycle period of the a.c. voltage barriers and wells swap positions, and each particle oscillates between these positions. Particle oscillations result in dynamic confinement \cite{Vasi13}. Fig.~\ref{468Stab} presents confinement regions for electrically charged microparticles levitated in 8, 12 and 16 electrode trap geometries, that are investigated in \cite{Mih16a, Mih16b}. The confinement regions are identified as follows: for an 8 electrode trap the trapping region is located between solid gray lines, for a 12 electrode trap the area is delimited by dark gray dash lines, while for a 16 electrode trap the zone is delimited by black dash-dot-dot lines. Beyond these regions multipole traps cannot confine particles. In case of low values of the electric charge, the a.c. field can no longer compensate the gravity force and particles flow across the trap. When the particles reach the right-hand area of the trapping region, the electric field pushes them out of the trap during one half-cycle of the oscillation.

\begin{figure}[bth] 
	\centering
	\includegraphics[scale=0.5]{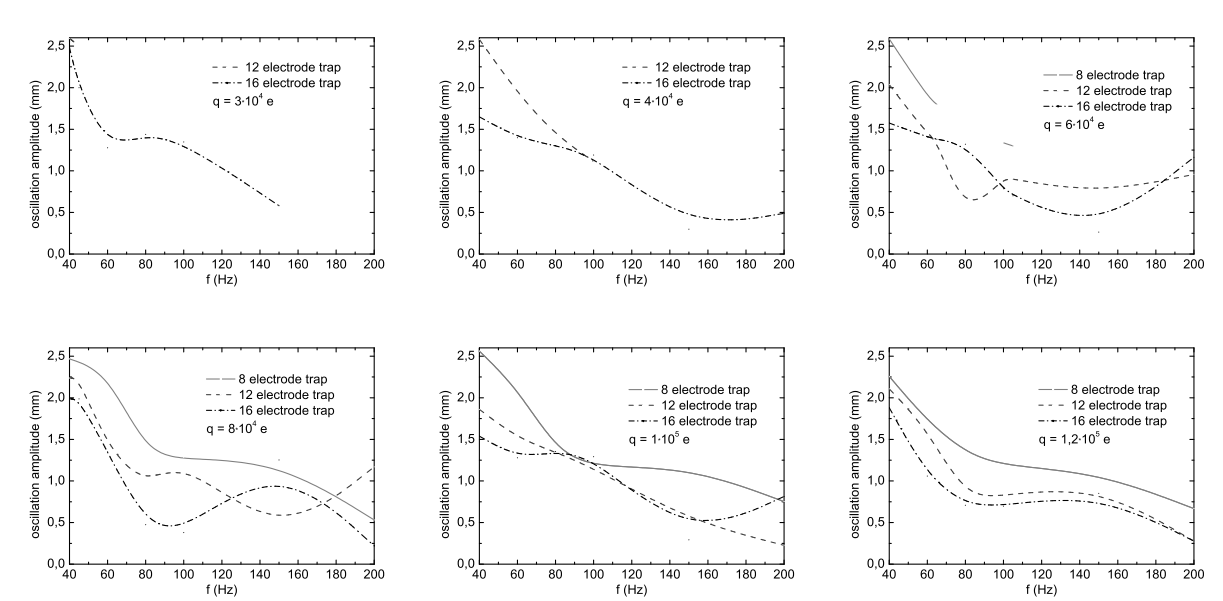}
	\caption{Dependence of the average oscillation amplitude on the number of trap electrodes, particle charge $q_p$ and a.c. voltage frequency $f$. Calculations are performed for the following parameter values: particle radius $r_p = 5 \ \mu$m, electric charge $q_p = 3 \times 10^4 e \div 1.2 \times 10^5 e$. Source: picture reproduced from \cite{Lapi16d} with kind permission of the authors.}
	\label{468ampl}
\end{figure}

The dependence between the particle oscillation amplitude and the number of electrodes is presented in Fig.~\ref{468ampl}. As illustrated in Fig.~\ref{468ampl}(a), for a rather low value of the electric charge, the dependency of the averaged amplitude of motion on the frequency $f$ exists only for 12 and 16 electrode traps, because the regions of particle confinement for these traps are wider than the one corresponding to an 8 electrode trap. The dependence of the oscillation amplitude on the number of trap electrodes is complex and it depends on the electric charge and inter-particle interaction, as earlier mentioned. The higher the frequency of the a.c. field, the lower the oscillation amplitude. For an 8 electrode trap a resonance effect at $60$ Hz is identified, for particles with an electric charge value of $q = 8 \times 10^4 e$.  

\begin{figure}[bth] 
	\centering
	\includegraphics[scale=0.41]{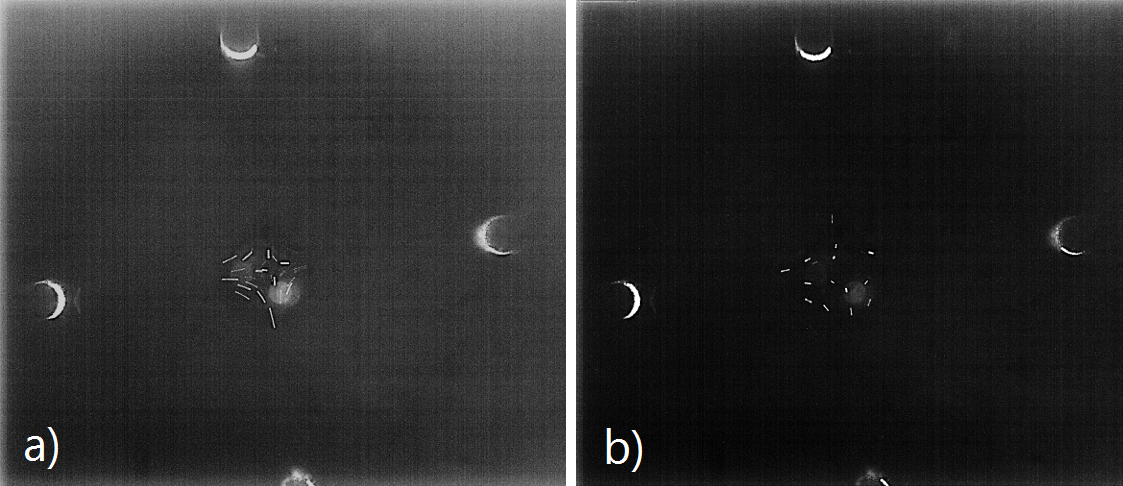}
	\caption{Dust structure in a quadrupole trap: $f = 50$ Hz (a), $f = 110$ Hz (b).}
	\label{quadro}
\end{figure}

\begin{figure}[bth] 
	\centering
	\includegraphics[scale=0.37]{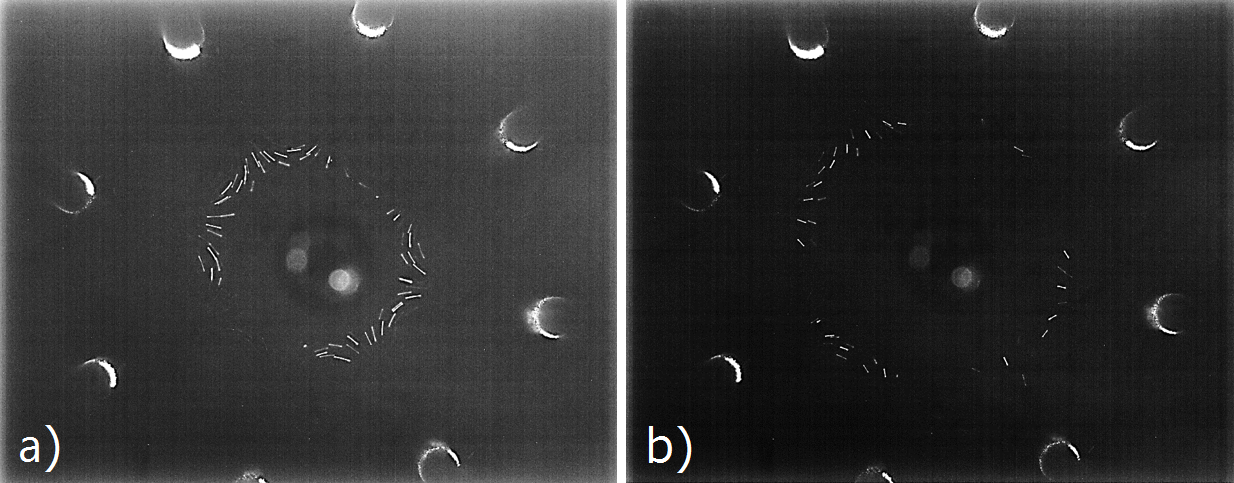}
	\caption{Dust structure in an octopole trap: $f = 50$ Hz (a), $f = 110$ Hz (b).}
	\label{octo}
\end{figure}

Experimental investigations are performed in quadrupole and octopole vertically oriented traps. Fig.~\ref{quadro} and Fig.~\ref{octo} show images of a horizontal cut of confined dusty structures, for various frequency values of the a.c. trapping voltage. The voltage amplitude is $4.25$ kV. The bright spot in the centre of the images represents the end electrode that provides particle levitation.

In case of a quadrupole configuration, for a frequency $f = 50$ Hz, particles are located near the symmetry axis of the trap. By increasing the frequency, the amplitudes of particle oscillations decrease while the inter-particle distances expand. When the frequency increases from 50 Hz up to 130 Hz, the oscillation amplitude drops off by 65\% (from $0.65$ mm down to $0.23$ mm). In case of an octopole trap particles are approximately arranged in a circle whose diameter increases with frequency (it expands from $16$ mm up to $27$ mm). The oscillation amplitude and the inter-particle distance depend on frequency, in a similar manner to the situation in a quadrupole trap. However, the dependence is less significant: by incrementing the frequency from 50 Hz to 130 Hz, the oscillation amplitude drops off only by 40\%, {\em e.g.} starting from $0.62$ mm down to $0.37$ mm.

\subsection{Perspectives and new technological approaches}

A recent paper introduces a multipole plasma trap that represents a novel approach based on using RF electric multipole fields to levitate charged particles of a plasma within a 3D volume. This new concept is presented in Ref. \cite{Hicks19a}, whereupon one of the trap electrodes is replaced with the aperture of a linear multipole stage which acts as a pipe by channelling plasma out of a source region into the trap volume. This linear multipole plasma transport can also be used to investigate 2D multipole confinement, whether an axial magnetic field is present or not.

\section{Mass spectrometry using electrodynamic traps}\label{Sec7}

Electrodynamic traps are clever tools, able to perform chemical characterization of microparticles \cite{Huang97, Jones97, Song06}. A 3D Paul trap can be used to perform both microparticle or even molecule levitation along with mass analysis \cite{Jons97}. Chemical characterization is then achieved by employing optical methods such as absorption, fluorescence and Raman spectroscopy \cite{Dem15}. After investigation by means of optical methods, the microparticle is analyzed using mass spectrometry. Current techniques for ion generation include laser ablation, desorption, or photoionization.      

Accurate control and manipulation of trapped ion species plays an important role in advancing many technologies, including ion-based quantum information processing (QIP) \cite{Bruz19, War20}, mass spectrometry (MS) \cite{March17a}, quantum metrology \cite{Wine11, Jord20, McCor19} and quantum engineering \cite{Zago11}. MS based techniques can also be used to identify the masses and relative abundances of Coulomb-crystallized ions levitated in a linear Paul trap \cite{Deb15}. In the field of life sciences and chemical analysis, MS using ion and charged particle traps is a versatile and powerful analytical instrument aimed at investigating chemical and biological species. Specimens with diameters ranging between $10$ nm $\div 20 \ \mu$m, i.e. DNA sequences, proteins, viruses, cells, bacteria and generally nanoparticles (NPs), can be diagnosed with high resolution and accuracy. In an attempt to further increase the resolution of MS, integration of advanced optical detection is considered a key approach \cite{Jiang11}. Similarly, optical spectroscopy of nanometre sized particles depends on the ion trap capacity. Gas phase NPs ranging from 3 nm to 15 nm in diameter can be contained and levitated using a Trapped Reactive Atmospheric Particle Spectrometer (TRAPS), with a maximum trap capacity of $5 \times 10^8$ particles and a residence time of 12 sec, for optical and X-ray spectroscopy studies \cite{Mein10}. Recent progress include ion traps that are developed with integrated fibre optics, able to efficiently collect fluorescence light from the ion \cite{VanDev10, Lee19}. Ion manipulation finds use in various applications such as miniaturized chemical detectors, vacuum technology, ultra-precise atomic clocks, QIP and secure communication schemes. Therefore, new approaches for particle (ion) control and manipulation represent a mandatory step towards implementing the next generation of advanced technologies \cite{Kolo07, Smith10}.

\begin{figure}[!ht] 
	\begin{minipage}[h]{0.55\linewidth}
		\center{\includegraphics[width=\linewidth]{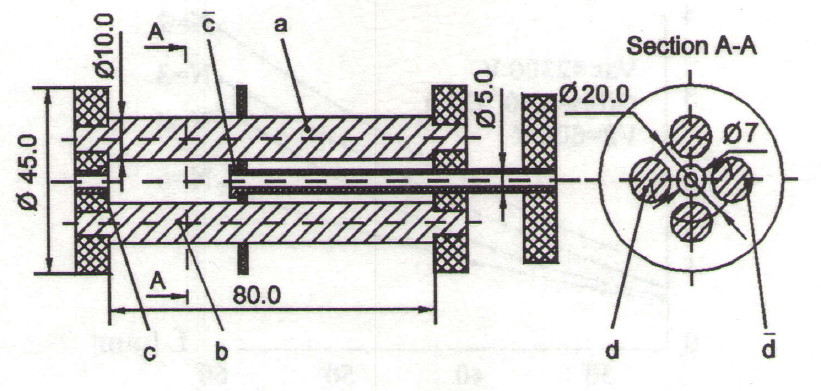}\\a)}
	\end{minipage}
	\begin{minipage}[h]{0.55\linewidth}
		\center{\includegraphics[width=\linewidth]{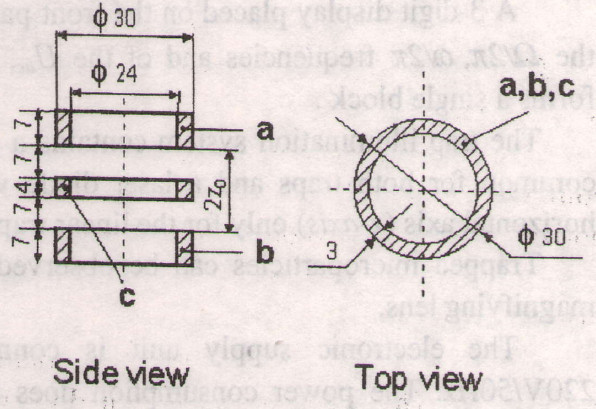}\\b)}
	\end{minipage}
	\caption{Sketch of (a) linear and (b) ring geometry Paul traps intended for levitating microparticles under SATP conditions. Image reproduced from \cite{Sto01} with kind permission of the publisher.}
	\label{LinTrap}
\end{figure}

A linear ion (Paul) trap  (LIT) uses a superposition of time varying, strongly inhomogeneous (a.c.) and d.c. electric potentials, to achieve a trapping field that dynamically confines ions and other electrically charged particles \cite{Major05, March05, Vini15, March17a}. When the a.c. trapping voltage frequency lies in the few Hz up to MHz or even GHz range, electrons, molecular ions and electrically charged micro and nanoparticles with mass larger than 10 {\em u} (atomic mass units), are confined \cite{Smith10}. Ion dynamics in a Paul trap is described by a system of linear, uncoupled equations of motion (Hill equations \cite{Mag66}), that can be solved analytically \cite{McLac64, Rand16, Vasi13, Lebe12}. The linear trap geometry can be used as a selective mass filter or as an actual trap by creating a potential well for the ions along the $z$ axis of the electrodes \cite{Sto01, Otto09}. In addition, the linear design results in increased ion storage capacity, faster scan times and simplicity of construction. A Paul trap runs in the mass-selective axial instability mode by scanning the frequency of the applied a.c. field \cite{Staff84, Nie08, Smith08}. Pictures of a linear and of an annular trap geometry intended for microparticle levitation under SATP conditions are presented in Fig.~\ref{LinTrap} \cite{Ghe98, Sto01}. 

Recent sampling and ionization methods extend mass spectrometry (MS) to many applications, such as investigation of biomolecules, aerosols \cite{Sue99}, explosives, petroleum and even microorganisms \cite{Peng04}. Late advances towards the miniaturization of mass spectrometers are presented in \cite{Ouya09, Peng11, Guo18}. The approach used consists either in minimizing the RF electronics and the vacuum system, or in using new materials such as polymers or ceramics. One of the challenges in chemical physics lies in the ability to control ion temperatures. To achieve this goal low-temperature photoelectron spectroscopy instruments are developed, aimed at investigating complex anions in the gas phase, including multiply charged ones and molecules \cite{Wang08}. Such an experimental setup consists of an electrospray ionization (ESI) source, a 3D Paul trap where ions are laser cooled, a ToF mass spectrometer (MS) and a magnetic-bottle photoelectron analyzer. The device enables a good control of ion temperatures in the $ 10 \div 350 $ K range.

\subsection{The quadrupole Paul trap as a mass spectrometer}

Since its early days, the Paul trap (3D QIT) \cite{Paul90, Major05, Paul58, Bla98, Doug09} has proven to be a versatile device that employs path stability as a means of separating ions according to their specific charge ratio. Mass analysis is mainly performed using either one of two techniques: (a) mass selective resonance detection, when presence of ions is detected by means of an external electronic circuit connected between the endcap electrodes, or (b) mass selective storage in which the positive ions are expelled over holes pierced in the endcap electrodes, either onto the first dynode of an electron multiplier or into a {\em channeltron}. Mass selective storage is based on external detection which lifts many of the issues associated with mass-selective detection \cite{March97, Kais91, March95b, Xu09, Gross17}. 

The resonance detection technique represents an original method to perform mass analysis using an ion trap \cite{Paul58, Bla98}. It employs supplying a small a.c. voltage of frequency $\omega_{ac}$ between the endcap electrodes, in addition to the d.c. and RF ion trapping voltages (denoted as $U_0$ and $V\cos\left(\omega t\right)$, respectively) applied between the endcap and ring electrodes. Any trapped ions that exhibit a mass to charge ratio $M/Q$ \cite{Kais91} in such a way that their fundamental frequency of motion in the axial direction, $\omega_z$, is equal to the perturbing field frequency value, $\omega_{ac}$, will resonate. Ions are neutralized at the endcap electrodes and the resulting current is detected \cite{Staff84, Fisch59, Schwa91, Guan93}. Broad-band nondestructive ion detection can be achieved in a QITMS by excitation of a cloud of trapped ions with different masses, followed by the recording of the ion image currents induced on a small detector electrode embedded in the surrounding endcap electrode \cite{Soni96, Nappi98}.  

The mass selective storage technique used to perform mass analysis of ions is devised by Dawson and Whetten \cite{Gross17, Daws69}. Mass spectrometers that rely on such technique have been experimentally tested and investigated \cite{March09, March05, Schwa91, Daws69}. Operation of a QIT based on this technique requires that the applied d.c. and RF trapping voltages are chosen in such a manner so as to narrow the range of $M/Q$ values trapped in the device, which results in tight confinement of ions with a particular $M/Q$ ratio. Ejection and detection of trapped ions is usually accomplished by applying a voltage pulse between the endcaps. The expelled ions reach an electron multiplier and the corresponding electric current is measured. To achieve a mass scan over a wide range of ion masses an experiment must be performed for every possible $M/Q$ value within the chosen range \cite{Staff84, Schwa91, Guna09}. The method of multiple scales can be used to characterize dynamics associated with early and delayed ejection of ions in mass selective ejection experiments in Paul traps, as demonstrated in \cite{Raja07}. The electric fields responsible for mass-selective axial ejection (MSAE) of ions trapped in a linear quadrupole ion trap are investigated in \cite{Lond03} using both analytical theory and computer modelling. As an outcome of the strong dependence of the axial field on radial displacement, trapped thermalized ions can be ejected axially from a linear ion trap in a mass-selective way when their radial amplitude is increased through a resonant response to an auxiliary signal.  

\begin{figure}[!ht] 
\begin{center}
	\includegraphics[scale=1.25]{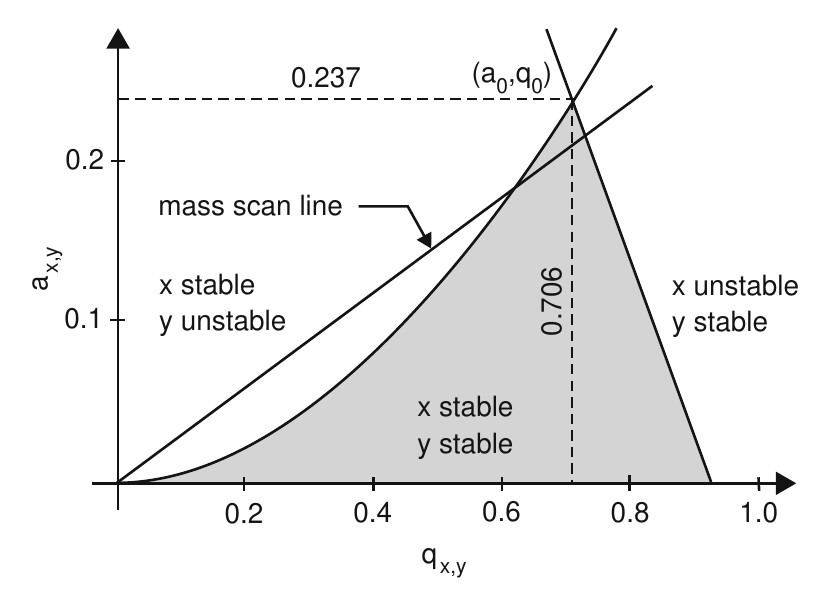}
	\caption{Scan line with constant ratio $a/q$ for operation of a Paul trap as a mass spectrometer. Picture reproduced from \cite{Werth09} by courtesy of Prof. G. Werth.}
	\label{MassScan}
\end{center}
\end{figure}

If a scan line is selected that crosses the stable region of a Paul trap near its upper corner, only a narrow range of masses will achieve stable confinement in the trap (see Fig.~\ref{MassScan}). The resolving power of the trap is enhanced as the scan line approaches the crest of the stable area.

The Quadrupole Ion Trap Mass Spectrometry (QITMS) is a promising technique able to perform mass analysis of micron-sized particles such as biological cells, aerosols and synthetic polymers \cite{Dale91, Batey14}. The trap can be operated in the mass-selective axial instability mode \cite{Nie08, Smith08}. Microparticle diagnosis can be achieved by operating the QIT as an EDB, for low frequency values of the a.c. field applied to the trap electrodes (typically less than 1 kHz) \cite{Doug09, March10, March95c}. Because of the low frequency, in order to achieve high mass measurement accuracy (better than 1 ppm) only one particle is analyzed at a time, over a time period ranging from seconds up to minutes. A specific charge $M/Q$ can be isolated in the ion trap by ejecting all other $M/Q$ particles when scanning various resonant frequencies. Moreover, an ion trap can be coupled to an Aerosol Mass Spectrometer (AMS) and thus investigate atmospheric (nano)particles \cite{Kul11, Kwa11, Ula11, Signo11, Cana07}. 

Conventional linear ion trap mass analyzers (LIT-MS) provide high ion capacity and show remarkable MS$^n$ capability. A recent experiment demonstrates successful measurement of Immunoglobulin G ions $\left( m/z \sim 150,000 \right)$, under conditions of controlled ion kinetic energy \cite{Patil16}. In order to achieve that, the experimental setup consists of a charge detector (CD) coupled to a rectilinear ion trap mass spectrometer (RIT-MS).  

In addition, state of the art results demonstrate how the capabilities of QIT mass spectrometers operated in the precursor and neutral loss scan mode can be extended using two approaches \cite{Szal20a}. In a first experiment, a triple resonance precursor ion scan is employed to enhance sensitivity, selectivity and molecular coverage. The method enhances the ion trap precursor ion scan using a second excitation frequency which selectively activates first generation (MS2) product ions as they are created, and yields second generation (MS3) product ions. Mass-selective ejection of the MS3 ions is achieved by means of a third auxiliary signal. The second capability demonstrated is {\em frequency tagging}, a method used to discriminate between ions ejected due to inherent instability and ions that are resonantly ejected by the product ion ejection frequency. In addition, when QIT MS are operated at constant trapping voltage, ions can be excited at their secular frequencies and all MS/MS experiments can be performed, including the two-dimensional tandem mass scan (2D MS/MS scan) in which all precursor ions and their subsequent product ions are both identified and correlated \cite{Szal20b}.

\subsection{Advantages of ion trap mass spectrometry}

Due to intrinsic features such as high sensitivity, compact size, somewhat large operating pressure and the unique capacity to achieve multistage tandem mass analysis (MS$^n$), QITs are largely used as mass analyzers \cite{March17b, March17a, Nie08, March10, Doug05}. First experimental setups were based on classical 3D Paul traps with hyperbolic-shaped electrodes \cite{Bla06, March05, Bla98, Doug09, Schwa91}. Practical realization of such geometries is frequently associated with mechanical imperfections and electrode misalignments \cite{Tak07, Wu15, Tian18} that degrade the strength of the quadrupole trapping field, which leads to the occurrence of higher order terms in the series expansion of the electric potential \cite{March05, Beat87, Aust10, Zhang11, Wang14}. Ion dynamics in a nonlinear resonance ion trap used to perform MS is characterized by a nonlinear Mathieu differential equation \cite{Wang93}. The performances of commercial quadrupole mass spectrometers (QMS) characterized by imperfections have been investigated with respect to an ideal hyperbolic 3D Paul trap geometry, using the computer simulation program SIMION 3D \cite{Bla98}. In addition, the case of a quadrupole mass spectrometer (QMS) in which a static magnetic field is applied axially in the $z$-direction along the length of the mass filter is presented in \cite{Syed10}. A technique that relies on the variation of the electric field in a cylindrical ion trap (CIT) mass spectrometer is presented in \cite{Sona13}. By employing this technique the CIT electrodes are split into a number of mini-electrodes, supplied at different voltages in order to achieve the desired field. The fundamental principles of QMS are discussed in \cite{Batey14}. 

An issue of high interest for MS experiments is related to the effects induced by supplying various d.c. magnitudes and polarities to only one of the endcaps of a 3D QIT. A monopolar d.c. field is obtained by supplying a d.c. potential to the exit endcap electrode, while the entrance endcap electrode is kept at ground potential. Control over the monopolar d.c. magnitude and polarity during time periods associated with ion accumulation, mass analysis, ion isolation, etc., leads to increased ion capture efficiency, increased ion ejection efficiency during mass analysis, and effective isolation of ions using lower a.c. resonance ejection amplitudes \cite{Koizu09, Prent11}. As an outcome of these remarkable experimental improvements the performance of a 3D ion trap used for MS experiments is greatly enhanced.

Late experiments show that decent performance can also be achieved using simplified trap geometries such as spherical \cite{Nosh14}, hybrid \cite{Arkin99}, linear \cite{Suda12}, planar \cite{Bahra19, Clark13, Song06, Aust10, Zhang11, Alda16} and cylindrical (toroidal) setups \cite{Li17, Tay12, Higgs16, Kot16, Kot17, Higgs18}, that exhibit the advantage of enhanced compactness while being easier to design and machine \cite{Tay12}. Reduced power operation of a mass analyzer, under conditions of minimum loss of spectral resolution and mass range, is beneficial when discussing in terms of portability. Given that the RF amplitude required to perform mass analysis scales with the square of analyzer dimensions, minimization of QITs represents a primary concern. The performance of a miniature, stainless steel, rectilinear ion trap (RIT), is investigated in \cite{Hend11}. Portable mass spectrometers usually employ cylindrical ion traps (CIT). A LIT exhibits sensibly improved trapping capacity and trapping efficiency with respect to a 3D QIT \cite{Major05, March09, March05}. A rectilinear ion trap geometry used in a miniaturized MS system is able to achieve a resolution of $\Delta M/Q = 0.6$ Th (FWHM -full width at half-maximum) and a mass range of up to $M/Q \sim 900$ \cite{Li14a}. Electrode misalignments in linear ion traps are investigated in \cite{Wu15} by employing SIMION software to model the case of a two-plate linear ion trap. Geometric misalignments in six degrees of freedom are analyzed with respect to the resolving power and ion detection efficiency. A new fabrication method, numerical simulations performed with SIMION and experimental results for micromachined CIT (m-CIT) arrays suggested for use in miniaturized mass spectrometers, are presented in \cite{Chaud14}. In addition, alignment of plates in a two-plate planar LIT is ideal to characterize misalignment effects, as it represents the most simple case that exhibits only six degrees of freedom (DOF) (three of them translational and the other three rotational) \cite{Tian18}. 

The scientific community carries out intensive efforts to miniaturize ion traps for portable MS applications. The most demanding technological challenge lies in limiting the energy consumption of the particular components of a mass spectrometer, such as the vacuum system, the ion source and the RF amplifier \cite{Jones97, Math06, Jau11}. Ref. \cite{Baig13} describes the development of a multichannel arbitrary waveform generator intended for QIP using ITs. A technique that achieves active stabilization of the harmonic oscillation frequency of a laser-cooled atomic ion levitated in a RF trap is demonstrated in \cite{Johns16}. The solution relies on sampling and rectifying the high RF voltage applied to the trap electrodes. Thus, the 1 MHz atomic oscillation frequency is stabilized to a value below 10 Hz. Use of this technique is expected to enhance the sensitivity of ion trap (IT) based MS and the fidelity of quantum operations in IT based QIP. A low power RF amplifier circuit for ion trap applications is presented and tested in \cite{Nori16}. The design and operation of a multichannel, low-drift, low-noise d.c. voltage source, particularly designed for supplying the electrodes of a segmented LIT is described in \cite{Beev17}.
 
Digital QMS(s) exhibit certain unique features and render more simple various ion handling operations as they provide an enhanced control over the frequency, duty cycle and amplitude of the trapping potentials. Ref. \cite{Brabe16} demonstrates how matrix solutions of the Hill differential equation can be used to explore the influence of the additional degrees of freedom on ion stability. In order to illustrate these effects, the paper provides the stability diagrams corresponding to a digital mass filter that employs asymmetric driving potentials.  

Nonlinear harmonics is the outcome of nonlinear fields in the trap, leading to the occurrence of the nonlinear resonance effect \cite{Zhou13}. Ion dynamics in presence of a quadrupole field with a weak superimposed octopole component \cite{Suda03} is shown to be characterized by the nonlinear Mathieu equation (NME), and it can be investigated by using the analytical harmonic balance (HB) method \cite{Xio16}. Chemical physics relies strongly on QITs as versatile investigation tools. ITs are currently used to investigate both external and internal dynamical systems. An interesting discussion focused on linear and nonlinear resonances in QITs is carried out in \cite{Sny16b, Sny17a}, with an emphasis on the effect of quadrupole field nonlinearity. Secular frequency scanning is investigated in \cite{Sny16a} for linear traps and different ion species.

The mass/charge range of a MS that operates either at the boundary of the first stability region of the Mathieu equation or in the resonance ejection mode \cite{Koizu09, Remes15}, is usually limited by the highest RF voltage value that can be supplied to the trap electrodes. High voltages are dangerous for miniature instruments as they might induce discharges onto the electrodes which are separated by small distances. To overcome this technological challenge, an alternative approach to mass range extension is used, based on a method of scanning the resonance ejection frequency nonlinearly in the form of an inverse Mathieu $q$ scan \cite{Sny17b, Sny17c}. The method is experimentally tested in case of a benchtop LTQ linear ion trap and for the Mini 12 miniature LITMS. In both situations an increase in mass range of up to 3.5 times is reported \cite{Sny17c}. In addition, an ion trap operated in the ac frequency scan mode can perform any arbitrary mass scan, as well as a sequence of scans, by using single ion injection \cite{Sny17d}.     

A novel method used for mass-selective removal of ions from a Paul trap by parametric excitation is described in Ref. \cite{Schmi20b}, based on applying an oscillating electric quadrupole field at twice the secular frequency, using pairs of opposing electrodes.

\subsection{Trap operation in the mass-selective axial instability mode}

\begin{figure}[bth]
	\centering
	\includegraphics[scale=0.85]{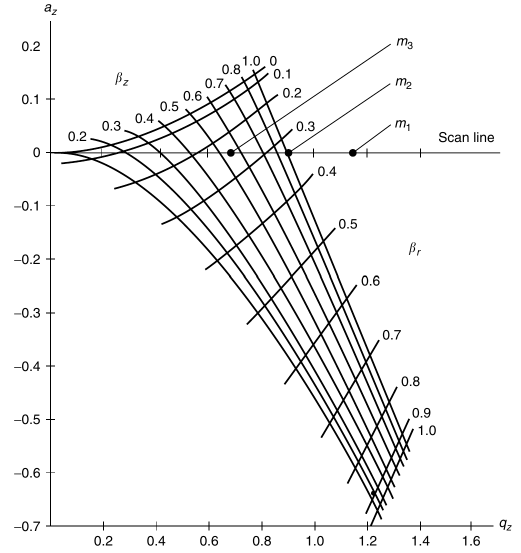} 
	\caption{Stability diagram for a quadrupole ion trap plotted in ($a_z, q_z$) space. The points marked $m_1$, $m_2$ and $m_3 \left(m_1 < m_2 < m_3\right)$, refer to the coordinates of three ions: $m_1$ is already ejected, $m_2$ is on the point of ejection, while the $m_3$ species is still trapped. Reproduced from R. E. March and J. F. J. Todd (Eds.), Practical Aspects of Ion Trap Mass Spectrometry, Vol. III, page 11, Figure 1.7. CRC Press, Boca Raton, FL, 1995. Copyright 1995 by CRC LLC. Reproduced with permission of CRC Press LLC in the format Journal/magazine via Copyright Clearance Center.}
	\label{axinstab}
\end{figure}

Operation of an ion trap in the mass-selective axial instability mode is reported for the first time in 1984 \cite{Staff84}. It is based on the fact that if the working point $\left(a_z, q_z\right)$ moves along a {\em scan line} which crosses either the $\beta_z = 0$ or $\beta_z = 1$ boundary, then the trapped ion will rapidly develop axial instability and get expelled onto an appropriately positioned detector. Experimentally, the simplest way to achieve axial instability lies in supplying the ring electrode with RF power only, which renders the scan line coincident with the $q_z$ axis \cite{March05, Staff84}. The method begins by creating ions using electron bombardment or photoionization, while maintaining the RF trapping voltage constant. Then, the RF voltage is ramped linearly in such a way that the $a_z, q_z$ values of the ions shift to the $\beta_z=1$ boundary at which point ions are expelled out of the trap whilst the $M/Q$ value increases, as illustrated in Fig.~\ref{axinstab}. 

Charge detection QITMS is demonstrated to be a remarkable technique suited for high-speed mass analysis of micron-sized particles such as biological cells and aerosols \cite{Smith08, Sue99}. Ref. \cite{Nie08} describes the technique and explains how the trap can operate in the mass-selective axial instability mode by scanning the frequency of the applied a.c. field. A calibration method is presented, aimed at finding the points of ejection in the stability diagram $\left(q_{eject}\right)$ for the individual particles that are investigated. Mass selective instability in a linear trap without the use of auxiliary waveforms is investigated in Ref. \cite{Reece19}, where it is demonstrated that digital ion traps enable duty cycle manipulation. Therefore, the duty cycle can be applied to enhance the resolution and sensitivity for mass-selected instability in a linear ion trap without the application of an auxiliary waveform.

\subsection{Specific charge measurement techniques based on ion trap mass spectrometry}\label{specharge}

We basically describe the method employed to measure the specific charge for different microparticle species levitated in various Paul trap geometries tested at INFLPR, operated under SATP conditions. The equation of motion for a particle of mass $M$ and charge $Q$ confined in the trap is \cite{Ghe98, Sto01}:

\begin{equation}
{\bf F} = Q {\bf E} - K{\bf r} + M{\bf g} + Q\bf{E_{dc}}  \ ,
\end{equation}
where ${\bf r} = \left( x, y, z\right)$ is the particle vector position, and $K \left(K > 0\right)$ is the coefficient describing the aerodynamic drag force. ${\bf E}$ represents the trap electric field, $M{\bf g}$ stands for the gravitational force and ${\bf E_{dc}} = V_2/\left(2 r_0\right)$ is the electric field produced by applying a static potential difference between the upper and lower rods (electrodes) of a quadrupole LIT separated at $2 r_0$, in order to shift the particles towards the trap centre. 

We introduce the parameters $\tau = \Omega t/2$ and $\Lambda = K / \left( M \Omega \right)$, where $\Omega$ is the RF drive frequency. We also denote $X = e^{\Lambda \tau}x, Y = e^{\Lambda \tau}y, \textrm{and} \ Z = e^{\Lambda \tau}z$. Then, the Mathieu equations that characterize  trapping are homogeneous along the $X$ and $Y$ axes, with well known solutions and stability domains. The Z-axis dynamics is described by an inhomogeneous Mathieu equation. In this case stability is obtained for (a) $i\mu $ real; (b) $\mu$ real and $|\mu| < \Lambda$; (c) $\mu -i$ real and $|\mu - i| < \Lambda$. Hence, the stability regions corresponding to solutions of the inhomogeneous equations of motion in presence of drag forces include not only the stability regions of the homogeneous equations but also part of the instability regions, which means they are extended. 

Typically, parametric excitation of trapped microparticle motion is achieved by applying a low amplitude and variable frequency additional a.c. voltage in series with the $V_{ac}$ trapping voltage \cite{Brou11}. When the supplementary a.c. field frequency is twice that of the secular motion, microparticles (located at the limit of the first Mathieu stability domain) resonantly absorb energy from the field whilst their secular motion amplitude exponentially increases \cite{Ghe96b}. If the trap contains a number of non-interacting particles, besides normal resonance at the secular motion frequency and parametric resonance at the double secular frequency, other weak resonances are observed \cite{Ved90} as a consequence of the presence of coupled terms for various combinations of the frequencies of motion.

\begin{figure}[!ht]
	\centering
	\includegraphics[scale=0.9]{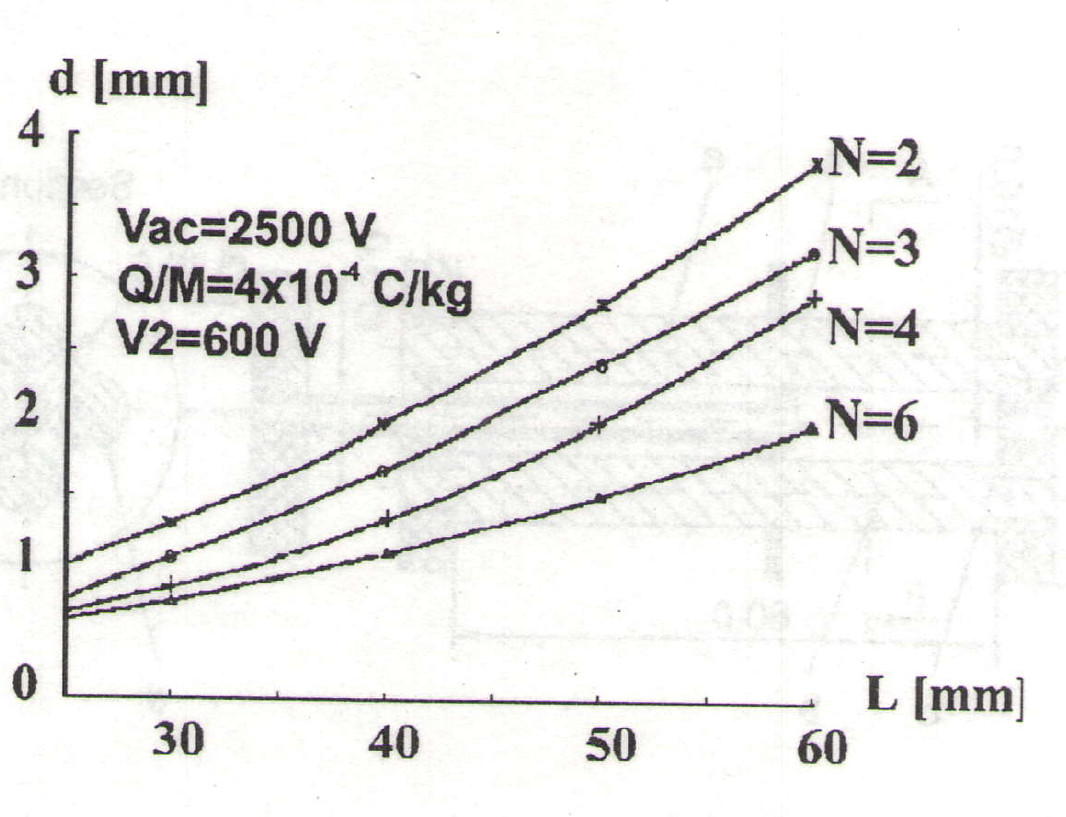} 
	\caption{Picture of the inter-particle distance $d$ for a variable linear trap geometry designed and tested in INFLPR \cite{Ghe98, Sto01}, presented in Fig.~\ref{LinTrap}, for SiC microparticles with dimensions ranging between $ 50 \div 1000 \ \mu$m. $N$ represents the number of microparticles while $L$ is the trap length. Reproduced from \cite{Sto01} with kind permission of the publisher.}
	\label{interpart}
\end{figure}

When the coupling parameter between particles (the ratio between the Coulomb and thermal energy) is greater than 175, solid effects occur in the levitated particles' system \cite{Major05, Boll84, Werth09, Dub99, Wine88, Werth05a}. As the inter-particle distance is larger for trapped microscopic particles than for trapped ions, it is much easier to observe such behaviour. In Fig.~\ref{interpart} we supply a picture of the inter-particle distance $d$ for a variable linear trap geometry designed and tested in INFLPR \cite{Ghe98, Sto01} (presented in Fig.~\ref{LinTrap}), for SiC microparticles with homogeneous and well-defined dimensions ranging between $ 50 \div 1000 \ \mu$m. $N$ stands for the number of microparticles while $L$ is the trap length. 

One up to thousands of microparticles can be levitated along the trap axis. Ordered structures (planar, zig-zag and volume structures) are reported in a LIT. The $Q/M$ specific charge ratio for SiC microparticles is estimated using the relation $Q / M = 2 g \ {\Delta z_0} /\Delta V_2$ (around $10^{-4}$ C/kg), for a static trapping voltage $V_0 = 500$ V \cite{Sto01}, where $\Delta z_0$ stands for the axial shift of the microparticle motion corresponding to a variation of the $V_2$ voltage denoted by $\Delta V_2$. The absolute electric charge is evaluated from the electrostatic equilibrium conditions for microparticles confined along the trap axis, considering the $V_2/2$ generated electrical forces and the inter-particle Coulomb repulsion. For a value $V_2 = 500$ V  and $L = 75$ mm, an electric charge value $Q$ of around $10^{-11}$ C is inferred. By knowing the specific charge value, the microparticle mass can be instantly evaluated. As the SiC density value is known, the microparticle radius \cite{Lapi18c} can be inferred. 

The specific charge was estimated for different trapped microparticle species. The values obtained are displayed in Tabel \ref{specific} \cite{Sto01}.

\begin{table}[h!]
\begin{center}
	\caption{Averaged values of the specific charge ratio for different microparticle species levitated in a linear Paul trap, presented in Fig.~\ref{LinTrap}}
	\label{specific}
	\begin{tabular}{l|c|c|c|r} 
		\textbf{Microparticle species} & \textbf{SiC} & \textbf{Anthracene} & \textbf{Alumina} & \textbf{Hydroxyl appatite} \\
		\hline
		$\bar{Q}/\bar{M} \times \ 10^{-4} \left[ \text{C/kg}\right]$ & $3.482$ &  $2.974$ & $4.147$ & $4.873$ \\
   \end{tabular}
\end{center}
\end{table}

A different diagnosis method employs parametric excitation of the microparticle motion, at a frequency $\omega/2\pi$.  When the value of $\omega$ is twice the secular motion frequency, parametric resonance occurs. Experimental measurements performed for $100 \ \mu$m diameter SiC microparticles, for $\Omega/2 \pi = 100$ Hz and $V_{ac} = 2.5$ kV, reveal parametric resonance at about 25 Hz. Hence, the specific charge of the SiC microparticles is evaluated as $5 \times 10^{-4}$ C/kg \cite{Ghe98}.

High resolution mass analysis of intact RNA, DNA and viruses is reported in Ref. \cite{Wang12}. The setup described achieves sampling of singly charged particles up to 200 nm in diameter at atmospheric pressure into vacuum, while also trapping large numbers $\left( > 10^6\right)$ at a point located in front of the end cap electrode of a linear quadrupole ion guide/trap. Measurements of charge density of tapered optical fibres using charged polystyrene particles confined in a linear Paul trap operating under SATP conditions are reported in \cite{Kami16}.

\section{Aerosols and nanoparticles}\label{Sec8}

According to Ref. \cite{Kul11} aerosols can be described as two-phase systems consisting of the suspended solid or liquid phase, levitated (immersed) in the surrounding gas phase. Despite their small size aerosols yield a major impact on global climate and health, but the underlying mechanisms and subsequent effects are still far from being explained. Typically, ambient aerosol particles are characterized by dimensions ranging from a few nm to a few hundred micrometers, while they consist of a wide variety of materials. About 90 \% of the aerosols found in the atmosphere are generated by natural sources. Volcanoes eject huge amounts of ash and volcanic debris in the troposphere along with sulphur dioxide and other gases. Sea spray is another important source of natural aerosols \cite{Sult17}, as well as the wind carrying desert sand particles or soil dust (including specific minerals) to either sea or land areas \cite{Kond06}. Both species of previously described aerosols are larger particles with respect to their human-made counterparts. Phytoplankton is also responsible for the emission of gases such as dimethyl sulfate. 

Solid and liquid aerosol particles that consist of a significant fraction of biological materials are called bioaerosols. Their dimensions vary between $0.1 \div 250 \ \mu$m. Airborne bacteria, viruses, toxins and allergens, or fungal spores, represent important types of bioaerosols \cite{Heo17}. Bioaerosols have a tremendous economic impact in the transmission of diseases of humans (e.g., flu, severe acute respiratory syndrome (SARS)), other animals, agricultural crops and other plants, while also causing asthma or allergies \cite{Srik08, Kim18}. Hence, detection and characterization of bioaerosols is an issue of large interest \cite{Hart20}. Moreover, as the world is confronted with an increasing wave of terrorist attacks, detection of bioaerosols or other aerosols associated with traces of explosives or chemical weapons is a matter of utmost importance. Hence, there is a stringent need for improved methods and instrumentation for rapidly characterizing harmful bioaerosols \cite{Kim08, Islam18}. Such instruments could also be used in studying and monitoring disease dissemination \cite{Ghosh15, Fuji17}. 

We also mention the contribution of cosmic aerosols that is negligible with respect to the aerosol sources described above. Space weather is a relatively new and important field of research as it represents a branch of space physics and aeronomy. It strongly influences radio communications, space missions, diagnostics of ionospheric and space plasmas, detection of pollutants and re-entry objects, weather prediction and the phenomenon of global warming. Recent scientific progress and results clearly show that nano- and micrometre-sized electrically charged particles coming from the interplanetary space and present in the Earth's atmosphere, can alter both the local properties and diagnostics of the interplanetary, magnetospheric, ionospheric and terrestrial complex plasmas. The various sources of charged dust particles and their effects on the near-Earth space weather are extensively presented in Ref. \cite{Pop11}. The remaining 10 \% of aerosols are considered anthropogenic or of human origin, as they are produced by a variety of sources. Even if less abundant than their natural counterparts, anthropogenic aerosols can prevail the air downwind of urban and industrial areas \cite{Kul11, Sande14, Tom17, Mao14, Marin19}. 

Aerosols play a prominent role in a wide range of scientific areas. We emphasize that aerosols are also intensively used in the pharmaceutical industry, while technological applications include spray drying or delivery of fuels for combustion. 

\subsection{Atmospheric aerosols. Sources. Classification. Physical and chemical properties} \label{atmaero}

One can distinguish between {\em primary} and {\em secondary} aerosol sources. The primary sources of aerosols are mostly of natural origin. Among them we enumerate oceans (sea salt aerosols), desert and semiarid regions, biological material, smoke resulted from burning of biological material, direct anthropogenic particle emissions such as soot, road dust, suspended particulate matter or interplanetary dust particulates. Secondary (indirect) aerosol sources, which represent the major source of particles below $1 \ \mu$m in radius, are generated by conversion of available natural and man made atmospheric trace gases into solid and liquid particles. Extraterrestrial particles are generated by meteor showers or comet debris that disintegrate on collision, get trapped in the Van Allen radiation belts and occur mostly over the Earth poles \cite{Rama18}. Atmospheric aerosols represent a mixture of solid or liquid particles suspended in air \cite{Mull08}. Similar to astrophysical dusty plasmas, atmospheric aerosols are made out of micro- and nanoparticles (NPs) \cite{Rama18, Sein16, Agran10}. 

Human industrial activity that leads to the release of anthropogenic aerosols in the atmosphere is responsible for adverse health effects, by producing haze in urban areas or even deposition of acids \cite{Kim08, Islam18, Via13, Kelly15}. Anthropogenic aerosols alter the radiation balance of the Earth (albedo) \cite{Pand95}. Although NPs represent the largest portion of ambient aerosol concentration, in fact they rarely account for a significant fraction of the total mass \cite{Sein16}. NPs can emerge from primary or secondary processes \cite{Via16}. Primary particles are generated directly at source while secondary ones occur from the condensation of gas-phase species. The main aerosol species that exist in the atmosphere include sea salt, mineral dust, sulfate, nitrate, and carbonaceous aerosols (e.g., black carbon - BC \cite{Liu18} and organic carbon) \cite{Rama18}.

Different micron sized organisms immersed in air are called airborne particles or bioaerosols, a category which mainly includes pathogenic and non-pathogenic, live or dead fungi and bacteria, viruses, allergens, pollen, etc., most of them held responsible for the increasing incidence of health issues of human beings and other living animals \cite{Heo17, Srik08, Kim18, Nowo16}. Unfortunately, pathogenic bioaerosols are also used as biological weapons, such as the bio-terrorist attacks that took place in 2001 using {\em Bacillus antracis} spores. Together with the pandemic outbreaks of (a) flue due to the influenza AH1N1 virus in 2009, or (b) due to the severe acute respiratory syndrome coronavirus 2 (SARS-CoV-2) in 2020, these events represent strong arguments that emphasize the significance of bioaerosol research \cite{Kim08, Ghosh15}. This is a matter of high importance as bioaerosols are considered to represent future weapons of mass destruction (WMD). Currently, there is a strong interest to investigate and explain the mechanisms responsible for the occurrence and transport of bioaerosol, with an aim to identify the associated health risks for population and limit the exposure. In-depth details on indoor bioaerosol sampling and dedicated analytical techniques are found in Refs. \cite{Ghosh15, Fuji17}, where it is demonstrated that apart from bioaerosol recognition future development of their control mechanisms represents a primary concern \cite{Mull08, Nowo16}. 

The complex picture associated to an aerosol, which is the outcome of the heterogeneity in the particle size, chemical composition, phase and mixing state, is furthermore complicated by the inherently rapid coupling with the surrounding gas phase \cite{Wills09}.

\subsection{Global climate and environment protection. Aerosols and quality of life}\label{globcli}

The direct relationship between the presence of aerosols in the atmosphere and factors such as the quality of life raises high scientific challenges, which renders chemical analysis of aerosol particles an issue of great interest \cite{Kul11, Scha12}. Ocean-derived microbes in sea spray aerosol (SSA) possess the ability to affect global climate and weather by acting as ice nucleating particles in clouds \cite{Sult17, Smirn17}. As the atmosphere is characterized by a large variety of organic compounds with complex structure and in many cases with intrinsic fragility, the study of atmospheric chemistry and global climate evolution uses different tools and methods in order to achieve such goal \cite{Rama18, Tom17, Bou15, Box15}. Many questions are still left unanswered together with the accompanying mechanisms that account for these effects. For example, the processes responsible for the formation of secondary organic aerosol (SOA) components \cite{Vol21} are not yet explained. Different types of particles can clump together to form hybrids that are difficult to discriminate. Changes in humidity or temperature may result in drastic changes in how certain aerosol species behave and interact with cloud droplets. Hence, there is a well-established interest towards developing new sensitive mass spectrometric methods with soft ionization techniques that can help in identifying such components among atmospheric particles, while also characterizing the underlying mechanisms. The main objective lies in building new tools for laboratories and ground-based aerosol monitoring stations, able to supply relevant data that are missing or are rendered incomplete using methods other than ion trap MS. 

\begin{figure}[bth]
	\centering
	\includegraphics[scale=0.5]{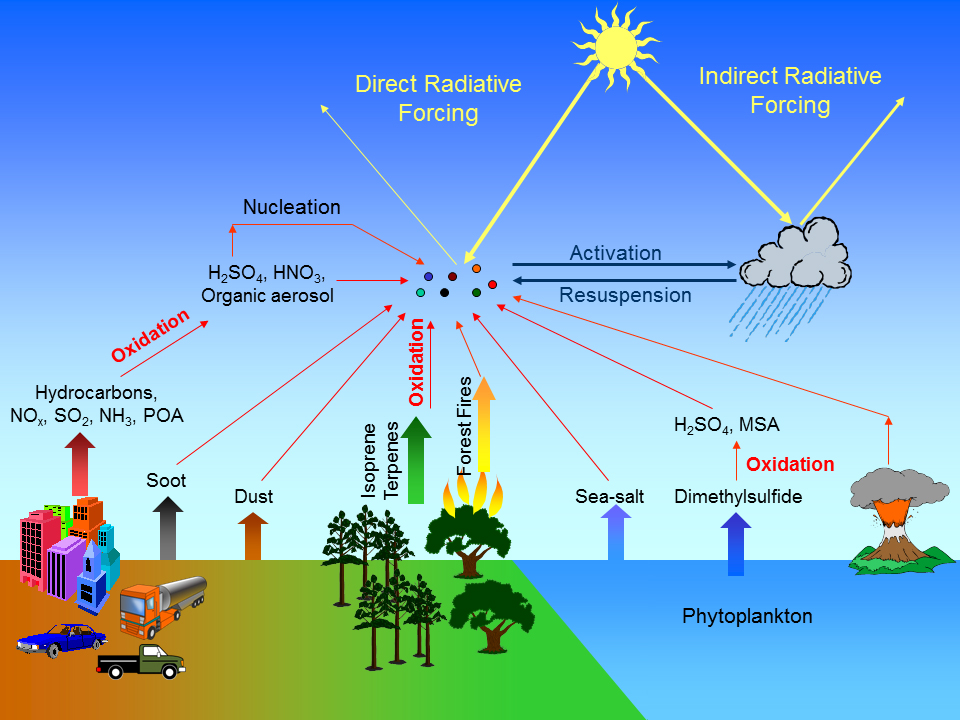} 
	\caption{Natural and anthropogenic aerosol sources. Aerosol particles alter global climate both individually and by performing as cloud condensation nuclei (CCN), thus influencing the energetic balance of the Earth (or albedo). Thus, gathering of accurate data to help understand the physical and chemical mechanisms characteristic to aerosols is a mandatory step to explain their effects on global climate. Picture reproduced from \cite{Kelly16}.}
	\label{Brook}
\end{figure}

Late studies demonstrate that presence of atmospheric aerosol particles in the atmosphere has significant direct and indirect effects on air quality, global climate and hydrological processes \cite{Kim08, Tom17, Smirn17, Kelly16}, things related to the phenomenon of global warming and the quality of life \cite{Nowo16, Bou15, Box15}. The impact produced by aerosol particles is the outcome of their physical and chemical properties \cite{Davis02, Davis97, Kond06, Mao14}. Aerosols are presumed to have a larger impact on climate compared to greenhouse gases \cite{Smirn17}, according to the Intergovernmental Panel on Climate Change (IPCC) Working Group (WG) I 4-th and 5-th Assessment Report (AR4 \& AR5) released in 2007 and 2013, respectively \cite{IPCC6}. Nevertheless, there still is a high uncertainty about it owing to the aerosol complex composition and a still incomplete picture needed to characterize the interactions between aerosols and global climate \cite{Kul11, Sein16, Kim08, Mao14, Kelly16}. We distinguish between two types of interaction: direct and indirect interactions. By indirect effect hydrophilic aerosols act as cloud condensation nuclei (CCN) affecting cloud cover and implicitly the radiation balance. Direct interactions account for the light scattering mechanism \cite{Berg17} on aerosols, resulting in cooling effects. On the other hand, aerosols containing black carbon (BC) \cite{Liou11, Liu19} or other substances absorb incoming light thus heating the atmosphere \cite{Kul11, Rama18}. According to the measurements performed, the direct radiative effect of BC would be the second-most important contributor to global warming after absorption by carbon dioxide (CO$_2$) \cite{Liu19}. In addition to scattering or absorbing radiation, aerosols can alter the reflectivity or albedo of the planet. The impact of aerosols and the associated interactions are shown in Fig.~\ref{Brook}.

In addition, nonspherical particles have a major impact on the Earth atmosphere as they influence processes such as radiative forcing \cite{Rama18, Liou11, Lee16}, photochemistry, new particle formation and phase transitions \cite{David18}. The scientific community is highly interested in explaining the physical properties and chemical composition of submicron aerosol particles, in an effort to better characterize radiative forcing \cite{Liou11, Lee16} and air quality \cite{Kul11, Sein16, Cana07, Kim08, Tom17, Kulma11}. Organic Aerosol (OA) represents the prominent fraction of non-refractory submicron particle mass (between 18 and 70 \%), while in tropical forest regions OA stands for almost 90 \% of the total fine aerosol mass. Thus, OA is the most abundant but least characterized fraction of atmospheric aerosol particles, as an outcome of its highly complex chemical composition \cite{Vogel13a, Marmu20}.   

There is a large interest towards minimizing the uncertainties associated with data collection when evaluating the impact of aerosols on global climate \cite{Lebe12, Islam18, Agran10, Lee16, Kirch08}. Different approaches and methods have been developed to analyze particles ranging from 10 nm to $10 \ \mu$m in diameter size, which consist of salts, soot, crustal matter, metals and organic molecules, often mixed together \cite{Nash06, Wang06}. Late investigations show that the largest uncertainty in radiative forcing of climate stems from the interaction of aerosols with clouds \cite{Rama18, Nowo16, Kelly16}. Thus, accurate characterization of aerosols and of their intrinsic properties is a mandatory step towards explaining many important processes in the atmosphere, while also characterizing the energy balance of the Earth \cite{Sein16, Tom17, Bou15}. The incoming solar radiation is modified when it passes through the atmosphere by two fundamental processes: light scattering and light absorption \cite{Liou11}. One can frequently find the term attenuation or light extinction, which includes both processes. The attenuation of light by these processes results in important climatic consequences. Aerosol particles both absorb and scatter light, with the efficiency of the processes being highly dependent on their size and shape, chemical composition, morphology and the wavelength of the incident radiation. Aerosol dimensions vary between $0.001 \div 100 \ \mu$m, but the maximum in scattering efficiency is found for aerosol particles with dimensions ranging from $0.1 \ \mu$m to $1 \ \mu$m \cite{Rama18}.

The European Space Agency (ESA) considers that Earth observation will result in {\em a reliable assessment of the global impact of human activity and the likely future extent of climate change} \cite{Earth}. Radiative forcing can not be estimated with precision at current time, that is why reliable assessment depends on the development of global models for aerosols and clouds that are well backed up by observations \cite{Kim08, Islam18, Lee16}. The ESA Living Planet Programme includes the EarthCARE (Earth Clouds, Aerosol and Radiation Explorer) mission which investigates the aerosols' interaction with clouds. The primary target of the EarthCARE mission (scheduled for launch in 2021) is to advance knowledge {\em of the role that clouds and aerosols play in reflecting incident solar radiation back into space and trapping infrared radiation emitted from Earth’s surface} \cite{Earth}. An enhanced understanding and better modelling of the relationship between clouds, aerosols and radiation, is considered to be a vital priority in climate research, weather prediction and quality of life assurance. The mission will employ state of the art high-performance lidar and radar technology to perform global observations of the vertical structure of clouds and aerosols \cite{Islam18}, while in the same time carrying out radiation measurements \cite{Kim08, Mull08, Hal18, Ans05, Come13, Come17}.  

The field application of an aerosol concentrator in parallel with the use of an atmospheric pressure chemical ionization ion trap mass spectrometer (APCI-IT-MS) represents a state of the art technique used to characterize aerosol particles, with an emphasize on OA \cite{Vogel13a}. Field measurements showing that atmospheric secondary organic aerosol (SOA) particles can be present in a highly viscous, glassy state, have triggered intensive investigations that approach the issue of low diffusivity of water in glassy aerosols \cite{Bast17}. Moreover, water diffusion experiments on highly viscous single aerosol particles levitated in an EDB demonstrate a typical shift behaviour of the Mie scattering resonances \cite{Bast18}. This represents a clear indicator on the changing radial structure of the particle which provides an experimental method and approach aimed at monitoring the diffusion process inside the particle.

Bioaerosols largely influence the Earth system, especially as an outcome of the interplay between the atmosphere, biosphere, climate and public health. As shown in Section \ref{atmaero}, airborne bacteria, fungal spores, pollen and other bioparticles, can induce or aggravate human, animal and plant diseases \cite{Nowo16}. In addition, bioaerosols may serve as nuclei for cloud droplets, ice crystals and precipitations, thus strongly influencing climate and rainfalls \cite{Tom17}. 

State-of-the-art experiments performed in the CLOUD chamber at CERN Geneva demonstrate that below $+5 \degree$C, nitric acid and ammonia vapours can condense upon freshly nucleated particles of about few nanometres diameter \cite{Wang20b}. In addition, at temperatures below $-15 \degree$C nitric acid and ammonia can nucleate directly through an acid-base stabilization mechanism which yields ammonium nitrate particles. As these vapours are quite abundant, the resulting particle growth rates can easily exceed 100 nanometres per hour due to gas-phase supersaturations that are expected to appear especially during wintertime in inhomogeneous urban settings.       

\subsection{Hazardous effects of aerosols and nanoparticles on humans. Fine particulate matter and coarse particle pollution}\label{hazard}

Late studies aimed towards assessing the impact of airborne NP exposure on human health result in extensive research activity directed towards better characterizing these particles and understanding which properties are most prominent in the context of health effects \cite{Kelly15}. Therefore, comprehension of the sources, chemical composition, physical structure and ambient concentrations of NPs has significantly enhanced in the last decade \cite{Sande14, Islam18, Via13, Via16}.

Airborne particles with dimensions $< 100$ nm are called ultrafine particles or NPs. Ultrafine aerosol concentrations in urban areas strongly depend on time and physical location \cite{Bzdek12}. Aerosol particles with diameter less than 10 $\mu$m enter the pulmonary bronchi, while those whose diameter is lower than 2.5 $\mu$m reach the pulmonary alveoli (a region where gas exchange takes place) and from here they are carried into blood \cite{Srik08}. Aerosols and NPs are known to inflict harmful effects on humans. There is a strong effort towards limiting the maximum concentration and enforcing safety levels for atmospheric aerosol micro and nanoparticles (including dust), most hazardous to human health \cite{Heo17, Kim18, Fuji17}. One can distinguish between two main classes: (i) fine particles with a diameter less than 2.5 $\mu$m (also called fine particulate matter or PM$_{2.5}$) which are the most dangerous, and (ii) larger particles with a diameter less than 10 $\mu$m but larger than 2.5 $\mu$m, namely the particulate matter PM$_{10}$ (also called coarse particles). In some fields of engineering PM$_{2.5}$ and PM$_{10}$ particles are called NPs \cite{Via13, Via16}. Important steps are being taken worldwide to limit pollution levels caused by PM$_{2.5}$ and PM$_{10}$ particles, in order to minimise their harmful effects on humans and biological tissue. In the EU and North America directives are enforced which regulate the PM$_{2.5}$ and PM$_{10}$ levels. EU Member States are compelled to set up {\em sampling points} in urban and rural areas. Besides particulate matter these sampling points must perform measurements on the concentration of sulphur dioxide, nitrogen dioxide and oxides of nitrogen, lead, benzene and carbon monoxide. Measurements are also performed to assess the impact of street characteristics and traffic factors, in order to evaluate the exposure of people to BC \cite{Dons13, Willi16}.

Strong evidence indicates that breathing in PM$_{2.5}$ over the course of hours to days (short-term exposure) and months to years (long-term exposure) might cause serious public health effects that include premature death, adverse cardiovascular and respiratory effects, or even harmful developmental and reproductive effects \cite{EPA21}. Lung cancer seems to be associated with the emission of NPs produced by diesel engines \cite{Kirch08}. Scientific data also indicates that breathing in larger sizes of particulate matter (coarse particles or PM$_{10}$) may also have public health consequences \cite{Kelly15}. In addition, particle pollution degrades public welfare by producing haze in cities or by constantly increasing the rate of allergies for population living in urban areas. People with obesity or diabetes are more vulnerable to increased risk of PM - related health effects. This is why constant monitoring of the various polluting agents is a major concern for health services and life quality assurance in most countries throughout the world \cite{EPA21, IRSST}. Aerosol characteristics exhibit large fluctuations in time as the anthropogenic part is quite substantial in urban areas, causing incidence of associated health effects to be sensibly higher. This is an issue of utmost interest as the largest fraction of world population is located around urban areas, and it is therefore severely exposed to degraded and degrading air quality effects. Current ambient air quality monitoring network is solely based on fixed monitoring sites, which does not always reflect the exposure impact and the associated effects on humans \cite{Srik08, Kim18, Tom17, Nowo16, Bou15}.

Investigations demonstrate that breathable aerosols with dimensions ranging between $1 \div 10 \ \mu$m are mostly dangerous for human health, as they are responsible for respiratory (allergic asthma and rhinitis, airway inflammation) and infectious human diseases \cite{Ghosh15}. Regular or ordinary human activities generate bio-aerosols. Airborne bio-aerosols are carried by air flows which means they can be inhaled or attached to human bodies \cite{Heo17, Kim18}. A mandatory requirement to enforce occupational safety and public health purposes lies in monitoring and controlling bio-aerosol concentration in the air.    

Recent research indicates that NPs are also associated with toxic effects on humans \cite{IRSST} as they are widely used by the cosmetics industry. Many sunscreens contain NPs of zinc oxide or titanium dioxide. There are manufacturers that have added C60 fullerenes into anti-aging creams because these particles can act as antioxidants. Strong evidence suggests that normally inert materials can become toxic and damaging when they are nano-sized. Evidence collected indicates that NP effects on human (living) tissue are extremely dangerous \cite{IRSST}. As many metals are toxic, the metallic content of the nanoparticulate burden represents an important factor to consider when attempting to assess the impact of NP exposure on human health \cite{Sande14}. Special care must be taken for people occupied in this area or exposed to nanomaterials, which strongly motivates the need to further analyze and characterize nanomaterial systems \cite{Wang06}. NPs can be classified into different classes based on their properties, shapes, or sizes. The main categories include fullerenes, metal NPs, ceramic NPs and polymeric NPs. NPs possess unique physical and chemical properties due to their high surface area and nanoscale size \cite{Khan17}.

\subsection{Investigation methods. Quantitative uncertainties. Mitigation}\label{InvestMeth}

Laboratory techniques aimed at investigating aerosol properties and associated physico-chemical processes usually involve the study of single particles, particle ensembles, or they collect aerosol properties out of bulk phase studies \cite{Kwa11, Pin11}. Techniques aimed at investigating single aerosol particles help in better characterizing the sources, the chemistry and the evolution of atmospheric aerosols. Single-particle techniques provide for: (1) measurement on the composition of single particles of roughly one picogram of mass even if they occur in very low concentration, (2) rapid measurement of temporal and spatial fluctuations of specific aerosol species, especially those that exhibit low concentration, and (3) information on the morphology and internal structure of single particles \cite{Pin11}. The techniques that are used to characterize single particles in atmospheric aerosols are: (a) laser-induced-fluorescence (LIF), (b) aerosol laser-ablation mass spectroscopy, and (c) two-dimensional (2D) angular elastic scattering.

On the other hand, quantitative uncertainties in the amount of aerosols and especially aerosol properties need to be mitigated. It is obvious that only more accurate measurements combined with complex computer modelling will provide the critical information that scientists need, with an aim to fully integrate aerosol impact into climate models and thus minimize uncertainty about climate evolution in the future. Aerosol MS has proven to be a very promising and versatile technique to perform diagnosis of such aerosol components \cite{Davis10, Kul11, Sein16, Sue99}. Aerosol mass spectrometers (AMS) generally use time-of-flight (ToF) mass spectrometers or linear quadrupole mass filters (QMF). Recent investigations demonstrate that an ion trap represents a very powerful instrument to perform chemical analysis of aerosol particles. 

LIDAR techniques provide important information regarding the particle effective radius for anthropogenic aerosols, in case of forest fire smoke or for mixtures of anthropogenic aerosols and biomass burning generated aerosols \cite{Hal18, Ans05, Come13}. Moreover, multiwavelength Raman lidar observations performed in Europe and worldwide supply important data regarding natural sources of pollution and human produced pollution in the free troposphere \cite{Mull08}.

The effect of aerosols is considered to be the major cause of uncertainty in the global climate radiation balance calculations. Aerosol extinction rests on the wavelength, size, concentration, composition, and to a lesser extent on the aerosol shape. Consequently, new methods and techniques are required to characterize and model these quantities. The size distribution of larger aerosols can be monitored by means of a multistatic lidar, at least in the spherical approximation. Aerosols that are small compared to the incident wavelength exhibit a Rayleigh-like scattering dependence as their size cannot be estimated using multistatic lidar techniques. Raman lidar measurements, Mie models of extinction and backscatter Angstrom ratios, demonstrate that small aerosols bring a relevant contribution to optical scattering, while also indicating that size information can be extracted from lidar gathered data.

Measurements performed on particle ensembles that represent selected narrow size ranges of accumulation mode particles ($1 \ \mu$m diameter), generally produce a rapid assessment of a statistically averaged property. This approach allows scientists to explore aerosol optical properties, chemical aging and phase behaviour \cite{Kul11, Wills09,  Bou15, Krieg12, Guan11}. 

The Earth atmosphere consists of absorbing aerosol particles, ranging from high-absorbing BC to less-absorbing mineral dust and brown carbon \cite{Liu18, Scha12, Liu19}. The absorption of aerosol particles \cite{Liou11} has a major impact on climate, while it also poses a high risk to public health \cite{Tom17, Bou15, Box15}. This explains the primary concern towards monitoring and characterizing the regional and global distribution of aerosol particulates. Characterization of temporal and spatial distribution of absorbing aerosol particles demands a complex and varied approach. Remote sensing using laser and radar ground stations along with satellite monitoring represents a mandatory approach to achieve extended spatial and temporal coverage. In situ airborne and surface measurements bring important supplementary data. To these, we add single-particle measurements that provide critical insights that would be lost when performing ensemble measurements \cite{Kul11, Tom17, Wills09, Bou15, Walt19}.

Cloud physics research has been long time performed in the laboratory using a wide range of techniques. Experimental studies of cloud particles (haze droplets, cloud droplets, ice crystals, etc.) may be classified in two main categories: (a) those that require large populations of particles such as cloud chambers \cite{Hagen89, Song94b}, and (b) methods that isolate individual particles \cite{Davis02, Davis97, Davis90, Shaw00} such as EDB \cite{Hart92, Lamb96}, optical tweezers \cite{Gies15, Wills09}, acoustic levitation \cite{Sto11, Andra18}, etc. Single-particle studies avoid population effects by containing the particles in a very well defined region of space under conditions of minimal perturbation \cite{Stra55, Wuerk59, Wint91, Ghe95a, Ghe98, Shaw00}. 

\begin{figure}[!ht] 
	\begin{center}
		\includegraphics[scale=1.5]{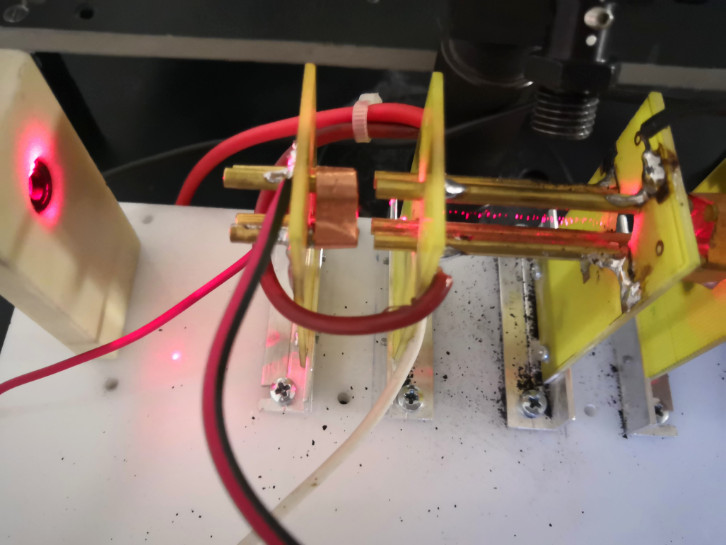}
		\caption{Levitation of solid aerosol particles of humic acid in a segmented Paul trap geometry under test at INFLPR. The trap operates under SATP conditions. Three independent sections of the trap can be observed in the images. Above the central section of the trap lies the collimation lens, located at one of the ends of an optical fibre that picks up the fluorescence signal and delivers it to a spectrometer. Work in progress.}
		\label{SegmTrapINFLPR}
	\end{center}
\end{figure}

In Fig.~\ref{SegmTrapINFLPR} we supply an image of a segmented quadrupole linear Paul trap under test in INFLPR, used to levitate anthropogenic aerosols, in our case humic acid or ammonium sulphate (NH$_4$)$_2$SO$_4$. Experiments are in progress and different trap geometries are currently under test, as well as a new microsystem based programmable voltage source that delivers three d.c. voltages ($0 \div 1000$ V) and an a.c. voltage source that delivers a 3.5 kV peak voltage at a variable frequency.  

Experiments demonstrate that a time series of optical resonance spectra of an evaporating, non-spherical, irregular aerosol particle levitated in an EDB, exhibits patterns which are associated to its evaporation kinetics \cite{Zardi09}. Simulated spectra of an evaporating, model aerosol particle, exhibit comparable features. These patterns can be used to characterize particle size variation with time. The complex physical and chemical mechanisms that govern the size, composition, phase, and morphology of aerosol particles in the atmosphere, represent a major challenge for scientists trying to characterize and model them. Measurements performed on single aerosol particles ($2 \div 100 \ \mu$m in diameter) levitated in electrodynamic, optical and acoustic traps, allow one to investigate individual processes under minimal perturbation and well controlled laboratory conditions \cite{Krieg12}. In addition, particle size measurements can now be performed with unprecedented accuracy (sub-nanometre) over a wide range of timescales. Hence, one can precisely identify the physical state of a particle while its chemical composition and phase can be determined under conditions of high spatial resolution.

Femtosecond spectroscopy opens new perspectives in bioaerosol sensing. On one side, the induced nonlinear light emission from the particles is remarkably peaked in the backward direction which is favourable for remote detection, and on the other side quantum control schemes allow discrimination from major interference such as soot and other organic non-biological particles \cite{Wolf10}. Optical trapping represents another versatile tool for investigating aerosol droplets \cite{Reich20}, where measurements techniques and methodologies include the observation of elastically-scattered light, measurement of a Raman or fluorescence fingerprint and application of conventional brightfield microscopy. An in-depth treatment of optical trapping and of the associated techniques can be found in Ref. \cite{Wills09}.   

Use of the light scattering mechanism to explore small airborne particles and reveal their intrinsic features represents a remarkable analytical tool in aerosol science. The scientific community makes intensive efforts to characterize scattering and absorption cross-sections of aerosol particles, as there is a large uncertainty in explaining the physico-chemical mechanisms responsible for the complex interaction between the atmospheric particles and solar radiation in Earth’s atmosphere \cite{Kul11, Tom17, Via16, Bou15, Box15}. The optical properties associated to an individual aerosol particle determine the amount of scattering and absorption that occurs during its interaction with electromagnetic radiation, and depend upon the size, shape, morphology, real and imaginary parts of the particle refractive index, and incident radiation wavelength. Scattering of electromagnetic radiation by particles can be divided into three main categories: (a) Rayleigh scattering, (b) Mie scattering, and (c) optical scattering. We can distinguish among the three regimes as the wavelength of the radiation is: (a) much larger, (b) of the same order, or (c) much smaller with respect to the particle size. The Rayleigh and optical scattering can be regarded as approximate solutions to the scattering problem, while the Lorenz-Mie theory supplies an analytical solution to investigate scattering of an electromagnetic plane wave by an isotropic, homogeneous sphere \cite{Bain18, Mie1908, Lock09}.   

The Lorenz-Mie theory is valid for all three scattering regimes, but its application is often restrained due to its high demands on computational time and the friendly features of the approximate models that are used to investigate Rayleigh or geometric regimes \cite{Wrie09}. A Lorenz-Mie calculation is based on the evaluation of the scattering coefficients. Ref. \cite{Bain18} presents an optical trapping setup intended for accurate studies of the physico-chemical processes, which enables performing single particle measurements. The paper demonstrates that Mie resonances can be used to determine the size and refractive index of optically trapped particles. State of the art results are focused on methods to infer the refractive index of aerosol particles from measured optical properties using refractive index retrievals (also known as inverse Mie methods). As stated in \cite{Rad18}, retrievals of the aerosol refractive index are based on two fundamental methods: (i) measurements of the extinction, absorption and/or scattering cross-sections or efficiencies of size- (and mass-) selected particles, for mass-mobility refractive index retrievals (MM-RIR), and (ii) measurements of the aerosol size distributions plus a combination of the extinction, absorption and/or scattering coefficients for full distribution refractive index retrievals (FD-RIR). The method is demonstrated for the study of pure aerosols or mixtures of ammonium sulfate and nigrosin aerosol \cite{Rad18}.

To summarize, we emphasize that the wide variety of atmospheric aerosol chemical composition and physico-chemical properties requires single-particle analysis in order to fully characterize the chemical evolution of particles and their effects on environment, clouds, global climate and health. Late progress and advances with respect to state-of-the-art techniques for individual particle analysis are presented in \cite{Sulli18}. The approach focuses on "online single-particle mass spectrometry, experiments performed with levitated isolated particles using an EDB or aerosol optical tweezers, and the use of electron, X-ray, and Raman spectromicroscopies for detailed analysis and chemical mapping of particles collected on substrates."

\subsection{High precision mass measurements for aerosols and nanoparticles}

Mass spectrometry (MS) provides high sensitivity coupled with a fast response time to probe chemically complex particles \cite{March10, March17a, March95c}. Off-line MS techniques require sample collection on filters but can provide detailed molecular speciation. In particular, off-line MS techniques using tandem MS experiments and high resolution mass analyzers provide improved insight into SOA formation and heterogeneous reaction pathways \cite{Pratt12a, Pratt12b}. MS techniques are used to perform real‐time measurements and thus acquire information about aerosol properties. The Aerosol Mass Spectrometer (AMS) uses aerodynamic lens inlet technology together with thermal vaporization and electron‐impact (EI) MS, to measure real‐time non‐refractory (NR) chemical speciation and mass loading as a function of the size of fine aerosol particles (with aerodynamic diameters ranging between $\sim 50 \ \textrm{and} \ 1,000$ nm). The original AMS is based on a QMS with EI ionization. A review of aerosol mass spectrometry is performed in Ref. \cite{Nash06}. More recent versions employ ToF mass spectrometers and produce full mass spectral data for single particles \cite{Cana07}. AMS are used to perform {\em in situ} measurements on gas vehicle emission and aerosol phase composition \cite{Chiri11}. 

A nanoaerosol mass spectrometer (NAMS) employed for real-time characterization of individual airborne NPs is described in Ref. \cite{Wang06}. The NAMS includes an aerodynamic inlet, a quadrupole ion guide, a quadrupole ion trap, and a ToF mass analyzer. Charged particles in the aerosol are drawn through the aerodynamic inlet, then focused through the ion guide and finally captured in the ion trap. Trapped particles are irradiated with a high-energy laser pulse to reach the “complete ionization limit”, where each particle is considered to be completely disintegrated into atomic ions. Within this limit the relative signal intensities of atomic ions supply the atomic composition. The method is firstly demonstrated with sucrose particles produced with an electrospray generator. A method to deconvolute overlapping multiply charged ions (e.g., C$^{3+}$ and O$^{4+}$) is outlined.

A novel Ion Trap Aerosol Mass Spectrometer (IT-AMS) for atmospheric particles is described in \cite{Kurt07}. Using this instrument the chemical composition of non-refractory component of aerosol particles can be measured quantitatively. The setup makes use of the well-characterized inlet and vaporization/ionization system of the Aerodyne Aerosol Mass Spectrometer (AMS). While the AMS uses either a linear quadrupole mass filter (Q-AMS) or a time-of-flight mass spectrometer (ToF-AMS) as the mass analyzer, the IT-AMS is based on a 3D QIT. The main advantages of an ion trap are the possibility to perform  MS$^n$-experiments along with ion/molecule reaction studies. The IT-AMS operates under full PC control and can be used as a field instrument due to its compact size. Experiments show that a mass resolving power larger than 1500 can be reached. This value is high enough to separate different organic species at $ M/Q \sim 43$. Calibrations performed with laboratory-generated aerosol particles indicate a linear relationship between signal response and aerosol mass concentration. These studies, together with estimates of the detection limits for particulate sulfate (0.65 $\mu$g/m$^3$) and nitrate (0.16 $\mu$g/m$^3$), demonstrate suitability of the IT-AMS to characterize atmospheric aerosol particles. An inter-comparison between an IT-AMS and a Q-AMS for nitrate in urban air yields a good agreement \cite{Kurt07}. 

Recent development and optimisation of an IT-AMS are reviewed in \cite{Fach17}, which leads to a more reproducible and robust mode of operation that enables the first field deployment of the instrument. Experimental results indicate that such setup is able to supply enhanced quantitative information on organics, nitrate and sulfate mass concentrations, while also providing reasonable detection limits ($ 0.5\div 1.4 \ \mu\textrm{g}\cdot \textrm{m}^{-3}$ for 1 h averages). Moreover, results obtained with an IT-AMS can directly be related to measurements performed by means of Aerodyne AMS.

\subsection{Aerosol mass spectrometry (AMS) investigations for organic and inorganic aerosols}

Particulate matter (PM) emissions from various sources represent a major source of concern that contribute to the degradation of air quality especially in urban environments, but not restricted to. Adverse effects of PM$_{10}$ and PM$_{2.5}$ particles on human health are discussed in Section \ref{hazard}. One of the methods used to evaluate their impact and determine the exposure level is based on using Optical Remote Sensing (ORS) techniques. The principle of the method relies on using path-integrated multi-spectral light extinction measurements in a vertical plane, by ORS instruments downwind of a passing PM source. The light extinction measurements assist in extracting path-averaged PM$_{2.5}$ and PM$_{10}$ mass concentrations, using inversion of the Mie extinction efficiency matrix for a wide range of size parameters \cite{Kim08}. Light extinction can be computed using the Lorenz-Mie theory \cite{Herg12} provided that one knows the optical properties and size distribution of PM. If the PM is non-absorbing then light extinction is the outcome of scattering alone. Hence, the refractive index will exhibit a non-zero real part and a zero imaginary part. The extinction efficiency depends on the particle diameter $d$ and electromagnetic radiation wavelength $\lambda$. In case of a broad spectral band, the extinction efficiency must be calculated for each combination of the particle size and the corresponding wavelength. Mass aerosol concentrations can be determined by means of the ORS method \cite{Kim08}.

Generally, properties of boundary-layer aerosols associated with local and regional emissions of particles and gases are very different with respect to those of free-tropospheric particles. Climate models show that particles persist in the boundary layer over the continents and they are precipitated within $2 - 4$ days, around a radius of up to $2000$ km with respect to their origin. Instead, free-tropospheric particles are often carried over large distances across different continents where the transport times are usually longer than $1$ week. These particles play a prominent role in the cloud formation processes, a feature that makes them important agents with respect to the aerosol indirect effect \cite{Kim08}. Multiwavelength Raman lidars have been used lately to characterize geometrical, optical, and microphysical properties of free-tropospheric pollution. Different aerosol types are monitored such as anthropogenic pollution, forest-fire smoke particles from North America and Siberia, arctic area pollution, and the drift of dust particles coming from the Sahara desert \cite{Kim08}. For a quite comprehensive overview of light detection and ranging (lidar) techniques aimed at characterizing desert dust one should refer to Ref. \cite{Mona12}. Ref. \cite{Come17} approaches fundamentals, technology, methodologies and state-of-the art characteristics of lidar systems used to retrieve aerosol information. Late results and directions of research can also be found in Ref. \cite{Hal18}. Recent advances and "state of the art science in techniques for individual particle analysis" are presented in Ref. \cite{Sulli18}.

\subsection{Analytical and numerical modelling of the problems of interest with respect to aerosol dynamics and elaboration of prediction models}

Detailed information about the extent and distribution of stratospheric aerosols is necessary to perform climate modelling or to validate aerosol microphysics models and investigate geoengineering. In addition, stratospheric aerosol loading has to be determined with sufficient precision so as to enhance the retrieval accuracy of key trace gases (e.g. ozone or water vapour) when explaining remote sensing measurements of scattered solar radiation. The most frequently used markers to characterize stratospheric aerosols are the aerosol extinction coefficient and the {\AA}ngstr{\"o}m coefficient. It is considered that a best approach is based on the use of particle size distribution parameters together with aerosol number density. Ref. \cite{Mali18} suggests a new retrieval algorithm to infer the particle size distribution of stratospheric aerosol from space-borne observations of scattered solar radiation in the limb-viewing geometry.

\subsection{Need for a complete system covering both non-refractory and refractory particles found in the atmosphere}

Single particle mass spectrometers (SPMSs) represent versatile instruments employed to infer the chemical composition and mixing state of aerosol particles in the atmosphere, under ever changing environmental conditions. Moreover, SMPSs can deliver high temporal resolution without prior sample preparation and produce {\em in situ} single particle structure data. SMPSs have demonstrated excellent abilities in detecting microbes and late experiments demonstrate that aerosol ToF mass spectrometers are able to identify aerosolized microbes in ambient sea spray aerosol \cite{Sult17}.

Ref. \cite{Sage18} demonstrates how mass spectrometry can be achieved with arrays of 20 multiplexed nanomechanical resonators, where each resonator is designed with a distinct resonance frequency which becomes its individual address. The paper also reports on mass spectra of metallic aggregates in the MDA range, characterized by more than one order of magnitude enhancement in the analysis time with respect to individual resonators. A 20 Nanoelectromechanical Systems (NEMS) array is probed in $150$ ms, under conditions of the same mass limit of detection as a single resonator \cite{Sage18}. Excellent agreement is reported with respect to a conventional ToF spectrometer operated in the same system.  

A laser ablation aerosol particle ToF mass spectrometer (LAAPTOF) used to investigate single aerosol particles is described in \cite{Shen19}. During the experiments seven major particle classes are identified. As an outcome of the precise particle identification and well-characterized overall detection efficiency (ODE) characteristic to this instrument, particle similitude can be conveyed into corrected number and mass fractions without requiring a reference instrument. An aerosol mass spectrometer (AMS) is also used, where the two MS exhibit good correlation regarding total mass for more than 85 \% of the measurement time, which represents a trace that non-refractory species measured by AMS may originate from particles consisting of internally mixed non-refractory and refractory components. The paper also reports on finding specific relationships between LAAPTOF ion intensities and AMS mass concentrations, for non-refractory compounds and particular measurement periods \cite{Shen19}. The approach used enables assignment of non-refractory compounds (measured by AMS) to different particle classes.

\section{Single particles confined in electrodynamic traps. Methods of investigation}\label{Sec9}

\subsection{Light scattering mechanisms}

Investigation of the light scattering mechanisms by small random particles did raise an enormous interest for science since the 1950s. Among the most prominent work in the domain we mention the books of van der Hulst \cite{Hulst81} and Bohren \cite{Bohr98}, to which we add Refs. \cite{Kul11, Rama18, Sein16, Tom17, Agran10, Bou15, Box15, Kokha08}. We distinguish between (a) elastic scattering where the photon energy does not change, and (b) inelastic scattering when both the photon energy and inner energy of the scattering particle modify. We can further classify elastic scattering as:

\begin{enumerate}
	\item Rayleigh scattering $\longmapsto$ photon scattering from small, uncharged but polarizable particles (such as atoms or molecules), whose dimensions are considerable smaller than the radiation wavelength. Polarization of scattering particles makes the scattered radiation to be emitted uniformly in all directions. Rayleigh scattering exhibits wavelength dependence where shorter wavelengths are scattered more than higher ones, 
	\item Mie or Debye scattering $\longmapsto$ photon scattering from relatively large particles or molecules with dimensions comparable or larger than the incident radiation wavelength. The resulting radiation is non-uniformly scattered and shows little wavelength dependence. Forward Mie scattering is stronger than backward scattering because the relative phase differences of contributions from different scattering locations on the particles dwindle,  
	\item Thomson scattering $\longmapsto$ photon scattering on charged particles.
\end{enumerate}     

Inelastic scattering can be also classified as:
\begin{itemize}
	\item Brillouin scattering $\longmapsto$ a mechanism which typically occurs in photon scattering from solid materials. It involves acoustical phonons with characteristic frequencies in the GHz region. The incident radiation wavelength is altered by the energy levels of sound waves or phonons in the solid material, but these shifts are quite small,
	\item Raman scattering $\longmapsto$ a mechanism where the frequency of the scattered radiation changes. It is used to perform diagnostic analysis. Raman scattering is typically weak and considerably less intense than the Rayleigh scattered light, which explains the special care taken in the design and operation of Raman spectrometers. In case of Raman scattering at gas molecules, the vibration and rotation states of the molecules change; as a general rule the molecules possess a higher energy after scattering which suggests a correspondingly lower photon energy of the scattered light (Stokes components). When molecules are previously excited, anti-Stokes components with increased optical frequency are reported. Analogously, Raman scattering can arise in solids involving phonons with frequencies in the THz region, also called optical phonons.
\end{itemize}	

We can also mention Compton scattering which is basically an inelastic form of Thompson scattering that occurs when the energy of the incident radiation starts to become comparable to the rest energy of the charged particle.

\subsection{Single particles levitated in particle traps}

As already shown in Section \ref{EDB}, an electrodynamic balance (EDB) represents a device used to levitate charged particles/droplets or ions, either in vacuum or under SATP conditions. Particular trap geometries such as ring shaped \cite{Nosh14} and hyperboloidal shaped electrodes \cite{Major05, March05, Singh17, Ghe94, Davis11}, segmented \cite{Pyka14} and planar geometries \cite{Bahra19, Clark13, Song06, Aust10, Zhang11, Alda16}, or even toroidal shaped electrodes \cite{Li17, Tay12, Higgs16, Kot16, Kot17, Higgs18}, are employed to generate a quadrupole electric field. Use of EDBs has pushed forward the domain of mass spectrometry while providing chemists and physicists an invaluable precision instrument to explore microparticles or NPs, and thus investigate their physico-chemical properties \cite{Kwa11, Pin11}.  

In the last two decades a considerable effort was invested to devise new methods and techniques to investigate chemical reactions and NPs at the single molecule level. Single molecule manipulation techniques include scanning probe microscopy (SPM) such as the Atomic Force Microscope (AFM) and the Scanning Tunneling Microscope (STM) \cite{Hey11}, or near-field scanning optical microscopy (NSOM) and variations of. However, these techniques raise technological challenges with respect to the highly complex tip-sample and sample-surface interactions. One of the first attempts to circumvent the issues related to SPM based techniques consists in developing a new single molecule control method which relies on a trap setup that is largely modified with respect to a conventional 3D QIT \cite{Seo03}. The method enables {\em in situ} investigation of single molecules and NPs by enabling ultra-sensitive detection; the device is called single nanoparticle ion trap (SNIT). 

Storage devices based on inhomogeneous, time-dependent electric fields, characterized by trap geometries such as quadrupole, linear octopoles, higher-order multipoles or series of ring electrodes, are described in \cite{Ger03}. Different species of charged particles can be levitated, starting with electrons via molecular ions and NPs. Manipulation of microparticles and NPs (including molecules) in vacuum represents one of the challenges of modern science, as it enables investigations on proteins, DNA segments, viruses and bacteria \cite{Jose10, Peng04, Wang12}, dusty plasmas \cite{Fort05, Morfi09}, or detection of tiny forces \cite{Li18, Haya18}. Measurements on single aerosol particles (ranging from $2$ to $100 \ \mu$m in diameter) levitated in electrodynamic, optical and acoustic traps, or deposited on a surface, allow individual processes and physical mechanisms to be studied in isolation under controlled laboratory conditions \cite{Krieg12}.

Non-destructive, optically detected mass measurements on single microparticles and NPs using electrodynamic (Paul) traps are performed in several groups around the world. A technical limitation of these setups comes from the fact that they use different particle injection methods. The major drawback when using such methods is caused by an imprecise control over the particle mass and range, which strongly affects the reproducibility and control of the experiment. Nanoparticle Mass Spectrometry (NPMS) is a very powerful and versatile tool to perform NP diagnosis. Very recent experiments suggest a novel technique called electrospray ionization (ESI) NP source \cite{Howd14}. The method is based on using a pair of radiofrequency (RF) ion guides that provide collisional cooling and focusing of the NP beam generated by the source, while also acting as prefilters (in fact, mass filters) and sorting species as a function of the specific mass-to-charge ($Q/M$) ratio.

Trapping of single particles or colloidal molecules in a planar aqueous Paul trap (PAPT) setup is described in \cite{Park12a, Park12b} and \cite{Guan11}, with a focus on control and localization of the  motion of a charged particle in a water environment. The random thermal noise due to Brownian motion in water is investigated, as it significantly affects the trapping process. Optical trapping of an ion and the associated advantages are presented in \cite{Schne10}. Trapping is demonstrated in regions that are located in the closest vicinity of the electrodes, while indicating that hybrid setups combining both optical and RF potentials can efficiently cover the full spectrum of composite confinements - from RF to optical. Experiments demonstrate the perturbing influence of the static electric potential which can easily restrain successful optical trapping. Different analytic models have been tested, suggesting that recoil heating stands for the most prominent heating effect, which is indicated by the experimental results \cite{Schne12a}. The ion lifetime in an optical dipole trap is constrained by photon scattering. State-of-the-art experimental techniques such as optical repumping may lead to a remarkable lifetime increase by three orders of magnitude \cite{Lamb17}. By employing these techniques, optical trapping and isolation of ions can be performed on a level that is comparable to neutral atoms. Such almost decoherence-free techniques enable scientists to accomplish isolation from the environment. This achievement represents a major breakthrough that opens new pathways towards novel regimes of ultracold interactions of ions and atoms, at collision energies considered previously inaccessible. This approach enables one to perform a novel class of experimental quantum simulations with ions and atoms in a variety of versatile optical trapping geometries, such as bichromatic traps or higher-dimensional optical lattices \cite{Lamb17}.

Ref. \cite{Wieb19} reports on single micron-sized melamine-formaldehyde particles levitated in the sheath of a RF-plasma and exposed to an intense laser beam, while trapped in optical tweezers. A reversible change in the particles' properties is observed such as a gain in particle charge, where the initial charge restores within minutes. A similar experiment reports on a cold-damping scheme used to cool one mode of the CM motion of an optically levitated NP in ultrahigh vacuum ($10^{-8}$ mbar), whose temperature drops off from room temperature down to an extremely low value of 100 $\mu$K \cite{Tebb19}.  

The outcome of laser phase noise heating on resolved sideband cooling, in the context of cooling the CM motion of a levitated NP in a high-finesse cavity, is analyzed in Ref. \cite{Meyer19}. As phase noise heating does not represent a fundamental physical constraint, the paper explores the regime where it becomes the main limitation in Levitodynamics. The interest in investigating this regime is highly motivated, given that it represents the main restriction in reaching the motional ground state of levitated mesoscopic objects under conditions of resolved sideband cooling. 

The dynamical backaction effect and the possibility to cool a levitated nanosphere in the absence of a cavity, or when the nanosphere is trapped outside a cavity and excites a continuum of electromagnetic modes is explored in \cite{Abba19}, where both Stokes and anti-Stokes processes are considered. Ground state cooling of the CM motion of a levitated silica nanoparticle (NP) based on coherent scattering into an optical cavity is described in \cite{Deli20, Deli19, Gonz19}. Excellent control of the dynamics of a levitated silica nanosphere which is transferred between two optical tweezers is demonstrated in \cite{Cala21}. Cooling of all three motional modes of a charged dielectric nanoparticle in a Paul trap is demonstrated in \cite{Dania21} where two methods are investigated, one of them based on electrical forces and the other on optical forces. A weighing metrology experiment of a single silica microsphere optically trapped and immersed in air is described in Ref. \cite{Hill20}, that demonstrates three different mass measurements based on fluctuations around thermal equilibrium. A direct comparison between parametric feedback cooling and velocity damping methods for a particle confined in a Paul trap is performed in \cite{Pen21}. All these systems provide novel protocols aimed at performing cavity optomechanics experiments with levitating nanoparticles, opening new perspectives for investigation of quantum phenomena and achieving accurate sensing devices.

Another field of large scientific interest is the study of micro and nanomechanical oscillators \cite{Gies15} and especially of their associated quantum mechanical motion, as experimental results are expected to provide novel insights into the boundary between the quantum and classical worlds. Cooling of a dielectric nanoscale particle trapped in an optical cavity is investigated in \cite{Bark10}. An experimental spectrometer setup based on two separate cryogenic ion traps that enable formation and characterization of solvated ionic clusters such as water, methanol and acetone around a protonated glycylglycine peptide, is described in \cite{Marsh15}. 

A novel technique demonstrates that {\em levitated electromechanics} can be successfully employed for electronic detection, cooling, and precision sensing of (single) microparticles and NPs \cite{Gold19}. The technique uses charged particles levitated in an ion trap, coupled to an RLC circuit. Sub-Kelvin temperatures are achievable with room temperature circuitry and it is estimated that trap operation in a cryogenic environment will enable ground-state cooling for micron-sized particles. It is also suggested that hybrid levitated opto-electromechanical setups are supposed to open new pathways towards deep cooling into the quantum regime, enabling scientists to engineer highly nonlinear states such as the squeezed states \cite{Gold19}. Ref. \cite{Ricci17} explores levitated nano-mechanical resonators and especially experimental aspects related to the precise control of nonlinear and stochastic bistable dynamics of a levitated NP in high vacuum. 

Ref. \cite{Ozd19} describes a novel design of a charge detection QIT-MS intended for the analysis of micron-sized dry inorganic and bioparticles, including red blood cells (RBCs) and  MCF-7 breast cancer cells of different sizes. The method suggested promises to be very fast in determining the mass of levitated microparticles. The system performs online analysis of various micro-sized particles up to $1 \times 10^{17}$ Da.

Characterisation of a charged particle nano-oscillator in a Paul trap is performed in \cite{Bull20}. The overall frequency stability and temperature stability are estimated using the Allan deviation, whilst the oscilator mass is determined with a 3 \% uncertainty.

Thus, optomechanical systems in the well-resolved-sideband regime are extremely promising candidates to investigate and illustrate various quantum phenomena, especially regarding backaction-evading measurements, mechanical squeezing, or engineering of nonclassical states of motion \cite{Qiu20, Dodon02}.

\subsection{Investigation of chemical and optical properties of trapped particles or microdroplets using optical techniques. General considerations}

Charge detection quadrupole mass spectrometry (CD-ITMS) driven by rectangular and triangular waveforms (rect-CD-ITMS and tri-CD-ITMS) has been developed in an attempt to achieve better characterization of microparticles. Experimental results indicate that the rect-CD-ITMS and tri-CD-ITMS can operate well and perform mass measurement of microparticles by using the frequency scan. By increasing the applied voltage and the signal-to-noise ratio (S/N) of the charge detector, the mass resolution can be further improved. In addition, rect-CD-ITMS and tri-CD-ITMS can be used to characterize red blood cells (RBCs) \cite{Xio12}.

On the other hand, handling and control of single cells and large biomolecules \cite{Sato96} is a prerequisite for various medical and biological techniques, starting with in vitro fertilization to genetic engineering. To achieve such goal, different techniques and devices for micromanipulation have been devised and tested: (a) optical tweezers \cite{Gies15, Ash86, Sato94, Svobo94, Mara13, Zheng13} that provide single particle manipulation but limited trapping capacity as an outcome of the strong focusing requirement; and (b) dielectrophoresis that enables massive manipulation under conditions of insufficient spatial resolution. A promising approach is demonstrated in \cite{Flor15} that associates advantages characteristic to both methods and allows optical trapping and manipulation of microparticles suspended in water due to laser-induced convection currents. Experimental results show that for low laser power ($0.8$ mW) particles are trapped at the centre of the beam, while at higher powers $\left( \sim 3 \ \text{mW} \right)$ particles arrange themselves in a ring that encircles the beam. Such a behaviour results out of the action of two competing forces: the Stokes and the thermo-photophoretic forces. Numerical simulations establish that thermal gradients achieve trapping. We also mention Ref. \cite{Kokha08} which is a comprehensive guide that approaches subjects such as the optical properties of aerosol particles, multiple light scattering and Fourier optics of aerosol media, or optical remote sensing techniques.  

Laser trapping of NPs is demonstrated in \cite{Leh15}, with an emphasize on colloidal metal NPs that exhibit remarkable features due to their particular interaction with electromagnetic radiation, caused by surface plasmon resonance effects. An experimental method that relies on the use of dynamic split-lens configurations with an aim to achieve trapping and spatial control of microparticles through the photophoretic force is presented in \cite{Liza18}. Optical trapping allows high precision investigations of many microphysical and chemical processes as it enables measurements at single-particle level. Such unique features are demonstrated in Ref. \cite{Bain18}, where the size and refractive index of single aerosol particles are determined using angular light scattering and Mie resonances.

\subsection{Light scattering measurements in particle traps}

As a rule when studying light scattering, photoemission from a particle surface, processes of evaporation and condensation of a drop, or chemical reactions with a participation of solid particles \cite{ Major05, Davis97, Ghe95a, Shaw00}, one preliminary charged particle has to be placed in the trap. Particle confinement is achieved at low gas pressures in the trap. The dynamics of charged dust particles and their capability to build up Coulomb structures in a Paul trap at low pressures are examined in Ref. \cite{Wuerk59}. A single trapped particle moves along Lissajous curves, depending on the parameters of the applied trapping voltage. As mentioned in \cite{Wuerk59} a trap with a single confined particle could serve as an analogue computer for solving the Mathieu equation. Large numbers of particles can be levitated in a trap, but in this case particles move chaotically depending on the frequency and amplitude of the trapping voltage applied to the electrodes.

Experiments demonstrate that alumina or SiO$_2$ spheres (grains) with diameter ranging between $0.1 \div 1 \ \mu$m can be simply detected by means of light scattering. Relevant information about the $Q/M$ ratio (also called specific charge) can be inferred by performing a Fourier analysis of the recorded signal which is modulated by the particle secular motion \cite{Ger03}. Recent techniques and results with respect to measurements of elastic scattering and absorption of aerosols are presented in \cite{Ula11}. Both spectroscopy and angular measurement of the intensity can be employed to infer the size of aerosol particles. Moreover, additional properties can be retrieved from the angular dependence of scattering.

Ref. \cite{Goue15} reviews laser based optical techniques used to characterize discrete particles embedded in {\em flows}. Measuring the absorption of a single aerosol particle represents a difficult task due to the complexity associated with the problem. Among the few techniques available there is none suitable for measuring single-particle absorption of coarse-mode nonspherical aerosols. Nevertheless, analysis of two-dimensional angular optical scattering (TAOS) patterns provides a possible solution to perform such task. TAOS patterns of single aggregate particles made out of spheres are reported in \cite{Zardi09} where optical resonance spectroscopy is employed to size evaporating solid, non-spherical particles. Single bi-sphere particles levitated in an EDB can be used to record 2D angular scattering patterns at different angles of the coordinate system of the aggregate with respect to the incident laser beam \cite{Krieg11}. Numerical simulations based on the Mackowski code are also used to perform a qualitative analysis of experimental results. The Brownian motion allows one to sort out high symmetry patterns for analysis. Ref. \cite{Walt19} reports on captured TAOS patterns with geometries similar to a previously designed instrument, that are simulated by employing a Multiple-Sphere T-Matrix (MSTM) code. By discriminating the speckle size and the integrated irradiance of these simulated TAOS patterns, one can distinguish between high-absorbing, weak-absorbing and non-absorbing particles, over a size range of $2 \ \mu$m to $10 \ \mu$m. The particle size can be estimated due to the presence of speckle in the scattering pattern. The next step lies in characterizing particle absorption using the integrated irradiance.

\subsection{Elastic scattering. Lorenz-Mie theory}

Some aspects of the Lorenz-Mie scattering theory \cite{Herg12} are presented in Section \ref{InvestMeth}. Many key technologies heavily rely on NPs and the area of applications is exponentially growing, starting from pygments used in paints and cosmetics, drug delivery, chemical and biological sensing, gas sensing, CO$_2$ capture, etc. \cite{Khan17}. Beyond their important applications related to multiple high technology areas (see Section \ref{hazard}), NPs also present a vivid interest for fundamental research, especially for the areas of complex plasmas and astrophysics.  

{\em In situ} analysis of NPs can be performed by employing the method of multiple-wavelength Rayleigh–Mie scattering ellipsometry. In Ref. \cite{Gebau03} the method is applied to characterize NPs levitated in low-pressure plasmas. Experiments demonstrate that size distribution and the complex refractive index can be determined with high precision. Moreover, the Rayleigh–Mie scattering ellipsometry also applies to achieve {\em in situ} analysis of NPs under high gas pressures and in liquids.

Interpretation of elastic scattering data based on Mie theory \cite{Herg12} enables accurate  measurements of the size and refractive index of microspheres. These measurements can be subsequently employed to analyze droplet evaporation and condensation processes for pure components and multicomponent droplets \cite{Davis10}. Moreover, the Rayleigh limit of charge on a droplet can be investigated using elastic scattering in order to determine the size and charge corresponding to droplet explosions. Phase function measurements (intensity versus scattering angle) and data gathered from morphology-dependent resonances (MDRs) provide alternative methods to determine the size and refractive index of droplets and microspheres \cite{David18, Smith08, Krieg12, Krieg11}, and thus characterize coated spheres \cite{Trevi09}. 
 
A QIT calibrated for microparticle MS that is employed to levitate dye-labelled polystyrene microspheres and study the associated fluorescence spectra, is presented in \cite{Trevi07}. The absolute mass and electric charge of the particle are resolved by measuring its secular oscillation frequencies. The microsphere radius is calculated by employing the Lorenz-Mie theory \cite{Mie1908, Lock09, Herg12, Mish09} to interpret the fluorescence emission spectrum dominated by optical cavity resonances \cite{Trevi07}. The laser-induced coalescence of two conjoined polystyrene spheres levitated in a QIT is investigated by monitoring optical morphology dependent resonances (MDRs) that show up in the fluorescence emission spectrum \cite{Trevi09}. Surface tension drives the heated bisphere into an individual sphere. At the end of the structural transformation phase, particle dimensions are inferred by analyzing the frequency shifts of the non-degenerate azimuthal MDRs. The relaxation time of the viscous sphere is then used to infer the polystyrene viscosity and temperature. An overview of light scattering experiments aimed at measuring one or more elements of the scattering matrix as functions of the scattering angle, for ensembles of randomly oriented small irregular particles in air, is supplied in \cite{Munoz11}.  

Elastic light scattering shows sensitivity to aerosol particle size, shape, complex refractive index and molecular density distribution within the particle. As it provides high finesse the technique is frequently used to achieve real-time information and perform {\em in situ} classification of the aerosol type. Elastic light scattering is especially recommended in discriminating hazardous bioaerosols amongst normal atmospheric background constituents \cite{Kul11, Tom17, Bou15, Box15}. Remarkable progress recorded due to improved theoretical models and enhancement of the computing power, enables performing numerical simulations that are able to infer light scattering patterns and mechanisms out of very complex and highly irregular systems. A new method that provides simultaneous measurements of back-scattering patterns and of images of single laser-trapped airborne aerosol particles is presented in \cite{Fu17}.

An algorithm used to calculate electric and magnetic fields inside and around a multilayered sphere is developed in \cite{Ladu17}. The algorithm includes explicit expressions for the Mie expansion coefficients inside the sphere. 

An EDB can be used to levitate highly viscous single aerosol particles for water diffusion experiments, as demonstrated in \cite{Bast18}. Particle growth and reduction of the shell refractive index are experimentally observed via redshift and blueshift behaviour of the Mie resonances, respectively. The particle radius as well as a core-shell radius ratio are inferred from the measured shift pattern and Mie scattering calculations. 

A system used to characterize individual aerosol particles based on stable and repeatable measurement of elastic light scattering is described in \cite{Lane18}. The technique relies on using a linear electrodynamic quadrupole (LEQ) particle trap that levitates charged particles injected using ESI. Particles are confined along the stability line which coincides with the trap centre. Optical interrogation is employed to investigate particles, the scattered light is collected and intensities are calculated based on the Lorenz-Mie scattering theory. Correlation of scattered light measurements for different wavelengths enable one to distinguish and categorize inhomogeneous particles \cite{Lane18}.  

Measurements performed towards determining the size and absorption of single nonspherical aerosol particles from angularly-resolved elastic light scattering are presented in \cite{Walt19}, where it is demonstrated how the analysis of two-dimensional angular optical scattering (TAOS) patterns provides a technique that enables measuring the absorption of a single aerosol particle. 

Ref. \cite{Soren19} reports investigations on the optical absorption cross section for homogeneous spheres that interact with electromagnetic radiation in the visible range, calculated using Mie equations. Four regimes can be distinguished: the Rayleigh Regime, the Geometric Regime, the Reflection Regime, and a Crossover Regime.

\subsection{Inelastic scattering. Raman and fluorescence spectroscopy}\label{RamanFl}

Inelastic scattering (Raman and fluorescence scattering) is employed to determine the chemical composition of microparticles and to study gas/particle and gas/droplet chemical reactions. In addition, inelastic scattering is also used to monitor polymerization of monomer microparticles. It is demonstrated that the temperature of a microparticle can be inferred by measuring the ratio of the anti-Stokes to Stokes scattering intensities. Moreover, inelastic scattering can be used to detect biological particles such as pollen and bacteria \cite{Davis10, Davis11}.

Characterization of ions produced as a result of the interaction between a high energy laser pulse and NPs is indispensable to perform quantitative determinations of the composition and size of NPs, by means of single particle mass spectrometry (SPMS). Ref. \cite{Zhou07} introduces a one-dimensional hydrodynamic model to explain the interaction mechanisms, demonstrating that the laser field is coupled to the non-equilibrium time-dependent plasma hydrodynamics of heated aluminium particles. The properties of ions generated during interaction with a strong laser pulse (532 nm wavelength, 100 mJ/pulse) are investigated for nanosecond pulses and particle dimensions ranging between $20 \div 400$ nm, that are most relevant for SPMS.  

Raman spectroscopy of single particles levitated in an EDB is described in Ref. \cite{Signo11}. The most prominent applications include the study of atmospheric aerosols. EDBs provide the advantage of enabling investigations of various atmospheric processes and phenomena that involve kinetic effects. Both static EDBs and the recently developed scanning EDBs (SEDBs) techniques are extremely suited to perform hygroscopic measurements and investigate phase transformations of levitated aerosol particles. Moreover, homogeneous reactions of Organic Aerosol Particles (OAP) can be investigated using this approach.  

The synergy between optical trapping and Raman spectroscopy \cite{Signo11} opens new pathways to explore, characterize and identify biological micro-particles. More precisely, optical trapping lifts the limitation imposed by the relative inefficacy associated to the Raman scattering process. As a result of applying this technique, the Raman spectroscopy can be employed to study individual biological particles in air and in liquid, providing the potential for particle identification with high specificity or to characterize heterogeneity of individual particles in a population \cite{Wang15}. Ref. \cite{Red15} is an introduction on the techniques used to integrate Raman spectroscopy together with optical trapping, in order to perform qualitative studies on single biological particles levitated in liquid and air. 

Time evolution of fluorescence and Raman spectra of single solid particles optically trapped in air is reported in \cite{Gong17}. Initially, the spectra exhibit strong fluorescence with weak Raman peaks, then fluorescence is bleached within seconds and only clean Raman peaks are reported. The experiment uses an optical trap assembled by using two counter-propagating hollow beams. Both absorbing and non-absorbing particles in the atmosphere are strongly trapped. The setup provides a novel method to investigate dynamic changes in the fluorescence and Raman spectra collected from a single optically trapped particle, under SATP conditions.

A custom-geometry LIT designed for fluorescence spectroscopy of gas-phase ions at ambient to cryogenic temperatures is described in Ref. \cite{Raja18}. Laser-induced fluorescence emitted by trapped ions is collected from between the trapping rods, normal to the excitation laser beam directed along the axis of the LIT. The original design allows an enhanced optical access to the ion cloud with respect to similar designs. As a general rule, studies of ions in the gas phase are associated with fluorescence spectroscopy due to the possibility to precisely control the solvation state of a chromophore. There are certain technical challenges to be solved when combining mass spectrometry with fluorescence spectroscopy, as ion density is low and scattering elements are persistent along the excitation laser beam path \cite{Raja18}.

Characterization of the physico-chemical properties of molecules implies studying their temporal reactions within a micro-sized particle in its natural phase. A recent experiment reports on measurements of temporal Raman spectra for different submicron positions of a laser-trapped droplet composed of diethyl phthalate and glycerol \cite{Kalu18}. Thus, laser-trapping Raman spectroscopy (LTRS) allows one to characterize single airborne particles, single cells, spores, etc., by supplying information on the particle size, shape, surface structure, extinction coefficient, refractive index, elastic scattering pattern, fluorescence and Raman shifts, etc.

Finally, we mention Ref. \cite{Chen19} that demonstrates a micro-nano mechanical sensor for weak gravity sensing field. In addition, progress recorded in this direction may be useful to characterize other short-range physics mechanisms such as Casimir forces. 

\subsection{Particle detection and investigation. Laser induced breakdown spectroscopy (LIBS).}

Laser Induced Breakdown Spectroscopy (LIBS) is a non-destructive atomic emission spectroscopy technique that employs a high energy laser pulse as an excitation source \cite{Dem15}. The laser atomizes and excites a small amount of material at the sample surface (in the nanogram to picogram weight range), developing a plasma plume that emits light at frequencies characteristic to the chemical elements present in the sample. As LIBS is basically a surface interrogation technique it may not identify species present in the inner side of a sample, unless the sample is homogenized prior to analysis. The sensitivity of LIBS lies in the ppm range, the analysis is performed in real-time and no preliminary sample preparation is typically required (unless homogenation is desired). LIBS yields atomic emission and plasma emission spectra that can be used to determine the elemental composition of single particles \cite{Pin11, Yua19}.

The influence of the water content, droplet displacement, and laser fluence on the laser-induced breakdown spectroscopy (LIBS) signal of precisely controlled single droplets, is investigated in Ref. \cite{Jarvi16}. NP analysis is extremely valuable for many applications. One of the disadvantages associated to optical excitation techniques stems from the bounded resolution which is the outcome of the diffraction limit. Single particle nanoanalysis was previously unattainable by means of LIBS, but state-of-the-art attogram-scale absolute limits of detection are achieved by employing optical trapping and levitation to isolate single particles \cite{Puro17}. The multielemental capabilities of this approach are demonstrated by subjecting two different types of nanometric ferrite particles to LIBS analysis \cite{Puro19}. Ref. \cite{Yua19} suggests a new one-point calibration method called single-sample calibration LIBS (SSC-LIBS) to improve the accuracy and stability of detecting major elements using only one standard sample. 

Finally, a tutorial review that presents the most recent advances and state of the art techniques used in LIBS is Ref. \cite{Andra20b}.

\section{Linear Paul traps as tools in laser plasma accelerated particle physics}\label{Sec10}

Strongly nonlinear longitudinal waves in plasma are introduced in \cite{Kriv92}, in a incessant quest to identify physical mechanisms for plasma accelerators of charged particles that allow one to achieve high acceleration rates and energy. 

For a very long time, one of the major issues in physics was to find a method to confine charged particles. As a combination of static fields cannot achieve that, a solution was devised in the 1950s when the concept of {\em alternating gradient (AG) focusing} was implemented \cite{Coura52, Coura58}. The basic idea lies in creating an array of electrostatic lenses arranged in an AG configuration. Particles can be accelerated or decelerated by applying an appropriate high-voltage switching sequence to the lenses. By periodically switching the orientation of two transverse fields a net focusing force is generated, oriented along both transverse coordinates. Thus, confinement of a charged particle beam in both transverse dimensions is achieved by combining two magnetic quadrupoles of opposite polarity into a doublet \cite{Kjae05, Dorf06, Fuku14, Fuku16}. The method is applied to confine ions in quadrupole mass filters, in Paul traps \cite{Paul90, Paul58, Stra55}, and in all type of particle accelerators \cite{Scha12}. Generally, the time-varying sinusoidal voltage applied to a quadrupole trap can exhibit an arbitrary waveform provided that a stable solution of the Hill (Mathieu) equation \cite{Mag66} that describes particle dynamics results. Hence, a square time-varying trapping voltage can be used. Two decades ago it was suggested to make use of the analogy between the dynamics of charged particles in the non-homogenous magnetic quadrupole field of accelerators or storage rings, and that of an electric quadrupole trap excited by square voltage pulses \cite{David00, Kjae01}. Instead of using typical sinusoidal voltages, controllable, pulsed RF voltages, are applied to the trap electrodes with an aim to investigate collective beam dynamics in various lattice structures \cite{Tak04, Kelli15}. Moreover, the collective dynamics of a one-component plasma (OCP) in a RF quadrupole trap is physically equivalent to that of a charged-particle beam that propagates throughout a periodic magnetic lattice. Experiments focused on using linear Paul traps to study space-charged-dominated beams are reported in the last two decades \cite{Tak04, Kjae05, Dorf06, Gils04, Chung07, Gils07}, while Refs. \cite{Kjae02a, Kjae02b} focus on emulations of beams in the ultra strongly coupled regime of Coulomb crystallization. 

In addition to significant contributions of the ion trap group from the Aarhus University, the group from the Hiroshima University has designed two different classes of trap systems, one of which uses a RF electric field (Paul trap) while the other one uses an axial magnetic field (Penning trap) to achieve transverse plasma confinement \cite{Fuku14, Oht10, Oka14}. These systems are called Simulator of Particle Orbit Dynamics (S-POD) \cite{Ito08}. Their intrinsic feature is the capability to approximately reproduce collective motion of a charged-particle beam propagating through long alternating-gradient (AG) quadrupole focusing channels using the Paul trap, and long continuous focusing channels using the Penning trap. Such an interesting feature allows one to investigate various beam-dynamics issues using compact setups without the need for large-scale accelerators. Linear and nonlinear resonant instabilities of charged-particle beams travelling in periodic quadrupole focusing channels are experimentally investigated using a compact non-neutral plasma trap \cite{Oht10}. The linear Paul trap system named S-POD is applied to investigate a collection of space-charge-induced phenomena. To emulate lattice-dependent effects, periodic perturbations are applied to the quadrupole electrodes which yields to additional resonance stop bands that shift depending on the plasma density. The loss rate of trapped particles is measured as a function of bare betatron tune with an aim to identify resonance bands in which the plasma becomes unstable. When an imbalance is created between the horizontal and vertical focusing, the instability bands split \cite{Fuku14}. Experimental results suggest that the instability band is somewhat insensitive to the phase of the quadrupole focusing element location within the doublet configuration, over a significant range of parameters. The Penning trap with multi-ring electrode geometry is employed to study beam halo formation driven by initial distribution perturbations \cite{Oka14}. A similar apparatus to the S-POD, the Intense Beam Experiment (IBEX), was designed at the Rutherford Appleton Lab (RAL)-UK. To use either one of these experimental setups to investigate beam dynamics under conditions of more intricate lattice configurations, novel diagnostic techniques have to be devised and tested for Paul traps. Such a technique is demonstrated in Ref. \cite{Marti18} where a new method is described to measure the beta function and the emittance at a given time in a Paul trap. A novel technique is also demonstrated in Ref. \cite{Ito19}, where direct measurement of low-order coherent oscillation modes in a LPT is performed by detecting image currents induced on the electrodes’ surfaces. The technique is based on picking up weak signals from the dipole and quadrupole oscillations of a plasma bunch, which are then subjected to Fourier analysis. Late experiments report on Paul-trap systems that allow positioning of fully isolated micrometer-scale particles with micrometer precision as targets in high-intensity laser-plasma interactions \cite{Ost19}. 

Besides quadrupole traps, multipole geometries are also intensively investigated. The experimentally tested configuration includes additional electrodes that enable one to control the strengths and time profiles of the low-order nonlinear fields that occur in the trap, independently of the linear focusing potential \cite{Fuku15}. Thin metallic plates are inserted in between the cylindrical electrodes (rods) of a quadrupole Paul trap. The size and arrangement of the extra electrodes are optimized by using a Poisson solver. Simple scaling laws are inferred, to perform a quick estimate of the sextupole and octopole field strengths as a function of plate dimensions. Particle tracking simulations performed clearly demonstrate the possibility to achieve controlled excitation of nonlinear resonances in a multipole Paul trap.

\subsection{Trapping of nanoparticles using dielectric containers and capillary tubes}

\begin{figure}[htb]
	\centering
		\includegraphics[scale=1.4]{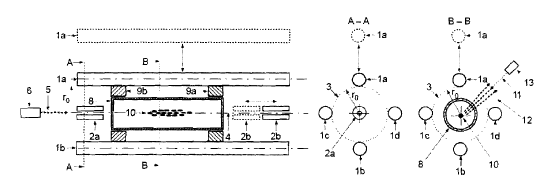}	
		\caption{Longitudinal and two transversal (corresponding to the A-A and B-B plans) cross-sections of the micro- and nanoparticle trapping setup in capillary dielectric tubes, based on a quadrupole linear Paul trap. Legend: 1a $\div$ 1d -- trap electrodes; 2a and 2b -- cylindrical endcap electrodes; 3 -- electrode cross section tangent to the circle of radius $r_0$; 4 -- quadrupole trap $z$ axis; 5 -- laser beam generated by the laser diode denoted as $6$;  8 -- dielectric container and electric charging system; 9a and 9b -- spacer rings; 10 -- trapped nanoparticles. Picture reproduced from \cite{Sto14} with kind permission from O. Stoican.}
	\label{dieltrp}
\end{figure}

A transportable setup intended for trapping an assembly of solid NPs (with dimensions in the micrometer or nanometer range) in a limited and well-defined region of space, without mechanical contact with the walls of an enclosure or any other supporting element, is described in Ref. \cite{Sto14}. The NPs are levitated in a dielectric container of random shape, fully closed or partially open, placed between the electrodes of a linear quadrupole electrodynamic (Paul) trap, as shown in Fig.~\ref{dieltrp}. There are several methods to achieve electrical charging of the NPs, as follows: (i) as an outcome of the physical contact due to the interaction with the dielectric wall of the container, (ii) by providing additional electrodes where high voltage pulses are applied, (iii) by sudden variation of the electric voltage supplied to the trap electrodes, or (iv) as a consequence of an ionizing radiation beam (generated by an external source) which acts upon the particles. The container, identified as (8) in Fig.~\ref{dieltrp}, is made of a dielectric material and it can be fully transparent, half-transparent or opaque, as it can be also filled with gas, with a mixture of gases, or it might be evacuated. Even capillary tubes can be used as containers. The system devised is able to confine conductive powders (either semiconductor, dielectric, or even radioactive), different types of germs, bacteria, viruses, seeds, or other microbiological material. The area of applications is quite large, as the system is also adequate for use in setups and equipments intended for the investigation, manipulation, measurement, and determination of the physico-chemical characteristics and properties of trapped microparticles or NPs. In addition it provides a large optical access for trapped microparticle species, while it can also be employed to investigate the effect of fields of force or of external radiation sources upon micron sized particles levitated in electrodynamic (Paul) traps. 

The NPs located inside the container are levitated in a well defined gaseous environment, whose pressure and chemical composition are controlled. Moreover, particles can also be trapped in vacuum. Such a system enables manipulation and investigation of toxic, radioactive, or biological  material, excluding the possibility to contaminate the human personnel or the surrounding environment. The recipient containing the solid particles which are trapped can be prepared at a different location, then safely carried to a location where the setup or the analysis and measurement system is to be used.    

We further explain operation of the experimental setup described in \cite{Sto14}, used to confine NPs in dielectric containers and capillary tubes. The system relies on a quadrupole electrodynamic trap that consists of four cylindrical, equidistantly spaced electrodes. The electrode cross section is a circular surface denoted as (3) in Fig.~\ref{dieltrp}, tangent to a circle of radius $r_0$. Moreover, two additional endcap electrodes are placed in a plane normal to the one holding the trap electrodes, at equal distance with respect to the trap ends. The trap geometry can exhibit a larger number of electrodes provided that this number is even. Fig.~\ref{dieltrp} shows a longitudinal section and two cross sections of the experimental setup. Two cylindrical endcap electrodes are located coaxially with respect to the trap axis (4). One or both of the endcap electrodes are perforated along the longitudinal axis, with an aim to allow illumination of the trapped NPs (denoted as 10) by means of a laser beam (5) generated by a cw laser diode (6). The laser diode is not required if the confined particles are observed directly or indirectly, by means of radioactive or X-ray emission. 

\begin{figure}[htb]
	\centering
		\includegraphics[scale=0.9]{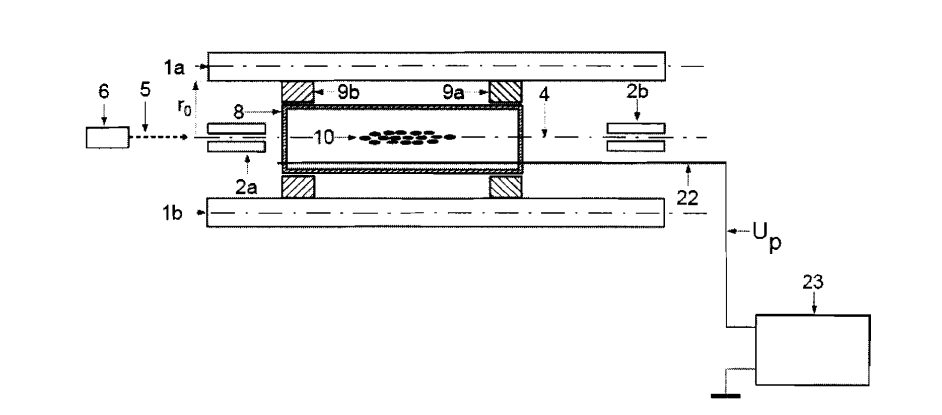}	
		\caption{View of experimental setup for trapping nanoparticles based on a dielectric container. Legend: 1a and 1b -- trap electrodes; 2a and 2b -- cylindrical endcap electrodes; 4 -- quadrupole trap $z$ axis; 5 -- laser beam generated by the laser diode denoted as $6$;  8 -- dielectric container; 9a and 9b -- spacer rings; 10 -- levitated nanoparticles; 22- conducting wire; 23 -- electronic module that delivers high voltage pulses of amplitude $U_p$. Picture reproduced from \cite{Sto14} with kind permission from O. Stoican.}
	\label{dieltrap2}
\end{figure}

Solid particles to be confined are located within a container made of a dielectric material, be it transparent, semi-transparent, or opaque. The container is located within the space delimited by the trap cylindrical electrodes (denoted as 1a $\div$ 1d) and by the trap endcap electrodes, denoted as 2a and 2b. The container vessel (8) may exhibit different shapes and geometries, its cross section might change along the longitudinal axis, it can either exhibit or not cylindrical symmetry, it can be open at both ends or it can be open at one end and closed in the opposite end, it can be fully closed or somewhat closed at both ends. The setup in Fig.~\ref{dieltrp} shows a dielectric container that exhibits a cylindrical shape while being fully closed at both ends. Either one or both of the endcap electrodes can change position and shift along the trap axis, which modifies the extent of the inner space delimited by the trap cylindrical electrodes. Fig.~\ref{dieltrp} also illustrates the possibility of shifting the (2b) electrode along the trap axis (4), where its new position is denoted by a dashed line. Thus, containing recipients of various lengths can be introduced in the inner inter-electrode space. The setup also allows for the possibility to detach one or more trap electrodes with an aim to remove or introduce the container that holds the particles to be confined \cite{Sto14}. In addition, Fig.~\ref{dieltrp} also illustrates the possibility to detach the (1a) electrode in an attempt to insert the dielectric container (8) in the inner inter-electrode space, while its new position is represented by a dashed line. The maximum dimensions of the dielectric container cross-section must be chosen so as to enable its location inside the inner volume delimited by the cylindrical electrodes (1a) $\div$ (1d), (2a) and (2b). Thus, even capillary tubes can be used as particle containers. Mechanically, the container (8) can be positioned by means of two spacer rings (9a) and (9b) (see Fig.~\ref{dieltrap2}) located at both ends. The radiation (12) emitted by trapped particles (10) is observed in a plane that is normal to the trap axis (4), and it can either be scattered light, radioactive emission, or X-radiation. The emitted radiation crosses the walls of the container (8) and it is observed by means of an instrument (denoted as 13) such as the objective of a microscope or of another optical instrument, a CCD camera, a photomultiplier, an array of photoelectric cells, a Geiger counter, etc. 

Analogous to other ion traps designed in INFLPR \cite{Mih16a, Sto01, Mih16b}, the power supply delivers a high a.c. voltage, $U_{ac}$ and two d.c. voltages, namely an $U_z$ voltage applied to the endcap electrodes to achieve axial confinement and a diagnosis voltage, $U_x$, applied between the upper (1a) and lower (1b) trap electrodes, whose electrical polarity can be reversed by means of an electronic switch. All the supply voltages are measured with respect to the (1a) electrode, electrically connected to the ground potential (taken as reference). The diagnosis voltage $U_x$ compensates gravity and enables the experimentalist to shift the particle position in the radial plane.      

Experimental parameters that allow stable trapping for particles with a particular value of the specific charge $Q/M$ depend on the NP dimensions, on the $U_x$ and $U_z$ voltages, as well as on the trapping voltage $U_{ac}$ amplitude ($V_0$) and frequency $f_0$. Generally, for a linear electrodynamic trap with $2n$ cylindrical electrodes, where $n \in N$ is a natural number ($n = 2, 3, \ldots $), when both the $U_x$ and $U_z$ voltages are zero and the polar coordinate $r$ is much smaller than the distance between the trap axis and the electrode surface $r_0$, the analytical expression of the multipole field generated in a point close to the trap axis is \cite{Asva09, Ger92}

\begin{equation}\label{diel1}
\Phi \left(r, \phi, t\right) = V_0 \left({\frac r{r_o}}\right)^n \cos \left(n\phi \right) \sin\left(\omega_0 t\right)  \ ,
\end{equation}    
where $r$ and $\phi$ are the polar coordinates of a specific point and $\omega_0 = 2 \pi f_0$. The experimental setup suggested in \cite{Sto14} can exhibit either a quadrupole or a multipole geometry with an even number of electrodes, while NPs are confined within a dielectric container. According to literature \cite{March97} trapping of particles is performed in optimum conditions if

\begin{equation}\label{diel2}
\frac QM < \frac{1.816 \ \pi^2 r_0^2\ f_0^2}{V_0} \ .
\end{equation}   

In order to find the optimum trapping point, depending on the dimensions and specific charge $Q/M$ of the nanoparticle species used, the a.c. voltage $U_{ac}$ amplitude $\left(V_0\right)$ and frequency $f_0$ can be modified along with the values of the trapping voltages, $U_x$ and $U_z$. Such an operation practically shifts the operating point in the Mathieu equation stability diagram with an aim to achieve stable trapping. Electrical charging of NPs can be achieved as follows: 

\begin{enumerate}
	\item as an outcome of the physical contact between particles confined in the dielectric container (8) without the need of extra electrodes or of another ionizing system,
	\item as a result of the interaction between NPs confined in the dielectric container (8) and its dielectric surface,
	\item by applying a high voltage electric pulse, $U_p$, across an electrode located within the dielectric container (8). According to Fig.~\ref{dieltrap2} the electrode can consist of a thin conducting wire (22) that crosses the container (8) parallel to the trap axis (4), located close to the inner surface, in order to severely reduce perturbations of the electric field distribution near the trap axis. Thus, the trapping electric field can be considered as approximately harmonic. The conducting wire (22) is connected to an electronic module (23) that delivers high voltage pulses of amplitude $U_p$ at the output,   
	\item an overvoltage can be applied for a short period of time, superimposed over the $U_x$ diagnosis voltage. To achieve that the secondary winding of a step-up transformer is connected to the (1b) electrode. The output of a pulse generator is fed into the primary winding of the step-up transformer, 
	\item by switching the electrical polarity of the d.c. voltage $U_x$, applied between the electrodes denoted by (1a) and (1b). It can be performed by either acting upon a mechanical switch or by means of an electronic switch, 
	\item NPs located within the container (8) can be acted upon for a specific period time by an ionizing radiation beam, such as ultraviolet (UV) radiation, X radiation, or radiation emitted by a radioactive source. The radiation interacts with the NPs that are to be trapped by ionizing and implicitly charging them.
\end{enumerate}

\subsection{Use of linear Paul traps for target positioning}\label{Target}

Theoretical and experimental investigations with respect to laser plasma acceleration of particles also use isolated mass-limited targets, that result in increased ion energies and peak spectral distributions of interest for a large number of applications \cite{Bul08, Klu10}. Different approaches used to realize mass-limited targets use micrometer sized water droplets, gas-cluster targets with dimensions in the $100$ nm range, and objects carried by thin structures with dimensions ranging between $100$ nm up to a few $\mu$m \cite{Ost16}. For all the cases listed the range of target dimensions is limited. In order to lift all target restrictions as well as unintended interactions, Paul traps are used in combination with high-power laser systems in an attempt to position isolated micron-sized targets in the laser focus with high accuracy. Novel Paul trap geometries allow sensibly higher laser energies and a significantly larger range in what concerns the diameter size of the target \cite{Ost16, Ost18}. The trap represents  a major progress with respect to recent Paul trap geometries used in similar experiments \cite{Sto14, Sok10}. The most notable difference lies in the fact that instead of friction between target particles and the surrounding gas, the Paul trap used in Refs. \cite{Ost16, Ost18} employs an ion gun to achieve charging. Thus, specific charge ratios $Q/M$ that are a few orders of magnitude higher are obtained, which leads to a minimization of the residual motion of trapped targets at comparable kinetic energies. The particle motion (and implicitly its kinetic energy) is not damped by friction with air but it is produced by an opto-electronic feedback loop that runs continuously, which allows trap operation at pressure values that go below $10^{-5}$ mbar \cite{Ost18}. Thus, diagnostics are not influenced by the background gas environment. The setup allows target positioning with an accuracy value that lies within the micron range, which is a considerable enhancement with respect to the typical FWHM diameter of high power laser foci (few-$\mu$m). 

Similar to the setup described in \cite{Sto14}, the transportable setup used in \cite{Ost18} uses a quadrupole Paul trap with cylindrical rods and provides a large optical access to the particle species of interest. There is a strong scientific interest towards using micron-scale particles as targets for high-intensity laser-plasma acceleration experiments. A tightly focused laser pulse interacts with the strongly coupled plasma that consists of levitated particles confined within a well defined region in space, located in the trap centre. Both the focusing optics and laser diagnosis packages should be provided for the experiment. The resulting ultra-short and intense X-ray, ion, and electron bursts, are expected to have important applications in physics and medical research. In comparison with metallic foil or bulk foil targets \cite{Groza19}, well isolated targets could exhibit major advantages especially in what regards the maximum energy of energetic ions, as well as their energy and source size distributions. The setup can provide targets of any solid material with diameters currently ranging from 500 nm up to $50 \ \mu$m. An almost interaction free environment is created as any particle interaction with nearby structures or background gas is sensibly reduced, while the positioning accuracy is controlled down to micron level precision \cite{Ost18}.  

The trap uses a cylindrical electrode geometry which yields to the occurrence of higher order terms of the electric potential, such as those corresponding to the 12-pole and 20-pole order. As the trapping field is far from being a pure quadrupole one (in fact it is anharmonic), the equations of motion in the radial plane ($x$ and $y$ directions) are coupled. Higher order terms lead to unstable trajectories and may even result in chaotic motion, as an outcome of the phenomenon of resonant heating \cite{Bla98}. The trap is operated with $q < 0.4$, namely in a region of the stability diagram where only very high order instabilities exist. In addition, due to the fact that the target levitates close to the trap centre, higher order instabilities are reduced to a minimum. Axial confinement along the $z$ axis is achieved by supplying additional d.c. voltages on the endcap electrodes of 3 mm diameter (located $10 \div 20$ mm apart), which leads to an additional motion along the trap axis. Anharmonic fields vanish in the trap centre and they are discarded in the discussion. The particle trajectory is measured electro-optically and filtered electronically. An electrical feed-back field is used with a proper phase-shift to turn the harmonic macromotion into the dynamics of a damped harmonic oscillator, which efficiently damps the macromotion \cite{Ost18}.  

\begin{figure}[htb]
	\centering
	\includegraphics[scale=0.65]{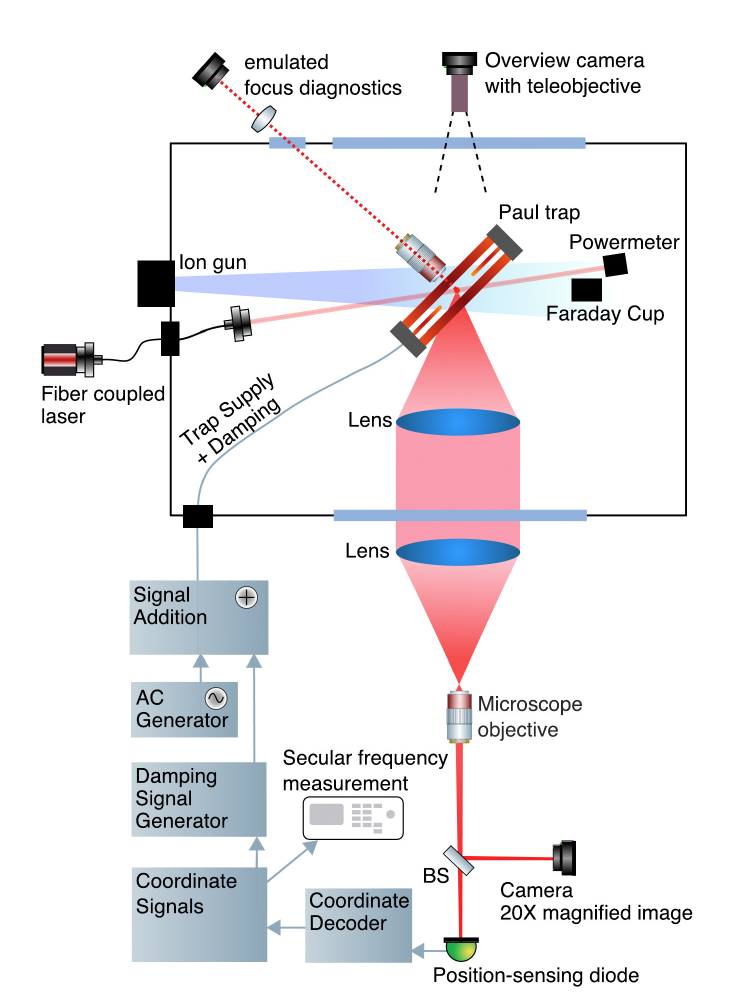}	
	\caption{Paul trap setup used for taget positioning. The setup comprises a vacuum chamber, the trap and power supplies, an ion gun to charge and a laser to illuminate the target particles, plus the optical measurement and feedback setup. Picture reproduced from \cite{Ost18} by courtesy of T. Ostermayr.}
	\label{PaulTarg1}
\end{figure}

The Paul trap based setup developed for laser-plasma experiments is shown in Fig.~\ref{PaulTarg1}. A vacuum chamber allows reaching pressures down to the $10^{-6}$ mbar range. Generally, laser-plasma accelerators (LPAs) operate at pressure values around $10^{-5}$ mbar to minimize the effects of the gaseous environment on the laser beam and to provide appropriate detection of laser-plasma accelerated particles. The Paul trap consists of four cylindrical copper rods of 5 mm diameter, equidistantly spaced around the trap centre. The trap radius is $r_0 = 8.1$ mm. Due to a large optical access provided by the geometry used, large-numerical-aperture optics can be employed, characteristic for experiments with ultrahigh-intensity lasers. All a.c. and endcap electrodes are positioned in precision manufactured ceramic seats to electronically isolate them with respect to each other. Moreover, each a.c. electrode and each endcap electrode is connected to its individual voltage supply. The a.c. trapping voltage can reach a maximum value of 3 kV with $\Omega = 2 \pi \cdot 5$ kHz, while the d.c voltage applied between endcap electrodes can rise up to 450 V.       
      
The setup includes a particle reservoir located above the trap centre, filled with a few milligrams commercial sample of monodisperse spherical particles used as target. By means of a mechanical loading system the particles fall into the trap through a 100 $\mu $m hole. An ion beam -- generated by an ion gun -- is fed into the trap and charges the particles. Thus, a few particles are confined in the trap. The endcap voltage is dropped off to a value of a few volts before trapping, which yields a strong Coulomb repulsion that expels most of the particles out of the trap leaving just a single one inside. Then, the ion gun is shut off and the pressure is reduced to a lower value. Simultaneously, the a.c. and d.c. trapping voltages are increased towards their maximal values to ensure tight confinement in the radial plan. A 660 nm laser diode (that delivers $50$ mW of output power) is used to visualize the particles. Two thirds of the output power are coupled into a single-mode fiber. An adjustable collimator lens performs laser beam focusing to a spot size between $0.25$ and $1$ mm FWHM (Full-Width at Half Maximum) in the trap centre. Three independent imaging systems collect the stray light from target. A position-sensing diode (PSD) is employed to record the CM motion of the particle's image. The frequency of the particle motion is inferred by means of a Fast Fourier Transform (FFT) algorithm. The coordinate signal is phase-shifted and delivered as an additional voltage to the trap electrodes, which leads to an effective damping of the particle motion in absence of a particular gas.
     
The key parts of the concept used are the optical setup and the optoelectronic damping, that are responsible for achieving experiments with well positioned targets under high-vacuum conditions, which eliminates the need for buffer gas that would otherwise affect the high-power laser and diagnostics for the laser plasma interaction. The setup represents a remarkable improvement as it can be customized to fit any interaction chamber geometry. Another critical issue lies in the fact that the system should be able to collect the small amount of stray light created by microscopic target particles. Since only the CM of the particle image is tracked by the PSD, a large aperture lens with comparably small depth of focus ($\sim 1 \div 2 \ \mu$m) is used to provide enhanced collection efficiency at the cost of image sharpness (which is almost irrelevant if only the CM is tracked), if the particle moves along the focal plane before reaching its final damped state in the  focus of the lens system. The fibre-coupled laser is fed into the vacuum chamber by means of a vacuum feedthrough, which results in reducing the influence of stray light to a minimum.

\subsection{Particle acceleration using Petawatt (PW) class lasers and strongly coupled plasmas levitated in electrodynamic traps or or optical levitation traps}\label{PWTargets}

Plasma acceleration is a technique to accelerate charged particles such as electrons, positrons, and ions, using an electric field associated with an electron plasma wave or other high-gradient plasma structures (like shock and sheath fields) \cite{Taji79, Joshi06, Hook13}. The plasma acceleration structures are created using ultra-short laser pulses or energetic particle beams that are matched to the plasma parameters \cite{Down14a, Down14b, Li14b}. These techniques offer a way to build high performance particle accelerators of much smaller size than conventional devices.  

It was suggested for the first time in 2009 to use compact Laser Plasma Accelerators (LPAs) with an aim to produce broadband radiation similar to the outer space and use it for radiation hardness testing of electronic parts \cite{Rosen11, Gan15}. LPAs offer cost-effective radiation generation which makes them extremely suited for testing radiation resistivity of electronic components. Besides that, LPAs have the potential to become versatile tools for reproducing certain aspects of space radiation under far better reproducibility than state-of-the-art technology \cite{Hook14}, among which we can enumerate existing linear particle accelerator, cyclotron (MeV energy protons), synchrotron (energetic protons in the GeV range), or electron synchrotron facilities \cite{Hook13}. Ground-based conventional particle accelerators cannot reproduce the exponential electron beam distribution spectrum that occurs in space, and a typical approach to ground-based radiation hardening assessment (RHA) involves investigating the impact of radiation at multiple monochromatic energies, and then extrapolating the results to other energies based on modelling and simulation. The primary deficiency of such ground-based approaches lies in the fact that naturally occurring space radiation environment energy spectrum is dramatically different compared to the output of a conventional Earth-bound accelerator. On Earth, accelerators based on classical technology do not generate particle energy spectra that follow the exponential distribution found to characterize particle energy spectra in space, but deliver instead nearly monoenergetic distributions.

Petawatt (PW) laser technology and the mechanism of plasma acceleration of charged particles (especially electrons and protons) find extremely important areas of applications for space technology. The advantage of plasma acceleration lies in the fact that the acceleration field is much stronger than the one generated by conventional RF accelerators. In case of RF accelerators the field exhibits an upper limit, determined by the threshold for dielectric breakdown of the acceleration tube. Such a phenomenon limits the acceleration gain over any given area, requiring large dimension accelerators in order to achieve high energies. Present interest is focused on using ultrashort high intensity laser pulses that are matched to the plasma parameters. Using such mechanisms it is possible to create very large electric field gradients and very compact structures with respect to classical accelerators. The basic idea lies in using LPA techniques with an aim to reproduce the cosmic radiation environment in terms of particle species (electrons and protons) \cite{Gan19}. This is an issue of large interest for the European Space Agency, in support of the Jupiter Icy Moon Explorer (JUICE) mission aimed at exploring Jupiter and three of its moons: Ganymede, Europa, and Callisto \cite{juice}. The basic idea is to demonstrate the concept of laser plasma acceleration of electrons and protons \cite{Groza19} up to energies of tens of MeV, or even larger than 100 MeV. The outcome lies in achieving an exponential energy distribution of the generated electrons that would be similar to the distribution met in the Jovian system. Nevertheless, while space radiation can be characterized as continuous, the LPA technique generates a discontinuous flux of charged particles, as emphasized in \cite{Gan15}.  

\begin{figure}[htb]
	\centering
	\includegraphics[scale=1.2]{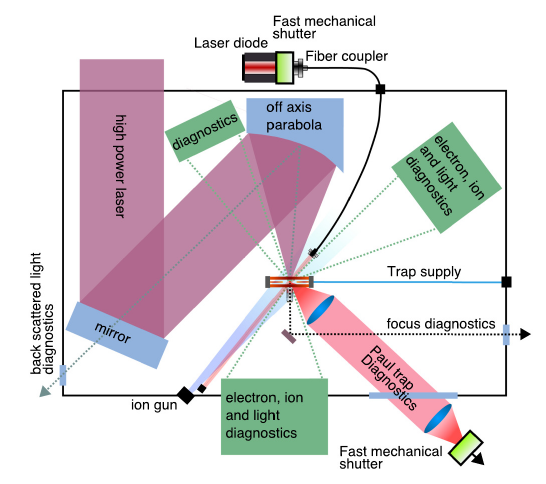}	
	\caption{Paul trap setup used in laser plasma acceleration experiment. Picture reproduced from \cite{Ost18} by courtesy of T. Ostermayr.}
	\label{PaulTarg3}
\end{figure}

As shown in Section \ref{Target} custom designed Paul trap systems allow accurate positioning of micrometer-sized particles with micron precision, which makes them extremely useful as targets in high intensity laser-plasma interactions. A setup intended for a laser plasma experiment is proposed in \cite{Ost18}. A sketch of the setup is reproduced in Fig.~\ref{PaulTarg3}. A high power laser beam is delivered via the beam-line system (BLS) and then focused on the trap using an off-axis parabolic (OAP) mirror. The laser focus is monitored via an optical system consisting of a microscope objective and a mirror. The black dotted line labelled as focus diagnostics denotes the optical path of the microscope. Electron, ion and X-ray detectors, can be placed in the plane of laser propagation, marked as green.   

Fast mechanical shutters are used to reduce the influence of stray light. Laser beam powers of up to 500 TW and energies up to 150 J are used. The secular frequency of levitated microparticles is measured by real-time tracking of the PSD signal. As the trap parameters ($r_0$, $\Omega$ and $V$) are known, by measuring the secular frequency one can determine the charge-to-mass ratio $Q/M$ and the confinement strength. Thus, the particle position can be established with remarkable accuracy. Repetitive measurements performed for $10 \ \mu$m diameter polystyrene particles yield a value of $0.29$ C/kg for the charge-to-mass ratio and an electric charge value $Q = 9.9 \times 10^5 e$ (for a secular frequency value $\omega_{sec}/2 \pi = 125$ Hz). Measurements performed on single levitated particles demonstrate reproducibility, stability, and high accuracy in determining the particle location \cite{Ost18}. The most important parameter that reflects the quality of damping is the residual motion of a levitated particle, which can be estimated using the {\em in situ} focus diagnostics. The method used is based on determining the CM and then plotting the 2D histogram of the occurrence of a specific particle position \cite{Ost18}. The very low values of the Full-Width at Half Maximum (FWHM) of the motion demonstrate that strong confinement is achieved, which leads to high accuracy in pinpointing the particle position along with long trapping times. Thus, CM tracking is shown to be the most accurate method to determine the residual motion.      
   
An experimental demonstration of the acceleration of a clean proton bunch is performed in Ref. \cite{Hilz18}. To achieve that, a microscopic and three-dimensional (3D) plasma confined near critical density is created, that evolves from a $1 \ \mu$m diameter plastic sphere which is levitated and positioned with micrometer precision in the focus of a Petawatt (PW) laser pulse. The emitted proton bunch is reproducibly observed with central energies ranging between $20 \div 40$ MeV and narrow energy spread (down to 25\%), showing almost no low-energetic background. To put this energy spread in perspective, typical LPAs using extended bulk foil targets produce beams with ultrabroad energy spectra decaying exponentially towards their maximum energy. Using 3D particle-in-cell (PIC) simulations the acceleration process is fully characterized, emphasizing the transition from organized acceleration to Coulomb repulsion. This reveals limitations of current high power lasers and viable paths to optimize laser-driven ion sources.

Finally, theoretical and experimental aspects of relativistic intense laser-microplasma interactions and potential applications are nicely presented in Ref. \cite{Ost19}. The book describes a Paul-trap based target system developed and tested so as to provide fully isolated, well defined and well positioned micro-sphere-targets, for experiments with well focused PW laser pulses. The interaction between the PW laser pulses and the micro-spheres converts these targets into microplasmas \cite{Dub99, Boll90b}, able to emit proton beams with kinetic energies higher than 10 MeV \cite{Ost19, Hilz18}. The spatial distribution and kinetic energy spectrum of the proton beam can be modified by changing the acceleration mechanism. The results span from broadly distributed spectra in cold plasma expansions to spectra with relative energy spread, that is reduced down to 25 \% in spherical multi-species Coulomb explosions and in directed acceleration processes. Both analytical modelling and  numerical simulations show good agreement with the experimental results. The 3D PIC simulations characterize the complete acceleration process and clearly emphasize the transition from regular acceleration to Coulomb repulsion \cite{Hilz18}. This is a strong argument towards employing microplasmas to engineer laser-driven proton sources. 

\begin{figure}[htb]
	\centering
	\includegraphics[scale=1.5]{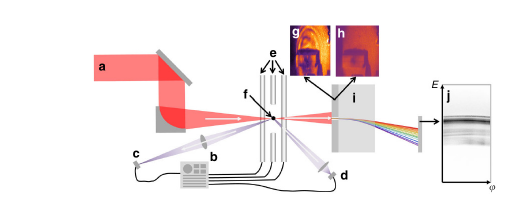}	
	\caption{Paul trap setup used in laser plasma proton acceleration experiment: (a) Incoming PHELIX laser beam, (b) Paul trap power supply, (c) 660 nm illumination laser, (d) electrooptical diagnostics for target damping and positioning, (e) Paul trap electrodes, (f) trapped target scatter screen for transmitted light with (g) and without (h) target, (i) magnetic spectrometer, (j) IP raw proton/ion data (without degraders). Picture reproduced from \cite{Hilz18} by courtesy of T. Ostermayr.}
	\label{PaulTarg4}
\end{figure}

A sketch of the experimental setup used in Ref. \cite{Hilz18} is shown in Fig. \ref{PaulTarg4}. Targets are extracted out of a reservoir by mechanical vibration. As particles fall freely inside the operating trap, electrical charging is achieved by means of an ion gun. The trapped particle amplitude is reduced using a two phase mechanism: buffer gas cooling at $10^{-4}$ mbar pressure, followed by electro-optical damping for a vacuum pressure lower than $10^{-5}$ mbar. Similar to the experiment described in \cite{Ost18} stray light is collected onto a PSD and a small fraction of it is imaged onto a CCD camera. The hollow spheres are created by covering polystyrene spheres with a Polymethylmethacrylate (PMMA) layer. Then, the polystyrene cores are subsequently dissolved away.

Other aspects are approached in Ref. \cite{Ost19}, such as future perspectives and directions of action for the study of laser-microplasma interactions using non-spherical targets. 

An optical levitation trap operated in vacuum, intended for use in high-intensity, high-energy laser interaction experiments, is reported in \cite{Price15}. Isolated microtargets consisting of $10 \ \mu$m diameter oil droplets are levitated in the trap that employs a a focused, vertically propagating, continuous wave (CW) laser beam. Capture and manipulation of the microtargets is achieved by photon momentum transfer at much longer working distances than commonly used by optical tweezer systems, for timescales that exceed one hour. A high speed ($10$ kHz) optical imaging and signal acquisition system is developed in order to pinpoint the levitated droplets position and dynamic behaviour under both atmospheric and vacuum conditions, within a $\pm 5 \ \mu$m spatial resolution. 

The trap performance is determined in a series of high-intensity $10^{17}$  W$\cdot$cm$^{-2}$ laser experiments aimed at evaluating the the X-ray source size and assessing the free-electron temperature of a single isolated droplet target, as well as measuring the emitted RF pulse. The results demonstrate that optically levitated microdroplets can be considered as a viable solution to perform high-intensity laser interaction and point source experimental investigations.

\section{Conclusions and perspectives}\label{Concl}

The paper is a review on the physics and applications of strongly correlated Coulomb systems levitated in electrodynamic (Paul) traps. After a short introduction into the physics of particle traps at the beginning of Section \ref{Sec2}, the issue of the anharmonic contributions to the electric potential generated by a linear RF trap is discussed with respect to recent investigations \cite{Pedre18b}. Particle dynamics is examined for a classical system of two particles levitated in a 3D Paul trap. A theoretical model is introduced with an aim to characterize regular and chaotic orbits for the system described, depending on the chosen control parameters \cite{Mih21}. The phase portraits illustrate ordered and chaotic orbits, while particle dynamics is shown to be ergodic. By studying the sign of the eigenvalues of the Hessian matrix one can analyze and classify the critical points, as well as the system stability. Recent advances in the development of nonlinear traps are mentioned, before introducing optical levitation and acoustic levitation techniques. The problem of instabilities in electrodynamic traps is also approached, with respect to the most relevant and recent contributions to the domain. Chaos is reviewed in Sec. \ref{chaos} where the dynamics of a charged particle confined in a quadrupole nonlinear Paul trap is investigated. Phase portraits, Poincar{\'e} sections, and bifurcation diagrams illustrate the onset of chaos, as well as the existence of strange attractors in the associated dynamics, which can be characterized as strongly nonlinear. 

Non-neutral, complex plasmas and Coulomb systems are investigated in Section \ref{Sec3}. Strongly coupled Coulomb systems encompass various many-body systems and physical conditions such as dusty (complex) plasmas, or non-neutral and ultracold plasmas \cite{Fort06a, Werth05a, Morfi09}. We emphasize that ion traps represent versatile instruments to investigate many-body Coulomb systems \cite{Schne12b, Yos15}, or dusty and non-neutral plasmas. The dynamical time scales associated with trapped microparticles lie in the tens of milliseconds range, while microparticles can be individually observed using optical methods. As the background gas is dilute particle dynamics exhibits strong coupling regimes characterized by collective motion \cite{Major05, Piel17, Melz05, Sav12}. Weakly-coupled plasmas and strongly-coupled Coulomb systems are presented in Sections \ref{weak} and \ref{strong}, respectively. Section \ref{Sec3} illustrates how a Paul trap represents a clever tool to investigate the physics of few-body phase transitions and thus gain new insight on such mesoscopic systems \cite{Schli96, Walth95}. 

Strongly coupled plasmas confined in quadrupole electrodynamic traps are reviewed in Section \ref{Sec4}. Stable 2D and 3D structures are discussed, including Coulomb structures \cite{Vasi13, Syr18, Lapi19b, Lapi15c} as well as Microparticle electrodynamic ion traps (MEITs) \cite{Libb18}, with an emphasize on latest techniques, experimental results and applications. Hamiltonians for systems of $N$ particles are used to describe ordered structures and collective dynamics for systems of charged particles (ions) levitated in quadrupole 3D traps with cylindrical symmetry. The Hamilton function for the system of $N$ particles confined in a combined quadrupole 3D trap is obtained \cite{Ghe00, Mih18, Mih21, Mih22a}. 
 
Waves in plasmas are investigated in Section \ref{waves} using particle-in-cell (PIC) simulations. Caustic solitary density waves are discussed for microparticles levitated in electrodynamic traps \cite{Syr19b, Syr19c}. Analytical and numerical simulations are performed in order to better explain the physical mechanisms and phenomena involved. The interaction between an acoustic wave and levitated microparticles is also investigated \cite{Sto11}. Section \ref{Sec6} presents the physics of multipole traps and their associated advantages, along with the latest experimental techniques and results \cite{Mih16a, Mih16b}. Analytical and numerical simulations are supplied that support experimental data and results.    

Sections \ref{Sec7} and \ref{Sec9} review mass spectrometry methods and techniques used to investigate single particles levitated in electrodynamic or optical traps \cite{Gong18, Puro19}. Emphasis is placed on mass spectrometry techniques \cite{Werth09, March05, March17a} and novel methods based on elastic and inelastic scattering of light, using Mie theory \cite{Bast18, Mie1908, Herg12} or Raman and fluorescence spectroscopy \cite{Signo11,  Wrie09, Mish09, Red15}. Recent experiments seem to indicate that inelastic scattering can be successfully applied to the detection of biological particles such as pollen, bacteria, aerosols, traces of explosives, or synthetic polymers \cite{Wang15}. 

Despite their micro and nanometre range size, aerosols have a major impact on global climate and health but the underlying mechanisms and subsequent effects are still far from being explained. Section \ref{Sec8} brings new perspectives and sheds new light on aerosol and nanoparticle diagnosis \cite{Sande14, Kulma11, Krieg12, Sulli18}, with a special emphasis on the optical properties associated with them \cite{Kul11, Rama18, Sein16, Islam18, Tom17}. Novel methods and techniques are described, aimed at investigating and characterizing these particles. 

Section \ref{Sec10} focuses on laser plasma accelerated physics and use of electrodynamic traps as targets for high power lasers. Two decades ago it was suggested to employ the analogy between the dynamics of charged particles in the non-homogenous magnetic quadrupole field of accelerators or storage rings and that of an electric quadrupole trap excited by square voltage pulses \cite{Kjae01}. Instead of using typical sinusoidal voltages, controllable, pulsed RF voltages are applied to the trap electrodes, with an aim to investigate collective beam dynamics in various lattice structures \cite{Kelli15}. Moreover, collective dynamics of a one-component plasma (OCP) in a RF quadrupole trap is physically equivalent to that of a charged-particle beam that propagates throughout a periodic magnetic lattice \cite{Fuku14, Oht10}. Such an interesting feature allows one to investigate various beam-dynamics issues using compact setups, without the need for large-scale accelerators. Theoretical and experimental investigations with respect to laser plasma acceleration of particles also use isolated mass-limited targets, that result in increased ion energies and peak spectral distributions of interest for a large number of applications \cite{Bul08, Klu10}. In order to lift all target restrictions as well as unintended interactions, Paul traps have been used in combination with high-power laser systems in an attempt to position isolated micron-sized targets in the laser focus with high accuracy \cite{Ost19, Hilz18}.

\section{Acknowledgements}

B. M. is grateful to Prof. G{\"u}nther Werth, Dr. Tobias Ostermeyr (LBNL) and Prof. Kenneth Libbrecht (Caltech), for their courtesy and valuable comments on the manuscript. B. M. gratefully acknowledges support from the Romanian Space Agency (ROSA) - Contract Nr. 136/2017 \textit{Quadrupole Ion Trap Mass Spectrometry} (QITMS) and Contract No. 16N/2019 Romanian Ministry of Research and Innovation (Romanian National Core Program LAPLAS VI). Part of the activities and work reported in Section \ref{Sec10} were supported through the European Space Agency (ESA) - Contract. No. 4000121912/17/NL/CBi. B. M. also acknowledges support along the many years from several colleagues at INFLPR, and especially O. Stoican, {\c S}t. R{\u a}dan, and Prof. V. Gheorghe who introduced me to the universe of ion traps and the wonderful physics associated with them.   

V. F., R. S., and L. S. acknowledge stimulating discussions with Professors V. E. Fortov and O. F. Petrov. Work has been carried out under financial support of the Russian Foundation for Basic Research (RFBR) via grant 18-08-00350.  


\bibliography{CCP22}

\end{document}